\documentclass[journal]{IEEEtran}

\IEEEoverridecommandlockouts
\usepackage{cite}
\usepackage{caption}
\usepackage{amsmath,amssymb,amsfonts}
\usepackage{algorithmic}
\usepackage{graphicx}
\usepackage{tabularx}
\usepackage{textcomp}
\usepackage{xcolor}
\usepackage{stfloats}
\usepackage{multirow}
\usepackage{amsmath}
\usepackage{booktabs}
\usepackage{array}
\usepackage{subfigure}
\usepackage{makecell}
\usepackage{array}
\usepackage{booktabs}
\usepackage{overpic}
\usepackage{url}
\usepackage{textcomp}
\usepackage{diagbox}
\usepackage{upgreek}
\usepackage{bm}
\usepackage[figuresleft]{rotating}

\def\BibTeX{{\rm B\kern-.05em{\sc i\kern-.025em b}\kern-.08em
		T\kern-.1667em\lower.7ex\hbox{E}\kern-.125emX}}
\usepackage[breaklinks,colorlinks,linkcolor=black,citecolor=black,urlcolor=black]{hyperref}
\begin{document}

\bstctlcite{setting}
\title{On the Road to 6G: \\Visions, Requirements, Key Technologies and~Testbeds}

\author{Cheng-Xiang Wang,~\IEEEmembership{Fellow,~IEEE}, Xiaohu You,~\IEEEmembership{Fellow,~IEEE}, Xiqi Gao,~\IEEEmembership{Fellow,~IEEE}, Xiuming Zhu, Zixin Li, Chuan Zhang,~\IEEEmembership{Senior Member,~IEEE}, Haiming Wang,~\IEEEmembership{Member,~IEEE}, Yongming Huang,~\IEEEmembership{Senior Member,~IEEE}, Yunfei Chen,~\IEEEmembership{Senior Member,~IEEE}, Harald Haas,~\IEEEmembership{Fellow,~IEEE}, John S. Thompson,~\IEEEmembership{Fellow,~IEEE},  Erik G. Larsson,~\IEEEmembership{Fellow,~IEEE}, Marco Di Renzo,~\IEEEmembership{Fellow,~IEEE}, Wen Tong,~\IEEEmembership{Fellow,~IEEE}, Peiying Zhu,~\IEEEmembership{Fellow,~IEEE}, Xuemin (Sherman) Shen,~\IEEEmembership{Fellow,~IEEE}, H. Vincent Poor,~\IEEEmembership{Life Fellow,~IEEE}, and Lajos Hanzo,~\IEEEmembership{Life Fellow,~IEEE}
\thanks{This work was supported by the National Key R\&D Program of China under Grant 2018YFB1801101, the National Natural Science Foundation of China (NSFC) under Grants 61960206006 and 62122020, the Key Technologies R\&D Program of Jiangsu (Prospective and Key Technologies for Industry) under Grants BE2022067, BE2022067-1, and BE2022067-5, the EU H2020 RISE TESTBED2 project under Grant 872172, the EU H2020 ARIADNE project under Grant 871464, the EU H2020 RISE-6G project under Grant 101017011, the U.S National Science Foundation under Grants CCF-1908308 and CNS-2128448, the Engineering and Physical Sciences Research Council project under Grants EP/W016605/1 and EP/X01228X/1, and the European Research Council's Advanced Fellow Grant QuantCom under Grant 789028. Thanks are also extended to Xichen Mao, Yinglan Bu, Wenxie Ji, Zihao Zhou, Yue Yang, Lijian Xin, Hengtai Chang, and Duoxian Huang, who have provided valuable assistance and advice during this work.}
\thanks{C.-X. Wang (corresponding author), X. H. You (corresponding author), X. Q. Gao, X. M. Zhu, Z. X. Li, C. Zhang, and Y. M. Huang are with the National Mobile Communications Research Laboratory, School of Information Science and Engineering, Southeast University, Nanjing 210096, China, and also with the Purple Mountain Laboratories, Nanjing 211111, China (email: \{chxwang, xhyu, xqgao, xm\_zhu, lizixin, chzhang, huangym\}@seu.edu.cn). 

H. M. Wang is with the School of Information Science and Engineering and the State Key Laboratory of Millimeter Waves, Southeast University, Nanjing 210096, China, and also with the Pervasive Communication Research Center, Purple Mountain Laboratories, Nanjing 211111, China (email: hmwang@seu.edu.cn).

Y. F. Chen is with the School of Engineering, the University of Warwick, Coventry CV4 7AL, U.K. (e-mail: yunfei.chen@warwick.ac.uk).

H. Haas is with the LiFi Research and Development Center, Department of Electronic and Electrical Engineering, University of Strathclyde, Glasgow G1 1XQ, U.K. (e-mail: harald.haas@strath.ac.uk).

J. S. Thompson is with the Institute for Digital Communications, School of Engineering, University of Edinburgh, Edinburgh EH9 3JL, U.K. (e-mail: john.thompson@ed.ac.uk).

E. G. Larsson is with  with the Department of Electrical Engineering (ISY), Link{\"o}ping University, 581 83 Link{\"o}ping, Sweden (e-mail: erik.g.larsson@liu.se).

M. Di Renzo is with Universit\'e Paris-Saclay, CNRS, CentraleSup\'elec, Laboratoire des Signaux et Syst\`emes, 3 Rue Joliot-Curie, 91192 Gif-sur-Yvette, France. (marco.di-renzo@universite-paris-saclay.fr)

W. Tong is with the Wireless Advanced System and Competency Centre, HUAWEI Technologies Co., Ltd., Ottawa, ON K2K 3J1, Canada (e-mail: tongwen@huawei.com).

P. Y. Zhu is with HUAWEI Technologies Canada Co. Ltd., Ottawa, ON K2K 3J1, Canada (e-mail: peiying.zhu@huawei.com).

X. Shen is with the Department of Electrical and Computer Engineering, University of Waterloo, Waterloo, ON N2L 3G1, Canada (e-mail: sshen@uwaterloo.ca).

H. V. Poor is with the Department of Electrical and Computer Engineering, Princeton University, Princeton, NJ 08544, USA (e-mail: poor@princeton.edu).

L. Hanzo is with the School of Electronics and Computer Science, University of Southampton, Southampton SO17 1BJ, U.K. (e-mail: lh@ecs.soton.ac.uk)}

}
\markboth{IEEE Communications Surveys \& Tutorials, vol.~xx, no.~xx, February~2023}%
{C.-X. Wang, \MakeLowercase{\textit{et al.}}: On the Road to 6G: Vision, Requirements, Key Technologies and Testbeds.}

\maketitle

\begin{abstract}
Fifth generation (5G) mobile communication systems have entered the stage of commercial development, providing users \textcolor{black}{with new services and improved user experiences as well as offering} a host of novel opportunities to various industries. However, 5G still faces many challenges. To address these challenges, international industrial, academic, and standards organizations have commenced research on sixth generation (6G) wireless communication systems. A series of white papers and survey papers have been published, which aim to define 6G in terms of requirements, application scenarios, key technologies, etc. \textcolor{black}{Although ITU-R has been working on the 6G vision and it is expected to reach a consensus on what 6G will be by mid-2023, the related global discussions are still wide open and the existing literature has identified numerous open issues.} This paper first provides a comprehensive portrayal of the 6G vision, technical requirements, and application scenarios, covering the current common understanding of 6G. Then, a critical appraisal of the 6G network architecture and key technologies is presented. Furthermore, existing testbeds and advanced 6G verification platforms are detailed for the first time. In addition, future research directions and open challenges are identified for stimulating the on-going global debate. \textcolor{black}{Finally, lessons learned to date concerning 6G networks are discussed.}
\end{abstract}


\IEEEpeerreviewmaketitle

\section{Introduction}
\begin{figure}[tb]
	\centerline{\includegraphics[width=0.49\textwidth]{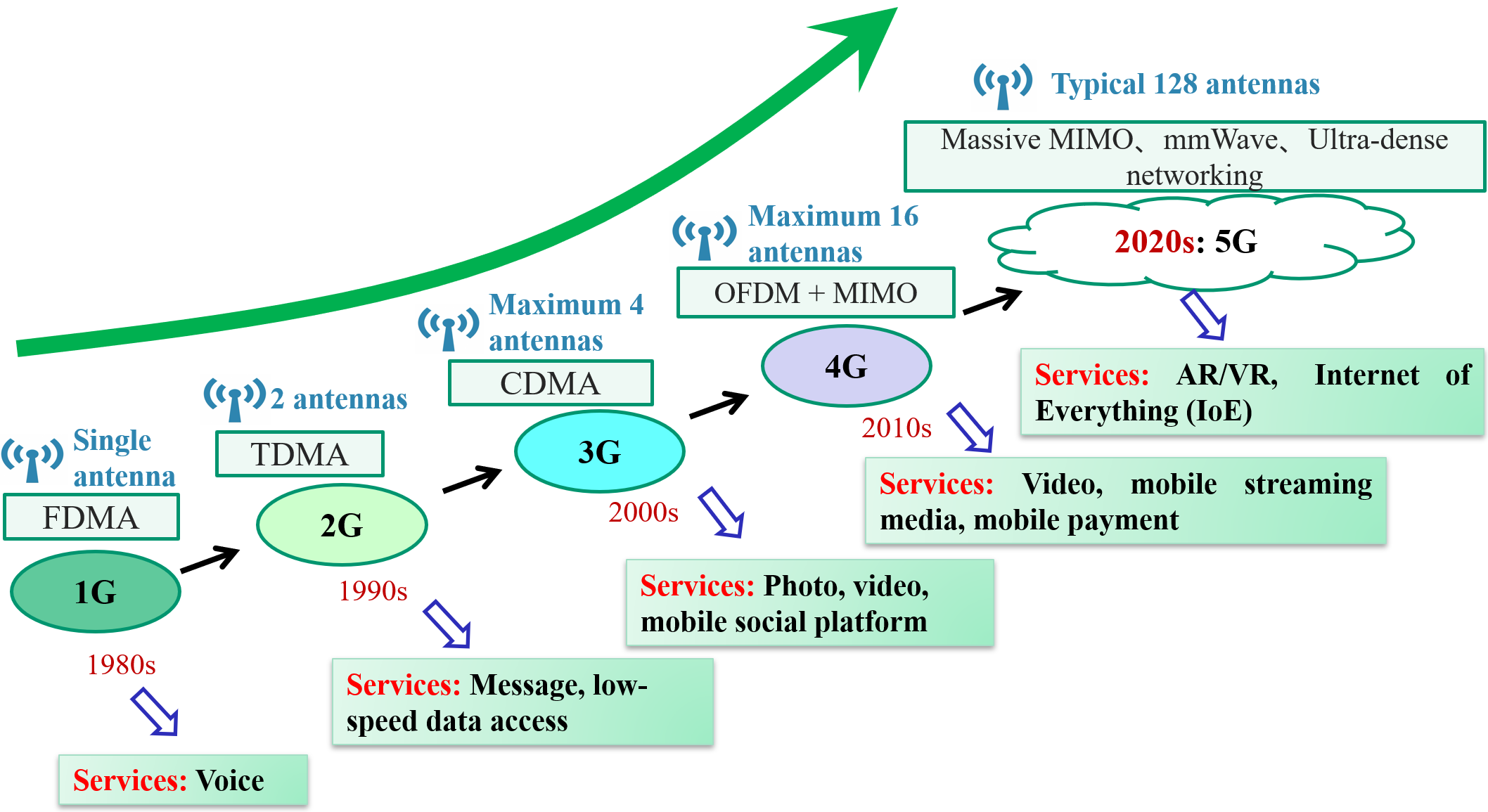}}
	\caption{\textcolor{black}{1G-5G: Antennas, multiple access technologies, and services.}}
	\label{fig_1G_5G}
\end{figure}

\IEEEPARstart{W}{ith} the rapid development of communication applications, communication technologies are undergoing revolutionary changes generation after generation. Up till now, the development of cellular mobile communication systems has undergone five generations. From the first generation (1G) analog communication systems to fifth generation (5G) digital communication systems, each generation incorporates higher frequencies, larger bandwidths, and \textcolor{black}{higher data rates. Starting from 2019}, 5G has been officially commercialized, employing sub-6 GHz and millimeter wave (mmWave) bands, with a peak rate of 20 Gbps. 
From the architecture's perspective, mobile communication systems have been evolving towards more antennas, more advanced multiple access technologies, and richer services, as shown in Fig. \ref{fig_1G_5G}. The 5G base stations exploit massive multiple-input multiple-output (MIMO)\cite{Lu2014_mMIMO}, mmWave, and ultra-dense networking (UDN) technologies\cite{Kamel2016_UDN}, supporting \textcolor{black}{up to 64 transceiver chains with more antenna elements. Currently, commercial 5G base station products using 128 antennas are mature, and HUAWEI is the first to release a massive MIMO base station with 384 antennas\cite{HUAWEI_384}}. In addition, 5G can support augmented reality (AR), virtual reality (VR), and Internet of Everything (IoE).

\subsection{5G Limitations and Challenges}
Although 5G offers significant improvements over fourth generation (4G) communication systems, it still has several limitations. Currently, \textcolor{black}{there are applications and services requiring better communication performance that is beyond 5G's capabilities, such as global coverage, ultra-high data rate transmission, ultra-low latency, ultra-dense connection, high precision positioning, ultra-reliable and safe connection, low power consumption, high energy efficiency (EE), as well as ubiquitous intelligence.} To address these limitations, several challenges need to be addressed. Global maritime communications and satellite communications in high-latitude regions need to be further explored to achieve global coverage. Ultra-high data rate transmission needs to be significantly improved so that the peak data rate can reach Tbps level, supporting \textcolor{black}{services such as ultra high-definition video and telemedicine}. At low transmission speed, the end-to-end (E2E) latency needs to be less than 1 millisecond, while at high speed, the latency should reach the microsecond level. The connection density should reach $10^8$ devices/${\rm km^2}$, meeting the needs of connecting dense crowds and industrial equipment. In addition, the positioning accuracy needs to be improved to achieve an outdoor centimeter level and an indoor sub-centimeter level for high precision positioning. A series of novel applications \textcolor{black}{such as tactile Internet, vehicle to everything (V2X), and wireless data centers}, have higher requirements for reliability. Energy consumption is also a key issue for many applications, and consequently the power consumption needs to be reduced and the network EE needs to be increased by 100 fold. Moreover, a great number of intelligent applications prompt communication systems to have a higher intelligence level. \textcolor{black}{By analyzing the gaps between 5G capabilities and future demands, Table {\ref{table_5G_limitations}} summarizes the limitations and challenges of 5G, which are expected to be explored and resolved in sixth generation (6G) communication systems. Detailed analysis of 5G and 6G KPIs will be introduced later.}

\renewcommand{\arraystretch}{1.2}
\begin{table*}[t]
    \centering
    \caption{Summary of 5G limitations and challenges.}
    \begin{tabular}{|c|c|c|c|c|}
        \hline
				\textbf{Key Indicator} & \textbf{Main Industrial Application} & \textbf{Industry Requirement} & \textbf{5G KPI} & \textbf{Future Research Challenge}  \\ \hline
				Global Coverage & \begin{tabular}[c]{@{}c@{}}Marine communication \\ Satellite communication\end{tabular} & \begin{tabular}[c]{@{}c@{}} Cover sea and \\ remote areas \end{tabular} &  \begin{tabular}[c]{@{}c@{}} Ocean coverage: only 5$\%$ \\ Land coverage: only 20$\%$ \end{tabular} & \begin{tabular}[c]{@{}c@{}} Global coverage: \\ space-air-ground-sea \end{tabular}  \\ \hline
        \begin{tabular}[c]{@{}c@{}} Ultra-high Data Rate \\ Transmission\end{tabular}& \begin{tabular}[c]{@{}c@{}} Ultra high \\ definition (HD) video, \\ holographic image \end{tabular} & \begin{tabular}[c]{@{}c@{}} Very high speed \\ transmission \end{tabular} & \begin{tabular}[c]{@{}c@{}} Transmission rate: $<$ 20 Gbps \\ User experienced \\ data rate: $\sim$ 100 Mbps \end{tabular} & \begin{tabular}[c]{@{}c@{}} Peak data rate \\ at Tbps level \\ User experienced \\ data rate: 1--10 Gbps \end{tabular} \\ \hline
        Ultra-low Latency & \begin{tabular}[c]{@{}c@{}} Automatic drive,\\ high-precision \\ industrial production \end{tabular} & \begin{tabular}[c]{@{}c@{}} High speed and low \\ latency transmission \end{tabular} & \begin{tabular}[c]{@{}c@{}} Delay: $<$ 1 ms at static \\ and low speeds, but cannot \\ be reached at high speeds \end{tabular} &  Sub-second ($<$ 1 ms) \\ \hline
        \begin{tabular}[c]{@{}c@{}} Ultra-dense \\ Connection \end{tabular} & \begin{tabular}[c]{@{}c@{}} Crowded shopping malls, \\ stations, fully automatic\\ production line \end{tabular} & \begin{tabular}[c]{@{}c@{}} Super dense population,\\ super dense equipment \end{tabular} &  $10^6$ devices/$\rm {km^2}$ & \begin{tabular}[c]{@{}c@{}} Connection density: up to \\ $10^8$devices/k${\rm m^2}$ \end{tabular} \\ \hline
        \begin{tabular}[c]{@{}c@{}} High Precision\\ Positioning \end{tabular} & \begin{tabular}[c]{@{}c@{}} Unmanned vehicle \\ positioning and navigation, \\indoor precise positioning \end{tabular} & \begin{tabular}[c]{@{}c@{}} Any circumstances of \\outdoor/indoor\\ precise positioning \end{tabular} & \begin{tabular}[c]{@{}c@{}} Outdoor: $\sim$ 10 m \\ Indoor: $\sim$ 1 m  \end{tabular} & \begin{tabular}[c]{@{}c@{}} Achieve outdoor meter \\ level, indoor centimeter \\ level positioning \end{tabular} \\ \hline
        Ultra-reliable/safe & \begin{tabular}[c]{@{}c@{}} Tactile Internet, V2X, \\ telemedicine, wireless data, \\center, wireless brain- \\machine interface \end{tabular} & Super reliable/safe & 99.999\%  & \begin{tabular}[c]{@{}c@{}}99.99999\% \end{tabular} \\ \hline
        \begin{tabular}[c]{@{}c@{}} Low Power \\ Consumption/High \\ Energy Efficiency \end{tabular} & \begin{tabular}[c]{@{}c@{}} Internet of bio-nano-things, \\ intermediate altitude \\ communications \end{tabular} & \begin{tabular}[c]{@{}c@{}} Reduce the power \\ consumption and improve \\ the energy efficiency as \\ much as possible \end{tabular} & \begin{tabular}[c]{@{}c@{}} Network energy \\ efficiency: $10^7$ bit/J \end{tabular} & \begin{tabular}[c]{@{}c@{}} Enhance the network energy \\ efficiency to $10^9$ bit/J  \end{tabular} \\ \hline
        \begin{tabular}[c]{@{}c@{}} Ubiquitous \\ Intelligent \end{tabular} & \begin{tabular}[c]{@{}c@{}} Digital twins, integrated \\ sensing and communication \\ (ISAC), AI applications \end{tabular} & \begin{tabular}[c]{@{}c@{}} Support a series of \\ intelligent applications \end{tabular} & Low & High \\ \hline

  \end{tabular}
  \label{table_5G_limitations}
\end{table*}

\subsection{Recent Developments}
The evolution of wireless communication systems is roughly an iterative process of every ten years. At present, the research on 6G is in the early stage of exploration, and a collection of countries and standardization organizations around the world have announced their plans for 6G research. 
In 2018, the Finnish government launched the first large 6G research program in the world. 
In the United States, the Federal Communications Commission (FCC) opened the terahertz (THz) spectrum for 6G research and proposed the idea of developing 6G based on ``mmWave + THz + satellite" in Mar. 2019. 
In Oct. 2020, the Alliance for Telecommunications Industry Solutions (ATIS) led the formation of the NextG Alliance, which is a trade organization specializing in the management of 6G development in North America.
In China, the Ministry of Science and Technology officially launched the 6G research in Nov. 2019, and the International Mobile Telecommunications 2030 (IMT-2030) promotion group was established to promote 6G technology research. 
In Apr. 2020, Japan's Ministry of Internal Affairs and Communications released Japan's 6G strategic plan. 
In South Korea, the time schedule for 6G was released in Jan. 2020, and it was expected to be commercialized in 2028. 
\textcolor{black}{In Apr. 2021, Germany announced an investment in 6G research, including a 6G Research Hub and a 6G Platform.}
\textcolor{black}{In Europe, the 6G Smart Networks and Services Industry Association (6G-IA) has been set up for next-generation networks and services.}
As an international organization for standardization, the International Telecommunications Union (ITU) released the initial schedule of 6G research in Feb. 2020. It is expected that the research on the 6G vision and corresponding technical propositions are likely to be completed by 2023. 
As exemplified in Table~{\ref{table_6G_projects}}, a series of pioneering projects focusing on next-generation wireless networks have been initiated around the world.
\begin{figure}[tb]
	\centerline{\includegraphics[width=0.5\textwidth]{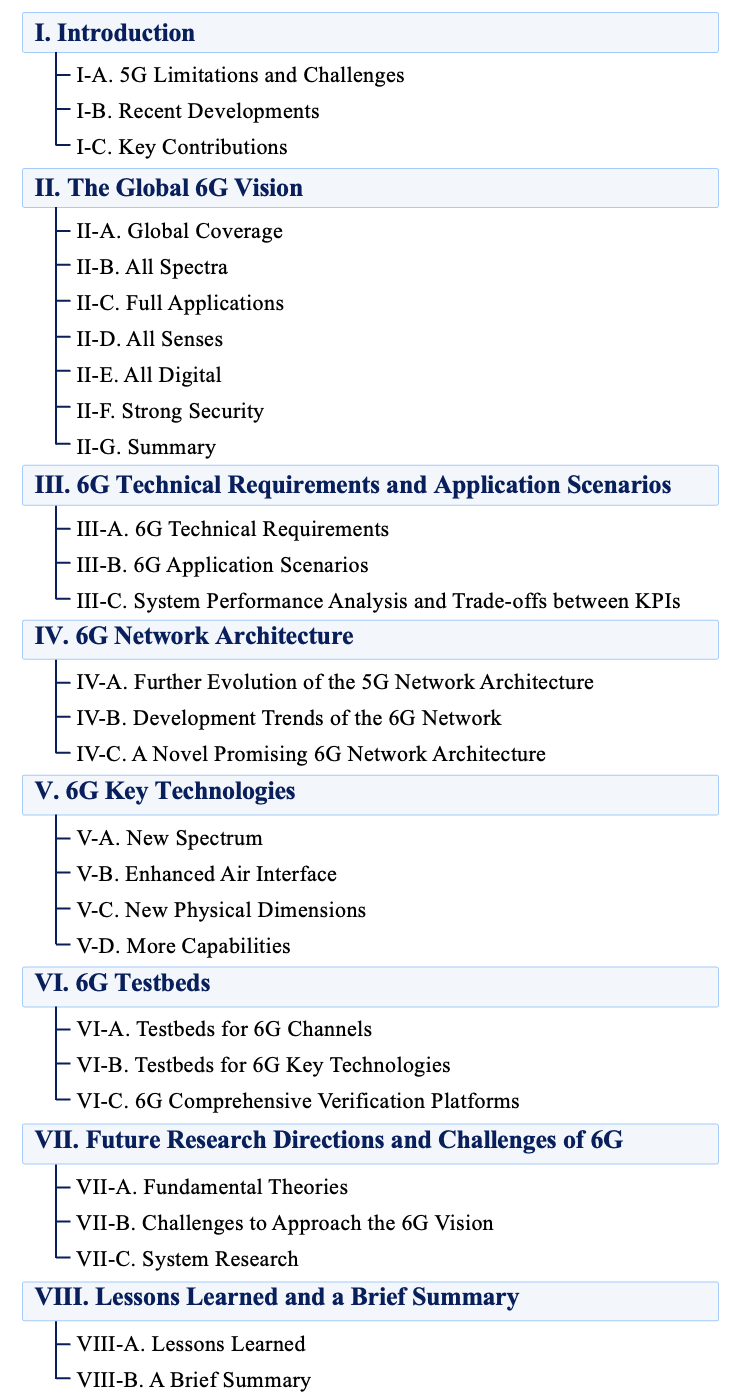}}
	\caption{\textcolor{black}{The outline of this survey paper.}}
	\label{fig_outline}
\end{figure}
\begin{table*}[t]
  \centering
  \caption{Summary of global 6G projects.}
  \begin{tabular}{|c|c|c|r|}
  \hline
   \textbf{Ref.} & \textbf{Year} & \textbf{Country} & \textbf{6G Project} \\ \hline
   \cite{Oulu} & 2018 & Finland & 6G-enabled wireless smart society and ecosystem (6Genesis) \\ \hline
   \cite{China_Project2019} & 2019 & China & Satellite communication technology integrated with 5G/6G \\ \hline
   \cite{AI_EU} & 2019 & EU & Artificial Intelligence Aided D-band Network for 5G Long Term Evolution \\ \hline
   \cite{China_Project2020} & 2020 & China & Research on the theory and key technologies of ultra-wideband photonic THz wireless transmission \\ \hline
   \cite{Space_Japan} & 2020 & Japan & Research and development on satellite-terrestrial integration technology for beyond 5G \\ \hline
   \cite{China_Project2021} & 2021 & China & 6G communication-aware-computing converged network architecture and key technologies \\ \hline
   \cite{China_Project2021} & 2021 & China & 6G ultra-low latency ultra-reliable large-scale wireless transmission technology \\ \hline
   \cite{SENTINEL} & 2021 & Germany & Project SENTINEL for flexible 6G networks consisting of non-terrestrial networks, THz, localization, etc. \\ \hline
   \cite{RISE-6G} & 2021 & EU & Reconfigurable intelligent sustainable environments for 6G wireless networks (RISE-6G) \\ \hline
    \cite{REINDEER} & 2021 & EU & Project REINDEER for smart connectivity platform creating hyper-diversity \\ \hline
    \cite{Hexa-X} & 2021 & EU & Project Hexa-X for 6G vision and intelligent fabric of technology enablers connecting human, physical, and digital worlds \\ \hline
    \cite{RINGS} & 2021 & USA & Resilient \& intelligent NextG systems (RINGS)\\ \hline
   \cite{China_Project2022} & 2022 & China & AI-driven 6G wireless intelligent air-interface transmission technologies \\ \hline
   \cite{China_Project2022} & 2022 & China & 6G research on smart and simple network architecture and autonomous technologies \\ \hline
   \cite{China_Project2022} & 2022 & China & Endogenous security and privacy protection technologies for 6G \\ \hline
   \cite{China_Project2022} & 2022 & China & New network architecture and transmission methods for 6G smart applications \\ \hline
    \cite{KT6G} & 2022 & Korea & Quantum cryptographic communication for aerospace and space applications\\ \hline
    \cite{Open6GHub} & \textcolor{black}{2022} & \textcolor{black}{Germany} & \textcolor{black}{Project Open6GHub for society and sustainability including adaptive 6G RAN technologies, connected intelligence, etc.}\\ \hline
    \cite{6GStart} & \textcolor{black}{2022} & \textcolor{black}{EU} & \textcolor{black}{6GStart: Starting the Sustainable 6G SNS Initiative for Europe} \\ \hline
    \cite{6GTandem} & \textcolor{black}{2022} & \textcolor{black}{EU} & \textcolor{black}{Project 6GTandem for dual-frequency distributed MIMO technologies}\\ \hline
    \cite{Hexa_X_II} & \textcolor{black}{2023-} & \textcolor{black}{EU} & \textcolor{black}{Project Hexa-X-II launched to address challenges in sustainability, inclusion, and trustworthiness}\\ \hline
	\cite{TERA6G} & \textcolor{black}{2023-} & \textcolor{black}{EU} & \textcolor{black}{TERAhertz integrated systems enabling 6G Terabit-per-second ultra-massive MIMO wireless networks (TERA6G)}\\ \hline
    \multicolumn{4}{|l|}{*EU: European Union, USA: United States of America} \\ \hline
  \end{tabular}
  \label{table_6G_projects}
\end{table*}

The above plans come with very active preliminary works on 6G. A series of white papers and related survey papers have been released, defining the envisioned requirements, application scenarios, key performance indicators (KPIs) of 6G, and discussing the network architecture as well as key enabling technologies, etc. 
For example, Finland's 6G flagship organization released the first 6G white paper\cite{Oulu} in Sept. 2019, proposing the vision of ``ubiquitous wireless intelligence" and focusing on the key drivers, challenges, and related research issues of 6G. After that, it also released a series of white papers\cite{Flagship_WPs}, covering networking, machine learning (ML) applications, business, edge intelligence, security, and other aspects of 6G. In \cite{RohdeS}, Rohde \& Schwarz released a white paper exploring the evolution of 5G to 6G and forecasting 6G key technologies. In \cite{Ericsson}, Ericsson focused on the vital role of artificial intelligence (AI) in the next generation of intelligent networks, identifying five distinct challenges regarding the applications of AI in 6G. In China, the China Center for Information Industry Development (CCID) took the lead in releasing a white paper\cite{CCID} on the 6G concept and vision in Mar. 2020, and conducted an investigation\cite{CCID2} on the global progress and development prospects of 6G in Apr. 2021. In addition, \textcolor{black}{HUAWEI\cite{HUAWEI}}, China Mobile\cite{ChinaMobile}, China Unicom\cite{Unicom}, UNISOC\cite{UNISOC}, DATANG Mobile\cite{DATANG}, have also carried out a series of initial research projects on 6G. The IMT-2030 promotion group in China released 6G white papers, defining business scenarios and potential key technologies\cite{IMT-2030}, as well as typical applications and KPIs \cite{IMT2030-APP_KPI} for 6G. In \cite{NTT-Docomo}, NTT DOCOMO carried out research on the further evolution of 5G and the requirements, use cases, and related technologies of 6G, pointing out that the technology of mobile communication system changes every ten years, and the cycle of creating new value for the markets on mobile communication business is about 20 years. In \cite{Samsung}, Samsung envisioned that 6G would provide users with ultimate experience through hyper-connectivity between humans and everything. Besides, several alliances in the communication industry have also carried out early works on 6G. The Next Generation Mobile Networks (NGMN) alliance analyzed the driving factors of 6G from social goals, market expectations, and necessities, and discussed the journey from 5G to 6G from the perspective of a 6G vision\cite{NGMNA}. The 5G Infrastructure Association (5GIA) released a white paper\cite{5GIA}, looking forward to 6G in terms of driving force, necessities, key technologies, and architecture. In addition, the 6G Alliance of Network AI (6GANA) has been actively exploring the realization of network AI and the construction of endogenous intelligent 6G networks\cite{6GANA}. 
\textcolor{black}{Meanwhile, there are a large number of 6G survey papers\cite{Huang2019_28, Zhang2019_32, Zhang2019_33, Zong2019_34, Chowdhury2020_6,Khan2020_8,Gui2020,Bariah2020_11,Saad2020_12,Viswanathan2020_13,Giordani2020_16,Liu2020_18,Chen2020_19,SEU,Bhat2021,Jiang2021_21,Gustavsson2021,Alwis2021,Tataria2021,Nguyen2022_Survey,Wang2022_6G,Shen2022_DT}, which provide outlooks on 6G from different perspectives such as vision, requirements, use cases, and key technologies. These forward-looking 6G review works have also inspired a variety of researches on potential 6G-oriented technologies.}

\renewcommand{\arraystretch}{1.2}
\begin{table*}[t]
	\centering
	\caption{A comparison of existing 6G survey papers.}
	\begin{tabular}{|l|l|c|c|c|c|c|c|c|}
		\hline
		\multirow{2}{*}{\textbf{Ref.}} & 
		\multirow{2}{*}{\textbf{Year}} & 
		\multicolumn{7}{c|}{\textbf{Content Coverage}} \\ \cline{3-9}
		{\ } & {\ } & \textbf{Vision} & \textbf{KPIs} & \textbf{\begin{tabular}[c]{@{}c@{}} Application \\ Scenarios \end{tabular}} & \textbf{\begin{tabular}[c]{@{}c@{}} Network \\ Architecture \end{tabular}} & \textbf{\begin{tabular}[c]{@{}c@{}} Key \\ Technologies \end{tabular}} & \textbf{Testbeds} & \textbf{Challenges} \\ \hline

		T. Huang \emph{et al.}\cite{Huang2019_28} & 2019 & \textcolor{black}{$\bigcirc$} & $\ $ & $\ $ & $\surd$ & $\surd$ & $\ $ & $\triangle$ \\ \hline	
		L. Zhang \emph{et al.}\cite{Zhang2019_32} & 2019 & $\surd$ & $\ $ & $\triangle$ & $\ $ & $\surd$ & $\ $ & $\triangle$ \\ \hline	
		Z. Zhang \emph{et al.}\cite{Zhang2019_33} & 2019 & $\surd$ & $\surd$ & $\triangle$ & $\surd$ & $\surd$ & $\ $ & $\triangle$ \\ \hline	
		B. Zong \emph{et al.}\cite{Zong2019_34} & 2019 & $\surd$ & $\ $ & $\triangle$  & $\surd$ & $\surd$ & $\ $ & $\triangle$ \\ \hline	
		M. Z. Chowdhury \emph{et al.}\cite{Chowdhury2020_6} & 2020 & \textcolor{black}{$\bigcirc$} & $\ $ & $\surd$ & $\triangle$ & $\surd$ & $\ $ & $\surd$ \\ \hline	
		L. U. Khan \emph{et al.}\cite{Khan2020_8} & 2020 & \textcolor{black}{$\bigcirc$} & $\surd$ & $\surd$ & $\triangle$ & $\surd$ & $\ $ & $\surd$ \\ \hline	
		G. Gui \emph{et al.}\cite{Gui2020} & 2020 & \textcolor{black}{$\bigcirc$} & $\surd$ & $\ $ & $\surd$ & $\surd$ & $\ $ & $\surd$ \\ \hline
		L. Bariah \emph{et al.}\cite{Bariah2020_11} & 2020 & $\surd$ & $\ $ & $\triangle$ & $\ $ & $\surd$ & $\ $ & $\surd$ \\ \hline
		W. Saad \emph{et al.}\cite{Saad2020_12} & 2020 & $\surd$ & $\surd$ & $\surd$ & $\triangle$ & $\surd$ & $\ $ & $\surd$ \\ \hline	
		H. Viswanathan \emph{et al.}\cite{Viswanathan2020_13} & 2020 & $\surd$ & $\surd$ & $\triangle$ & $\surd$ & $\surd$ & $\ $ & $\triangle$ \\ \hline	
		M. Giordani \emph{et al.}\cite{Giordani2020_16} & 2020 & \textcolor{black}{$\bigcirc$} & $\ $ & $\surd$ & $\surd$ & $\surd$ & $\ $ & $\triangle$ \\ \hline	
		G. Liu \emph{et al.}\cite{Liu2020_18} & 2020 & $\surd$ & $\surd$ & $\surd$ & $\surd$ & $\ $ & $\ $ & $\triangle$ \\ \hline	
		S. Chen \emph{et al.}\cite{Chen2020_19} & 2020 & $\surd$ & $\surd$ & $\triangle$ & $\surd$ & $\surd$ & $\ $ & $\triangle$ \\ \hline	
		X.-H. You \emph{et al.}\cite{SEU} & 2021 & $\surd$ & $\surd$ & $\surd$ & $\surd$ & $\surd$ & $\ $ & $\triangle$ \\ \hline	
		J. R. Bhat \emph{et al.}\cite{Bhat2021} & 2021 & \textcolor{black}{$\bigcirc$} & $\surd$ & $\surd$ & $\surd$ & $\surd$ & $\ $ & $\surd$ \\ \hline	
		W. Jiang \emph{et al.}\cite{Jiang2021_21} & 2021 & $\surd$ & $\surd$ & $\surd$ & $\triangle$ & $\surd$ & $\ $ & $\triangle$ \\ \hline	
		C. D. Alwis \emph{et al.}\cite{Alwis2021} & 2021 & \textcolor{black}{$\bigcirc$} & $\surd$ &$\surd$ & $\triangle$ & $\surd$ & $\ $ & $\surd$ \\ \hline	
		H. Tataria \emph{et al.}\cite{Tataria2021} & 2021 & \textcolor{black}{$\bigcirc$} & $\surd$ & $\surd$ & $\surd$ & $\surd$ & $\ $ & $\triangle$ \\ \hline	
		D. C. Nguyen \emph{et al.}\cite{Nguyen2022_Survey} & 2022 & \textcolor{black}{$\bigcirc$} & $\surd$ & $\surd$ & $\ $ & $\surd$ & $\ $ & $\surd$ \\ \hline	
		Z. Wang \emph{et al.}\cite{Wang2022_6G} & 2022 & $\surd$ & $\ $ & $\surd$ & $\ $ & $\surd$ & $\ $ & $\triangle$ \\ \hline		
		X. Shen \emph{et al.}\cite{Shen2022_DT} & 2022 & \textcolor{black}{$\bigcirc$} & $ \  $ & $ \  $ & $ \surd $ & $ \surd $ & $ \  $ & $ \surd $ \\ \hline
		\textbf{This survey paper} & 2022 & $\surd$ & $\surd$ & $\surd$ & $\surd$ & $\surd$ & $\surd$ & $\surd$ \\ \hline
		\multicolumn{9}{|l|}{\begin{tabular}[l]{@{}l@{}} Note:  \end{tabular}} \\  
		\multicolumn{9}{|l|}{\begin{tabular}[l]{@{}l@{}} \textcolor{black}{1.	For the ‘Vision’ column, the $\surd$ symbol indicates that an overall vision for 6G was proposed in the reference, the $\bigcirc$ symbol means that the} \\ \textcolor{black}{authors in the reference shown their outlook for several aspects of 6G, but an overall 6G vision was missing in the reference.} \end{tabular}} \\ 
		\multicolumn{9}{|l|}{\begin{tabular}[l]{@{}l@{}} 2. For other columns, the $\surd$ symbol indicates that this aspect is covered in detail in the reference, the $\triangle$ symbol means that this aspect is only \\ mentioned briefly or with other contents but not discussed comprehensively in a single section in the reference, and the blank means that this aspect \\ is not covered at all in the reference.  \end{tabular}} \\ \hline
		
	\end{tabular}
	\label{Table_paper_comp}
\end{table*}

\subsection{Key Contributions}
\textcolor{black}{With the rapid development of the mobile communications industry, there is an urgent need to solve the limitations of 5G and continue to develop 6G. Based on a series of existing forward-looking 6G works, we provide a comprehensive discussion and summary of 6G. This paper aims to put forward the definition of 6G covering the current common understanding of 6G and to investigate the most recent developments in 6G thoroughly. A brief comparison of this survey paper with existing 6G survey papers is given in Table~{\ref{Table_paper_comp}}. The novelty and contributions of this survey can be summarized as follows.}
\begin{enumerate}
  \item {The global 6G vision, KPIs, and application scenarios are critically appraised, covering the current common threads of 6G research.}
  \item {\textcolor{black}{The development trends, research status, and standardization progress} of the 6G network architecture as well as key technologies are examined. Furthermore, a promising 6G network architecture is proposed.}
  \item {Crucially, the existing 6G testbeds concerning the 6G-style wireless channels, the pivotal 6G components, and 6G verification platforms are reviewed for the first time.}
  \item {Commencing from the global 6G vision, a suite of open research directions and key challenges of 6G are discussed, concluding with the lessons learned from the critical evaluations of vast body of literature cited.}
\end{enumerate}

{ \em The rest of the survey is organized based on the rationale of outlining the rich state-of-the-art relying on 600+ authoritative citations and crisp summary tables, followed by identifying the critical knowledge gaps. The existing concepts are then critically appraised in terms of their pros and cons as well as trade-offs, paving the way for addressing the open research problems by following a range of promising avenues, whilst avoiding~pitfalls.}

\textcolor{black}{An illustration of the outline of this survey paper
  is shown in Fig.~\ref{fig_outline}.}  To elaborate a little further
in the above-mentioned spirit, Section~II introduces the 6G vision and
compares it with relevant papers, illustrating the comprehensive nature
of the emerging 6G vision.  Section~III details the 6G KPIs and
application scenarios. The relationships between these 6G KPIs and
compelling application scenarios as well as the inevitable trade-offs
between 6G KPIs are also discussed.  In Section~IV, the evolution of
the 6G network architecture is scrutinized and a novel 6G network
architecture is proposed.  Section~V introduces the whole spectrum of
key 6G enabling technologies relying on a four-pronged attack. \textcolor{black}{The standardization progress of the 6G network architecture and key technologies are summarized at the beginning of Section~VI.} A
unique distinguishing aspect of this treatise is that in Section~VI
the pioneering 6G-style wireless channel simulators and sounders, the
key technology evaluation testbeds and verification platforms are
examined. Future research directions and challenges on the road
to making the 6G vision a reality are outlined in
Section~VII. Finally, the key lessons learned from the critical
appraisal of the literature and our conclusions are offered in
Section~VIII.

\section{The Global 6G Vision}
6G is expected to be different from 5G in several aspects. From the perspective of application requirements, 5G extended the ``Mobile Internet" in 4G to the ``IoE". Based on 5G, 6G will continue to enhance the mobile Internet and IoE but will also deeply integrate them with AI and big data to realize the intelligent IoE. As for technical requirements, compared with 5G, 6G will pursue wider coverage, higher rates, more connections, ultra-low latency, ultra-high positioning accuracy, \textcolor{black}{integration of communications and sensing}, more intelligence, \textcolor{black}{more security, and better substitutability. Note that 6G will be a network that goes beyond communication.}
These application requirements and technical requirements lead us to envision what the 6G will be like.

While 5G is being rolled out globally, a number of research initiatives have proposed ideas for the 6G vision. In Sept. 2019, the 6G flagship led by the University of Oulu published the world's first 6G research white paper and proposed the vision of ``ubiquitous wireless intelligence" for 6G\cite{Oulu}. In Mar. 2020, the white paper on 6G concept and vision published by the CCID envisaged 6G to open up a unified network of ubiquitous intelligent connection between virtual and real worlds\cite{CCID}. In Nov. 2020, UNISOC published a white paper, hoping 6G can achieve the link between macro and micro, the fusion of virtual and reality, the mapping between digital and physical, \textcolor{black}{the crossing of past and future, and the matching between technologies and requirements, emphasizing the importance of AI}\cite{UNISOC}. In Dec. 2020, it was expected by DATANG Mobile that the 6G capability of basic communication, intelligence, wireless sensing, network security, and network computing power would be enhanced\cite{DATANG}. 
\textcolor{black}{In Apr. 2021, the NGMN alliance analyzed driving factors of 6G and presented the overall vision of 6G in terms of 6G attributes and key design considerations\cite{NGMNA}.}
In June 2021, the IMT-2030 promotion group envisaged the 6G vision of ``intelligent connection of everything and digital twins"\cite{IMT-2030}. 
In Aug. 2021, HUAWEI looked forward that 6G would realize the ``Intelligence-of-everything", the interconnection of people and things would eventually evolve into an intelligence interconnection\cite{HUAWEI}. \textcolor{black}{In addition to 6G white papers that have been published by different organizations, the 6G vision has also been put forward in numerous works\cite{Zhang2019_32,Zhang2019_33,Zong2019_34,Bariah2020_11,Saad2020_12,Viswanathan2020_13,Liu2020_18,Chen2020_19,SEU,Jiang2021_21,Wang2022_6G}. 
Three characteristics of 6G were envisioned in \cite{Zhang2019_32}, i.e., mobile ultra-broadband, super Internet of Things (IoT), and AI. The authors of \cite{Zhang2019_33} believed that 6G will enable the blueprint of connecting everything, full-dimensional coverage, technology cross-integration, and human-like intelligent autonomy. A six-F trend set was proposed for 6G in \cite{Zong2019_34} as ``full spectra, full coverage, full dimension, full convergence, full photonics, full intelligence". In \cite{Bariah2020_11}, the ``human-thing intelligence" interconnectivity and tactile communication were envisioned for 6G. Saad \emph{et al.}\cite{Saad2020_12} emphasized the importance of technology convergence, and proposed a holistic 6G vision in terms of applications, trends and technologies. The authors of \cite{Viswanathan2020_13} identified that 6G is for the interconnection of physical, biological, and digital worlds. 
China Mobile \cite{Liu2020_18} summarized the 6G vision as ``digital twin and ubiquitous intelligence". It was looked ahead in \cite{Chen2020_19} that the human society would become a ubiquitous intelligent mobile society in the 6G era. In \cite{SEU}, You \emph{et al.} in Southeast University (SEU) and Purple Mountain Laboratories (PML) put forward the 6G vision of ``global coverage, all spectra, full applications, and strong security". Based on extensive investigations, the author in \cite{Jiang2021_21} pointed out the disruptive transformations from previous generations to 6G. In \cite{Wang2022_6G}, Wang \emph{et al.} envisioned the vision of ``intelligent connection of everything, digital twin". }

\begin{figure*}[tb]
	\centerline{\includegraphics[width=0.9\textwidth]{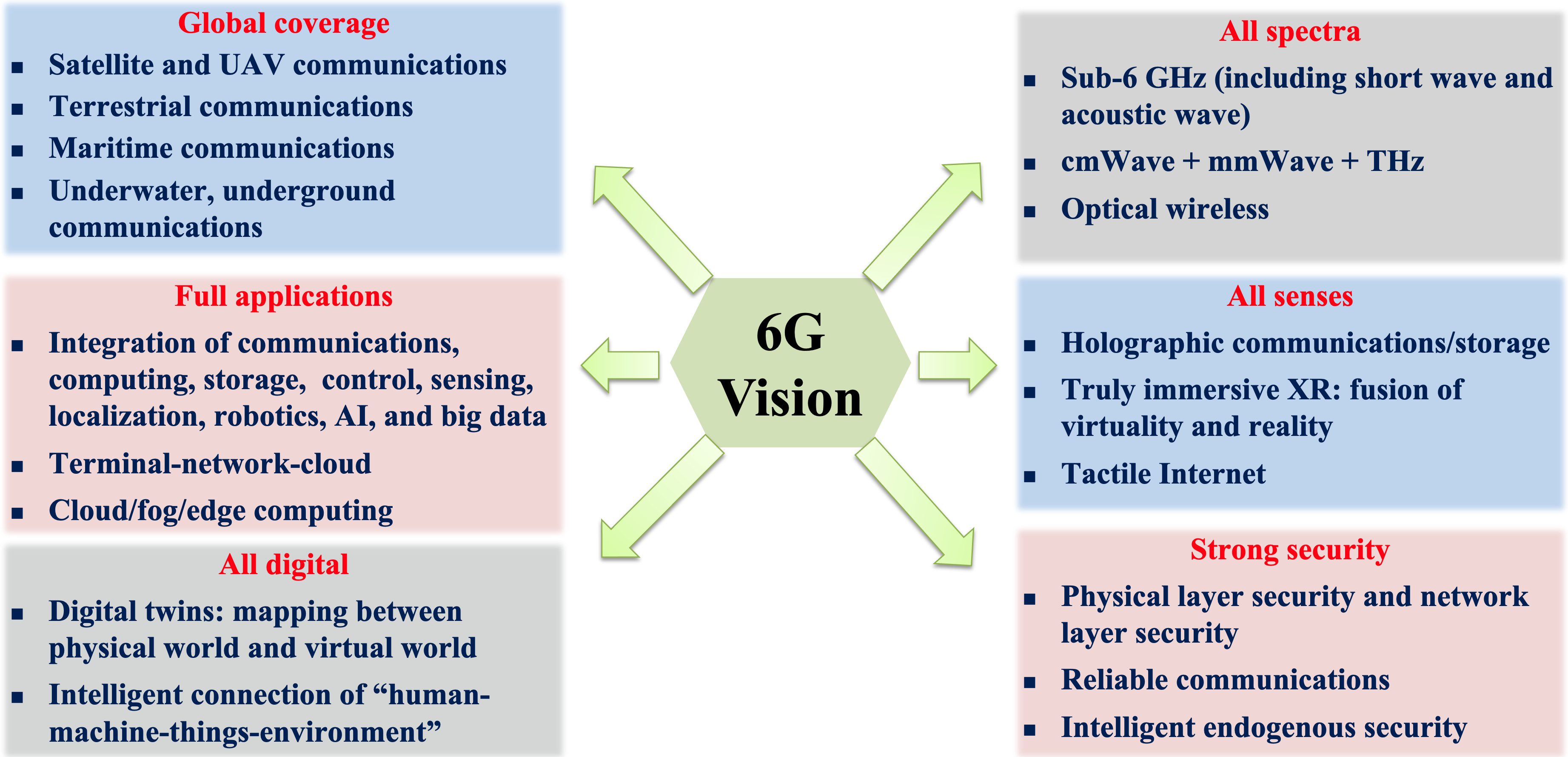}}
	\caption{6G vision: Global coverage, all spectra, full applications, all senses, all digital, and strong security.}
	\label{fig_6G_vision}
\end{figure*}

Overall, the 6G vision can be descripted as ``global coverage, all spectra, full applications, all senses, all digital, and strong security", as shown in Fig. \ref{fig_6G_vision}. In order to achieve global coverage, 6G will expand from terrestrial communication to an integrated space-air-ground-sea communication network. In order to meet the application requirements of huge traffic and huge connection, the full spectra will be fully mined by 6G communication networks, including sub-6 GHz, mmWave, THz, and optical bands. In order to serve various vertical industries, communications, computing, storage, control, sensing, positioning, AI, and big data will be deeply integrated, giving birth to full application scenarios such as ISAC. 6G will also provide users with a full sensory experience through holographic communications and storage, immersive extended reality (XR), tactile Internet, and other applications. Based on the digital twin, 6G will realize the mapping between the digital world and the physical world, and finally realize the intelligent connection of ``human-machine-things-environment". Strong security is embodied by taking the security into account when designing the communication network, known as \textcolor{black}{network endogenous security\cite{SCIS_Wu2022}}, including physical layer security and network layer security. Combined with AI, 6G will also achieve intelligent endogenous security.

\subsection{Global Coverage}
\textcolor{black}{Currently}, communication services are largely limited to terrestrial mobile communications, while remote areas and special scenarios are still blind zones. In order to achieve ubiquitous global coverage, 6G will expand from terrestrial communications to space-air-ground-sea communications, integrating satellite communications\cite{Kodheli2021}, unmanned aerial vehicle (UAV) communications\cite{Mozaffari2019}, terrestrial ultra-dense communication networks, maritime communications\cite{Yu2020}, underwater communications\cite{Jouhari2019}, and underground communications\cite{Saeed2019}. By \textcolor{black}{integrating various communication networks}, 6G will achieve seamless three-dimensional (3D) ubiquitous coverage and connectivity, providing multiple communication services. The global coverage can provide communication services to users with outside the coverage of terrestrial communication networks, \textcolor{black}{such as the remote area Internet access, enhanced on-board communications\cite{Jiang2021_21}, and communications in underground mines}. In the event of terrestrial communication network outage due to disasters (e.g., flood and fire), space-air networks can provide fast, stable, and high-quality emergency services. The full-space deep coverage can also realize the full-space environment monitoring, deep-sea exploration, and other services. 

In June 2017, the Sat5G Alliance was established in Europe to explore the integration of satellites and 5G networks\cite{Liolis2019}. In July 2019, the ITU-R M.2460-0 report discussed key challenges of integrating satellite systems into the next generation of mobile communication systems\cite{ITU-R-M.2460-0}. The 3rd Generation Partnership Project (3GPP) also carried out a series of non-terrestrial network (NTN) standardization work. The 3GPP TR 38.811 standard\cite{3GPP-TR-38.811} and the 3GPP TR 38.821 standard\cite{3GPP-TR-38.821} were published, aiming at exploring the astro-earth fusion communication architecture. With the continuous development of technologies and the deepening of standardization work, 6G will eventually achieve global coverage with the help of space-air-ground-sea integrated communication networks.

\subsection{All Spectra}
As new high-data rate communication services and applications continue to evolve, the demand for wireless mobile traffic is increasing exponentially. Existing communication systems are facing the challenge of spectrum congestion in the radio frequency (RF) band, which is insufficient for higher rate services. 
The 5G has already made nearly full use of sub-6 GHz and started to explore the mmWave band\cite{Rappaport2017}. In the 6G era, in order to meet the technical requirements of huge traffic and huge connection, all spectra including sub-6 GHz, \textcolor{black}{centimeter wave (cmWave)}, mmWave, THz\cite{Tekbiyik2019}, and optical wireless bands\cite{Chowdhury2020} will be fully exploited. The THz band has the advantages of huge bandwidth and ultra-high data rate\cite{Wang2020_6GCMM-survey}, providing strong support for the 6G wireless data center\cite{Ahearne2019}, nano Internet\cite{Akyildiz2010}, ultra-short distance communications, and other new application scenarios. The optical bands used for optical wireless communications (OWCs) include infrared (IR), visible light, and ultraviolet (UV) bands, with nearly thousands of THz of unused spectral resources. Moreover, the visible light band has a number of advantages, including being green and economical, no spectrum regulation, high security, no electromagnetic (EM) interference\cite{Ghassemlooy2015}. In the scenario where RF communications are limited, OWCs have great application potential and have spawned a series of optical communication technologies, including \textcolor{black}{visible light communications (VLC), light fidelity (LiFi), optical camera communications (OCC), free space optical (FSO) communications, and light detection and ranging (LiDAR)\cite{Chowdhury2018}}.

As the first effort on the THz standardization work towards 6G, the IEEE 802.15.3d standard was released for the THz band in the 300 GHz\cite{Petrov2020}. In terms of the standardization for OWCs, existing work includes the IEEE 802.11 standard related to diffuse IR communications\cite{802.11}, the IEEE 802.15.7r1 aiming at short range OWCs\cite{VLCA2018}, and the IEEE 802.15.13 working on multi-Gigabit/s OWCs\cite{IEEE-802.15.13}. In addition, the IEEE 802.11bb task group had been working on integrating the light medium in the base IEEE 802.11 standards\cite{Tgbb}. Based on the sub-6 GHz band and mmWave band of 5G, 6G will explore higher frequency bands as needed, and finally realize the deep mining of the full spectra. Multiple frequency bands will coexist and fuse with each other, to enable different services.

\subsection{Full Applications}
With the diversification of services and the continuous development of communication systems, 6G will generate massive data. New technologies, such as AI and big data, will be fully utilized to explore the intelligent potential of 6G networks and realize a series of intelligent applications. On one hand, these novel technologies will help the development of more advanced and intelligent communication systems, providing new ideas and paradigms in the research of wireless channel modeling, network multiple access, rate control, caching and reloading, secure and stable connections\cite{Luong2019,Wang2020_6GCMM-survey}. On the other hand, the ubiquitous intelligent 6G networks will also provide intelligent applications, such as smart cities, smart agriculture, and smart transportation. In addition, 6G will not only provide communication services but also other services that combine communications, computing, storage, control, sensing, localization, and robotics, giving birth to a series of diversified service applications, such as the ISAC\cite{Cui2021,Liu2020,Barneto2021}.

In recent years, there has been continuous research interest in wireless AI, which aims to integrate AI with wireless communication networks. In June 2019, the 3GPP TR 23.791 defining the functional specifications for the data collection and analysis in automated cellular networks was released\cite{3GPP-TR-23.791}. Several organizations, including the open-RAN (O-RAN) alliance, are also working on the combination of AI and wireless communication networks\cite{Shafin2020,O_RAN}. Compared with the wireless AI, the development of ISAC is still at a relatively preliminary stage. It is hoped that various applications will fuse deeply with each other, making joint contributions to the realization of a smarter and more comprehensive 6G network.

\subsection{All Senses}
From 1G to 5G, the interconnection between people has evolved into the interconnection between people and things. In 6G, with the support of a variety of communication technologies, users will be provided with a full sensory experience through holographic communications and storage, immersive XR, tactile Internet, and other applications. People's perception of reality is obtained through various sensory organs. With the support of 6G providing reliable communications with large bandwidth and low latency, all human's five sense information, including sight, hearing, touch, taste and smell will be enabled through 6G communication networks and fully reproduced in front of the user, achieving the fusion of virtuality and reality\cite{UNISOC}. It is envisioned that 6G communication systems will realize multi-sensory interconnections, providing a wide range of applications in the entertainment, skill learning, medical health, and other fields\cite{Holland2019,Kolovou2021}.

In June 2019, Qualcomm combined its technical advantages in the XR field and mmWave frequency bands to launch the world's first XR platform that integrated 5G and AI, i.e., the Snapdragon XR2 platform\cite{Qualcomm_XR2}. Many technology companies around the world, \textcolor{black}{e.g., Qualcomm, Microsoft, Apple, and HUAWEI}, have conducted research on XR products. In addition, the IEEE 1918.1 TI Standards WG\cite{IEEE-1918.1} has also been committed to the standardization of the tactile Internet, discussing and studying the definition, KPI requirements, application scenarios, architecture, interfaces, and other technical aspects of the tactile Internet\cite{Holland2019}. It is believed that, under the comprehensive promotion of future communication technologies and application requirements, 6G can finally provide users with a full sensory experience in a variety of scenarios.

\subsection{All Digital}
Thanks to the advancement of communications, sensing, computing, storage, as well as the development of big data and AI, in the 6G era, the digital twin technology will further develop and evolve. The physical reality will be precisely digitized, and the digital and physical worlds will map to and influence each other\cite{UNISOC}. The digital world is more than a digital mapping of the physical world. Inferences and predictions of the digital world can correspond to those of the physical world, accurately reflecting and predicting the physical world in real time\cite{IMT-2030} and serving as a reference for decision-making in the physical world. Through the interconnection of the physical world and the virtual digital world, 6G will realize the intelligent connection of ``human-machine-things-environment" including the ``environment" of the virtual world\cite{CCID}, driving a series of applications, \textcolor{black}{such as human body digital twins and digital twin cities\cite{Ivanov2020}}. 

At present, a number of standardization organizations have carried out standardization works on digital twins\cite{DT_WhitePaper}, e.g., IEEE and ISO/IEC JTC1. At the same time, innovative high-tech enterprises, such as Microsoft\cite{Microsoft_DT} and Siemens\cite{Siemens_DT} have also explored digital twin technologies and products. It is believed that the 6G era will be a new era in which the digital virtuality and physical reality are deeply integrated.

\subsection{Strong Security}
With the continuous development of communication networks, security has become an important issue. In addition to traditional security problems, such as virus and distributed denial of service (DDoS) attacks, 6G will face a series of new security threats, for instance, large-scale data breaches and learning-empowered attacks with the development of new application scenarios, new technologies, and huge user information explosion\cite{Nguyen2021}. On the basis of designing the network to provide various new services, 6G also needs to take security into consideration during its design to realize strong security, including physical layer security and network layer security. The quantum communication technology\cite{Zhang2017}, blockchain technology\cite{Dorri2017}, and other potential security technologies will promote to constitute an endogenous security mechanism to ensure that the 6G network is credible, manageable and controllable. Combined with AI, 6G will also realize the intelligence endogenous security, aiming at the independent identification and solution of network security problems\cite{Siriwardhana2021,Al-Garadi2020}. 

The communication security has always been an important focus of research. The ISO/IEC JTC1 SC27 standardization group published a series of standards for information security management, supply chain security management, network virtualization security, including ISO/IEC 27005, ISO/IEC 27036, and ISO/IEC 27033-7. The 3GPP also published 3GPP TR 33.813\cite{3GPP-TR-33.813} for network slicing enhanced security research, and 3GPP TS 33.501\cite{3GPP-TS-33.501} for 5G system security architecture and processes, etc. Besides, there are various standardization organizations working on the 6G security, \textcolor{black}{e.g., IEEE and ITU-T, as summarized in \cite{Porambage2021}} The security is an indispensable part of the 6G research. It is reasonable and necessary to envision that the future 6G network will be a highly secure and trusted network.

\subsection{Summary}
In summary, 6G will provide global coverage and be a network offering a series of novel applications, \textcolor{black}{such as intelligent applications, full-sensory applications, and digital twin applications.} This will be enabled by deep mining of full spectra and the integration of a range of new technologies. In addition, 6G will be a network with endogenous security by taking security into account.

\textcolor{black}{A comparison of relevant papers on the 6G vision is given in Table \ref{Table_literature_6G_Vision}. It can be seen that the proposed consistent description of the overall 6G vision in this paper is comprehensive, covering most prospects on the 6G in current academic and industrial efforts. Besides, there are some new research interests in the prospects of strong security, full sensory applications, and digital twins. They will be covered in this work too.}


\renewcommand{\arraystretch}{1.2}
\begin{table*}[t]
	\centering
	\caption{\textcolor{black}{A comparison of relevant papers on the 6G vision.}}
	\begin{tabular}{|l|c|c|c|c|c|c|}
		\hline
		\textbf{Ref.} & \textbf{Global Coverage} & \textbf{Full Applications} & \textbf{Strong Security} & \textbf{All Spectra} & \textbf{All Senses} & \textbf{All Digital} \\ \hline
		6G Flagship\cite{Oulu} & $\surd$ & $\surd$ & $\triangle$ & $\ $ & $\triangle$ & $\ $ \\ \hline
		CCID\cite{CCID} & $\surd$ & $\surd$ & $\ $ & $\ $ & $\triangle$ & $\surd$ \\ \hline
		UNISOC\cite{UNISOC} & $\surd$ & $\surd$ & $\triangle$ & $\surd$ & $\surd$ & $\surd$ \\ \hline
		DATANG Mobile\cite{DATANG} & $\surd$ & $\surd$ & $\surd$ & $\ $ & $\triangle$  & $\triangle$ \\ \hline
		NGMN Alliance\cite{NGMNA} & $\surd$ & $\ $ & $\surd$ & $\ $ & $\ $  & $\surd$ \\ \hline
		IMT-2030\cite{IMT-2030} & $\surd$ & $\surd$ & $\surd$ & $\surd$ & $\surd$ & $\surd$ \\ \hline
		HUAWEI\cite{HUAWEI} & $\surd$ & $\surd$ & $\surd$ & $\surd$ & $\triangle$ & $\surd$ \\ \hline
		L. Zhang \emph{et al.}\cite{Zhang2019_32} & $\ $ & $\surd$ & $\ $ & $\ $ & $\surd$ & $\ $ \\ \hline
		Z. Zhang \emph{et al.}\cite{Zhang2019_33} & $\surd$ & $\surd$ & $\triangle$ & $\triangle$ & $\surd$ & $\ $ \\ \hline
		B. Zong \emph{et al.}\cite{Zong2019_34} & $\surd$ & $\surd$ & $\ $ & $\surd$ & $\ $ & $\ $ \\ \hline
		L. Bariah \emph{et al.}\cite{Bariah2020_11} & $\triangle$ & $\surd$ & $\ $ & $\surd$ & $\surd$ & $\ $ \\ \hline
		W. Saad \emph{et al.}\cite{Saad2020_12} & $\surd$ & $\surd$ & $\ $ & $\surd$ & $\surd$ & $\ $ \\ \hline
		H. Viswanathan \emph{et al.}\cite{Viswanathan2020_13} & $\ $ & $\surd$ & $\triangle$ & $\ $ & $\surd$ & $\surd$ \\ \hline
		G. Liu \emph{et al.}\cite{Liu2020_18} & $\triangle$ & $\surd$ & $\triangle$ & $\triangle$ & $\triangle$ & $\surd$ \\ \hline
		S. Chen \emph{et al.}\cite{Chen2020_19} & $\surd$ & $\surd$ & $\surd$ & $\surd$ & $\ $ & $\ $ \\ \hline
		X.-H. You \emph{et al.}\cite{SEU} & $\surd$ & $\surd$ & $\surd$ & $\surd$ & $\triangle$ & $\triangle$ \\ \hline
		W. Jiang \emph{et al.}\cite{Jiang2021_21} & $\surd$ & $\surd$ & $\surd$ & $\surd$ & $\triangle$ & $\surd$ \\ \hline
		Z. Wang \emph{et al.}\cite{Wang2022_6G} & $\triangle$ & $\surd$ & $\ $ & $\triangle$ & $\triangle$ & $\surd$ \\ \hline
		\textbf{This survey paper} & $\surd$ & $\surd$ & $\surd$ & $\surd$ & $\surd$ & $\surd$ \\ \hline
		\multicolumn{7}{|l|}{\begin{tabular}[l]{@{}l@{}} Note: The $\surd$ symbol indicates that this aspect of outlook is clearly outlined in the vision of the reference, the $\triangle$ symbol means that this aspect \\ of outlook is only mentioned but not clearly outlined in the vision of the reference, and the blank denotes that this aspect of outlook is not shown \\ in the reference. \end{tabular}} \\ \hline
		
	\end{tabular}
	\label{Table_literature_6G_Vision}
\end{table*}

\section{6G Technical Requirements and Application Scenarios}
Different application scenarios envisaged in the 6G vision have different performance requirements for 6G communication systems. In this section, the 6G technical requirements and application scenarios will be discussed in detail.

\subsection{6G Technical Requirements}
ITU-R considered eight parameters to be KPIs of International Mobile Telecommunications 2020 (IMT-2020), while with the rapid development of mobile communication networks, these indicators will not be sufficient to cater for the disruptive use cases and applications in 2030 and beyond. 
The eight KPIs \cite{ITU-R_M.2083-0} used to assess 5G are still valid for 6G, but the values need to be updated due to the development of technologies and the emergence of new applications. Besides, new indicators are required for the evaluation of new services in 6G, including positioning, sensing, security, and intelligence. 

\begin{figure*}[tb]
	\centerline{\includegraphics[width=0.9\textwidth]{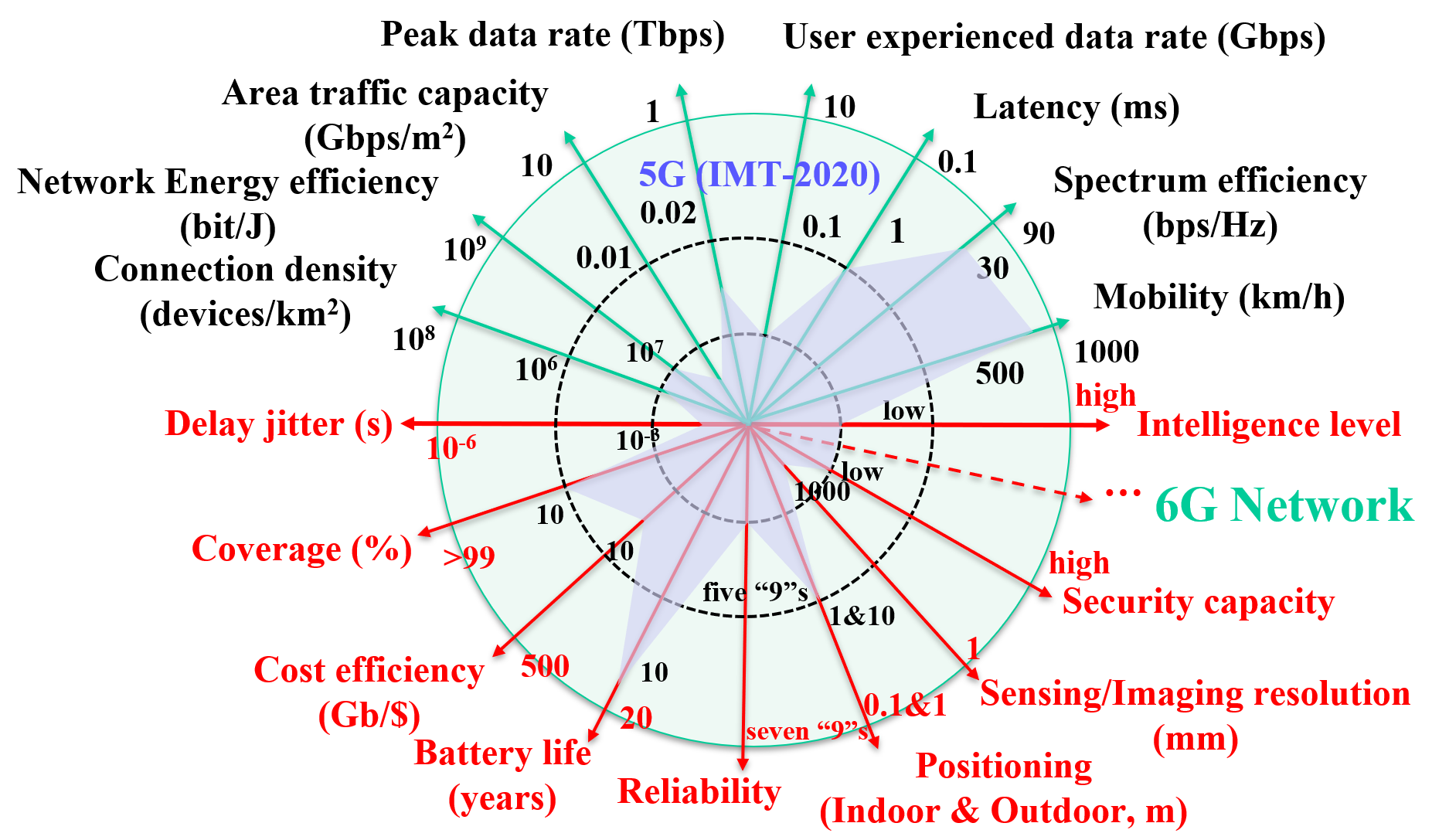}}
	\caption{6G KPIs.}
	\label{fig_6G_KPI}
\end{figure*}

In order to fill this gap, a number of quantitative and qualitative KPIs were proposed by different institutions and scholars \cite{ITU-R_M.2083-0,Oulu,NTT-Docomo,CCID,Samsung,SEU,vivo,UNISOC,Unicom,5GIA,HUAWEI,Akyildiz2020_1,Zhang2019_33,Khan2020_8,Gui2020,Saad2020_12,Viswanathan2020_13,Liu2020_18,Chen2020_19,Bhat2021,Jiang2021_21,Alwis2021,Tataria2021,Nguyen2022_Survey}. A comparison of related works on 6G KPIs is given in Table~\ref{Table_6G_KPI}. The table shows that the number of KPIs proposed in the existing papers is limited and cannot capture all the important characteristics of 6G. 
Besides, the analysis of KPIs in the literature is not thorough. In \cite{Oulu,NTT-Docomo,vivo,UNISOC,Unicom,Viswanathan2020_13,Bhat2021,Jiang2021_21}, the need for certain categories of indicators, such as security and intelligence, was suggested but corresponding KPIs were not defined.
For some KPIs, the reference values given by different papers vary greatly, such as delay jitter and network EE\cite{CCID,vivo,Akyildiz2020_1}, which may cause confusion. Some new KPIs were proposed, but there were no reference values given in the existing 6G survey papers, such as cost efficiency, security capacity, and intelligence level \cite{SEU,vivo}. Therefore, this paper aims to propose more comprehensive and reasonable KPIs and define their reference values based on extensive research and analysis.
\textcolor{black}{Fig. \ref{fig_6G_KPI} presents the proposed 17 KPIs for 6G, and those marked in red are not considered in 5G standardization. They are compared with the indicators for 5G, providing readers with an intuitive comparison. In Table \ref{Table_5Gvs6G_KPI}, the classification, definitions, references, as well as comparison of 5G and 6G KPIs are given. For 5G KPIs, corresponding references are labeled in the table. The quantitative values of 6G KPIs are obtained through the analysis of a number of literatures, which are explained in the text.}
Next, the proposed KPIs will be categorized into four classes and discussed as follows.

\renewcommand{\arraystretch}{1.2}
\begin{table*}[t]
	\centering
	\caption{\textcolor{black}{A comparison of relevant papers on the 6G KPIs.}}
	\begin{tabular}{|l|c|c|c|c|c|c|c|c|c|c|c|c|c|c|c|c|c|c|}
		\hline
		\multirow{2}{*}{\textbf{Ref.}}&
		\multirow{2}{*}{\rotatebox{270}{\textbf{Total number of KPIs}}} & 
		
		\multicolumn{4}{c|}{\textbf{Data Rate \& Delay}} & 
		\multicolumn{3}{c|}{\begin{tabular}[c]{@{}c@{}} \textbf{Capacity} \\ \textbf{\& Coverage}\end{tabular}} &
		\multicolumn{3}{c|}{\begin{tabular}[c]{@{}c@{}} \textbf{Service} \\ \textbf{Efficiency}\end{tabular}} &
		\multicolumn{7}{c|}{\textbf{\textcolor{black}{Diversified Service Evaluation}}} \\ \cline{3-19}
		{\ } & {\ } & \rotatebox{270}{\textbf{Peak data rate (Tbps)}} & \rotatebox{270}{\textbf{User experienced data rate (Gbps)}} & \rotatebox{270}{\textbf{Latency (ms)}} & \rotatebox{270}{\textbf{Delay jitter (us)}} & \rotatebox{270}{\textbf{Area traffic capacity ($\mathbf{Gbps/m^2}$)}} & \rotatebox{270}{\textbf{Connection density ($\mathbf{devices/km^2}$)}} & \rotatebox{270}{\textbf{Coverage (\%)}} & \rotatebox{270}{\textbf{Spectrum efficiency}} & \rotatebox{270}{\textbf{Network energy efficiency}} & \rotatebox{270}{\textbf{Cost efficiency}} & \rotatebox{270}{\textbf{Mobility (km/h)}} & \rotatebox{270}{\textbf{Battery life (years)}} & \rotatebox{270}{\textbf{Reliability}} & \rotatebox{270}{\textbf{Positioning (Indoor \& Outdoor, cm)}} & \rotatebox{270}{\textbf{Sensing/Imaging resolution (mm)}} & \rotatebox{270}{\textbf{Security capacity}} & \rotatebox{270}{\textbf{Intelligence level}}
		\\ \hline
		
		ITU-5G\cite{ITU-R_M.2083-0} & 8 & $\surd$ & $\surd$ & $\surd$ & $\ $ & $\surd$& $\surd$ & $\ $ & $\surd$& $\surd$& $\ $ & $\surd$ & $\ $ & $\ $ & $\ $ & $\ $ & $\ $ & $\ $ \\ \hline
		
		6G Flagship\cite{Oulu} & 8 & $\surd$ & $\ $ & $\surd$ & $\ $ & $\surd$& $\surd$ & $\ $ & $\ $ & $\surd$& $\ $ & $\ $ & $\surd$ & $\surd$ & $\surd$ & $\ $ & $\bigcirc$  & $\ $ \\ \hline
		
		NTT\cite{NTT-Docomo} & 7 & $\surd$ & $\ $ & $\surd$ & $\ $ & $\surd$& $\surd$ & $\surd$& $\ $ & $\ $ & $\ $ & $\ $ & $\bigcirc$ & $\surd$ & $\surd$ & $\bigcirc$ & $\bigcirc$ & $\ $ \\ \hline
		
		CCID\cite{CCID} & 9 & $\surd$ & $\surd$ & $\surd$ & $\ $ & $\surd$& $\surd$ & $\ $ & $\surd$& $\surd$& $\ $ & $\surd$ & $\ $ & $\ $ & $\surd$ & $\ $ & $\ $ & $\ $ \\ \hline
		
		Samsung\cite{Samsung} & 7 & $\surd$ & $\surd$ & $\surd$ & $\ $ & $\ $ & $\surd$ & $\ $ & $\surd$& $\surd$& $\ $ & $\ $ & $\ $ & $\surd$ & $\ $ & $\ $ & $\ $ & $\ $ \\ \hline
		
		SEU \& PML\cite{SEU} & 12 & $\surd$ & $\surd$ & $\surd$ & $\ $ & $\surd$& $\surd$ & $\surd$& $\surd$& $\surd$& $\surd$ & $\surd$ & $\ $ & $\ $ & $\ $ & $\ $ & $\surd$ & $\surd$ \\ \hline
		
		VIVO\cite{vivo} & 12 & $\surd$ & $\surd$ & $\surd$ & $\surd$ & $\ $ & $\surd$ & $\surd$& $\surd$& $\surd$& $\surd$ & $\surd$ & $\ $ & $\surd$ & $\surd$ & $\ $ & $\bigcirc$ & $\ $ \\ \hline
		
		UNISOC\cite{UNISOC} & 10 & $\surd$ & $\surd$ & $\surd$ & $\ $ & $\surd$& $\surd$ & $\ $ & $\surd$& $\surd$& $\ $ & $\surd$ & $\ $ & $\surd$ & $\surd$ & $\ $ & $\ $ & $\bigcirc$ \\ \hline
		
		China Unicom \cite{Unicom} & 11 & $\surd$ & $\surd$ & $\surd$ & $\ $ & $\surd$ & $\surd$ & $\ $ & $\surd$ & $\surd$ & $\ $ & $\surd$ & $\ $ & $\surd$ & $\surd$ & $\surd$ & $\bigcirc$ & $\bigcirc$ \\ \hline
		
		5GIA\cite{5GIA} & 6 & $\ $ & $\surd$ & $\surd$ & $\ $ & $\surd$& $\surd$ & $\ $ & $\ $ & $\surd$& $\ $ & $\ $ & $\ $ & $\ $ & $\surd$ & $\ $ & $\ $ & $\ $ \\ \hline
		
		HUAWEI\cite{HUAWEI} & 11 & $\surd$ & $\surd$ & $\surd$ & $\ $ & $\surd$& $\surd$ & $\surd$& $\ $ & $\surd$& $\ $ & $\ $ & $\surd$ & $\surd$ & $\surd$ & $\surd$ & $\ $ & $\ $ \\ \hline
		
		IMT-2030\cite{IMT2030-APP_KPI} & 14 & $\surd$ & $\surd$ & $\surd$ & $\surd$ & $\surd$& $\surd$ & $\surd$& $\surd$ & $\surd$& $\bigcirc$ & $\surd$ & $\ $ & $\surd$ & $\surd$ & $\surd$ & $\ $ & $\surd$ \\ \hline
		
		I. F. Akyildiz \emph{et al.}\cite{Akyildiz2020_1} & 10 & $\surd$ & $\surd$ & $\surd$ & $\surd$ & $\surd$& $\surd$ & $\ $ & $\surd$& $\surd$& $\ $ & $\surd$ & $\ $ & $\surd$ & $\ $ & $\ $ & $\ $ & $\ $ \\ \hline
		
		\textcolor{black}{Z. Zhang \emph{et al.}\cite{Zhang2019_33}} & 8 & $\surd$ & $\surd$ & $\surd$ & $\ $ & $\surd$& $\surd$ & $\ $ & $\surd$& $\surd$& $\ $ & $\surd$ & $\ $ & $\ $ & $\ $ & $\ $ & $\ $ & $\ $ \\ \hline
		
		L. U. Khan \emph{et al.}\cite{Khan2020_8} & 7 & $\surd$ & $\ $ & $\surd$ & $\ $ & $\surd$& $\ $ & $\ $ & $\surd$& $\surd$& $\ $ & $\surd$ & $\ $ & $\surd$ & $\ $ & $\ $ & $\ $ & $\ $ \\ \hline
		
		G. Gui \emph{et al.}\cite{Gui2020} & 8 & $\surd$ & $\ $ & $\surd$ & $\ $ & $\surd$& $\surd$ & $\ $ & $\surd$& $\surd$& $\ $ & $\ $ & $\ $ & $\ $ & $\ $ & $\ $ & $\surd$ & $\surd$ \\ \hline
		
		\textcolor{black}{W. Saad \emph{et al.}\cite{Saad2020_12}} & 5 & $\surd$ & $\ $ & $\surd$ & $\ $ & $\ $& $\ $ & $\ $ & $\surd$ & $\surd$ & $\ $ & $\ $ & $\ $ & $\surd$ & $\ $ & $\ $ & $\ $ & $\ $ \\ \hline
		
		H. Viswanathan \emph{et al.}\cite{Viswanathan2020_13} & 7 & $\surd$ & $\ $ & $\surd$ & $\ $ & $\surd$& $\surd$ & $\surd$ & $\ $ & $\ $ & $\ $ & $\ $ & $\bigcirc$ & $\surd$ & $\surd$ & $\bigcirc$ & $\ $ & $\ $ \\ \hline
		
		G. Liu \emph{et al.}\cite{Liu2020_18} & 10 & $\surd$ & $\surd$ & $\surd$ & $\ $ & $\surd$& $\surd$ & $\ $ & $\surd$& $\surd$& $\ $ & $\surd$ & $\ $ & $\surd$ & $\surd$ & $\ $ & $\ $ & $\ $ \\ \hline
		
		S. Chen \emph{et al.}\cite{Chen2020_19} & 11 & $\surd$ & $\surd$ & $\surd$ & $\ $ & $\surd$& $\surd$ & $\surd$& $\surd$& $\surd$& $\ $ & $\surd$ & $\ $ & $\surd$ & $\surd$ & $\ $ & $\ $ & $\ $ \\ \hline
		
		J. R. Bhat \emph{et al.}\cite{Bhat2021} & 14 & $\surd$ & $\surd$ & $\surd$ & $\surd$ & $\surd$ & $\surd$ & $\surd$& $\surd$ & $\surd$& $\ $ & $\surd$ & $\surd$ & $\surd$ & $\surd$ & $\surd$ & $\bigcirc$ & $\ $ \\ \hline
		
		W. Jiang \emph{et al.}\cite{Jiang2021_21} & 11 & $\surd$ & $\surd$ & $\surd$ & $\ $ & $\surd$& $\surd$ & $\surd$& $\surd$& $\surd$& $\ $ & $\surd$ & $\ $ & $\surd$ & $\surd$ & $\ $ & $\bigcirc$ & $\ $ \\ \hline
		
		\textcolor{black}{C. D. Alwis \emph{et al.}\cite{Alwis2021}} & 8 & $\surd$ & $\surd$ & $\surd$ & $\ $ & $\surd$& $\surd$ & $\ $ & $\surd$& $\surd$& $\ $ & $\surd$ & $\ $ & $\ $ & $\ $ & $\ $ & $\ $ & $\ $ \\ \hline
		
		
		
		H. Tataria \emph{et al.}\cite{Tataria2021}& 8 & $\surd$ & $\surd$ & $\surd$ & $\ $ & $\surd$& $\surd$ & $\ $ & $\ $ & $\surd$& $\ $ & $\surd$ & $\ $ & $\surd$ & $\ $ & $\ $ & $\ $ & $\ $ \\ \hline
		
		D. C. Nguyen \emph{et al.}\cite{Nguyen2022_Survey}& 7 & $\surd$ & $\ $ & $\surd$ & $\ $ & $\surd$& $\surd$ & $\surd$ & $\surd$& $\surd$& $\ $ & $\ $ & $\ $ & $\ $ & $\ $ & $\ $ & $\ $ & $\ $ \\ \hline

		\textbf{This survey paper} & 17 & $\surd$ & $\surd$ & $\surd$ & $\surd$ & $\surd$& $\surd$ & $\surd$& $\surd$& $\surd$& $\surd$ & $\surd$ & $\surd$ & $\surd$ & $\surd$ & $\surd$ & $\surd$ & $\surd$ \\ \hline
		\multicolumn{19}{|l|}{\begin{tabular}[l]{@{}l@{}} Note: The $\surd$ symbol indicates that this KPI is defined in the reference. The $\bigcirc$ symbol means that this KPI is not defined, but it is\\ mentioned that there is a requirement for indicator of this type. The blank denotes that this KPI is not covered in the reference. \end{tabular}} 
		\\ \hline
	\end{tabular}
	\label{Table_6G_KPI}
\end{table*}

\renewcommand{\arraystretch}{1.2}
\begin{table*}[t]
	\centering
	\caption{\textcolor{black}{Comparison of 5G and 6G KPIs.}}
	
	\begin{tabular}{|cc|c|c|c|c|}
		\hline
		\multicolumn{2}{|c|}{\textbf{KPI}}                 & \textbf{Definition} & \textbf{5G} & \textbf{6G} & \textbf{Enhancement} \\ \hline
		\multicolumn{1}{|c|}{\multirow{4}{*}[-25pt]{\begin{tabular}[c]{@{}c@{}}\textbf{Data rate \&} \\\textbf{Delay} \end{tabular}}} & \multicolumn{1}{m{2.5cm}|}{\textbf{Peak data rate}}  & \multicolumn{1}{m{4.3cm}|}{Maximum achievable data rate under ideal conditions per user/device} & 20 Gbps \cite{ITU-R_M.2083-0} & 1 Tbps  & 50$\times$ \\ \cline{2-6}
		
		\multicolumn{1}{|c|}{}                  & \multicolumn{1}{m{2.5cm}|}{\textbf{User experienced data rate}}  & \multicolumn{1}{m{4.3cm}|}{The data rate that is available ubiquitously across the coverage area to a mobile user/device} & 100 Mbps \cite{ITU-R_M.2083-0}  & 10 Gbps & 100$\times$ \\ \cline{2-6} 
		
		\multicolumn{1}{|c|}{}                  & \multicolumn{1}{l|}{\textbf{Latency}} & \multicolumn{1}{m{4.3cm}|}{The time from when the source sends a packet to when the destination receives it} & 1 ms \cite{ITU-R_M.2083-0} & 0.1 ms & 10$\times$  \\ \cline{2-6} 
		
		\multicolumn{1}{|c|}{}                  & \multicolumn{1}{l|}{\textbf{Delay jitter}} &  \multicolumn{1}{m{4.3cm}|}{The latency variations in the system} & 1 ms \cite{Nasrallah2019} & 1 ${\upmu}$s  & 1000$\times$ \\ \hline
		
		\multicolumn{1}{|c|}{\multirow{3}{*}[-10pt]{\begin{tabular}[c]{@{}c@{}}\textbf{Capacity \&} \\\textbf{Coverage} \end{tabular}}}  & \multicolumn{1}{l|}{\textbf{Area traffic capacity}} & \multicolumn{1}{m{4.3cm}|}{Total traffic throughput served per geographic area} & 10 Mbps/${\rm m^2}$  \cite{ITU-R_M.2083-0} & 10 Gbps/${\rm m^2}$ & 1000$\times$ \\ \cline{2-6} 
		
		\multicolumn{1}{|c|}{}                  & \multicolumn{1}{l|}{\textbf{Connection density}} & \multicolumn{1}{m{4.3cm}|}{Total number of connected and/or accessible devices per unit area} & $10^6$ devices/$\rm {km^2}$ \cite{ITU-R_M.2083-0} &  $10^8$ devices/$\rm {km^2}$ & 100$\times$\\ \cline{2-6} 
		
		\multicolumn{1}{|c|}{}                  & \multicolumn{1}{l|}{\textbf{Coverage}} & \multicolumn{1}{m{4.3cm}|}{The coverage percentage of network service} & 10\% \cite{KPI_Wu2020} & 99\% & 10$\times$ \\ \hline
		
		\multicolumn{1}{|c|}{\multirow{3}{*}[-17pt]{\begin{tabular}[c]{@{}c@{}}\textbf{Service} \\ \textbf{efficiency}\end{tabular}}} & \multicolumn{1}{l|}{\textbf{Spectrum efficiency}} & \multicolumn{1}{m{4.3cm}|}{Average data throughput per unit of spectrum resource and per cell} & 30 bps/Hz \cite{Akyildiz2020_1} & $\geq$90 bps/Hz & $\geq$3$\times$ \\ \cline{2-6} 
		
		\multicolumn{1}{|c|}{}                  & \multicolumn{1}{m{2.5cm}|}{\textbf{Network energy efficiency}} & \multicolumn{1}{m{4.3cm}|}{The quantity of information bits transmitted to/received from users, per unit of energy consumption} & $10^7$ bit/J \cite{vivo} & $10^9$ bit/J & 100$\times$ \\ \cline{2-6} 
		
		\multicolumn{1}{|c|}{}                  & \multicolumn{1}{l|}{\textbf{Cost efficiency}} & \multicolumn{1}{m{4.3cm}|}{The ratio of user's data consumption benefit and its data traffic cost} & 10 Gb/\$ \cite{vivo} & 500 Gb/\$ & 50$\times$ \\ \hline
		
		\multicolumn{1}{|c|}{\multirow{7}{*}[-33pt]{\begin{tabular}[c]{@{}c@{}}\textbf{Diversified} \\ \textbf{service evaluation}\end{tabular}}} & \multicolumn{1}{l|}{\textbf{Mobility}} & \multicolumn{1}{m{4.3cm}|}{Maximum speed at which a defined quality of service (QoS) and seamless transfer between radio nodes can be achieved} & 500 km/h  \cite{ITU-R_M.2083-0} & 1000 km/h & 2$\times$ \\ \cline{2-6} 
		
		\multicolumn{1}{|c|}{}                  & \multicolumn{1}{l|}{\textbf{Battery life}} & \multicolumn{1}{m{4.3cm}|}{The life span of IoT equipment batteries} & 10 years \cite{KPI_Zhong2019} & 20 years & 2$\times$ \\ \cline{2-6} 
		
		\multicolumn{1}{|c|}{}                  & \multicolumn{1}{l|}{\textbf{Reliability}} & \multicolumn{1}{m{4.3cm}|}{The percentage of packets successfully received under a certain upper delay limit} & 99.999\% \cite{Samsung} & $\geq$99.99999\% & $\geq$100$\times$ \\ \cline{2-6} 
		
		\multicolumn{1}{|c|}{}                  & \multicolumn{1}{l|}{\textbf{Positioning}} & \multicolumn{1}{m{4.3cm}|}{The accuracy of positioning for indoor and outdoor} & 1 m \& 10 m \cite{CCID}& 10 cm \& 1 m & 10$\times$ \\ \cline{2-6} 
		
		\multicolumn{1}{|c|}{}                  & \multicolumn{1}{m{2.5cm}|}{\textbf{Sensing/Imaging resolution}} & \multicolumn{1}{m{4.3cm}|}{The resolution of sensing and imaging} & 1 m\cite{KPI_Barneto2019} &  1 mm & 1000$\times$ \\ \cline{2-6} 
		
		\multicolumn{1}{|c|}{}                  & \multicolumn{1}{l|}{\textbf{Security capacity}} & \multicolumn{1}{m{4.3cm}|}{The rate of trustworthy information from the transmitter to the receiver under the threat of eavesdroppers } & Low & High & -- \\ \cline{2-6} 
		
		\multicolumn{1}{|c|}{}                  & \multicolumn{1}{l|}{\textbf{Intelligence level}} & \multicolumn{1}{m{4.3cm}|}{The intelligent level of communication system} & Low & High & -- \\ \hline
		
	\end{tabular}
	\label{Table_5Gvs6G_KPI}
\end{table*}

\subsubsection{Data rate \& Delay}

Due to the emergence of a number of new services, higher rate and lower delay are inevitable for 6G communications. Three indicators are proposed to evaluate 6G performance in this regard. They are peak data rate, user experienced data rate, and latency. For 6G networks, the peak data rate can be 1 Tbps with the help of new technologies, such as THz and OWCs, \textcolor{black}{which can increase by more than 50 times over 5G\cite{Samsung, UNISOC, HUAWEI, Akyildiz2020_1,  Liu2020_18, Bhat2021, Jiang2021_21, Tataria2021, Nguyen2022_Survey}}. The user experienced data rate, which is defined as the maximum rate GB/userthat can be guaranteed with a probability of more than 95\% when required, \textcolor{black}{will achieve 10 Gbps \cite{vivo, 5GIA, HUAWEI, Liu2020_18, Chen2020_19, Bhat2021}. }
Latency, defined as the minimum delay of air interface access, is expected to be \textcolor{black}{0.1 ms for some specific applications, such as intelligent driving, tele-surgery, and industrial control \cite{SEU, Oulu, NTT-Docomo, CCID, Samsung, vivo, UNISOC, Unicom, HUAWEI,Akyildiz2020_1, Khan2020_8, Viswanathan2020_13, Liu2020_18, Chen2020_19, Bhat2021, Jiang2021_21}}. 
Besides, delay jitter is an important indicator to quantify the latency variations in communication systems. \textcolor{black}{Thanks to the development of deterministic networks\cite{DetNet_WP}, it can achieve 1 us for 6G \cite{SEU, Oulu, NTT-Docomo, CCID, Samsung, vivo, UNISOC, Unicom, HUAWEI, Akyildiz2020_1, Khan2020_8, Viswanathan2020_13, Liu2020_18, Chen2020_19, Bhat2021, Jiang2021_21} \cite{HUAWEI,Akyildiz2020_1,Bhat2021}.}

\subsubsection{Capacity \& Coverage}
In order to support growing demands for next-generation new scenarios, such as smart factory and smart city, higher air traffic capacity and connection density are considered as essential conditions. \textcolor{black}{The area traffic capacity is expected to achieve 10 Gbps/${\rm m^2}$ \cite{NTT-Docomo, CCID, Unicom, HUAWEI, Khan2020_8, Liu2020_18, Bhat2021}, and the connection density is expected to be $10^8$ devices/${\rm km^2}$ \cite{SEU, Oulu, vivo, Liu2020_18}.}
Providing users with broader and seamless coverage is another important requirement of 6G networks. For 6G, 3D global coverage is expected to extend the 5G two-dimensional (2D) terrestrial coverage\cite{Jiang2021_21}. The 5G network can only cover land and some offshore areas, \textcolor{black}{only 20$\%$ of land and 5$\%$ of ocean \cite{KPI_Wu2020}, approximately equal to 10\% of the whole earth.} \textcolor{black}{The coverage percentage of 6G network is expected to be more than 99\% \cite{Chen2020_19, Nguyen2022_Survey}.} In high-value areas, 6G coverage needs to further improve EE and user experience, while in low-value areas, it is important to reduce coverage costs\cite{vivo}.

\subsubsection{Service efficiency}
In order to meet the need for sustainable development, we propose three efficiency-related indicators, taking the service efficiency at multiple levels into account. 
Owing to the development of advanced spectrum management technologies, \textcolor{black}{the spectrum efficiency (SE) of 6G systems can triple that of 5G, reaching 90 bps/Hz\cite{SEU, UNISOC, Khan2020_8, Chen2020_19, Jiang2021_21, Nguyen2022_Survey}.} Network EE refers to the number of bits transmitted over 1 Joule. According to \cite{vivo}, the consumption of a typical 5G base station is about 1-2 kW for an average data rate of 10 Gbps. Therefore, network EE is about $10^7$ bit/J for 5G. \textcolor{black}{For 6G, it is expected to increase by about 100 times in order to provide more efficient services with the same or lower energy consumption\cite{SEU, vivo, HUAWEI}.}
As a complement to SE and EE, economic efficiency was introduced to evaluate the effectiveness of the trade-off  \cite{KPI_Ku2013,KPI_Patcharamaneepakorn2016,KPI_Zhang2018}. In \cite{IMT-2020-5G}, a similar concept named cost efficiency was proposed as the ratio of user's data consumption benefit and its data traffic cost. \textcolor{black}{The trade-off between cost efficiency and EE for 5G system was analyzed in\cite{zi_energy_2016}. However, it was not until the  advance of 6G system that the importance of cost efficiency was gradually appreciated.} For 6G, cost efficiency will help equipment providers and service teams to consider the profitability of the communication industry while providing quality services to consumers. \textcolor{black}{In \cite{vivo}, it was proposed that spending on communication for a user is expected to be less than 1\% of gross domestic product (GDP) per capita. As a rough estimate, the cost efficiency of 5G is about 10 Gb/\$, with \$10,000 GDP per capita and 100 GB/user/month communication cost. The cost efficiency for 6G will become 500 Gb/\$, based on a doubling of GDP per capita and a 100-fold increase in service traffic.}

\subsubsection{Diversified service evaluation}
6G networks will largely enhance and extend 5G applications \cite{SEU}. With the increase of applications, \textcolor{black}{some indicators will be introduced to evaluate the diversified service quality for 6G communication systems.}
High-speed mobile communication scenarios, such as high-speed aircrafts and ultra-high-speed trains, place high demands on mobility. \textcolor{black}{The mobility of transceivers in 6G communication systems can be more than 1000 km/h, which is a huge increase compared with that in 5G\cite{CCID, vivo, UNISOC, Unicom, Akyildiz2020_1, Khan2020_8, Liu2020_18, Chen2020_19, Bhat2021, Jiang2021_21, Tataria2021}.}
In \cite{Oulu}, it was pointed out that limited batteries would be an evident obstacle for the development of IoT communication. Besides, in \cite{HUAWEI}, it was stated that scenarios such as smart city and smart home would have high requirements on the battery life of sensing devices, and it would be necessary to improve the battery life to ensure the stable availability of sensing devices. \textcolor{black}{The longest battery life for smart scenarios is expected to be 20 years\cite{Oulu, HUAWEI, Bhat2021}.}
The development of the IoT places high demands on the reliability of the communication system, that is, the correctness of the transmitted information under a certain upper delay limit. \textcolor{black}{From Table \ref{Table_6G_KPI}, it can be seen that reliability is not included in \cite{ITU-R_M.2083-0} but has attracted considerable attention for 6G.} Reliability requirements of 6G networks are expected to be application specific. For stringent scenarios such as strengthened ultra-reliable and low latency communications (uRLLC), \textcolor{black}{only one erroneous bit is permitted in 10 million transmitted bits\cite{Akyildiz2020_1, Viswanathan2020_13, Bhat2021, Tataria2021}.}
Sensing positioning and imaging are important foundations for smart connections of all things. 
\textcolor{black}{6G positioning accuracy is expected to reach 1 m for outdoor scenarios and 10 cm for indoor scenarios\cite{Oulu, CCID, vivo}. 6G sensing/imaging resolution is expected to reach 1 mm\cite{HUAWEI}.}
In addition, two important indicators, security capacity and intelligence level, are proposed to qualitatively evaluate the service capability of 6G systems. It is believed that 6G communication systems can achieve endogenous security through new technologies, such as quantum communication and blockchain, so that the system can have a higher security capacity. \textcolor{black}{Related indicators have been considered in \cite{SEU,Oulu,NTT-Docomo,vivo,Unicom,Gui2020,Bhat2021,Jiang2021_21}.} Besides, with the accelerated permeation of AI, the intelligence level of the 6G communication system is expected to be greatly \textcolor{black}{enhanced\cite{SEU,UNISOC,Unicom,Gui2020}.} \textcolor{black}{These two indicators have not been quantitatively studied yet.}

\subsection{6G Application Scenarios}
\begin{figure*}[tb]
	\centerline{\includegraphics[width=0.95\textwidth]{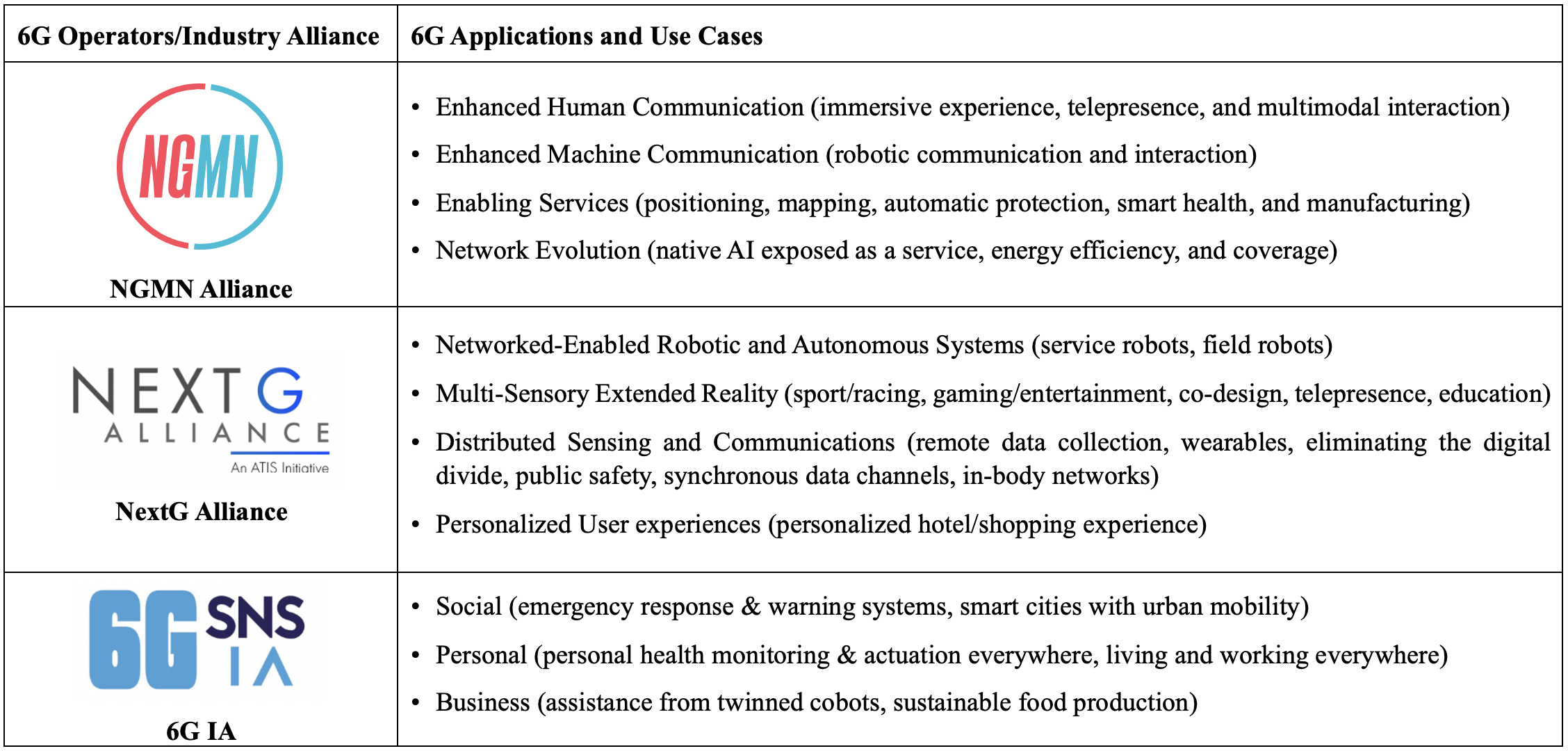}}
	\caption{\textcolor{black}{Prospects of 6G operator/industry alliances for 6G application scenarios\cite{NGMNA-Applications,NextG-Applications,6GIA-Applications}.}}
	\label{fig_6G_use_cases}
\end{figure*}
In the 5G era, there are three main application scenarios, i.e., enhanced mobile broadband (eMBB), massive machine type communications (mMTC), and uRLLC, aiming to meet high requirements for large bandwidth and high data rate, large connection density, high reliability and low latency, respectively. 
A number of relevant papers have demonstrated their outlooks for 6G application scenarios. \textcolor{black}{As exemplified in Fig. \ref{fig_6G_use_cases}, relevant operators and industry alliances have envisioned several applications and use cases for 6G\cite{NGMNA-Applications,NextG-Applications,6GIA-Applications}. It can be seen that the industry and operators are highly interested in typical scenarios in different fields, such as immersive applications related to the personal field, robotics, automation, and remote data collection in the commercial field.}
Since diverse application scenarios have different requirements on KPIs of communication systems, it is reasonable to classify application scenarios using KPIs, as for 5G.
However, most of these works only introduced several possible application scenarios in 6G, but did not give a classification for these scenarios, including 6G white papers\cite{CCID,Samsung,UNISOC,DATANG,Unicom,IMT-2030,HUAWEI} and a series of 6G survey papers\cite{Bariah2020_11,Viswanathan2020_13,Giordani2020_16,Chen2020_19,Tataria2021}. At the same time, classifications of application scenarios in several works were not related to 6G KPIs, e.g., applications for intelligent life, intelligent production, and intelligent society in \cite{Liu2020_18}, immersive, intelligent, and ubiquitous applications in \cite{Wang2022_6G}. Besides, classifications using 6G KPIs in other works are not comprehensive. For example, the authors of \cite{Chowdhury2020_6} \textcolor{black}{and \cite{Bhat2021}} ignored a series of novel applications such as space-air-ground-sea integrated networks and digital twin applications. You \emph{et al.}\cite{SEU} ignored new scenarios that combine characteristics of typical scenarios defined in 5G. Jiang \emph{et al.} \cite{Jiang2021_21} ignored the enhancement of 5G applications. Therefore, a comprehensive investigation of potential 6G application scenarios with a reasonable classification is urgently required.

\begin{figure*}[tb]
	\centerline{\includegraphics[width=0.95\textwidth]{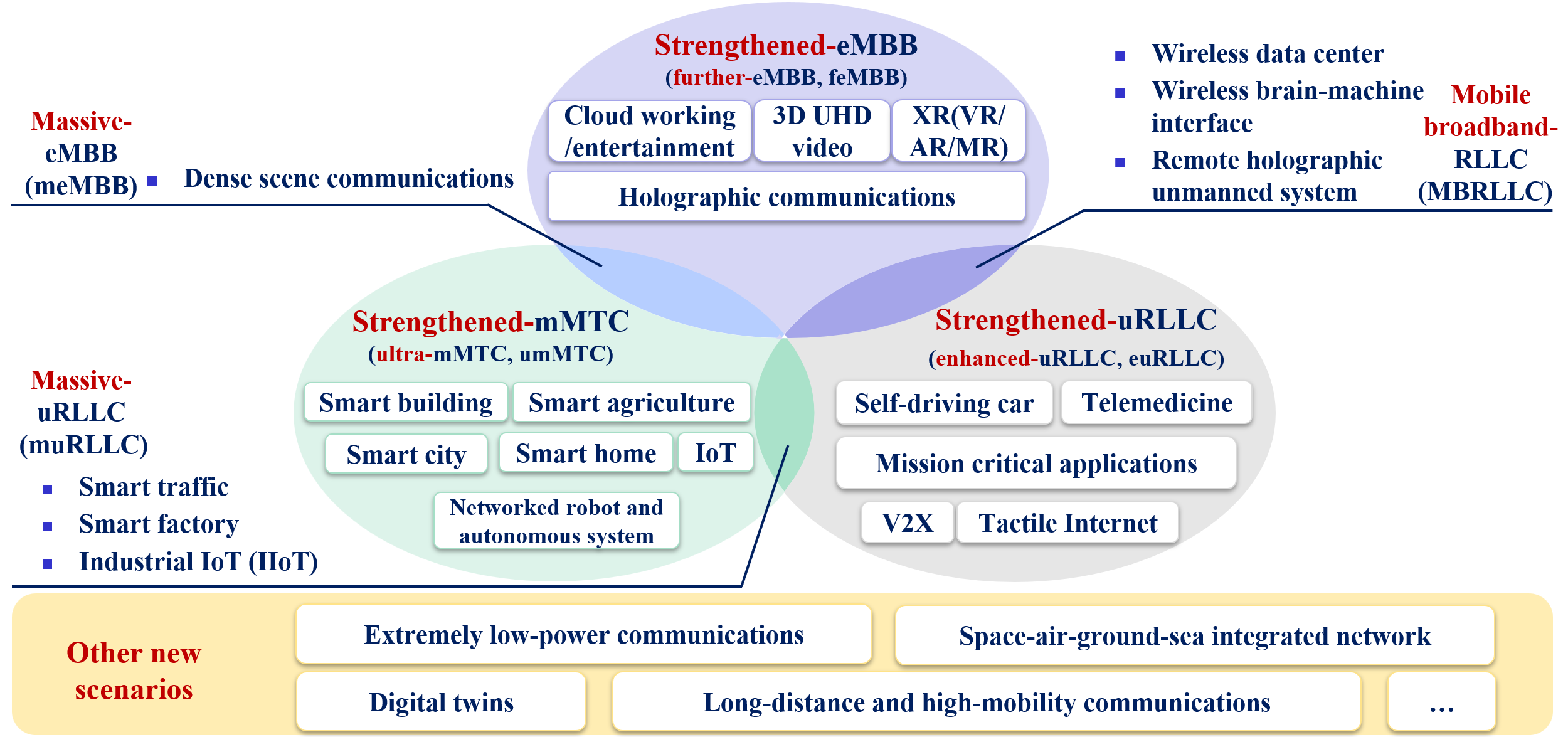}}
	\caption{Potential 6G application scenarios.}
	\label{fig_6G_application_scenarios}
\end{figure*}

Generally, 6G will continue to enhance and expand the above application scenarios to achieve further-eMBB (feMBB), ultra-mMTC (umMTC), and enhanced-uRLLC (euRLLC). In 2030, these three scenarios will not only meet traditional KPIs, such as data rate, connection density, and delay in communication systems, but also new KPIs in imaging, positioning, sensing, security capacity, and intelligence level, etc. Moreover, with the integration and development of technologies, 6G will also develop several potential application scenarios that combine characteristics of different kinds of scenarios. In addition to the three scenarios strengthened from 5G, it is envisioned that 6G will also provide additional application scenarios including: 1) massive eMBB (meMBB), aiming at high data rate, large bandwidth, and large connection density, e.g., dense scene communications; 2) mobile broadband reliable and low latency communication (MBRLLC), for scenarios with high data rate, large bandwidth, low latency, and high reliability, e.g., wireless data centers and wireless brain-machine interfaces; 3) massive uRLLC (muRLLC), requiring low delay and high reliability while having large connections for large-scale machine-type communication in special scenarios, e.g., smart transportation, smart factory, and industrial IoT. Moreover, there could be other potential scenarios, such as extremely low-power communications, digital twin applications, space-air-ground-sea integrated networks, as well as long-distance and high-mobility communications. Our vision for 6G application scenarios is shown in Fig. \ref{fig_6G_application_scenarios}. 
Next, several representative new application scenarios envisioned in the 6G era will be discussed.

\subsubsection{Wireless data center}
With the development of science and technology, various industries have produced a massive amount of data, which plays a decisive role in supporting modern society. In addition, technologies, such as AI and big data, are also expected to dig deeper into the hidden information embedded in the data to achieve more intelligent applications. In addition to focusing on technological development, these massive amount of data also need enough space to store. At present, data centers are mainly wired, with high complexity, high maintenance costs, high power consumption, and large space\cite{Bariah2020_11}. The demand for wireless data centers arises at the moment. However, due to the challenges of implementing ultra-high data rate transmission in existing wireless communication systems, they cannot meet the needs of data centers for data storage, transmission, and scheduling. With the further in-depth research of THz and the cloud data center, 6G is expected to take advantage of the ultra-large bandwidth of THz to achieve ultra-high data rate transmission\cite{Wang2020_6GCMM-survey,THz_Chen2019}, realizing the next-generation cloud-based wireless data center\cite{Ahearne2019,Bariah2020_11}. The data will be stored on multiple cloud servers, and THz will serve as the transmission medium to support deployments and operations of wireless data centers.

\subsubsection{Tactile Internet}
\textcolor{black}{As summarized in Fig.\ref{fig_1G_5G}, communication services in previous generations mainly focused on the digitization of visual and auditory information.}
With the advent of the 6G era, users' tactile information can also be collected, digitized, and transmitted over the network, finally forming the tactile Internet. The basic characteristics of the tactile Internet include the implementation of perception and synchronization actions, which can be used to transfer ``skills"\cite{Holland2019}. Tactile robots will act as a multi-modal avatar of human beings through the tactile Internet\cite{Haddadin2019}. The users' tactile information can be transmitted through the network, and remote robots complete corresponding actions according to the users' command. After acquiring the real-time status remotely, users can interact and control accordingly. The transmission of tactile information has very high requirements on the network delay and reliability. In the future, doctors can remotely guide robots to perform physical examinations through the tactile Internet, and remote robots can help miners to complete mining in high-risk areas. With the possible realization of ultra-low latency and ultra-high reliability communications, remote robotic surgery\cite{Kolovou2021} may also become possible.

\subsubsection{Digital twin applications}
At present, both academia and industry are exploring the connection between the physical real world and the digital world. With the further development and evolution of the digital twin technology, the physical reality will be more precisely digitized. The digital world and the physical world will map to and influence each other, achieving simultaneous development\cite{UNISOC,Szabo2019,Nguyen2021}. What's amazing is that, with the help of AI and other intelligent technologies, \textcolor{black}{users' operations and predictions in the digital world can correspond to the conditions in the physical world}, which provides a reference for decision-making in the physical world. Digital twins are expected to be used in many fields\cite{Barricelli2019,Vukovic2021,Ivanov2020,SEU,DATANG}. Typically, the city digital twin system\cite{Ivanov2020} is envisaged to include interconnected digital twins of urban infrastructure, transportation networks, urban ecological environment, power systems, and other systems, providing a series of functions, such as the environmental monitoring, emergency warning, and risk prediction. The realization of the city digital twin will help the rapid and effective decision-making of emergency plans and evaluation on urban design management plans. Another example is the digital twin body area network\cite{SEU}, which can simulate a virtual human body through 6G and information and communications technology (ICT). It can track around the clock, predict diseases in advance, and simulate the operations and medication on the virtual human body. The efficacy of drugs can be predicted which can speed up drug development, reduce costs, and improve the quality of human life. The combination of digital twin body area network and cloud, fog, sensor layer computing and other technologies will also help epidemic management, such as corona virus disease 2019 (COVID-19), including infection source search, drug development and other aspects. The realization of these novel application scenarios relies on smarter and more advanced 6G.

\subsubsection{Wireless brain-machine interface applications}
The wireless brain-machine interface is a method to help users control electronic devices through their brains, and is often used to provide intelligent communication between users and smart home applications or medical equipment\cite{Tripathy2016,Jafri2019}. It transmits the users' brain signals to the electronic device, and these signals are analyzed and converted into the command and operations for the device\cite{Chowdhury2020_6}. The wireless brain-machine interface is initially used in medical scenarios, providing a new method of transmitting information on patients with neurological injuries and diseases, such as paralysis and Parkinson's, to bring qualitative changes to the quality of their life. Recently, BrainGate has implemented a high-bandwidth, high-precision, and low-power wireless brain-machine interface system with which two paralyzed testers can ``perform" clicking and typing on a tablet computer using their brain signals only\cite{Simeral2021}. Also, the wireless system has been tested and its performance is very close to that of wired computer peripherals. The authors of \cite{Liu2021} proposed a fully integrated wireless sensor brain-machine interface, which can transmit key somatosensory signals, fingertip strength, and limb joint angles to the brain, providing a new solution for transmitting somatosensory feedback for the next generation of neuroprosthetics. In addition to applications in medical and health care scenarios, the wireless brain-computer interface in 6G may also help users better communicate with the environment and other users who use wireless brain-computer interface-supported devices\cite{Saad2020_12}. In the future, smarter 6G with higher speed, higher reliability, lower latency, higher perception accuracy, and lower power consumption will help the realization of wireless brain-machine interfaces, providing impetus for the realization of a smarter life.

\subsubsection{Holographic communication}
In many science fiction movies, there have been scenarios where the protagonist communicates with another person's virtual image, which is expected to become a reality in the 6G era. People can see and interact with holographic images of others, as if people were communicating face to face in the same place. On the basis of traditional 2D video communication, holographic communication can present high-precision real-time 3D images, requiring very large bandwidth, low latency, and high-precision resolution. Information including the 2D resolution, color and 3D tilt, angle and position will be transmitted through the communication network\cite{Jiang2021_21}. In addition to real-time holographic communications in the real world, holographic communication will also bring users the experience of connecting with the past and the future\cite{UNISOC}. Whether they are people who have passed away, things that have disappeared, or things that have not yet appeared, they can all be preserved and constructed in a holographic manner. Users will be able to communicate with these past and future holographic images, and get a new experience in the past and future worlds.

\subsubsection{Emergency rescue communication}
From 1G to 5G, the terrestrial mobile communication system has achieved higher coverage, larger bandwidth, faster speed, lower delay, and denser network. However, when an area suffers from large-scale natural disasters, such as earthquakes, floods, mudslides or other severe accidents caused by humans, the terrestrial communication network in the area may be completely paralyzed. People who need help cannot send out distress signals  in time, and the external rescue task will also be hindered. Besides, there is not enough communication network coverage in several scenarios, such as oceans and deserts. In the event of accidents and emergencies, the golden 72 hours will be the key to saving people's lives. With the further realization of 3D full space coverage in 6G, UAV\cite{Pirzada2020,Jin2020,Feng2020} and satellite communication networks\cite{Shin2017,Pirzada2020} will respond quickly and be deployed on demand, providing emergency communications to help quick search and rescue. Taking the golden rescue time into account, it is required to quickly provide a large-bandwidth network deployment with sufficient coverage area. At the same time, because the non-terrestrial networks have high battery requirements, it is hoped that the power consumption can be as low as possible and the emergency communication system can support for a longer period of time.

\subsubsection{Immersive XR}
XR is the collective term for VR, AR, and mixed reality (MR). The VR enables users to interact with another completely virtual digital world, while the AR provides users with an interactive experience of virtual objects in real physical worlds. The MR integrates the VR and AR, providing interactions for users with the real world, digital VR world and completely virtual objects\cite{Chowdhury2020_6}. Although XR has some practical applications in the 5G era, it is still in the initial stage, similar to the video services of the mobile Internet\cite{Jiang2021_21}. With the continuous development of computing, communication, sensing, imaging, storage, and other technologies, 6G will achieve a fully immersive XR. The virtual five senses including sight, hearing, touch, smell, and taste will be digitalized and transmitted. The video image resolution will be higher, the color will be more real, and the delay will be lower, giving users a more realistic and immersive feeling. In the future, 6G's ultra-high data rate, ultra-large bandwidth, low latency, high reliability, high imaging resolution, high sensing capabilities, and other characteristics will make immersive XR applications possible, which can be used \textcolor{black}{in a series of scenarios such as entertainment, telemedicine, and remote industrial control.}

\subsection{System Performance Analysis and Trade-offs between KPIs}
Due to the development of communication technologies and the enrichment of service requirements, 6G will provide a more comprehensive and high-quality communication experience and will exceed 5G in system performance on all fronts as shown in Table \ref{Table_5Gvs6G_KPI}. To evaluate 6G more thoroughly, the proposed 17 KPIs can be used to analyze system performance in terms of network transmission performance, network access performance, network efficiency, and QoS.

Although 6G will have a step change in these 17 KPIs, the KPIs often cannot be improved simultaneously in the same system due to hardware \textcolor{black}{impairments}, actual propagation environment and other limitations \cite{chen_fundamental_2011}. Moreover, performance improvements in communication systems often come at the cost of energy consumption. According to\cite{yu_power_2021}, the energy consumption of 5G is more than three times that of 4G, and thus it can be expected that the energy consumption of 6G may exceed that of 5G. As a result, considering the resource constraints, maintaining optimum performance in all aspects is superfluous for a practical communication system. As mentioned previously, 6G systems will be extended to support more abundant and various application scenarios. In addition to communication, 6G is expected to offer new services, \textcolor{black}{e.g., digital twins, AI, computing, localization, and sensing.} Since 6G has a wide range of applications but system resources are difficult to meet all the ultra-high performance demands simultaneously, the trade-offs of system performance metrics in each scenario and application are inevitable. In\cite{rasti_evolution_2022}, the authors summarized the technical requirements in different 6G application scenarios, including feMBB, umMTC, euRLLC, and MBRLLC. For example, it was proposed that latency, jitter and reliability are more significant in euRLLC scenarios than peak rate and connectivity. The different requirements for KPIs appear in other 6G scenarios as well, and this is the theoretical foundation for analyzing the performance trade-offs. 
\textcolor{black}{In this survey paper, the 6G technical requirements for different application scenarios are summarized in Table~\ref{table_KPI_AS}. For technical requirements with existing quantitative ranges, e.g., data rate, latency, and connection density, we investigate the quantitative requirements of these KPIs in different application scenarios. Note that each type of scenario does not require all indicators to reach the extreme. So, we only provide quantitative specific values for these indicators with high attention. However, for the emerging security capacity and intelligence level that have not been quantitatively studied yet, we only provide qualitative importance measure. As can be seen, there are great differences between 6G requirements of KPIs in different application scenarios. For example, digital twins impose high demands on data rate, latency, area traffic capacity, connection density, positioning, resolution, and intelligence level. Extremely low-power communications require high level of connection density and energy efficiency.}
A similar but simplified table was proposed in\cite{Jiang2021_21} and the table listed different levels of demand for KPIs for different technologies. Hence, trade-offs between KPIs based on their intrinsic relationships and application scenarios have been a hot topic since 4G and many scholars have conducted research on this fundamental issue.

Among all the KPIs, SE and EE have been the most concerned ones because every communication system upgrade is accompanied by a magnitude increase in these two metrics. For communication system operators, it is optimal to enhance both SE and EE. Unfortunately, in actual system deployment, EE decreases with SE enhancement, which means that it is essential to achieve the optimal operation points in terms of SE and EE for better overall performance\cite{wu_green_2014}. \textcolor{black}{In\cite{ku_spectral-energy_2013, zhang_performance_2018}, the trade-offs between SE and EE for the relay system were investigated} while in\cite{you_energy_2021} authors analyzed the same trade-off for RIS-aided system. Based on the trade-off between SE and EE, \textcolor{black}{in\cite{ruan_power_2019} authors investigated power allocation in satellite-vehicular networks} and the hybrid time division multiple access (TDMA) non-orthogonal multiple access (NOMA) system design was proposed in\cite{wei_spectral-energy_2022}. Furthermore, as a complementary KPI to SE and EE, cost efficiency has also received wide attention. The SE, cost efficiency, and their relationships were discussed for cellular networks in\cite{mishra_spectral_2019}. \textcolor{black}{In\cite{patcharamaneepakorn_quadrature_2018}, a quadrature space-frequency index modulation (IM) scheme was proposed and the SE-EE-cost efficiency-economic efficiency trade-off performance was investigated. An adaptive transmission scheme for integrated satellite-terrestrial networks was invented in\cite{ruan_energy_2018} and the trade-off among SE, EE, symbol error rate, and economic efficiency based on the proposed scheme was analyzed.}

In addition to the trade-off between SE and EE, the study of trade-offs between other KPIs is equally instructive for communication system design. The data rate, latency, and their trade-offs were investigated for secondary cellular networks and UAV networks\cite{chen_capacity_2018,wei_capacity_2019}. Furthermore, the delay and EE trade-offs also raise concerns. The trade-offs for D2D communication and maritime wireless networks were analyzed in\cite{luo_learning_2019,yang_efficient_2021}. Reliability, as one of the most focused KPIs, has similarly been extensively studied. Its trade-offs with security, latency, and cost efficiency have been validated\textcolor{black}{\cite{cao_security-reliability_2021,zeng_reliability_2021,chen_high_2019, song_cost-reliability_2017}}. As to the trade-offs in different application scenarios, in\cite{zhang_statistical_2021} the authors analyzed the delay and error-rate performances and solved the delay-violation probability minimization problem for the muRLLC scenarios in 6G cell-free massive MIMO systems. In\cite{wu_resource_2020}, the authors emphasized the importance of EE and QoS in designing resource management and network architecture for space-air-ground systems. The performance analysis of tactile internet and the trade-offs for URLLC massive MIMO systems were discussed in\cite{tarneberg_utilizing_2017}. 
In short, the trade-offs between 6G KPIs for optimal system performance have been widely recognized and the research on SE, EE, capacity, reliability and other metrics has been fruitful. However, due to the increased complexity of 6G systems and the abundant application scenarios, there is still a need for further research on 6G system performance analysis and trade-offs between KPIs.

\begin{sidewaystable*}[thp]
    \renewcommand\arraystretch{1.1}
    \footnotesize
    \caption{\textcolor{black}{6G requirements of KPIs in different application scenarios.}}
\begin{tabular}{|l|llll|lll|lll|lllllll|}
\hline

& \multicolumn{4}{c|}{\textbf{Data rate \& Delay}}                                                                                                           & \multicolumn{3}{c|}{\textbf{Capacity \& Coverage}}                                                           & \multicolumn{3}{c|}{\textbf{\begin{tabular}[c]{@{}c@{}}Service efficiency \\ (Compared to that of 5G)\end{tabular}}}                             
& \multicolumn{7}{c|}{\textbf{Service quality of new paradigms}}                                                                                                                                                                         \\                                                                                                		\cline{2-18}
  \begin{tabular}[c]{@{}l@{}} \textbf{Application} \\ \textbf{Scenarios} \end{tabular}      & \multicolumn{1}{l|}{\rotatebox{90}{\textbf{Peak data rate}}} & \multicolumn{1}{l|}{\rotatebox{90}{\textbf{User experienced data rate}}} & \multicolumn{1}{l|}{\rotatebox{90}{\textbf{Latency}}}          & \multicolumn{1}{l|}{\rotatebox{90}{\textbf{Delay jitter}}}        & \multicolumn{1}{l|}{\rotatebox{90}{\textbf{Area traffic capacity ($\mathbf{Gbps/m^2}$)}}}   & \multicolumn{1}{l|}{\rotatebox{90}{\textbf{Connection density ($\mathbf{devices/km^2}$)}}}      & \multicolumn{1}{l|}{\rotatebox{90}{\textbf{Coverage (\%)}}} & \multicolumn{1}{l|}{\rotatebox{90}{\textbf{Spectrum efficiency improvement}}}                                           & \multicolumn{1}{l|}{\rotatebox{90}{\textbf{Network energy efficiency improvement \ }}}   & \multicolumn{1}{l|}{\rotatebox{90}{\textbf{Cost efficiency improvement}}} & \multicolumn{1}{l|}{\rotatebox{90}{\textbf{Mobility (km/h)}}}          & \multicolumn{1}{l|}{\rotatebox{90}{\textbf{Battery life (years)}}} & \multicolumn{1}{l|}{\rotatebox{90}{\textbf{Reliability}}}                      & \multicolumn{1}{l|}{\rotatebox{90}{\textbf{Positioning}}}                          & \multicolumn{1}{l|}{\rotatebox{90}{\textbf{Sensing/Imaging resolution}}} & \multicolumn{1}{l|}{\rotatebox{90}{\textbf{Security capacity}}}    & \multicolumn{1}{l|}{\rotatebox{90}{\textbf{Intelligence level}}} \\ \hline
 

feMBB                                                                                           & \multicolumn{1}{l|}{\begin{tabular}[c]{@{}l@{}}$\geq$ 1 Tbps\\ \cite{Zhang2019_33, Tataria2021}, \\ \cite{Giordani2020_16, IMT2030-APP_KPI} \end{tabular}} &
\multicolumn{1}{l|}{\begin{tabular}[c]{@{}l@{}}$\geq$ 1--10 \\ Gbps \\ \cite{IMT2030-APP_KPI, Giordani2020_16}, \\ \cite{Zhang2019_33, DATANG} \end{tabular}} & 
\multicolumn{1}{l|}{\begin{tabular}[c]{@{}l@{}}0.1--10 ms\\ \cite{Tataria2021, Alwis2021}, \\ \cite{DATANG, IMT2030-APP_KPI} \end{tabular}} & 
\multicolumn{1}{l|}{\begin{tabular}[c]{@{}l@{}} $\leq$ 10 us \\ \cite{Nasrallah2019} \end{tabular}} & 
\multicolumn{1}{l|}{\begin{tabular}[c]{@{}l@{}}1 \\ \cite{Zhang2019_33} \end{tabular}} &
\multicolumn{1}{l|}{--} & --  & \multicolumn{1}{l|}{5x}   & 
\multicolumn{1}{l|}{\begin{tabular}[c]{@{}l@{}}10-100x \\ \cite{Zhang2019_33} \end{tabular}} & 5x & 
\multicolumn{1}{l|}{800} & \multicolumn{1}{l|}{--}  & \multicolumn{1}{l|}{--} & 
\multicolumn{1}{l|}{\begin{tabular}[c]{@{}l@{}} 10 cm\end{tabular}} & 
\multicolumn{1}{l|}{\begin{tabular}[c]{@{}l@{}}1 mm\end{tabular} }  & 
\multicolumn{1}{l|}{M} & H \\ \hline

umMTC & 
\multicolumn{1}{l|}{\begin{tabular}[c]{@{}l@{}} \textgreater 24 Gbps \\ \cite{Alwis2021}\end{tabular}} & 
\multicolumn{1}{l|}{\begin{tabular}[c]{@{}l@{}}1 Gbps \\ \cite{Giordani2020_16}\end{tabular}} & 
\multicolumn{1}{l|}{\begin{tabular}[c]{@{}l@{}}\textless \textless 5 ms \end{tabular}} &     
\multicolumn{1}{l|}{\begin{tabular}[c]{@{}l@{}}-- \end{tabular}} &
\multicolumn{1}{l|}{\begin{tabular}[c]{@{}l@{}}0.1-10 \\ \cite{IMT2030-APP_KPI}\end{tabular}} & 
\multicolumn{1}{l|}{\begin{tabular}[c]{@{}l@{}}$10^7$--$10^8$ \\ \cite{NTT-Docomo, Alwis2021}, \\ \cite{Giordani2020_16}\end{tabular}} & 
\multicolumn{1}{l|}{\begin{tabular}[c]{@{}l@{}}-- \end{tabular}} &  
\multicolumn{1}{l|}{\begin{tabular}[c]{@{}l@{}}5x \\ \cite{Giordani2020_16}\end{tabular}} &
\multicolumn{1}{l|}{\begin{tabular}[c]{@{}l@{}}10-100x \\ \cite{Giordani2020_16, Zhang2019_33}\end{tabular}} & 
\multicolumn{1}{l|}{\begin{tabular}[c]{@{}l@{}}5x \end{tabular}} & 
\multicolumn{1}{l|}{\begin{tabular}[c]{@{}l@{}}100 \end{tabular}} & 
\multicolumn{1}{l|}{\begin{tabular}[c]{@{}l@{}}10--20 \end{tabular}} & 
\multicolumn{1}{l|}{\begin{tabular}[c]{@{}l@{}}-- \end{tabular}} & 
\multicolumn{1}{l|}{\begin{tabular}[c]{@{}l@{}} 10--100 cm\end{tabular}} &
\multicolumn{1}{l|}{\begin{tabular}[c]{@{}l@{}}1 cm\end{tabular}} & 
\multicolumn{1}{l|}{\begin{tabular}[c]{@{}l@{}}H \end{tabular}} & 
\multicolumn{1}{l|}{\begin{tabular}[c]{@{}l@{}}H \end{tabular}} \\ \hline

euRLLC &
\multicolumn{1}{l|}{\begin{tabular}[c]{@{}l@{}}\textgreater 1 Gbps\\-- 1 Tbps \\ \cite{Alwis2021, Zhang2019_33}\end{tabular}} & 
\multicolumn{1}{l|}{\begin{tabular}[c]{@{}l@{}}$\approx$ 100 Mbps\\--1 Gbps \cite{IMT2030-APP_KPI}, \\ \cite{Tataria2021,Zhang2019_33}\end{tabular}} & 
\multicolumn{1}{l|}{\begin{tabular}[c]{@{}l@{}}\textless 0.1--3 ms\\  \cite{UNISOC, Alwis2021}, \\ \cite{Giordani2020_16, DATANG} \end{tabular}} &     
\multicolumn{1}{l|}{\begin{tabular}[c]{@{}l@{}}1--100 us \\ \cite{Bhat2021} \end{tabular}} &
\multicolumn{1}{l|}{\begin{tabular}[c]{@{}l@{}}-- \end{tabular}} & 
\multicolumn{1}{l|}{\begin{tabular}[c]{@{}l@{}}-- \end{tabular}} & 
\multicolumn{1}{l|}{\begin{tabular}[c]{@{}l@{}}-- \end{tabular}} &  
\multicolumn{1}{l|}{\begin{tabular}[c]{@{}l@{}}5x \\ \cite{Giordani2020_16}\end{tabular}} &
\multicolumn{1}{l|}{\begin{tabular}[c]{@{}l@{}}--\end{tabular}} & 
\multicolumn{1}{l|}{\begin{tabular}[c]{@{}l@{}}-- \end{tabular}} & 
\multicolumn{1}{l|}{\begin{tabular}[c]{@{}l@{}}100 \end{tabular}} & 
\multicolumn{1}{l|}{\begin{tabular}[c]{@{}l@{}}-- \end{tabular}} & 
\multicolumn{1}{l|}{\begin{tabular}[c]{@{}l@{}}$\geq$99.99\\999\% \\ \cite{Giordani2020_16, IMT2030-APP_KPI}, \\ \cite{Samsung, Bhat2021}\end{tabular}} & 
\multicolumn{1}{l|}{\begin{tabular}[c]{@{}l@{}}50 cm \\ \cite{HUAWEI}\end{tabular}} &
\multicolumn{1}{l|}{\begin{tabular}[c]{@{}l@{}}1 mm\end{tabular}} & 
\multicolumn{1}{l|}{\begin{tabular}[c]{@{}l@{}}H \end{tabular}} & 
\multicolumn{1}{l|}{\begin{tabular}[c]{@{}l@{}}M \end{tabular}} \\ \hline

meMBB &
\multicolumn{1}{l|}{\begin{tabular}[c]{@{}l@{}} $\geq$ 1 Tbps\end{tabular}} & 
\multicolumn{1}{l|}{\begin{tabular}[c]{@{}l@{}}$\geq$ 1 Gbps\end{tabular}} & 
\multicolumn{1}{l|}{\begin{tabular}[c]{@{}l@{}}1--10 ms\end{tabular}} &     
\multicolumn{1}{l|}{\begin{tabular}[c]{@{}l@{}}--\end{tabular}} &
\multicolumn{1}{l|}{\begin{tabular}[c]{@{}l@{}}1--10\end{tabular}} & 
\multicolumn{1}{l|}{\begin{tabular}[c]{@{}l@{}}$10^8$\end{tabular}} & 
\multicolumn{1}{l|}{\begin{tabular}[c]{@{}l@{}}-- \end{tabular}} &  
\multicolumn{1}{l|}{\begin{tabular}[c]{@{}l@{}}5x \\\cite{Zhang2019_33}\end{tabular}} &
\multicolumn{1}{l|}{\begin{tabular}[c]{@{}l@{}}10x \cite{Zhang2019_33} \end{tabular}} & 
\multicolumn{1}{l|}{\begin{tabular}[c]{@{}l@{}}5x \end{tabular}} & 
\multicolumn{1}{l|}{\begin{tabular}[c]{@{}l@{}}80--90 \end{tabular}} & 
\multicolumn{1}{l|}{\begin{tabular}[c]{@{}l@{}}-- \end{tabular}} & 
\multicolumn{1}{l|}{\begin{tabular}[c]{@{}l@{}}-- \end{tabular}} & 
\multicolumn{1}{l|}{\begin{tabular}[c]{@{}l@{}}-- \end{tabular}} &
\multicolumn{1}{l|}{\begin{tabular}[c]{@{}l@{}}-- \end{tabular}} & 
\multicolumn{1}{l|}{\begin{tabular}[c]{@{}l@{}}M \end{tabular}} & 
\multicolumn{1}{l|}{\begin{tabular}[c]{@{}l@{}}M \end{tabular}} \\ \hline

muRLLC & 
\multicolumn{1}{l|}{\begin{tabular}[c]{@{}l@{}}0.1--1 Gbps\\\cite{Alwis2021} \end{tabular}} & 
\multicolumn{1}{l|}{\begin{tabular}[c]{@{}l@{}} 100 Mbps \\ \cite{IMT2030-APP_KPI} \end{tabular}} & 
\multicolumn{1}{l|}{\begin{tabular}[c]{@{}l@{}}\textless 1 ms\\ \cite{CCID} \end{tabular}} &     
\multicolumn{1}{l|}{\begin{tabular}[c]{@{}l@{}}1--100 us \\ \cite{IMT2030-APP_KPI, DATANG}, \\ \cite{Tataria2021} \end{tabular}} &
\multicolumn{1}{l|}{\begin{tabular}[c]{@{}l@{}}-- \end{tabular}} & 
\multicolumn{1}{l|}{\begin{tabular}[c]{@{}l@{}}$10^6$--$10^8$ \\ \cite{IMT2030-APP_KPI}, \\ \cite{Alwis2021, Khan2020_8} \end{tabular}} & 
\multicolumn{1}{l|}{\begin{tabular}[c]{@{}l@{}}-- \end{tabular}} &  
\multicolumn{1}{l|}{\begin{tabular}[c]{@{}l@{}}5x \\ \cite{Giordani2020_16}\end{tabular}} &
\multicolumn{1}{l|}{\begin{tabular}[c]{@{}l@{}}-- \end{tabular}} & 
\multicolumn{1}{l|}{\begin{tabular}[c]{@{}l@{}}-- \end{tabular}} & 
\multicolumn{1}{l|}{\begin{tabular}[c]{@{}l@{}}20--60 \end{tabular}} & 
\multicolumn{1}{l|}{\begin{tabular}[c]{@{}l@{}}10--20 \\ \cite{KPI_Lu2017}, \\ \cite{IIoT_battery} \end{tabular}} & 
\multicolumn{1}{l|}{\begin{tabular}[c]{@{}l@{}}$\geq$ 99.99\\999\% \\  \cite{Tataria2021, Giordani2020_16},\\ \cite{Samsung, DATANG} \end{tabular}} & 
\multicolumn{1}{l|}{\begin{tabular}[c]{@{}l@{}} 10--50 cm \end{tabular}} &
\multicolumn{1}{l|}{\begin{tabular}[c]{@{}l@{}}1 mm--\\1 cm \\ \cite{IMT2030-APP_KPI, DATANG}\end{tabular}} & 
\multicolumn{1}{l|}{\begin{tabular}[c]{@{}l@{}}H \end{tabular}} & 
\multicolumn{1}{l|}{\begin{tabular}[c]{@{}l@{}}H \end{tabular}} \\ \hline

MBRLLC & 
\multicolumn{1}{l|}{\begin{tabular}[c]{@{}l@{}}$\geq$ 1 Tbps \end{tabular}} & 
\multicolumn{1}{l|}{\begin{tabular}[c]{@{}l@{}}0.1--1 Gbps\\ \cite{IMT2030-APP_KPI} \end{tabular}} & 
\multicolumn{1}{l|}{\begin{tabular}[c]{@{}l@{}}\textless 1 ms \\ \cite{IMT2030-APP_KPI} \end{tabular}} &   
\multicolumn{1}{l|}{\begin{tabular}[c]{@{}l@{}}1--100 us \end{tabular}} &  
\multicolumn{1}{l|}{\begin{tabular}[c]{@{}l@{}}1--10 \end{tabular}} &
\multicolumn{1}{l|}{\begin{tabular}[c]{@{}l@{}}-- \end{tabular}} & 
\multicolumn{1}{l|}{\begin{tabular}[c]{@{}l@{}}-- \end{tabular}} & 
\multicolumn{1}{l|}{\begin{tabular}[c]{@{}l@{}}5x \\\cite{Alwis2021}\end{tabular}} &  
\multicolumn{1}{l|}{\begin{tabular}[c]{@{}l@{}}-- \end{tabular}} & 
\multicolumn{1}{l|}{\begin{tabular}[c]{@{}l@{}}-- \end{tabular}} &
\multicolumn{1}{l|}{\begin{tabular}[c]{@{}l@{}}10 \end{tabular}} & 
\multicolumn{1}{l|}{\begin{tabular}[c]{@{}l@{}}-- \end{tabular}} & 
\multicolumn{1}{l|}{\begin{tabular}[c]{@{}l@{}}$\geq$ 99.99\\999\%\end{tabular}} & 
\multicolumn{1}{l|}{\begin{tabular}[c]{@{}l@{}}10 cm \end{tabular}} & 
\multicolumn{1}{l|}{\begin{tabular}[c]{@{}l@{}}1 mm--\\1 cm\end{tabular}} & 
\multicolumn{1}{l|}{\begin{tabular}[c]{@{}l@{}}H\end{tabular}} &
\multicolumn{1}{l|}{\begin{tabular}[c]{@{}l@{}}H\end{tabular}} \\ \hline

SAGSIN & 
\multicolumn{1}{l|}{\begin{tabular}[c]{@{}l@{}}$\geq$ 1 Gbps \end{tabular}} & 
\multicolumn{1}{l|}{\begin{tabular}[c]{@{}l@{}}0.1--10 Gbps \\ \cite{IMT2030-APP_KPI} \end{tabular}} & 
\multicolumn{1}{l|}{\begin{tabular}[c]{@{}l@{}}40 ms --\\ 0.7 s\end{tabular}} &     
\multicolumn{1}{l|}{\begin{tabular}[c]{@{}l@{}}-- \end{tabular}} &
\multicolumn{1}{l|}{\begin{tabular}[c]{@{}l@{}}-- \end{tabular}} & 
\multicolumn{1}{l|}{\begin{tabular}[c]{@{}l@{}}-- \end{tabular}} & 
\multicolumn{1}{l|}{\begin{tabular}[c]{@{}l@{}}99 \end{tabular}} &  
\multicolumn{1}{l|}{\begin{tabular}[c]{@{}l@{}}3x \\ \cite{KPI_SAGSIN_SE} \end{tabular}} &
\multicolumn{1}{l|}{\begin{tabular}[c]{@{}l@{}}10x \end{tabular}} & 
\multicolumn{1}{l|}{\begin{tabular}[c]{@{}l@{}}3x \end{tabular}} & 
\multicolumn{1}{l|}{\begin{tabular}[c]{@{}l@{}}$\geq$ 1000 \\  \cite{IMT2030-APP_KPI} \end{tabular}} & 
\multicolumn{1}{l|}{\begin{tabular}[c]{@{}l@{}}5--15 \end{tabular}} & 
\multicolumn{1}{l|}{\begin{tabular}[c]{@{}l@{}}-- \end{tabular}} & 
\multicolumn{1}{l|}{\begin{tabular}[c]{@{}l@{}}50--100 cm \\ \cite{IMT2030-APP_KPI,HUAWEI} \end{tabular}} &
\multicolumn{1}{l|}{\begin{tabular}[c]{@{}l@{}}1 cm\\ \cite{IMT2030-APP_KPI} \end{tabular}} & 
\multicolumn{1}{l|}{\begin{tabular}[c]{@{}l@{}}H \end{tabular}} & 
\multicolumn{1}{l|}{\begin{tabular}[c]{@{}l@{}}H \end{tabular}} \\ \hline

Digital twins & 
\multicolumn{1}{l|}{\begin{tabular}[c]{@{}l@{}}0.1--1 Tbps\\ \cite{IMT2030-APP_KPI,Samsung} \end{tabular}} & 
\multicolumn{1}{l|}{\begin{tabular}[c]{@{}l@{}}$\geq$1 Gbps\\ \cite{IMT2030-APP_KPI} \end{tabular}} & 
\multicolumn{1}{l|}{\begin{tabular}[c]{@{}l@{}}0.1--1ms\\ \cite{UNISOC,DATANG} \end{tabular}} &     
\multicolumn{1}{l|}{\begin{tabular}[c]{@{}l@{}}-- \end{tabular}} &
\multicolumn{1}{l|}{\begin{tabular}[c]{@{}l@{}}0.1--10 \\ \cite{IMT2030-APP_KPI, DATANG}\end{tabular}} & 
\multicolumn{1}{l|}{\begin{tabular}[c]{@{}l@{}}$10^7$--$10^8$\\ \cite{IMT2030-APP_KPI,DATANG} \end{tabular}} & 
\multicolumn{1}{l|}{\begin{tabular}[c]{@{}l@{}}-- \end{tabular}} &  
\multicolumn{1}{l|}{\begin{tabular}[c]{@{}l@{}}2x\\ \cite{Samsung} \end{tabular}} &
\multicolumn{1}{l|}{\begin{tabular}[c]{@{}l@{}}10x \end{tabular}} & 
\multicolumn{1}{l|}{\begin{tabular}[c]{@{}l@{}}-- \end{tabular}} & 
\multicolumn{1}{l|}{\begin{tabular}[c]{@{}l@{}}100 \end{tabular}} & 
\multicolumn{1}{l|}{\begin{tabular}[c]{@{}l@{}}20\\ \cite{Bhat2021}\end{tabular}} & 
\multicolumn{1}{l|}{\begin{tabular}[c]{@{}l@{}}-- \end{tabular}} & 
\multicolumn{1}{l|}{\begin{tabular}[c]{@{}l@{}}10--100 cm \end{tabular}} &
\multicolumn{1}{l|}{\begin{tabular}[c]{@{}l@{}}1 mm\end{tabular}} & 
\multicolumn{1}{l|}{\begin{tabular}[c]{@{}l@{}}M \end{tabular}} & 
\multicolumn{1}{l|}{\begin{tabular}[c]{@{}l@{}}H \end{tabular}} \\ \hline

LDHMC & 
\multicolumn{1}{l|}{\begin{tabular}[c]{@{}l@{}}30 bps\\--10 Gbps\end{tabular}} & 
\multicolumn{1}{l|}{\begin{tabular}[c]{@{}l@{}}--\end{tabular}} & 
\multicolumn{1}{l|}{\begin{tabular}[c]{@{}l@{}} \textgreater 1 s\end{tabular}} &     
\multicolumn{1}{l|}{\begin{tabular}[c]{@{}l@{}}-- \end{tabular}} &
\multicolumn{1}{l|}{\begin{tabular}[c]{@{}l@{}}-- \end{tabular}} & 
\multicolumn{1}{l|}{\begin{tabular}[c]{@{}l@{}}-- \end{tabular}} & 
\multicolumn{1}{l|}{\begin{tabular}[c]{@{}l@{}}-- \end{tabular}} &  
\multicolumn{1}{l|}{\begin{tabular}[c]{@{}l@{}}5x \\ \cite{brito_brazil_2020}\end{tabular}} &
\multicolumn{1}{l|}{\begin{tabular}[c]{@{}l@{}}-- \end{tabular}} & 
\multicolumn{1}{l|}{\begin{tabular}[c]{@{}l@{}}-- \end{tabular}} & 
\multicolumn{1}{l|}{\begin{tabular}[c]{@{}l@{}}$\geq$ 1000 \\  \cite{Alwis2021,Giordani2020_16}, \\ \cite{Zhang2019_33} \end{tabular}} & 
\multicolumn{1}{l|}{\begin{tabular}[c]{@{}l@{}}-- \end{tabular}} & 
\multicolumn{1}{l|}{\begin{tabular}[c]{@{}l@{}}-- \end{tabular}} & 
\multicolumn{1}{l|}{\begin{tabular}[c]{@{}l@{}}1 m\end{tabular}} &
\multicolumn{1}{l|}{\begin{tabular}[c]{@{}l@{}}1 cm \end{tabular}} & 
\multicolumn{1}{l|}{\begin{tabular}[c]{@{}l@{}}M	 \end{tabular}} & 
\multicolumn{1}{l|}{\begin{tabular}[c]{@{}l@{}}M	 \end{tabular}} \\ \hline

ELPC & 
\multicolumn{1}{l|}{\begin{tabular}[c]{@{}l@{}}1--50 Mbps\\ \cite{Alwis2021}\end{tabular}} & 
\multicolumn{1}{l|}{\begin{tabular}[c]{@{}l@{}}--\end{tabular}} & 
\multicolumn{1}{l|}{\begin{tabular}[c]{@{}l@{}}1 ms--1 s \\ \cite{Alwis2021}\end{tabular}} &     
\multicolumn{1}{l|}{\begin{tabular}[c]{@{}l@{}}-- \end{tabular}} &
\multicolumn{1}{l|}{\begin{tabular}[c]{@{}l@{}}-- \end{tabular}} & 
\multicolumn{1}{l|}{\begin{tabular}[c]{@{}l@{}}$10^7$ \\  \cite{Zhang2019_33} \end{tabular}} & 
\multicolumn{1}{l|}{\begin{tabular}[c]{@{}l@{}}-- \end{tabular}} &  
\multicolumn{1}{l|}{\begin{tabular}[c]{@{}l@{}}-- \end{tabular}} &
\multicolumn{1}{l|}{\begin{tabular}[c]{@{}l@{}}10--100x \\ \cite{Zhang2019_33} \end{tabular}} & 
\multicolumn{1}{l|}{\begin{tabular}[c]{@{}l@{}}-- \end{tabular}} &  
\multicolumn{1}{l|}{\begin{tabular}[c]{@{}l@{}}5--10 \end{tabular}} &
\multicolumn{1}{l|}{\begin{tabular}[c]{@{}l@{}}battery\\ free \\ \cite{KPI_Zhu2019} \end{tabular}} & 
\multicolumn{1}{l|}{\begin{tabular}[c]{@{}l@{}}-- \end{tabular}} & 
\multicolumn{1}{l|}{\begin{tabular}[c]{@{}l@{}}1 nm \\ (in-body) \end{tabular}} & 
\multicolumn{1}{l|}{\begin{tabular}[c]{@{}l@{}}-- \end{tabular}} & 
\multicolumn{1}{l|}{\begin{tabular}[c]{@{}l@{}}M \end{tabular}} & 
\multicolumn{1}{l|}{\begin{tabular}[c]{@{}l@{}}M \end{tabular}} \\ \hline

\multicolumn{18}{|l|}{*SAGSIN: space-air-ground-sea integrated network, LDHMC: long-distance and high-mobility communications, ELPC: extremely low-power communications, H: high importance, M: medium importance.} \\ \hline
\end{tabular}
\label{table_KPI_AS}
\end{sidewaystable*}

\section{6G Network Architecture}
\begin{figure*}[t]
	\centerline{\includegraphics[width=0.95\textwidth]{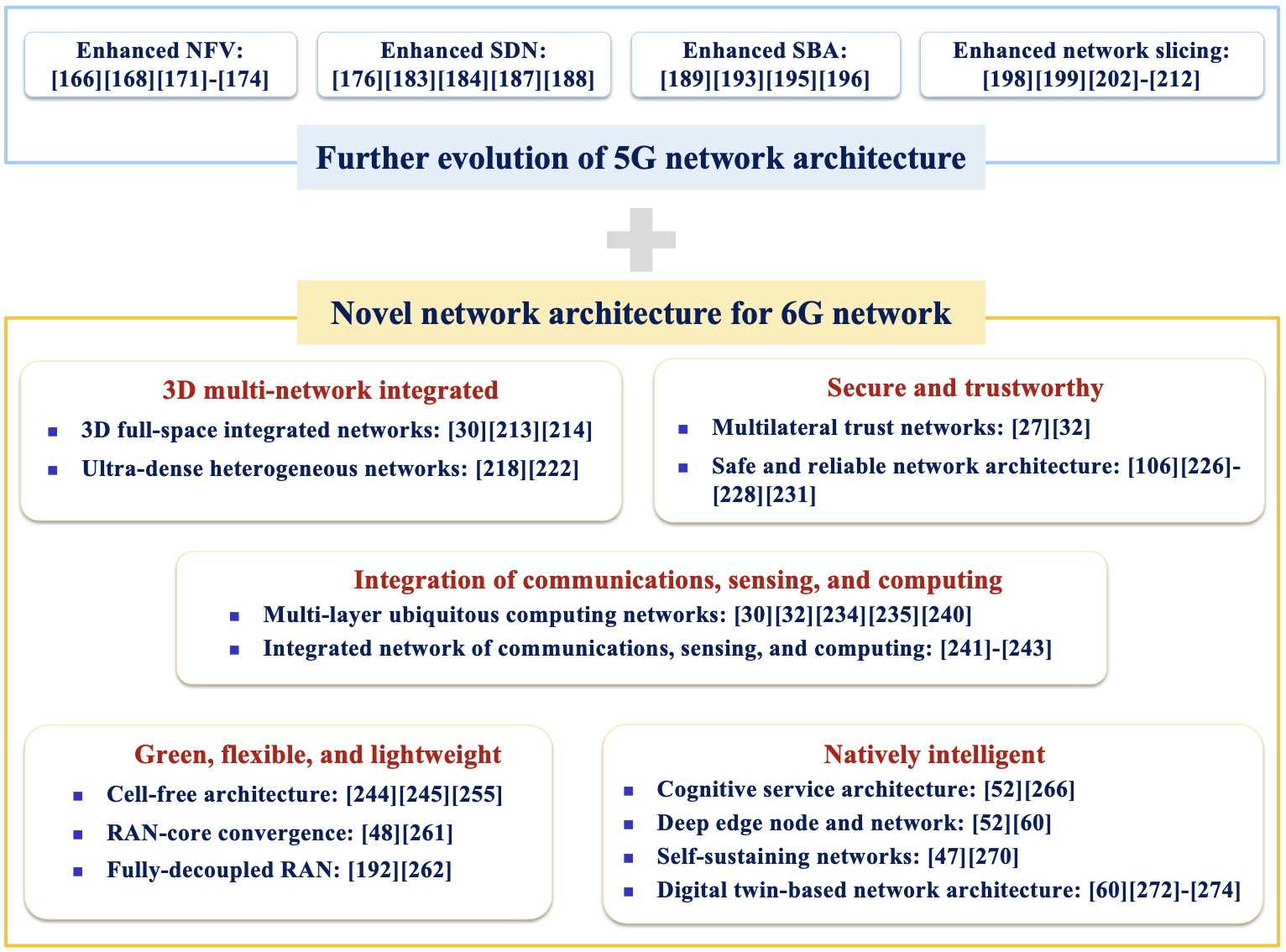}}
	\caption{\textcolor{black}{Development trends of the 6G network architecture and representative references.}}
	\label{fig_6G_Network_Architecture}
\end{figure*}

With the continuous development of application requirements and technical requirements, the three major application scenarios in current 5G are expected to be enhanced and expanded in 6G. Novel application scenarios, such as digital twins, integration of communication, computing, and sensing, distributed AI applications, are also envisaged in 6G. These enhanced and novel application scenarios put forward higher requirements on the KPIs of the communication system. In addition to the significant enhancement of 5G KPIs, communication systems are also required to acquire new capabilities on positioning, sensing, imaging, intelligence level, \textcolor{black}{etc., as shown in Fig. \ref{fig_6G_KPI}.} It is difficult to improve these KPIs only by using new air interface technologies. Hence, it is necessary to revolutionize the entire communication network at the architecture level, to enable the network to provide diverse applications and reduce the cost and energy consumption. 

From 1G to 5G, the communication network architecture is developing in the direction of modularization, softwareization, virtualization, and cloudification. 
Key network architectures in 5G include the network slicing, network functions virtualization (NFV), software defined network (SDN), and service-based architecture (SBA), making the network more flexible and bringing improvements in multiple aspects including application services and costs. However, there are still many challenges in deploying these network architectures in real networks. For the sake of application requirements, technical requirements, and cost considerations, the 6G network architecture will integrate novel network architectures and technologies on the basis of the further evolution of the 5G network architecture, moving towards the following five directions: 1) \textcolor{black}{3D multi-network integrated}; 2) secure and trustworthy; 3) \textcolor{black}{integration of communications, sensing, and computing}; 4) green, flexible, and lightweight; 5) natively intelligent. Development trends of the 6G network architecture are summarized in Fig. {\ref{fig_6G_Network_Architecture}}. \textcolor{black}{In addition, we have specified representative references to help readers understand the corresponding concept for each key point of 6G network architecture in Fig. {\ref{fig_6G_Network_Architecture}}.}
Some existing survey papers did discuss several potential architectural components, but they did not analyze the evolution trend or propose any overall architecture for 6G, e.g., \cite{Chowdhury2020_6,Khan2020_8,Saad2020_12,Chen2020_19,SEU,Jiang2021_21,Alwis2021}. What's more, the overall 6G network architecture introduced in existing works \cite{Huang2019_28,Zhang2019_33,Gui2020,Viswanathan2020_13,Giordani2020_16,Liu2020_18,Bhat2021,Tataria2021} cannot cover all the design concepts. For example, Huang \emph{et al.} \cite{Huang2019_28} ignored enhanced 5G architecture, flexible, lightweight, and energy-efficient networks, deep fusion of resources, and network security.

In this section, we will first review and summarize the further evolution of the 5G network architecture, then introduce the development trends of the 6G architecture one by one comprehensively, and finally we will propose a potential novel comprehensive network architecture for the 6G communication network.

\subsection{Further Evolution of the 5G Network Architecture}
\subsubsection{NFV}
Traditional fixed hardware-based networks have high deployment cost and low network flexibility. NFV is a revolutionary network architecture that transforms fixed networks to software-based programmable networks. Its design concept is to decouple software and hardware\cite{Han2015}. With the help of virtualization technologies, multiple virtualized servers can be deployed on one or more physical hardware resources. Different servers can be configured on demand to execute different network functions, thus providing network support for diverse applications. Using the NFV to virtualize the network can significantly reduce the equipment, operation, and maintenance costs of hardware, improve the efficiency of network operation and management, shorten the development cycle of network services, and make the network flexible and scalable\cite{SEU,Pattaranantakul2018}. The architecture of NFV mainly consists of three components, i.e., the NFV infrastructure (NFVI), the virtual network functions (VNFs), and the NFV management and orchestration (NFV-MANO)~\cite{SEU}.

In November 2012, the European Telecommunications Standards Institute (ETSI) cooperated with several telecom network operators to form the Industry Specification Group for NFV\cite{U.Rehman2019}. 
In recent years, a large number of NFV-related studies have emerged.
Related works on NFV can be divided into three categories according to their focus\cite{U.Rehman2019,Yi2018}, i.e.,
the conceptual design of the integration of NFV with various other networks, the research on resource allocation, orchestration, and other algorithms for NFV, and the literature review of various aspects of NFV. 

Recently, there are also plenty of research works on the enhancement of NFV. In order to minimize the decrease of NFV in the throughput and latency performance compared with fixed network infrastructures, the demand for optimization and acceleration technologies comes into being.  Linguaglossa \emph{et al.} \cite{Linguaglossa2019} presented a comprehensive overview and derived corresponding guidelines on a range of acceleration technologies, including low-level hardware acceleration and high-level software acceleration solutions. 
The reinforcement learning was used in \cite{Lin2021} to reduce the performance degradation of the flow scheduling in heterogeneous environments due to recently emerging programmable accelerators. In order to achieve effective resource management and allocation, improve the operating efficiency, and meet high latency requirements of applications, Kianpisheh \emph{et al.} \cite{Kianpisheh2021} proposed a joint access control and resource allocation algorithm based on parallel VNF processes. The experience network intelligence (ENI) working group established by ETSI is committed to using AI and ML technologies to adjust and optimize the VNFs in dynamic requirements and environments\cite{Jiang2021_21}. Network security is also an important issue that needs to be considered when employing NFV. Pattaranantakul \emph{et al.} \cite{Pattaranantakul2018} presented a set of recommendations for protecting NFV-based services based on the established threat classification and analysis, and discussed the latest security countermeasures.
A joint research on the mapping between virtual networks and hardware devices in NFV, routing strategies, and security defense issues were studied comprehensively in \cite{Xue2022}.

In addition to the aforementioned software optimization and acceleration, power allocation, access management, security, etc., NFV also faces research challenges \textcolor{black}{such as the system complexity, cloud-native NFV-MANO \cite{Breitgand2021}, network programming and automation\cite{U.Rehman2019}, service quality and data privacy in crowdsourced edge-based NFV\cite{Rahman2021}, and edge cloud virtualization technologies\cite{Valsamas2021}.} There is still a long way to go to enhance and deploy NFV in real beyond 5G (B5G) and 6G networks.

\subsubsection{SDN}
Different from decoupling software and hardware in NFV, the SDN is based on the idea of decoupling the control plane and the forwarding plane, aiming to separate network control functions from forwarding functions\cite{SEU,Das2020,Jimenez2021}. The SDN architecture mainly includes the application plane, control plane, and data (forwarding) plane\cite{Das2020}. In SDN, a software-defined logical network is abstracted on the physical network. Network control functions are centralized to the SDN controller which interfaces with upper-layer applications and instructs lower-layer network infrastructures that only retain forwarding functions to complete forwarding operations. SDN makes each layer relatively independent. This has a series of advantages, such as programmability, flexibility, low costs, and high efficiency\cite{SEU,Rawat2019,Jiang2021_21}.

Since the concept of SDN was proposed, research on various aspects of SDN implementation and enhancement has made a lot of progress. Limited by the heterogeneity and complexity of optical devices, the transport network may be the last part to fully realize SDN. A comprehensive survey on the current status and future directions of transport SDN was given in \cite{Alvizu2017}. SDN has also been studied to integrate with various other communication networks, e.g., smart grid communication\cite{Rehmani2019}, underwater wireless sensor networks\cite{Luo2018}, satellite networks\cite{Tu2020}, and vehicular networks\cite{Gao2020}. In \cite{Rawat2019}, the SDN was proposed to integrate with edge computing and blockchain to improve the efficiency and security of wireless network virtualization. As for more in-depth algorithm study, Das \emph{et al.} \cite{Das2020} reviewed comprehensively on the SDN controller placement problem, which is a key network design consideration that affects network latency, resiliency, EE, and load balancing. The controller synchronization in distributed SDN was quantified and analyzed in \cite{Zhang2021}. In addition, the characteristics of logically centralized control, global view of the network, and dynamic resource allocation in SDN bring some opportunities to achieve the network intelligence. Xie \emph{et al.} \cite{Xie2019} provided an overview of research issues and challenges in applying ML technologies to SDN. Note that, while SDN has the advantages of programmability, flexibility and openness, it will also lead to new security issues that can be easily overlooked. A detailed review and analysis on typical security issues and solutions in SDN were conducted in \cite{Jimenez2021} according to the STRIDE threat model\cite{Jimenez2021_19}.

Moreover, in order to take full advantage of SDN, the challenges of deploying SDN in 6G include switch forwarding schemes in hybrid SDN\cite{Hussain2021}, service virtualization and flow management\cite{Manogaran2021}, energy-efficiency optimization\cite{Tu2020}, handover schemes\cite{Gao2020} and freshness-aware age optimization\cite{Gao2021} for multipath transmission control protocol (TCP) in SDNs.

\subsubsection{SBA}
With the development of mobile communications, in order to support various applications, the core network needs to perform more and more functions and become more and more complex. However, different applications may only require part of these network functions. Based on technologies including cloud computing, virtualization, and micro-services, the 5G core network adopts the SBA, which has been accepted by 3GPP\cite{3GPP-TS-29.500_SBA}. The network elements providing each service are separate modules that are connected together to provide the services of the core network. Different service network functional modules are used for deployment according to the needs of application services. \textcolor{black}{Note that SBA realizes the modularization of 5G core network functions in the form of service function chains by means of NFV.}
The reference SBA for 5G core network and details of the modular functions can be found in \cite{SEU_192, Do2018,SEU}. \textcolor{black}{Similar to the design concept of decoupling control functions in SDN, 5G core network functions in SBA are divided into two classes, i.e., control plane functions and user plane functions (UPFs)\cite{Yu2019}. }In addition to the flexible and modular features originally brought by SBA, the architecture also inherits the advantages of underlying technologies\cite{SEU,Rudolph2019}, including on-demand computing of cloud computing, flexible and efficient resource management of virtualization, flexibility, fine-grained property, and independent scalability of micro-services. The security concepts, techniques, and challenges in the SBA were discussed in \cite{Rudolph2019}. In \cite{Buyakar2019}, the authors prototyped an SBA core network in an NFV environment and proposed a load balancing strategy that can significantly reduce the delay of the control plane. 

However, using the SBA only on the core network is not enough to maximize the benefits of SBA. While studying how to actually deploy SBA in the core network, research interests also focus on the further evolution of SBA in 6G networks. 

The future 6G communication network will be extended on the basis of using SBA in the 5G core network, and will further realize the SBA for the E2E network. In \cite{Zeydan2022}, Zeydan \emph{et al.} explored the potential of applying the SBA to the radio access network (RAN), and introduced the design concepts and implementation details of service-based RAN. The definition of network services at RAN side, the interface between service modules, and other details need to be further studied. 
What's more, a holistic E2E SBA, which extended the design concept of SBA to the access network and user plane, was proposed and applied to the integrated air-space-ground network in \cite{Wang2022_SBA}. The evolution and simulation of related protocols of the holistic SBA were also discussed and analyzed. In order to apply SBA to a variety of networks more effectively, SBA needs to be enhanced according to specific situations.
Looking forward to the 6G vision, a more flexible holistic SBA with full coverage and full applications is a potential candidate for the next-generation network architecture.

\subsubsection{Network slicing}
In the 5G era, the communication system has three scenarios: eMBB, uRLLC and mMTC. Network requirements of these scenarios are quite different. In order to support different application services, the communication network needs to be flexible enough to provide various services of high quality to the greatest extent while considering the cost. Taking these factors into consideration, the network slicing based on NFV, SDN, cloud and edge computing\cite{Khan2020_8_NS} emerged. The core concept of the network slicing is to multiplex independent logical networks virtualized on the same physical network infrastructure to support different application services\cite{SEU,Khan2020,Debbabi2020}. In addition to services to customers (ToC) directly, 5G network slices offer a range of services to business (ToB), such as cloud games, power, medical, ports, and industry. So far, there have been several surveys \cite{Khan2020,Khan2020_8_NS,Khan2020_16,Khan2020_17,Debbabi2020} on state-of-art network slicing. For instance, the authors of \cite{Khan2020} reviewed the current network slicing research in terms of taxonomy, requirements, and research challenges. Debbabi \emph{et al.} \cite{Debbabi2020} investigated the architecture of network slicing and focused on the analysis and overview from the perspective of~algorithms. 

As a key network paradigm of 5G, network slicing brings many advantages. However, the E2E integrity, slice specialization level, and intelligence level of 5G network slices are still limited. 

In the future, the network slicing will mainly develop further in three directions. Firstly, research interests are focusing on implementing holistic E2E network slicing systems. In \cite{Dogra2021}, a next-generation wireless communication network architecture containing slices at three levels: cloud, RAN, and application level, was proposed. Khan \emph{et al.} \cite{Khan2021} proposed an E2E network slicing framework including RAN and core network slicing for 5G vehicular Ad-Hoc networks. The authors of \cite{Li2020_NS} investigated an E2E network slicing system architecture including RAN, transport network, and core network. The software simulation and real hardware demo of the proposed architecture were also demonstrated in \cite{Li2020_NS}. Secondly, specialized/tailored network slicing is also one of the directions of evolution. It was envisioned in \cite{Viswanathan2020_13} that with further development of slicing and virtualization in the future, network slices can become highly specialized. More specifically, Cao \emph{et al.} \cite{Cao2022} proposed the TailoredSlice-6G algorithm, which can realize the tailored resource allocation of slices in 6G networks to provide tailored slices. \textcolor{black}{In \cite{Sherman_new_3}, a two-level soft customized RAN slicing scheme was proposed, which can satisfy QoS requirements of uRLLC and eMBB services simultaneously.} Lastly, there is a growing research interest in achieving intelligent network slicing using AI\cite{Shen2020}, digital twins\cite{Wang2022_NS}, and deep reinforcement learning (DRL) \cite{Mei2021,Nassar2022,Suh2022}. In \cite{Shen2020}, an AI-assisted next-generation RAN functional architecture based on network slicing was proposed, with AI-assisted network topology, network protocol, and resource management. Based on a new graph neural network model, Wang \emph{et al.} \cite{Wang2022_NS} proposed a scalable digital twin of network slicing which can accurately reflect the network behavior, predict E2E slice performance in unknown environments, and provide intelligent network slicing management. The authors of \cite{Mei2021} proposed a novel hierarchical DRL framework that incorporated the modified deep deterministic policy gradient (DDPG) and the double deep Q-network algorithm to maximize the long-term QoS of services and the SE of network slices. In \cite{Nassar2022} and \cite{Suh2022}, the DRL was explored to adaptively learn optimal slicing strategies and find optimal resource allocation strategies, respectively. In addition, challenges such as slice isolation, dynamic slice creation and management\cite{SEU}, and multi-tenant networks\cite{Reyhanian2022} are also research topics that have attracted great interest in network slicing. It can be expected that various application network slices in the future will lead to a qualitative leap in our lives.

\subsection{Development Trends of the 6G Network}
\subsubsection{\textcolor{black}{3D multi-network integrated}}
5G networks, as well as previous generations of communication networks, have mainly focused on deploying network access points to provide connectivity for communication devices on the ground. However, communication in remote areas is limited by low wireless coverage. In addition, communication systems that only rely on terrestrial communication networks have poor robustness and cannot provide timely communication in the event of various disasters. The 6G network will be a \textcolor{black}{3D full-space network} deeply integrated with the ultra-dense terrestrial heterogeneous communication network.

\paragraph*{\textcolor{black}{3D full-space integrated network}} 
The future 3D full-space integrated communication network is a 3D layered, integrated, and cooperative network, which is built on terrestrial networks but extended to space-based networks, aerial networks, maritime networks, underwater networks,\cite{UNISOC,Guo2022_SAGS}, \textcolor{black}{and underground networks.} With various extended networks, the \textcolor{black}{6G full-space integrated network} can be flexibly configured and has the advantage of high resilience\cite{IMT-2030}. For instance, a space-based network can be deployed as an extended backhaul network that helps terrestrial base stations access the core network, or as a node with base station functions\cite{UNISOC}. Besides, the space-based extended network can also be enabled to carry part of functions of the terrestrial core network, such as the access and mobility management function (AMF), the user plane function (UPF), and the session management function (SMF)\cite{Chen2020_19,UNISOC}. In addition to these advantages of high flexibility and resilience, the \textcolor{black}{3D full-space integrated network} architecture also has the superiorities of improving communication coverage, rapid deployment and reducing network operating costs in edge areas\cite{Giordani2020_16}, and on-demand dynamic resource allocation brought by the network globality\cite{Chen2020_19}.

At present, the research on the integrated 3D space-air-ground-sea network is ongoing and some progress has been made\cite{Guo2022_SAGS,Zhao2021,Pang2020,Mao2021,Kato2019}. However, the construction of the 3D full-space network architecture still faces a series of challenges\textcolor{black}{, which will be discussed later.}

\paragraph*{Ultra-dense heterogeneous network (UDHN)}
In 5G, in order to meet various needs of the network, ultra-dense networks have been studied to integrate various enabling technologies\cite{Adedoyin2020}. With the development of diverse communication technologies and various networks, as well as the increasing demand for communication density in various applications, ultra-dense heterogeneous networking will still be one of the key development trends of 6G. Various networks will continue to deepen the integration and combine up-to-date technologies to form a multi-layer UDHN, which can improve the overall QoS of the network and reduce costs\cite{Chowdhury2020_6}.

In recent years, research challenges arising from UDN and the heterogeneity of various networks have been the focus of research. More recently, there have been several new advances in addressing these research issues. In \cite{Jo2020}, a self-optimization scheme for coverage and system throughput was proposed for UDHNs. The adaptive cell selection method in UDHNs was studied in \cite{Alablani2021}. Sun \emph{et al.}\cite{Sun2021} investigated coordinated multiple points (CoMP) handover schemes for UDHNs considering user movement trends. In addition, with the continuous evolution of AI, new intelligent methods, such as ML and deep learning (DL) were also used to solve typical problems in ultra-dense networks\cite{Sharma2020,Zhang2021_ML,Kim2021}.

\subsubsection{\textcolor{black}{Secure and trustworthy}}
While it is critical to innovate the network architecture to meet requirements of various applications in the 6G communication network, \textcolor{black}{security and trustworthiness are also important aspects that cannot be ignored.} On one hand, the fusion of communication technologies with data technologies and industrial operation technologies, as well as the marginalization and virtualization of facilities will lead to a more blurred 6G network security boundary. Therefore, the traditional security trust model can no longer meet the requirements of the 6G security and trustworthiness\cite{IMT-2030}. On the other hand, with the change of the network architecture and the emergence of new services and new terminals, 6G networks will face novel security threats\cite{UNISOC,SEU}, \textcolor{black}{e.g., the data privacy issue, security risks of models and algorithms, as well as software or system vulnerabilities.} The new 6G network architecture should be based on a more inclusive multilateral trust model, taking security issues into consideration at the beginning of the network design, to achieve the endogenous \textcolor{black}{security and trustworthiness}.

\paragraph*{Multilateral trust network}
The current communication system mainly adopts the centralized bridge trust model\cite{HUAWEI}. However, the requirements and importance of security in different application scenarios are usually different. Porambage \emph{et al.}\cite{Porambage2021} summarized the key requirements for security in major 6G applications comprehensively. The diversity of 6G scenarios makes the traditional security trust model face many challenges. It is necessary to establish a multilateral trust model to cover the trustworthiness in different situations, and to endogenously carry a more robust, smarter, and scalable security mechanism\cite{IMT-2030,HUAWEI}. A network that supports multilateral trust includes three modes of trust models: bridge, consensus, and endorsement.
The core of multilateral trust is the decentralized consensus of all parties, including the mode of centralized authorization bridge and third-party endorsement. A detailed introduction of the three trust models can be found in \cite{HUAWEI}. The three trust modes are interrelated, making the multilateral trust model more inclusive.

\paragraph*{Safe and reliable network architecture}
In Fig. \ref{fig_6G_Network_Architecture}, we have summarized the evolution and development direction of the 6G network architecture in multiple aspects. It is worth noting that the evolving communication network architecture is also facing novel security threats. In \cite{Porambage2021}, the authors analyzed the security threats in several potential 6G network architectures \textcolor{black}{including RAN-core convergence, intelligent network management and scheduling, edge intelligence, and specialized 6G networks.} While designing the 6G network architecture, potential 6G security technologies are introduced to ensure the network security, such as blockchain technologies\cite{Li2020,S2021}, quantum communications\cite{Hamamreh2019}, and physical layer security technologies\cite{Al-Mohammed2021}. The implementation of standardized open network interfaces \textcolor{black}{with high modularity} in O-RAN also provides some new ideas for improving network security\cite{O-RAN_Security}.
The network security will gradually evolve from traditional security protection to the endogenous security with self-adaptive, autonomous and self-growing security capabilities\cite{UNISOC}. The basic concepts, problems, properties, structures, and functional applications were introduced in \cite{Jin2021}. Moreover, Wang \emph{et al.}\cite{Wang2021_Security} analyzed and applied the endogenous security principle from both theoretical and simulation perspectives. In addition to challenges of security technologies, the integrated evolution design of security architecture and network architecture is also the key to realizing an endogenous secure and reliable network architecture\cite{IMT-2030}. It is necessary to take both communication and security into account and make the optimal trade-off between cost and benefit.

\subsubsection{\textcolor{black}{Integration of communications, computing, and sensing}}
The 5G network architecture introduced mobile edge computing to reduce service latency and backhaul costs, and to alleviate the traffic pressure. To achieve the vision of full applications and to meet the requirements of lightweight and dynamic computing, the communication, computing, and sensing functions in the 6G era will be deeply integrated. Each network node will have the functions of data transmission, computing, and sensing, providing better services for various 6G application scenarios.

\paragraph*{Multi-layer ubiquitous computing network}
The emergence of cloud computing and edge computing has improved the network performance and supported a series of novel services and applications. On one hand, cloud computing centralizes resources and management in the cloud, providing terminal devices and users with flexible and on-demand resource allocation, less management burden, flexible pricing models, and convenient application and service provisioning\cite{SEU_211}. 
On the other hand, edge computing satisfies the network requirements of time-critical applications. The introduction of fog computing connects the cloud and edge computing to form an integrated multi-layer computing network that can flexibly handle computing tasks in various networks\cite{SEU, SEU_211, Ghosh2020}. With the continuous evolution of computing technologies, in the 6G era, network nodes including cloud computing centers, access networks, bearer networks, core networks, and terminals will have certain computing resources and capabilities. These diverse computing powers will be connected and coordinated in a networked manner, forming a multi-layer cloud-fog-edge-terminal ubiquitous computing network, which can realize on-demand scheduling and efficient sharing of computing services\cite{IMT-2030,UNISOC}.

\textcolor{black}{Currently, several prospective studies have been initialized on this topic. In \cite{ComputingNetwork_ChinaMobile,ComputingNetwork_ChinaUnicom}, China Mobile and China Unicom have carried out conceptual research on the computing network.} Based on cloud-fog-edge collaborative networking, Refat \emph{et al.}\cite{Refat2020} proposed a flexible mobile grammar teaching tool. In \cite{Fantacci2020}, the performance evaluation and optimization of three-layer cloud-fog-edge computing infrastructure were investigated based on queuing theory. A mobility-driven real-time cloud-fog-edge collaboration framework was proposed in \cite{Ghosh2020}, which can efficiently deliver processed information to user devices based on user mobility prediction and intelligent decision-making. Furthermore, research problems in the computing network have also aroused interests. In order to achieve better resource allocation in the computing network, the authors of \cite{Han2021} proposed a bandwidth allocation method based on utility optimization. The sleep mechanism of base stations was studied in \cite{Alnoman2021}, while the joint cloud-edge computing model was used to improve the system computing performance. In addition, the ITU-T study group 13 adopted the standard ITU-T--Y.2501\cite{ComputingNetwork_ITU-T}, defining the computing power network framework and architecture. At present, the industry is transforming from the division scheme to the collaboration scheme of computing and networks, and is integrating computing and networks\cite{IMT-2030}. It is worth noting that there are still many challenges in multi-layer ubiquitous computing networks\cite{UNISOC}, including theoretical research on cloud-fog-edge computing, demand for computing resources in different scenarios, and perception and measurement of computing powers, as well as the computing network operation management and control, multi-party cooperation, and operation mode from the perspective of real network deployment.


\paragraph*{Integrated networks of communications, computing, and sensing}
In addition to data transmission and computing capabilities, each node in the 6G network will have certain sensing capabilities to meet various new 6G application services that require sensing, imaging, positioning and other capabilities, such as high-precision positioning, mapping, and reconstruction, and gesture/activity recognition\cite{HUAWEI}. Therefore, the 6G network will evolve into an integrated network of communications, sensing, and computing, which is defined as a network with both physical-digital spatial perception and ubiquitous intelligent communications and computing capabilities\cite{INCSC_WP}. The multi-layer cloud-edge-terminal computing network will provide on-demand real-time scheduling and efficient sharing of computing resources, serving ISAC business functions. The architecture of the integrated network is composed of three layers, including distributed terminals, edge network, and core cloud network. The communications, sensing, and computing resources will be deeply integrated and mutually beneficial, providing efficient services for new intelligent applications such as intelligent transportation, UAV networks, space-air-ground-sea integrated networks, environmental detection, and metaverse. 

At present, the research on the integrated network of communications, sensing, and computing is still in its early stage, mainly focusing on concepts and requirements. In 2021, Zhang \emph{et al.} proposed the concept of joint communications, sensing, and computation enabled 6G intelligent networks and outlined the application requirements and network architecture\cite{Yan2021,Feng2021}. In 2022, the China Institute of Communications released the first white paper\cite{INCSC_WP} on the integrated network of communications, sensing, and computing, which defined the application scenarios and requirements, analyzed the enabling technologies, and predicted the direction of evolution towards intelligence. \textcolor{black}{Related challenges will be addressed later.}

\subsubsection{Green, flexible and lightweight}
Achieving green, flexibility and lightweight is a continuous trend in the development of the network architecture. This is also the key to allocating network resources dynamically and flexibly, enhancing the network efficiency, and reducing network deployment and operation costs. Beyond the flexible network slicing of 5G, the network architecture of 6G will become greener, more flexible, and more lightweight, using cell-free architecture, the RAN-Core convergence architecture, and the fully-decoupled RAN architecture and other promising techniques.

\paragraph*{Cell-free/less architecture}
Since 5G, the performance of the communication system has been greatly limited by the boundary effect of the traditional cellular architecture, which refers to the phenomenon of poor communication quality and strong interference from other users at the edge of the cell. The traditional cellular network structure requires complex co-processing and high deployment costs, which makes current technologies extremely limited in their ability to mitigate boundary effects. In addition, issues such as load balancing, interference management, and handover overhead, are also thorny problems in traditional cellular architectures\cite{SEU}. To address these challenges, cell-free (or cell-less) network architecture with massive MIMO was proposed and quickly became the focus of research\cite{Ngo2017,demir_foundations_2021,Papazafeiropoulos2020,Papazafeiropoulos2021,Zhang2021_CF,Makhanbet2022,Ye2022,Datta2022,Vu2020,Wang2022,Alonzo2021,Ammar2021}. A similar user-centric no cell architecture has been shown \textcolor{black}{in 2016} by HUAWEI\cite{HUAWEI-UCNC}. In the cell-free network architecture, areas are no longer divided based on cellular grids. Massive MIMO antenna arrays and access points are geographically distributed in a large area and controlled by unified central processing units (CPUs), jointly serving user terminals with the same resources. \textcolor{black}{The cell-free massive MIMO architecture is promising in next-generation systems due to its high network coverage, low cost, high macro-diversity gain, low path loss, as well as huge SE and EE\cite{Papazafeiropoulos2020,SEU}}.

\textcolor{black}{Since the cell-free architecture was proposed in \cite{Ngo2017}, research has been widely conducted to validate its advantages. In\cite{ngo_total_2018}, the authors analyzed the SE and EE for cell-free massive MIMO system and the results demonstrated that the cell-free architecture would provide the same QoS in a greener way. A similar conclusion was also revealed in\cite{papazafeiropoulos_towards_2021}. Moreover, due to the fact that cellular-free networks serve users through access points, the network deployment could be more flexible\cite{demir_foundations_2021,Ammar2021}. As a result, its potential advantages have led to a large number of studies on cell-free networks in recent years.} Focusing on the impact of hardware impairments, Papazafeiropoulos \emph{et al.}\cite{Papazafeiropoulos2021} investigated the performance of scalable cell-free systems with low computational complexity. The effects of phase drifts and noise in non-ideal hardware on cell-free systems were studied in \cite{Zhang2021_CF}. In order to improve the transmission reliability in dynamic cell-free massive MIMO networks, the authors of \cite{Makhanbet2022} proposed a learning-based energy-delay-aware power control strategy. Ye \emph{et al.} \cite{Ye2022} studied channel estimation methods in cell-free systems, and proposed a high-precision channel covariance matrix estimation scheme with fingerprint-based localization. In addition to theoretical studies in cell-free systems, researches have also focused on the \textcolor{black}{integration of cell-free networks with other technologies}. In \cite{Datta2022}, Datta \emph{et al.} proposed and analyzed full-duplex cell-free massive MIMO systems that can take full throughput and EE advantages of full-duplex communications. Based on cell-free massive MIMO networks, a novel federated learning framework in wireless environments was proposed in \cite{Vu2020}. So far, the cell-free architecture has been applied to many new communication scenarios, including mmWave communications\cite{Wang2022}, VLC \cite{Beysens2021}, satellite and UAV communications \cite{Liu2021_CF}, and communications in indoor factory environments \cite{Alonzo2021}. To make the cell-free/less architecture into reality, there are still many issues to be explored. The main research challenges and opportunities in cell-free massive MIMO networks were reviewed in \cite{Ammar2021} comprehensively, including issues such as the fronthaul link, estimation of channel state information (CSI), and resource allocation.

\paragraph*{RAN-Core convergence}
Currently, the core functions of the communication network are becoming decentralized, while higher-level RAN functions are becoming centralized. In order to make the 6G network more lightweight and flexible, the Bell Labs \cite{Viswanathan2020_13,Redana_Bell-Labs_Report} proposed a novel network architecture concept of RAN-Core convergence, which means that part of the RAN architecture (centralized high-level RAN) and part of the core network (sinking edge core network) can be unified to a single entity. On one hand, the complexity of the network and the cost of transmission can be reduced. On the other hand, the scalability of the network elements on the RAN side will be increased. Note that issues such as network decomposition in the RAN-Core converged network architecture and the coordination of protocol suites between different networks require further study\cite{Viswanathan2020_13}.

\paragraph*{Fully-decoupled RAN architecture}
From 4G to 5G, the architecture of the core network has been designed to separate the control plane from the user plane, which could improve the flexibility of the network. \textcolor{black}{However, on the RAN side of 5G, each base station still needs to be equipped with control functions and functions in the user plane (e.g., data transmission functions)\cite{Yu2019_10}.}
So, there is still room for network architecture optimization. In \cite{Yu2019_10,Yu2019}, a novel architecture that fully decouples the RAN side was proposed, and its advantages in terms of EE, reliability, mobility, and flexibility were discussed. In the fully-decoupled RAN architecture, base stations are divided into control base stations and data base stations, which perform control and data transmission functions separately. Control base stations are usually macro cell base stations with a large range, while data base stations are micro-cell base stations. By multiplexing the second generation (2G)/3G network infrastructure or spectrum resources, terminals can transmit control signals through low-frequency control channels and control base stations, which improves spectrum utilization and makes network more flexible. Besides, the design concept of this architecture also takes the resource allocation and power consumption into account. \textcolor{black}{Note that full decoupling is not achieved only by separating control base stations and data base stations.} The data base stations for uplink and downlink transmission are completely separated. In this way, control base stations can coordinate spectrum resources more accurately, and the interference between users will also be reduced. In addition, separating the uplink and downlink base stations and deploying fewer downlink base stations to serve larger areas can reduce the power consumption of the network and terminals, which can make the network greener and more energy efficient. Details about the design concept of the fully decoupled RAN architecture can be found in \cite{Yu2019}. Recently, Zhao \emph{et al.}\cite{Zhao2021_FD-RAN} investigated the uplink joint base station reception issue in the fully decoupled RAN architecture, and designed an effective parallel uplink base station selection strategy based on SE maximization. However, the current research on the fully decoupled RAN architecture is still in the early stage.

\subsubsection{Natively intelligent}
It is widely believed that the 6G will be more intelligent. In the 5G era, there have already been studies to improve the intelligence level of the network. The 5G core network has added the network data analysis function (NWDAF), which improves the data collection and analysis capabilities of the network. However, due to limited data sources, lack of data privacy protection and support for external AI services, the NWDAF cannot provide native AI support for the network. With the fast development of AI technologies (including DL\cite{DL_Network2019}, reinforcement learning, and federated learning) and the enhancement of the comprehensive capabilities of network nodes, such as communications, computing, and sensing, 6G networks will support native AI intelligence, which has two different aspects\cite{IMT2030_Network}, i.e., AI for Network (AI4Net) and Network for AI (Net4AI). On one hand, novel AI technologies are used in network planning, maintenance and optimization, enabling self-operation, self-maintenance, and self-repair capabilities of the network. On the other hand, the network with native intelligence will be able to provide more intelligent AI application services for users. Currently, more and more research interests are focused on natively intelligent network architecture, such as cognitive service architecture, deep edge nodes and networks ($\rm{DEN^2}$), self-sustaining networks (SSNs), and digital twin-based network architecture. \textcolor{black}{These technologies will help to achieve natively intelligent 6G networks.}

\paragraph*{Cognitive service architecture}
From 5G to 6G, diverse application scenarios have emerged. This means changing service scenarios, personalized user needs, and changing business requirements. These factors, in turn, require 6G networks to be flexible enough. Although the modular 5G core network with the SBA can be deployed using different network modules based on the needs of services, the SBA in the 5G core network uses rough configuration and lacks real-time sensing and dynamic adaptability to changes in service requirements\cite{SEU}. In \cite{SEU}, a novel cognitive service architecture, which has two main features, i.e., the ability to accurately identify target behaviors, scene semantics and user characteristics, as well as the unified service description method, was proposed. More recently, the design concept and implementation details of the cognitive service architecture were described in detail in \cite{Li2021_CSA}. In order to realize the cognitive function of the network, traditional network services are upgraded to cognitive services, and real-time perception and AI reasoning capabilities are added in addition to basic network functions. On one hand, the upgraded SBA can perceive network status such as the request flow, resource and topology status, and operation and maintenance events in real time. On the other hand, the intelligent ability of AI can be used to realize the online feature matching and local reasoning functions of the network. The updating of the cognitive capabilities of network functions and interfaces is accomplished by updating the knowledge graph of cognitive services. In addition, in the cognitive service architecture, AI is also introduced to enable the intelligent resource scheduling function module for the 6G core network to improve the overall performance of the network. More details about the cognitive service architecture can be found in \cite{Li2021_CSA}. Especially, in addition to investigating the design concept of the cognitive service architecture, Li \emph{et al.} \cite{Li2021_CSA} also conducted a case study of the cognitive service architecture through the session establishment process. The results illustrated that the cognitive service architecture can improve the performance of the system by simplifying the process of network function interaction with the knowledge graph of network services. In order to realize the cognitive service architecture, it is necessary to focus on various enabling technologies such as unified network semantics, polymorphic interfaces supporting cognitive services, service continuity assurance, and general platforms for the computing network integration.

\paragraph*{$DEN^2$}
In the future, 6G will serve novel industry scenarios that require extreme performance and local data processing. To this end, 6G is expected to deploy communication services and intelligence at the edge, to gradually realize pervasive intelligence of the network. The $\rm{DEN^2}$ was proposed in \cite{SEU}. Its essential design concept is large-scale networking consisting of collaborative and controllable deep edge node entities \textcolor{black}{which provide communication services, intelligence, computing, etc.} \textcolor{black}{A similar idea, called connected AI, was proposed in\cite{Shen2022_DT}.} A schematic diagram of the architecture of $\rm{DEN^2}$ was given in \cite{SEU}, including the architecture of $\rm{DEN^2}$ operation, $\rm{DEN^2}$ control and management, and networking architecture. From the perspective of promoting the intelligence of deep edge nodes, the key function of $\rm{DEN^2}$ is to support native AI, including data access, storage, processing, inference, and knowledge distribution\cite{SEU}. \textcolor{black}{Through the networking of deep edge nodes, $\rm{DEN^2}$ can promote the integration of resources such as communication and computing and make full use of these resources, further enhancing the intelligence level of 6G networks.}

\paragraph*{SSN}
In order to improve the efficiency of network deployment, management, and maintenance and reduce the cost of network operation, the self-organizing network (SON) pursuing network management deployment and management automation has been proposed, which has recently attracted much attention\cite{Zhang2020_SON,Youssef2021,Palacios2018}. However, a potential paradigm shift for 6G is from SON to the SSN as SON is limited to only adapting its functionality to specific environmental conditions\cite{Saad2020_12}. The main feature of SSN is the ability to self-sustain and permanently maintain the performance of the network in highly dynamic and complex environments (including unknown environments). 

So far, research on SSN is still in its infancy, and several studies have focused on SSN at the RAN side. In traditional network slicing, the network only performs operations for specific scenarios, and manual intervention is always required to solve unforeseen network situations and problems \cite{Mei2020}. To address this challenge, in \cite{Mei2020}, Mei \emph{et al.} proposed an intelligent self-sustaining RAN slicing framework. Driven by AI technologies, the self-sustaining RAN slice architecture combines self-management of multi-granularity network resources, self-optimization and self-learning of slice control performance, and adaptive control strategies under unpredictable network conditions. It can autonomously maintain high QoS performance of various services under different network conditions. In addition, the authors of \cite{Mei2020} also tried to apply the proposed self-sustaining RAN slicing framework to vehicular networks, and the case study illustrated the advantages of flexibility, self-learning, and fast automatic adjustment of the self-sustaining RAN slicing framework. We can believe that with the further development of AI technologies and the improvement of network intelligence, the future 6G network will develop from the RAN side and gradually realize an E2E SSN.

\paragraph*{Digital twin-based network architecture}
The 6G network will be a ubiquitous network with endogenous security, providing the ultimate network experience, supporting diverse application scenarios, and covering all scenarios. In order to realize the 6G vision, the network architecture needs to be innovated in many aspects. However, traditional network optimization and innovation often rely on the physical networks, which take a long time and have high implementation costs. In order to make the network evolve more efficiently, the concept of digital twin\cite{Shen2022_DT_74,Wu2021_DTNSurvey,Ahmadi2021,X.Nguyen2021_DT} has been applied to the innovation and evolution of communication network in recent years. This brings new opportunities and methods to improve various KPIs of 6G networks, such as SE, EE, intelligence, and security. The digital twin, which is defined as a physical product, a virtual product, and the connection between them, was first proposed by Michael Grieves in 2003 \cite{Shen2022_DT_74}. With the continuous advancement of modeling and simulation technologies, digital twins are gradually being applied to various industries such as manufacturing, aviation, healthcare, and 6G networks \cite{Wu2021_DTNSurvey}. In \cite{Ahmadi2021}, the authors discussed the relationship between the digital twin and 6G. On one hand, more advanced and intelligent 6G can promote the realization and application of digital twins in various industries. On the other hand, the intelligent digital twin can also facilitate the design, deployment, operation, and maintenance of 6G networks. Until now, the research on applications of digital twins in 6G networks is still in the early stage, and there are various definitions or implementations\cite{Shen2022_DT}. The widely recognized design concept of the digital twin-based network architecture refers to a virtual digital twin network that is constructed on the real physical network, and they are interactively mapped in real time. The twin network realizes the mapping and control of the physical network through closed-loop simulation and optimization. The digital twin network can provide guidance for the deployment, management, and operation of the real 6G network, and improve the autonomy and automation level of the 6G network\cite{IMT-2030,X.Nguyen2021_DT,Ahmadi2021}. 

The digital twin-based network architecture has attracted much attention, and a large number of studies have emerged in recent years. In \cite{Yu2019_Cybertwin,Yu2019_10}, the authors proposed a cybertwin-based network architecture in which digital cybertwins of the end users hosted at the edge of the network can offer three major functions: communication assistance, network behavior logs, and digital asset ownership. On the basis of this work, Li \emph{et al.}\cite{Li2021_DT} investigated the joint virtual network topology design and embedding in the cybertwin-based 6G core networks. \textcolor{black}{To facilitate user-centric networking, Shen \emph{et al.}\cite{Shen2022_DT} proposed a digital-twin-based network architecture integrating network slicing and AI. This framework could achieve fine-grain and flexible network management.} In \cite{Wang2021_DT_NFV}, a virtual digital twin instance of a physical network was established to capture the dependencies between anomalies and faults in NFV environments in real time. The simulation results demonstrated the effectiveness and advantage of digital twins to assist in analyzing root causes of anomalies in NFV environments. Naeem \emph{et al.}\cite{Naeem2021} used the digital twin to assist in the optimal allocation of network slice resources. They utilized graphs to build digital twins of network slices and tried to use graph neural networks to learn complex relationships of network slices. The network state based on the digital twin can finally be forwarded to the deep distributed Q network proposed in \cite{Naeem2021} to learn the optimal network slicing strategy. In addition, some other works focused on issues in digital twin-assisted edge networks, such as the application of digital twins in the offloading of intelligent computing tasks of IoT devices and the selection of mobile edge servers\cite{D.Duy_2022,Liu2022_DT}, the communication efficiency and data privacy protection of digital twin edge networks\cite{Lu2021_DT_security}, and the optimization issues on edge association\cite{Lu2021_DT_TII,Lu2021_DT_JIOT}.

It can be seen that the current research on digital twins in 6G networks is still at the preliminary stage, mainly involving the application of digital twins in certain parts of the 6G network. Due to the complexity of the 6G network, the research on the holistic 6G network architecture using the digital twin is indeed a very huge and complicated task. 
It is worth noting that while the digital twin accelerates the development of 6G networks, the construction of a real-time digital twin network system also has higher requirements on the data rate, reliability, delay, and other KPIs of the network. Therefore, the 6G network and the digital twin will integrate, promote each other, and develop~together.

\subsection{A Novel Promising 6G Network Architecture}
\begin{figure*}[tb]
	\centerline{\includegraphics[width=0.95\textwidth]{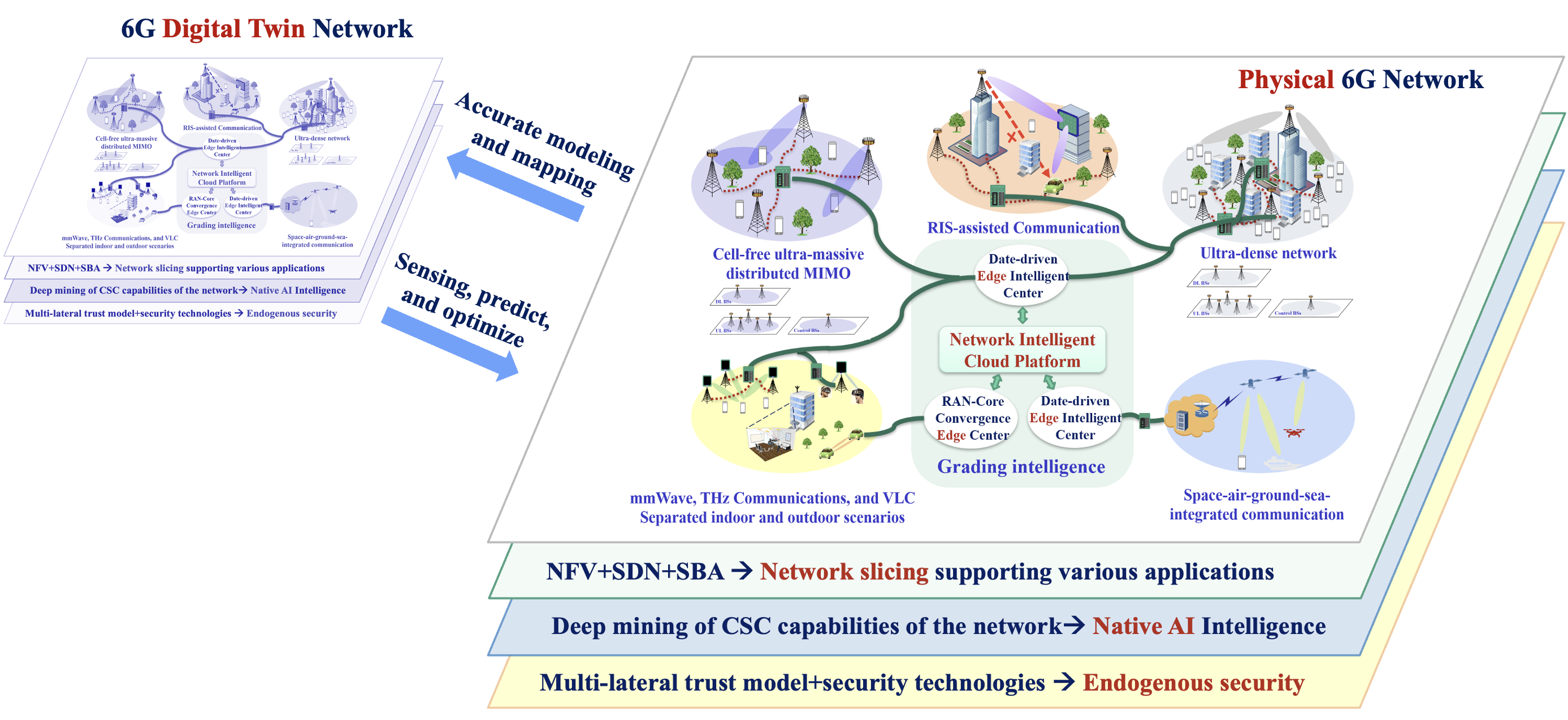}}
	\caption{A novel promising 6G network architecture (CSC: Communications, sensing, and computing).}
	\label{fig_6G_network_architecture}
\end{figure*}


In order to realize the 6G vision, satisfy the 6G KPIs and serve diverse new application scenarios, the 6G network architecture will undergo a comprehensive innovation. On the basis of further evolution of the 5G network architecture, the 6G network architecture will develop towards five directions as shown in Fig. \ref{fig_6G_Network_Architecture}. \textcolor{black}{In this section, we propose a novel comprehensive network architecture for the 6G communication network, as illustrated in Fig. \ref{fig_6G_network_architecture}. The architecture connects a number of potential future network technologies/components into a single framework. Since it has become a consensus that the future 6G network will be a heterogeneous integration of various networks, the architecture concept we proposed is similar to the concept in several white papers and projects, e.g., \cite{5GIA}. The innovation is that the proposed comprehensive architecture takes more potential components into consideration, including the hierarchical intelligent multi-layer ubiquitous computing network, new flexible architectures such as the cell-free architecture, and new technologies to enhance the network intelligence such as the digital-twin based network architecture. Next, the proposed 6G network architecture will be introduced.}

\textcolor{black}{First of all, with the development of computing, storage, and other technologies, the 6G network will become more intelligent, flexible, efficient, and ubiquitous. Cloud, fog, and edge computing will be dependent and complementary to each other, providing the networks with on-demand real-time scheduling and efficient sharing computing resources, which are essential to AI technologies. Based on this, an integrated multi-layer and hierarchical intelligent network, including a network intelligent cloud platform and data-driven edge intelligent centers is expected to form.} Among them, the fog computing acts as a bridge, connecting the centralized cloud and distributed edge networks. The mobile edge computing (sinking of network functions) in 5G will continue to evolve, making the 6G network decentralized. Edge intelligent centers will support a series of communication networks. The cell-free architecture integrated with ultra-massive MIMO antennas will break the boundary effect in traditional cellular architectures, and bring a series of improvements to the network, including the SE and EE. \textcolor{black}{Interestingly, reconfigurable intelligent surface (RIS) has the ability to actively control the wireless channel, which will be introduced detailedly in Section~\ref{sec:RIS}.} It has great potential for coverage enhancement and capacity improvement of future wireless networks. The communication densities in various application scenarios are gradually increasing, and the 6G network will be an UDHN. In addition, full spectra resources including mmWave, THz, and VLC will be deeply exploited and utilized to build a variety of networks to serve full coverage scenarios, i.e., space-air-ground-sea integrated networks\cite{VLC_Hass2020}. In particular, several RAN-Core converged edge computing centers which integrate high-level RAN functions with partially sinking edge functions, can reduce the delay of transmission and can better support applications with low latency requirements, such as automatic driving and telemedicine. In the 6G network architecture, the RAN side will be further decoupled and the control plane of the base station will be further separated from the user plane, so that network resources can be used more flexibly and efficiently. 

Secondly, the NFV, SDN, and SBA will continue to develop and evolve, providing the source and impetus for the realization of overall E2E network slicing. What's more, network elements in 6G networks will have additional computing and sensing capabilities. By deeply mining the communication, sensing, and computing capabilities, large amount of training data and distributed computing power will be possible. Thus, some architectures which can improve network intelligent level, such as cognitive service architecture, ${\rm DEN^2}$, and SSNs, will become a reality sooner. Further, 6G networks will gradually achieve native intelligence. From the perspective of network security, 6G networks will use a multilateral trust model, and combine various security technologies such as blockchain technology and physical layer security to achieve the endogenous security of the network. Note that edge intelligent centers in 6G networks will be improved in various aspects compared with those in 5G networks. The data-driven 6G edge intelligent centers will deeply integrate resources including communications, sensing, computing, and AI, with functions of task-centric dynamic service orchestration. In addition, these edge intelligent centers will contribute to 6G's native intelligence and endogenous security.

Finally, with the help of AI technologies and communication-sensing-computing integrated networks, digital twins will help push forward the evolution of the 6G network architecture. The real 6G network will be accurately modeled to construct the corresponding digital twin network, and the two will map to each other in real time. The twin network can track the changes of the real network, and predict the performance of the network optimization scheme through closed-loop simulation and optimization, providing optimization guidance for the deployment, management, and operation of the real 6G network. With the development of various enabling technologies, it is promising that the proposed comprehensive 6G network architecture will finally become a reality.

\section{Key 6G Technologies}

\begin{figure*}[tb]
	\centerline{\includegraphics[width=0.97\textwidth]{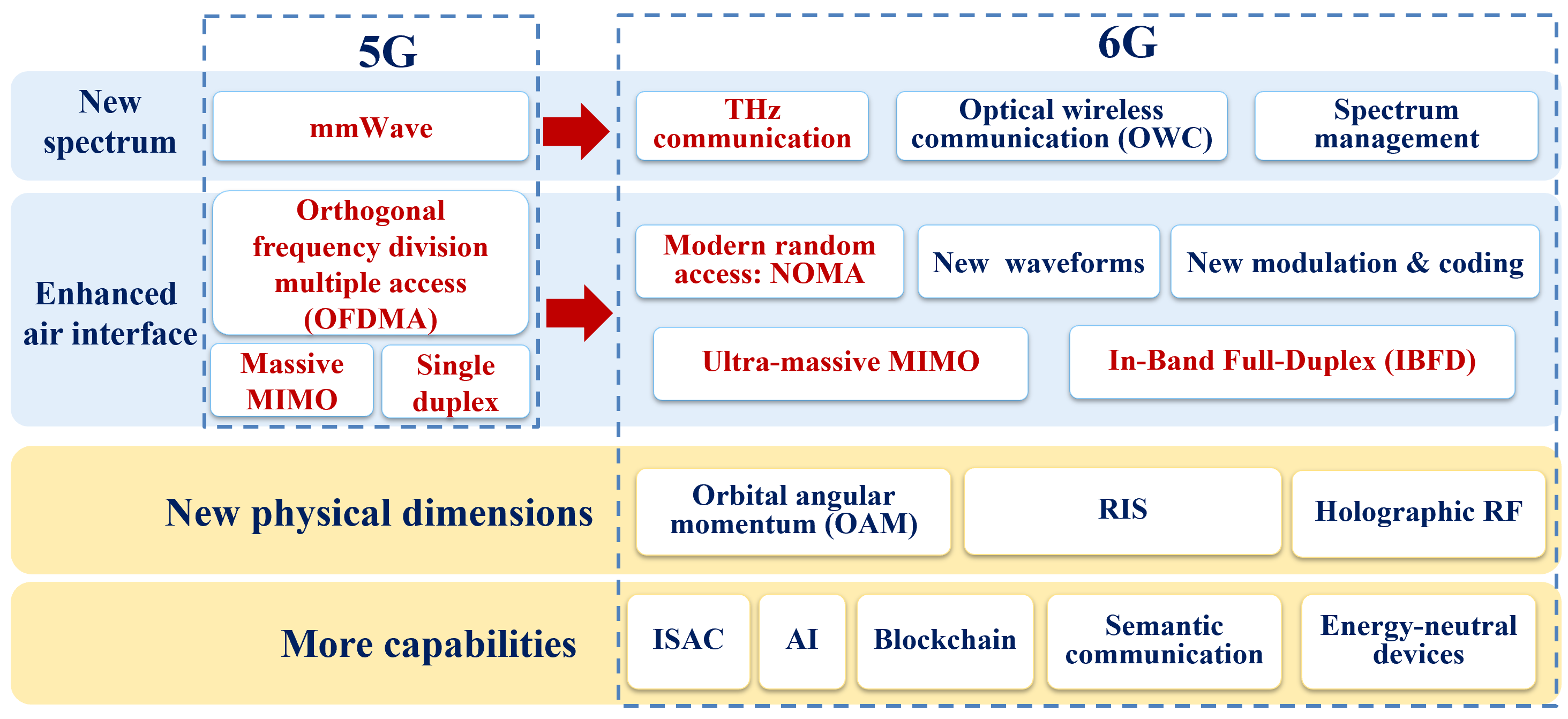}}
	\caption{Potential 6G key technologies.}
	\label{fig_6G_Tech}
\end{figure*}

\renewcommand{\arraystretch}{1.25}
\begin{table*}[!t]
    \centering
	\caption{\textcolor{black}{Representative surveys for 6G technologies.}}
	\begin{tabular}{|l|l|l|}
		\hline
		\textbf{Technology}                                      & \textbf{Ref.} & \textbf{Key points}                                                                                  \\ \hline
		\multirow{7}{*}{\textbf{THz}}                            & \cite{THz_Huq2019}             & Standardization, scenarios, application, future research directions and open issues of THz           \\ \cline{2-3} 
		& \cite{THz_Elayan2020}          & THz generation methods, channel models, applications, standardization activities, and future outlook \\ \cline{2-3} 
		& \cite{THz_Tekbiyik2019}        & Challenges, novelties and standardization of THz                                                     \\ \cline{2-3} 
		& \cite{THz_Sarieddeen2020}      & THz sensing, imaging, and localization applications                                                  \\ \cline{2-3} 
		& \cite{THz_Sarieddeen2021}      & THz-specific signal processing                                                                       \\ \cline{2-3} 
		& \cite{THz_Do2021}              & THz Line-of-Sight MIMO communication                                                                 \\ \cline{2-3} 
		& \cite{THz_Ghafoor2020}         & MAC protocols for THz                                                                                \\ \hline
		\multirow{4}{*}{\textbf{OWCs}}                           & \cite{OWC_Chowdhury2018}       & Different promising optical wireless technologies                                                    \\ \cline{2-3} 
		& \cite{OWC_AlKinani2018}        & OWC channel research                                                                                 \\ \cline{2-3} 
		& \cite{Chowdhury2020}           & Optical wireless hybrid networks                                                                     \\ \cline{2-3} 
		& \cite{OWC_Matheus2019}         & Concepts, architecture, and challenges of VLC                                                        \\ \hline
		\multirow{3}{*}{\textbf{Spectrum management}}            & \cite{SPE_Wang2011}            & Fundamentals, architecture, applications, and important issues of CR                    \\ \cline{2-3} 
		& \cite{SPE_Liang2020}           & A systematic view for symbiotic radio                                                                \\ \cline{2-3} 
		& \cite{SPE_Bhattarai2016}       & Ongoing Initiatives, challenges, and a roadmap of dynamic spectrum sharing                           \\ \hline
		\multirow{4}{*}{\textbf{New waveforms and   modulation}} & \cite{NEW_Wei2021e}            & Motivation, features, challenges, and applications of OTFS                                           \\ \cline{2-3} 
		& \cite{NEW_Darwazeh2018}        & The history of SEFDM development                                                                     \\ \cline{2-3} 
		& \cite{ISAC_Liu2020}            & ISAC-specific waveform design                                                                        \\ \cline{2-3} 
		& \cite{NEW_Basar2017}           & Advances and research directions for IM                                                \\ \hline
		\multirow{2}{*}{\textbf{New coding}}                     & \cite{ZC_coding_1}             & Evolution of channel coding                                                                          \\ \cline{2-3} 
		& \cite{coding_Shao2019}         & Decoder ASIC implementations of Turbo, LDPC, and Polar                                               \\ \hline
		\multirow{2}{*}{\textbf{Modern random access}}           & \cite{NOMA_Chen2018}           & Design, standardization progress and use cases of NOMA                                               \\ \cline{2-3} 
		& \cite{NOMA_Ding2017}           & NOMA research, innovations, applications, and challenges                                             \\ \hline
		\multirow{3}{*}{\textbf{Ultra-massive MIMO}}             & \cite{IMT2030_MIMO}            & An overall research report on ultra-massive MIMO                                                     \\ \cline{2-3} 
		& \cite{Zhang2019}               & Benefit, signal processing techniques, and research challenges of cell-free massive MIMO             \\ \cline{2-3} 
		& \cite{Ammar2022}               & Current status and future directions of user-centric cell-free massive MIMO network architecture     \\ \hline
		\multirow{2}{*}{\textbf{IBFD}}                           & \cite{IBFD_Liu2015}            & Research status and future challenges of in-band full-duplex relaying                                \\ \cline{2-3} 
		& \cite{IBFD_Kolodziej2019}      & Techniques and systems survey of IBFD                                                                \\ \hline
		\multirow{3}{*}{\textbf{OAM}}                            & \cite{OAM_Chen2020}            & Generation, detection, and emerging applications of OAM waves                                        \\ \cline{2-3} 
		& \cite{OAM_Yang2021}            & The generation, detection techniques, and application of THz-OAM beams                               \\ \cline{2-3} 
		& \cite{OAM_Liu2019}             & Insights and design guidelines of OAM-based sensing systems                                          \\ \hline
		\multirow{6}{*}{\textbf{RIS}}                            & \cite{Marco_Basar2019a}        & Solutions, research issues, related communication-theoretic models, and performance limits of RIS    \\ \cline{2-3} 
		& \cite{Marco_Liu2021h}          & Principles, performance evaluation, beamforming design and resource management of RIS                \\ \cline{2-3} 
		& \cite{Marco_DiRenzo2020}       & Research, applications, and challenges of RIS-empowered smart radio environments                     \\ \cline{2-3} 
		& \cite{Marco_DiRenzo2020a}      & The key differences and similarities between RISs and relays                                         \\ \cline{2-3} 
		& \cite{IMT2030_RIS}             & An overall research report on RIS                                                                    \\ \cline{2-3} 
		& \cite{Marco_Pan2021}           & Advantages, principles, applications, and research directions of RIS                                 \\ \hline
		\textbf{Holographic radio}                               & \cite{OULU_BC}                 & Realization and signal processing of holographic radio                                               \\ \hline
		\multirow{4}{*}{\textbf{AI}}                             & \cite{IMT2030-Wireless_AI_WP}  & An overall research report on wireless AI                                                            \\ \cline{2-3} 
		& \cite{AI_Zhang2020a}           & AI-based 5G and B5G algorithm, implementation, and optimization                  \\ \cline{2-3} 
		& \cite{AI_OShea2017a}           & Applications and open challenges of DL for the physical layer                                        \\ \cline{2-3} 
		& \cite{AI_Zhang2019}            & Techniques and challenges of ML for Internet congestion control                                      \\ 	\hline
		\multirow{6}{*}{\textbf{ISAC}}                           & \cite{IMT2030_ISAC}            & An overall research report on ISAC                                                                   \\ \cline{2-3} 
		& \cite{ISAC_Zhang2021}          & Motivation, methodologies, challenges, and opportunities of realizing perceptive mobile network      \\ \cline{2-3} 
		& \cite{ISAC_Zhang2021_signal}   & Signal processing techniques for joint communication and radar sensing                               \\ \cline{2-3} 
		& \cite{ISAC_Cui2021}            & Applications, trends, and challenges of ISAC for ubiquitous IoT                                      \\ \cline{2-3} 
		& \cite{ISAC_Ma2020}             & Dual-function radar-communications strategies and their relevance to autonomous vehicles             \\
		& \cite{SCIS_Xiao2022}           & An overview research on integrated localization and communication             \\
		 \hline
		\multirow{6}{*}{\textbf{Blockchain}}                     & \cite{BC_Wang2021}             & Fundamentals and recent efforts of Blockchain-enabled wireless communications                        \\ \cline{2-3} 
		& \cite{BC_Weiss2019}            & The application of blockchain to radio spectrum management                                           \\ \cline{2-3} 
		& \cite{BC_Shen2022}             & Blockchain solutions to address the challenges in data management                                    \\ \cline{2-3} 
		& \cite{BC_Ray2021}              & Blockchain for IoT-based healthcare                                                                  \\ \cline{2-3} 
		& \cite{BC_Mollah2021}           & Blockchain for the IoV                                                              \\ \cline{2-3} 
		& \cite{BC_Zhuang2021}           & Blockchain for cybersecurity in smart grid                                                           \\ \hline
		\textbf{Semantic communication}                          & \cite{SEM_Luo2022}             & The latest DL andE2E communication based semantic communications                                     \\ \hline
		\multirow{2}{*}{\textbf{Energy-neutral devices}}   
		& \cite{boyer2013backscatter}    & A tutorial survey of backscatter modulation   \\ \cline{2-3} 
		& \cite{CJIT_Xu2021a}  &  Resource allocation in backscatter communication networks  \\ \hline

	\end{tabular}
	\label{table_tech_survey}
\end{table*}

The ambitious 6G vision gives us an exciting blueprint for future communications systems. On the basis of using all the spectra and providing users with global coverage, the connotation of the communication system will be further expanded to realize the intelligent services that integrate communication, sensing and computing with security assurance. 
In this regard, 5G key technologies are no longer sufficient to support the aforementioned 6G vision. There have been several works on potential key technologies for 6G. In \cite{SEU, Chowdhury2020_6, Bariah2020_11, Saad2020_12}, \textcolor{black}{the authors listed several possible techniques without classification, which may be confusing and unable to provide readers with a holistic understanding of technological development directions. Some articles only investigated a few technologies \cite{Zong2019_34, Bhat2021},} or did not provide sufficient insights into the proposed techniques \cite{Zhang2019_33}. In \cite{ITU-R-M.1295-E}, ITU-R reported future technology trends towards 2030 and beyond, while it was a draft version without literature review. In addition, some proposed technologies have advantages while their applicable scope is limited, so that they are less likely to become key technologies in the future 6G, such as nanoscale communications\cite{Khan2020_8} and molecular communications \cite{Saad2020_12}. Based on existing research and the latest development of related technologies, we point out 16 potential 6G key technologies and divide them into four evolution directions, as shown in Fig. \ref{fig_6G_Tech}. \textcolor{black}{For readers' reference, we also summarize several representative surveys for 6G technologies in Table~\ref{table_tech_survey}.}

A portion of potential 6G technologies are further evolutions of 5G key technologies. 5G increased data rate through mmWave technology, while 6G is expected to introduce THz, OWCs, and advanced spectrum management technologies to meet the rapidly increasing demand for data service. 5G key air interface technologies, such as orthogonal frequency division multiple access (OFDMA), massive MIMO and half duplex, will be further enhanced in 6G. For example, modern random access technologies like NOMA, ultra-massive MIMO and in-band full-duplex (IBFD) are expected to achieve further enhancement in the capacity and efficiency of the communication system. In addition, academia and industry are actively exploring new physical dimensions, such as orbital angular OAM, RIS, and holographic RF, to achieve a revolutionary breakthrough in the way of data transmission. What's more, there are  a number of new technologies that are expected to be used to increase the capabilities of communication networks. For example, ISAC can give communication systems the opportunity to have converged sensing capabilities. AI is expected to provide disruptive and intelligent solutions for all layers of the communication system. Blockchain is promising to provide a guarantee for the security and reliability of communication systems. Semantic communication is promising to explore a new way of information extraction to break the transmission bottleneck of classic communication systems.

In this section, the concepts, applications, developments and challenges of these potential 6G key technologies will be discussed.

\subsection{New Spectrum}
\subsubsection{THz}
\

It is predicted that mobile data traffic will increase fivefold by 2024\cite{THz_Huq2019}. Under the 6G vision mentioned before, the rapid growth of video services and the emergence of new applications, such as VR/AR, autonomous driving and IoTs, have led to the increasing demand for high data rate transmission and low latency services\cite{THz_Rappaport2019}. Most existing 5G technologies are stuck in the mmWave band and can only achieve average rates of up to 1 Gbps\cite{THz_Huq2019}. Facing non-negligible spectrum congestion problems, 5G communication systems are insufficient to meet the rapidly increasing demand for 6G data services. 

THz (0.1--3 THz) is the last unexplored spectrum gap between mmWave and optical frequency ranges. THz is characterized by high frequency, large bandwidth, high path loss, severe molecular absorption, abundant diffuse scattering and extremely narrow beam. Although there is still some distance from practical applications, THz is regarded as one of the most promising technologies for 6G because of its ability to provide robust support for ultra-high data rate services\cite{THz_Elayan2020}.

THz technology is expected to play a significant role in applications including communication, sensing, imaging, and positioning: 1) The THz communication system is promised to support high data rate communication services from indoor to outdoor, such as HD holographic video conferences, the ultra-high resolution video formats, the downloading of HD film files, the VR technique, vehicle-to-vehicle (V2V) communications, wireless fronthaul and backhaul links, and space applications like inter-satellite communications \cite{THz_Chen2019,THz_Elayan2020,THz_Sarieddeen2020,THz_Huq2019}; 2) Because of the short wavelength, high link directionality, and small antenna aperture, THz is less vulnerable to free space diffraction, and can be applied to improve communication security\cite{THz_Chen2019,THz_Sarieddeen2020}; 3) THz technology is expected to be utilized in nanoscale devices, such as health, military, environmental pollution monitoring and ultra-high-speed on-chip communication\cite{THz_Chen2019,THz_Tekbiyik2019}; 4) THz technology is able to support applications beyond communication, such as spectrometers for explosive detection, gas sensing, security body scanning, imaging in rescue and surveillance and positioning with centimeter level accuracy \cite{THz_Sarieddeen2020, THz_Rappaport2019}.

In order to make THz applications a reality, many technical bottlenecks still need to be overcome. 
First, hardware devices suitable for THz need to be developed, such as electronic or photonic devices that can generate high-frequency EM waves, broadband and directional antennas, and THz amplifiers\cite{THz_Tekbiyik2019,THz_Liu2020e}. Current complementary metal oxide semiconductor (CMOS) technologies cannot deal with frequencies greater than 300 GHz. Besides, it is essential to fully understand the unique characteristics of the THz channel and establish an accurate and general channel model for communication system design and analysis \cite{THz_Wang2021}. THz-specific air interface techniques and MAC protocols need to be explored further \cite{THz_Sarieddeen2021,THz_Do2021,THz_Ghafoor2020}. What's more, it is necessary to deal with the problem of short communication distance caused by the THz band's severe propagation losses and power limitation \cite{THz_Sarieddeen2020,THz_Sarieddeen2019}. Last, health and safety issues of THz need to be analyzed and evaluated \cite{THz_Huq2019}.

In 2008, IEEE 802.15 established the THz Interest Group. In 2019, FCC announced a 10-year open test for 95 GHz--3 THz. In 2022, X. You \emph{et al.} proposed a photonics-aided transparent fiber-THz-fiber transmission system, breaking the publicly reported world record of THz communication system real-time transmission rate to reach more than 100 Gbps transmission for the first time \cite{THz_Zhang2022d,THz_Zhang2022c}. Global research on THz is steadily developing. It can be expected that breakthroughs in THz radiation and detection technology will bring significant changes to human life and social development in the near~future.

\subsubsection{OWCs}
\

Beyond the THz spectrum, OWCs provide high density broadband communication service with the advantages of ultra-low latency, inherent physical layer security, zero EM interference, abundant free unlicensed spectrum, relatively low costs, and simplicity of deployment, serving as a prominent complementary to RF-based wireless communication systems\cite{Bariah2020_11, Giordani2020_16, OWC_Chowdhury2018, OWC_Koonen2018, Oulu, OWC_Strinati2019}.

The optical band consists of IR (760 nm--1mm wavelength), visible light (360--760 nm), and UV (10--400 nm). With simple structures and low-cost equipment, the IR communication system is suitable for long-distance data transmission, but it is vulnerable to atmospheric effects such as fog\cite{Bariah2020_11}. Visible light can provide communication services along with illumination, and it can also be used as a source of energy. \textcolor{black}{Solar cells can be used for simultaneous energy harvesting and high speed data reception \cite{OWC_Tav2021}.} The concept of light-based IoT was proposed in \cite{OWC_Xu2019,OWC_Perera2021}. \textcolor{black}{New generation of lighting devices are entering the market which are based on blue laser that excite phosphor.In\cite{OWC_Lee2022}, a dual wavelength (blue converted to white light plus IR) was presented and the aggregate data rate was 26 Gbps.} Compared to other light bands, UV has lower background noise and higher atmospheric scattering, making it promising for communications for non-line-of-sight (NLoS) links\textcolor{black}{\cite{OWC_He2019}}. However, UV light may have a negative impact on health and safety, which needs to be fully evaluated before practical application\textcolor{black}{\cite{OWC_Deh2022}}.

In recent years, increasing attention has been paid to optical wireless, which is considered promising from ultra-short distance to ultra-long distance communications\cite{OWC_Chowdhury2018}. The main optical wireless technologies can be divided into five categories: VLC, LiFi, OCC, FSO, and LiDAR. VLC has great application potential in indoor, underwater, and vehicular communications, as well as localization systems\cite{OWC_Matheus2019}. Complementing wireless fidelity (WiFi), LiFi can provide illumination and multiuser communication services at the same time. There is a growing interest in OCC because of its applications in V2X communications, indoor positioning,  digital signage and VR. FSO communication systems are used for high-data-rate communication in data centers, deep space communication and underwater systems\textcolor{black}{\cite{OWC_Hamza2019, OWC_Kaz2022}}. LiDAR serves as an engaging optical remote sensing technology, and it has great potential to be used for transportation, airborne as well as autonomous vehicular~communication.

Three orders of magnitude larger spectrum resources are available in the optical bands than the RF bands. However, the utilization of spectrum resources is significantly limited by the electrical bandwidth of optoelectronic devices. In recent years, the development of high performance optoelectronic devices has been a topic of great interest in optical wireless research. Exceptionally fast organic light emitting diode (OLED) with bandwidth of hundreds of MHz were designed and applied in VLC systems, achieving data rates of over 1 Gbps \cite{OWC_Yoshida2020}. A silicon photomultiplier was shown to achieve a bit error rate (BER) of $10^{-3}$ at a data rate of 3.45 Gbps \cite{OWC_Matthews2021}.
Channel measurements and modeling are crucial for OWC channels understanding and link design. A comprehensive review of OWC channel research was published in 2018 \cite{OWC_AlKinani2018}. OWC channel is not isotropic. Therefore, it is essential to consider the effect of device orientation on the channel gain \cite{OWC_Soltani2019,OWC_Arfaoui2021}.
In order to prevail over the effect of random orientation and blockage, multi-directional transmitter with adaptive spatial modulation was proposed in \cite{OWC_Soltani2019}. It was suggested that time domain spatial modulation would be a feasible option for next-generation orthogonal frequency division multiplexing (OFDM) based optical spatial modulation \cite{OWC_Yesilkaya2019}. A NOMA scheme for the beam steering and user clustering scenario was proposed to exploit the space diversity of VLC systems, and it can provide 10 Mbps sum rate gain for each NOMA user pair \cite{OWC_Eroglu2021}. \textcolor{black}{Moreover, the security of OWC has likewise been extensively investigated. In\cite{OWC_Su2021}, a physical layer security technique was proposed for multi-user VLC systems.}

\subsubsection{Spectrum management}
\

To address the shortage of spectrum resources, in addition to exploring the unused spectrum at higher frequencies, it is also important to improve the utilization of the limited spectrum. Facing the fact that traffic demand is highly dynamic and environment-dependent, the under-utilization of frequency bands needs to be tackled urgently. \textcolor{black}{Effective spectrum management based on cognitive radio (CR)\cite{SPE_Wang2011}, symbiotic radio (SR)\cite{SPE_Liang2020}, and dynamic spectrum sharing technique\cite{SPE_Bhattarai2016} is seen as an important method to improve the SE and EE of 6G communication systems\cite{CCID,SEU,Samsung,Chen2020_19,Jiang2021_21}.}

The concept of CR was first introduced in 1999 by Joseph Mitola \cite{SPE_Mitola1999}. In 2003, FCC recommended that any radio with adaptive spectrum awareness should be referred to as CR. Haykin\cite{SPE_Haykin2005} proposed a brain-empowered CR technology and defined it as an intelligent wireless communication system that can sense the external environment to adjust accordingly. Zhang \emph{et al.}\cite{SPE_Zhang2019} proposed CR-based vehicular networks that apply deep Q-learning to deal with highly dynamic topology due to changes in vehicle distribution available spectrum.
As one of the most recent evolutions of CR, SR leverages CR and ambient backscatter communication (AmBC) technologies to embed information in the ambient RF signal for mutually beneficial spectrum sharing, allowing secondary systems to share the spectrum, energy and infrastructure of the primary system with high efficiency\cite{SPE_Liang2020}. AmBC-based SR technology is promising in passive IoT, assisting spectrum and energy efficient communication design\cite{SPE_Zhang2019a,SPE_Long2020}.
Intelligent and dynamic spectrum sharing has been an active research topic in recent years. Sharma \emph{et al.}\cite{SPE_Sharma2018} proposed that full duplex wireless technology, making concurrent sensing and transmission possible, can improve the spectrum utilization efficiency via dynamic spectrum sharing. Naparstek and Cohen\cite{SPE_Naparstek2019} proposed a distributed dynamic spectrum access technique based on deep Q-learning. Blockchain-empowered dynamic spectrum sharing is promising to improve distribution, security and automation, and AI is expected to enhance the performance of pattern recognition and decision-making in dynamic spectrum sharing \cite{SPE_Jacob2020,SPE_Hu2021}.

\subsection{Enhanced Air Interface}
\subsubsection{New waveforms and modulation}
\

6G has more diverse and complex application scenarios to support its ``global coverage, full applications, strong security, all spectra, all senses, and all digital" vision. 6G will achieve a Tbps-level data rate, supply dense connections, provide a wider range of coverage, and pursue more intelligent and safer services. These improvements pose new challenges to waveform design and modulation.

The unique characteristics and requirements of application scenarios need to be considered when designing waveforms, which are closely related to the performance of the communication system. During the development of the 5G standard, multi-carrier systems, which have high SE but also a high peak-to-average ratio, such as OFDM, are mainly used. More existing waveforms are detailed in \cite{HUAWEI}. In order to flexibly adapt to the possible application scenarios of the 6G communication system, new waveform designs are expected to provide better performance in a targeted manner. Compared with low frequency, the potential high frequency scenarios of 6G lead to more challenges, such as large transmission path loss and the need for efficient high-frequency broadband power amplifiers. In \cite{NEW_Nitsche2014}, a single-carrier system with low peak to average power ratio (PAPR) was studied to address these challenges. For high mobility scenarios, waveforms in transform domain, such as orthogonal time frequency space (OTFS), can describe information such as delay and doppler more accurately \cite{NEW_Wei2021e}. As for high throughput scenarios, systems such as 
spectrally efficient frequency domain multiplexing (SEFDM)\cite{NEW_Darwazeh2018} and overlapped x domain multiplexing (OVXDM)\cite{NEW_Li2018} can be used to obtain higher SE. \textcolor{black}{ISAC technology (which will be introduced later) imposes new requirements on waveform design, expecting simultaneous communication and sensing with the same waveform\cite{ISAC_Liu2020}. }

Modulation has a great influence on the effectiveness and reliability of the communication system. Currently, quadrature amplitude modulation (QAM) modulation is widely used and is adopted by the long-term evolution (LTE) and new radio (NR) standards. 
In recent years, some other modulation techniques have attracted attention due to their advantages in shaping gains, PAPR, and robustness, including selected QAM, irregular QAM, constellation interpolation, multidimensional modulation, and IM\cite{HUAWEI,NEW_Basar2017}. 

\subsubsection{New coding}
\

Efficient channel coding technology can improve the capacity, reliability and quality of the services in the communication systems. 
Guided by Shannon’s theory, error-correcting codes (ECCs) realize a leap from algebraic coding to probabilistic coding, which had great success in improving the capacity, reliability, and quality of service in communication systems\cite{ZC_coding_1}. By introducing randomness and sparsity in coding and propagating soft messages based on factor graphs in decoding, advanced probabilistic codes can approach or even achieve the Shannon limit. Among them, the most representative ECCs are Turbo codes\cite{ZC_coding_2}, low-density parity-check (LDPC) codes\cite{ZC_coding_3}, and polar codes\cite{ZC_coding_4}, which are the standard codes for 4G data channels, 5G data channels, and 5G control channels, respectively. \textcolor{black}{Though their de-facto decoding algorithms and implementations are different\cite{coding_Shao2019}, they are all derived based on Bayes’ theorem and competitive for 6G ultra-high speed and ultra-low power consumption requirements, which impel a unified decoding framework for complex and variable scenarios in 6G communication systems.}

As linear block codes, the three ECCs can be decoded by a belief propagation (BP) decoder that employs the famous sum-product algorithm on a bipartite Tanner factor graph\cite{ZC_coding_5} but is only beneficial to LDPC codes due to their high sparsity. By adding auxiliary state variables in the factor graph, the sum-product algorithm becomes the Bahl, Cocke, Jelinek, and Raviv (BCJR) algorithm for Turbo codes\cite{ZC_coding_5,ZC_coding_6}, which propagates messages on trellis graphs in both forward and backward directions. With belief pushing in a successive message-passing schedule, the BP decoder evolves to the classical successive cancellation (SC) decoder for polar codes\cite{ZC_coding_7}, with which polar codes are proved to achieve Shannon capacity when the code length is infinite\cite{ZC_coding_4}. Limited by 5G control channels, the rate of channel polarization degrades for finite code lengths, and neither SC nor BP decoding can meet the performance requirement. The increase of codeword search space is necessary, which results in algorithms such as SC list/flip\cite{ZC_coding_8,ZC_coding_9}, and BP list/flip decoding\cite{ZC_coding_10,ZC_coding_11}. Though originated from the same decoding rule, the encoding schemes of Turbo, LDPC, and polar codes can be further enhanced, such as the generator polynomial of Turbo codes\cite{ZC_coding_12} and the information set of polar codes\cite{ZC_coding_13}, thus facilitating the unified decoding factor graphs and simplified decoding algorithms to improve the EE of the decoder in 6G. 
The approximation for exponential operations in message-passing has promoted the circuit realization of these three ECC decoders, which results in the windowed Max-log-BCJR decoder for 4G LTE Turbo codes\cite{ZC_new_1}, adaptive min-sum-BP decoder for 5G NR LDPC codes\cite{ZC_new_2}, and node-based SC decoder for 5G NR polar codes\cite{ZC_new_3}. A uniform design of ECCs at the circuit level is a key technology of 6G, such as Turbo/LDPC decoders\cite{ZC_coding_25} and LDPC/Polar decoders\cite{ZC_coding_26}. 
Driven by ultra-low latency and ultra-reliability of 6G communication systems, a short code length will be adopted for ECCs, where the randomness, sparsity, and channel polarization of the above three ECCs are debilitated, thus degrading the good performance of their de-facto decoding algorithms. One alternative approach is to employ the near-maximum likelihood decoding schemes for classical algebraic coding, which are also uniform schemes for any linear block codes, such as ordered statistics decoding (OSD)\cite{ZC_coding_27} and the recently proposed capacity-achieving guessing random additive noise decoding (GRAND)\cite{ZC_coding_28}. Another promising solution for short-length scenarios is to use concatenated codes, like Arıkan’s new polarization-adjusted convolutional (PAC) codes \cite{ZC_coding_29}. By fully exploiting the massive antennas in MIMO systems, a spatiotemporal 2-D coding scheme concatenates codes from the time domain to the space domain, to improve the reliability and transmission rate in a short decoding latency~\cite{ZC_coding_30}.

\subsubsection{Modern random access}
\

LTE employs OFDMA and 5G NR uses optimized OFDM-based waveforms and multiple access, both of which are orthogonal multiple access (OMA) technologies. The connection density of 6G communication systems will increase by tens of times compared to that of 5G. NOMA is recognized as the most promising modern random access technology for 6G, meeting the needs of low costs, high reliability, low latency, massive connectivity, and high throughput in the complex and variable scenarios of 6G communication systems~\cite{IMT-2030,HUAWEI,SEU,UNISOC,Jiang2021_21}.

In \cite{NOMA_Saito2013}, NTT DOCOMO first introduced the concept of NOMA and demonstrated that NOMA technology can improve the capacity and cell-edge user throughput performance. In contrast to traditional OMA technology, the core idea of NOMA is to encourage multiple terminals to reuse the same radio resources in the time, frequency and/or code. NOMA actively introduces interference information at the transmitter side and demodulates it using a successive interference cancellation receiver. NOMA is expected to improve spectrum efficiency, increase system capacity, reduce system latency due to scheduling and queuing, and ease reliance on accurate CSI and feedback quality at the cost of complexity \cite{NOMA_Chen2018,NOMA_Makki2020}.

NOMA schemes can be divided into three categories, including power domain NOMA such as multi-user overlay coding, code domain NOMA such as sparse code multiple access, and interleave based NOMA such as interleave division multiple access \cite{NOMA_Ding2017}.

In \cite{NOMA_Vaezi2019}, the combination of NOMA and emerging wireless technologies such as massive MIMO, mmWave, cognitive and collaborative communication, VLC, physical layer security, energy harvesting, and wireless caching was comprehensively summarized. It was shown that NOMA with these technologies can further improve the performance of future communication networks, such as scalability and greenness. In \cite{NOMA_Zeng2019_secure}, artificial noise was exploited to secure the confidential information of massive MIMO-NOMA networks, and thus, to maximize the sum secrecy rates and EE of the system. Recently, a novel framework of NOMA-assisted RIS was proposed for the deployment and passive beamforming design and was shown to further improve EE \cite{NOMA_Liu2021_RIS}. In  \cite{NOMA_Li2021}, NOMA and ambient backscatter were combined as two promising technologies for developing SE and EE systems and the reliability and security of these systems were investigated.

In general, NOMA has shown a non-negligible potential and received a lot of attention from academia and industry \cite{NOMA_Chen2018}. However, NOMA was not finally adopted in 5G due to some technical reasons and debates. The following challenges have to be addressed before the practical application of NOMA. First, high-performance, low-complexity multi-user interference cancellation algorithms should be explored. Besides, the enhancement of the security and trust shall be taken into account. What's more, a common and unified 6G NOMA framework needs to be developed \cite{NOMA_Makki2020}.

\subsubsection{Ultra-massive MIMO}
\

As one of the key technologies for 5G, the initial idea of massive MIMO was first proposed by Marzetta from Bell Labs in 2009 \cite{Marzetta2010} and has received much attention because of its ability to improve the SE significantly. When the number of antennas serving each user is greater than 10, it can be considered as massive MIMO. In 6G, larger antenna arrays will be exploited, using hundreds or even thousands of antennas, which is known as ultra-massive MIMO. It has the ability to achieve higher SE and EE, wider and more flexible network coverage, and higher positioning accuracy in more diverse frequency range \cite{IMT2030_MIMO}.
Ultra-massive MIMO has an exciting application prospect because of its unique characteristics. Firstly, further expansion of the antenna scale can provide spatial beams with very high spatial resolution and processing gain, thus improving the multiplexing capability and interference suppression of the network. It is promising to improve SE and reduce energy consumption of the system. Besides, the ultra-massive MIMO array has the ability to adjust beams in three dimensions, thus can provide non-terrestrial coverage. What's more, the ultra-massive MIMO array has extremely high spatial resolution, which can improve positioning accuracy in complicated wireless communication environments and achieve accurate 3D positioning.

In summary, important issues and trends in ultra-massive MIMO are listed as follows. 1) With the further increase of the antenna scale, the near-field effect and wideband effect will be more prominent. The approximation of plane wavefront no longer holds and the spherical wavefront needs to be considered due to the utilization of large antenna arrays \cite{THz_Wang2021}. The spatial- and frequency-wideband effects lead to the channel sparsity in the angle domain and the delay domain \cite{Wang2018}. For the near-field and wideband effects, plenty of research has been conducted in the fields of channel modeling, channel estimation, beam assignment, codebook design, and beam training\cite{Yu2016,Myers2022,Wei2022}; 2) \textcolor{black}{In order to utilize more abundant spectrum resources, ultra-massive MIMO is expected to utilize higher frequency bands, such as mmWave and THz. For ultra-massive MIMO in higher frequency bands, research is being carried out, focusing on integrated circuit design, channel characteristics, modulation techniques and so on \cite{THz_Sarieddeen2019,Huang2020_mmWave,MIMO_Zhao2021a};} 3) In addition to the traditional centralized active antenna array, ultra-massive MIMO is expected to take a more flexible and diverse approach for implementation. \textcolor{black}{By using RIS (which will be introduced later), instead of traditional active antennas, network coverage, multi-user capacity, and signal strength can be significantly improved \cite{He2020,Jamali2021,MIMO_Wang2022};} 4) The distributed ultra-massive antenna system can deploy a large number of distributed antennas over a wide geographical area to build cell-free network, which is conducive to achieving consistent user experience, obtaining high SE, and reducing the transmission energy consumption of the system \cite{Chen2018,Zhang2019,Ammar2022}; 5) The introduction of AI for ultra-massive MIMO technology helps to achieve intelligence in multiple aspects such as channel estimation, channel sounding, beam management, and user detection. How to meet real-time requirements and obtain training data needs to be addressed \cite{Wei2021,Albreem2022}; 6) Ultra-massive MIMO is also expected to be combined with space-air-ground-sea integrated networks. It will bring great performance gain for a series of expanded application scenarios such as satellite communications\cite{Gao_new_3}, skywave communications\cite{Gao_new_1}, and underwater acoustic communications\cite{Gao_new_4}.

\subsubsection{IBFD}
\

Different from the commercially available frequency-division duplex (FDD) and time-division duplex (TDD), IBFD technology enables a radio to transmit and receive in the same frequency band at the same time, which can theoretically double the spectrum efficiency, expand wireless transmission capacity, and enable a more flexible and efficient network access. IBFD is one of the key technologies being explored for future wireless communications\cite{IMT-2030,UNISOC,Giordani2020_16}.

IBFD has a long history and has been used in the design of continuous wave radar systems since the middle of the last century \cite{IBFD_Hara1963}. However, due to technical limitations, there has been no further practical application. It is only in recent years that IBFD has reignited research interest. In \cite{IBFD_Liu2015}, IBFD relaying was investigated as a typical application. In \cite{IBFD_Liu2017}, a full-duplex technique was proposed to enable simultaneous communication between multiple ambient backscatter nodes. In \cite{IBFD_Yilan2019}, an IBFD architecture using monostatic antenna was presented and detailed laboratory tests were conducted to investigate its performance. In \cite{IBFD_Hanawal2020}, frequency hopping technique and an additional operation mode called transmission-detection were proposed to improve the throughput of IBFD nodes under jamming attacks. In \cite{IBFD_Hassani2021}, IBFD was applied in a radar-communication system enabling joint communication and opportunistic wireless sensing for the first time.

In order to make the practical application of IBFD possible, the primary challenge is to develop in-band self-interference cancellation (SIC) techniques with moderate implementation complexity and cost. SIC technique can be divided into electronic SIC and optical SIC. In \cite{IBFD_Kolodziej2019}, the electronic SIC techniques applied to sub-6 GHz were investigated in detail and the total isolation performance resulting from combining different approaches was analyzed. In \cite{IBFD_Nawaz2018}, a 2.4 GHz dual-polarized microstrip patch antenna with extremely high insertion isolation for IBFD transceivers based on a shared antenna structure was presented. In \cite{IBFD_Komatsu2020}, an iterative estimation and cancellation technique for nonlinear IBFD transceivers was presented. In \cite{IBFD_Komatsu2021}, a theoretical analysis technique for IBFD systems using parallel Hammerstein self-interference cancellers in digital-domain was developed. As the bandwidth increases, the difficulty of implementing self-interference suppression will gradually rise. There are more challenges in applying IBFD technology to the THz and OWC bands of 6G systems. OSIC has gradually attracted research interest because of its large bandwidth and high accuracy. In \cite{IBFD_Chang2018}, a full RF characterization of an integrated microwave photonic circuit for SIC was conducted as the first test for an ``RF-IN and RF-OUT" photonic integrated circuit. In \cite{IBFD_Wang2022}, a photonics-assisted frequency conversion and SIC approach was proposed and experimentally demonstrated for IBFD communication.

\subsection{New Physical Dimensions}

To cope with the booming development of massive IoT, 6G communication is expected to achieve higher data rate using the existing spectrum resources. 
In addition to relying on traditional air interface technologies, such as multiple antennas, modulation, coding, and duplexing, finding new physical dimensions and transmission carriers to achieve revolutionary breakthroughs can also help to improve SE further.
OAM, RIS, and holographic radio are the most promising ones among~them.

\subsubsection{OAM}
\

OAM is an inherent physical quantity of EM waves. It is a dimension in addition to frequency, phase and space, providing a new dimension for modulation in wireless communications. 
EM waves with OAM, also known as vortex EM waves, have an angular momentum phase wavefront instead of the traditional plane wavefront. Specific antennas are used to generate orthogonal modes, each associated with a different orbital angular momentum mode carrying different information. Thus multiple OAM modes can coexist and transmit data simultaneously over a single communication link. Taking advantages of the orthogonal characteristics of different OAM modes, it is promising to achieve a high spectrum-efficiency and increase the channel capacity without any additional frequency band \cite{IMT-2030,SEU,HUAWEI,Chen2020_19}.

Since Allen \emph{et al.} \cite{OAM_Allen1992} discovered that the optical vortex with a spiral wavefront can carry OAM in 1992, research on OAM has been extended to both radio and acoustic fields. R. Chen \emph{et al.} \cite{OAM_Chen2020} presented a summary of generation and detection of optical, radio, and acoustic OAM. Recently, OAM technology has attracted attention to explore new dimensions in not only mmWave band but also THz region \cite{OAM_Yang2021}.

OAM technology has huge potential to provide high data rates service in free-space optical, optical fiber, radio communications, as well as acoustic communication systems \cite{OAM_Yousif2019,OAM_Chen2020}. In addition, the combination of OAM and MIMO communication is promising to achieve higher capacity and SE \cite{HUAWEI}, which can be classified into two types\cite{OAM_Ref_book}. One is that the signal is transmitted from OAM antennas and received by conventional antennas, and vice versa. The diversity of OAM modes is utilized to decrease spatial correlation function, thus increasing capacity and SE. Theoretical and numerial results proved that the OAM-based MIMO system equals to the conventional MIMO system with larger element spacing, making it possible to bring higher SE\cite{OAM_Zheng2016}. In \cite{OAM_Lei2021}, an OAM-based MIMO communication system with two OAM modes was proposed and its throughput was improved upon traditional MIMO by up to 30.50\%. The performance of this kind of OAM-based MIMO system in the multipath scenario was analyzed in \cite{OAM_Chen22020}. It was found that in the small angular spread scenario, the capacity of OAM-based MIMO system was superior to that of the conventional MIMO system. Another is that the signal is transmitted and received both by OAM antennas. The most common OAM antenna configuration is the uniform circular array (UCA). Although the UCA-based OAM is a subset of the conventional MIMO\cite{OAM_Edfors2012} and thus its SE is still constrained by the upper bound of MIMO, it has low complexity and is effective in the line-of-sight (LoS) scenario due to orthogonality between OAM modes compared with the conventional MIMO system. Y. Yagi \emph{et al.} \cite{OAM_Yagi2021} demonstrated over 200 Gbps transmission using a dual-polarized OAM-based MIMO multiplexing with UCAs. Besides, OAM multiplexing can be combined with other multiplexing technologies.
T. Hu \emph{et al.} \cite{OAM_Hu2019} proposed a time-switched OFDM-OAM MIMO to achieve a very high sum-rate and spectrum efficiency with low computational complexity. Moreover, analysis in the multipath and misalignment scenario were performed as well. In \cite{OAM_Liang2020}, a hybrid orthogonal division multiplexing scheme with phase difference compensation, incorporating both OAM and OFDM, was proposed to achieve high capacity in sparse multipath environments. A. A. Amin \emph{et al.} \cite{OAM_Amin2020} integrated OAM-MIMO multiplexing system with NOMA, enhancing the channel capacities of the downlink for multiple users.
OAM can enhance radar techniques. Therefore, OAM-based sensing systems have been introduced as a new microwave-sensing technology \cite{OAM_Liu2019}.

A number of challenges remain before OAM technology can be implemented in practice. Beam divergence and misalignment severely reduce the transmission distance of OAM EM waves. Besides, reflection and refraction can destroy the orthogonality of OAM waves. Thus, it is still an open problem for OAM applications in NLoS scenarios. As a basis for system analysis, channel measurements and modeling are still lacking for OAM. For future commercialization, component process, antenna design, and signal processing are the key technical difficulties to be overcome. 

\subsubsection{RIS}
\
\label{sec:RIS}

RIS is a surface composed of a large number of programmable 2D meta-materials of sub-wavelengths, each of which is capable of dynamically, intelligently, and independently manipulating incident signal to obtain the expected reflected signal or transmission signal. It is thus expected to form EM fields with controllable amplitudes, phases, polarizations and frequencies to enhance the communication performance.
Compared with the transmitters with conventional structures, RIS technology is low-cost, low-energy, and easy to deploy, which can significantly increase network transmission rate, enhance signal coverage, and improve frequency, energy, and cost efficiency~\cite{IMT-2030,Chowdhury2020_6,Marco_Basar2019a,Marco_Liu2021h}.

RIS is one of the promising technologies for upcoming 6G networks, with promising applications in creating smart ratio environment\cite{Marco_DiRenzo2020, Marco_Renzo2019}, improving massive connectivity,  enhancing coverage, avoiding coverage holes, replacing relays\cite{Marco_DiRenzo2020a}, boosting cell-edge transmission rate, achieving green communication, assisting EM environment sensing or high precision positioning\cite{IMT-2030}, improving communication reliability\cite{Gui2020}, enhancing wireless body sensor networks\cite{Bariah2020_11}, and metasurfaces holographic technologies\cite{HUAWEI, Marco_Huang2020e}. More details of the applications of RIS can be found in \cite{IMT2030_RIS, Marco_Pan2021}.

In recent years, a lot of research for RIS has been carried out.
In 2011, Generalized Snell's law was proposed and the development of EM metasurface was greatly enriched\cite{RIS_Yu2011}. In 2014, T. J. Cui \emph{et al.} proposed the concept of coding metamaterials, digital metamaterials and programmable metamaterials\cite{RIS_Cui2014}.
A lot of channel research has been conducted for RIS and becomes a baseline for further theoretical studies and practical applications. In \cite{RIS_Tang2021}, three free-space path loss models for RIS-assisted wireless communications were developed for far-field, near-field and near-field broadcasting cases. In addition, experimental measurements were carried out to further validate the proposed models. A physical and widely applicable RIS channel model was proposed for mmWave frequencies and its corresponding open-source SimRIS Channel Simulator was introduced in \cite{RIS_Basar2021}. 
Numerous experiments and tests have been conducted to explore the capabilities of the RIS prototype system to improve system efficiency, increase throughput, enhance coverage, etc.
In \cite{RIS_Huang2019}, RIS-based resource allocation methods were developed for downlink multiuser multiple-input single-output (MISO) systems, which can provide up to 300\% higher EE compared to using conventional relaying.
E. Basar \emph{et al.} proposed the concept of RIS-assisted IM for massive MIMO wireless networks, having the potential to provide considerably high SE at low signal to noise ratio (SNR) \cite{NEW_Basar2020}.
The trade-off between EE and SE for RIS-assisted MIMO uplink communication systems was studied in \cite{RIS_You2021}. In \cite{RIS_Chen2021}, a dual-polarized RIS was proposed to realize low cost ultra-massive MIMO transmission architecture towards future networks.
By leveraging RIS in a downlink NOMA system, the rate performance was improved significantly \cite{RIS_Yang2021}.
In \cite{RIS_Huang2020}, advances in DRL were leveraged to optimize the joint design of transmit beamforming matrix at the base station and the phase shift matrix at the RIS.
The effectiveness of coverage enhancement of deploying RISs in a mmWave cellular network was clarified by M. Nemati \emph{et al.} \cite{RIS_Nemati2020}. In\cite{Marco_Sihlbom2021}, the system-level simulation was conducted to validate the fact that RIS could improve outdoor and indoor coverage and ergodic rate.
J. Yuan \emph{et al.} introduced multiple RISs to a downlink MISO CR system, increasing the achievable rate of secondary users significantly \cite{RIS_Yuan2021}.

Multiple technical challenges remain for the future development and wide application of RIS, such as constraints of hardware capabilities, baseband algorithms, architectures of wireless networks, and networking methods need to be considered. Also, the cost and energy consumption of RIS devices in high frequency band, as well as the deployment scale and methods of RIS systems still need further works.

\subsubsection{Holographic radio}
\

Holographic radio leverages the holographic interference of EM waves to achieve dynamic reconstruction of EM space with real-time precision control. Utilizing a spatially continuous microwave aperture, it is a new method to achieve spatial multiplexing. Holographic radio is able to meet the demands for ultra-high SE, ultra-high traffic density, ultra-high capacity. It contributes to the convergence of imaging, sensing and wireless communication to support the intelligence of EM space\cite{IMT-2030,Saad2020_12}. Holographic radio is also known as holographic MIMO, which refers to the ultimate form of multi-antenna system with finite aperture\cite{HOL_Bjornson2019,HUAWEI}.

Rather than viewing unwanted signals as a harmful phenomenon, holographic radio is promising to exploit the interference as a useful resource for enhancement of EE\cite{Zong2019_34}. Besides, holographic radio can obtain the RF spectral hologram of the RF transmitting sources by utilizing holographic interference imaging, thus can save the overhead in CSI or channel estimations\cite{OULU_BC}. In the near future, holographic radio will give full play to its potential in applications such as smart factories, high-precision positioning, precise wireless power supply, and data transmission for a massive number of IoT devices\cite{IMT-2030}.

The realization of continuous aperture antenna array is one of the most primary technical difficulties of holographic radio. Currently, there are two approaches to realize a continuous microwave aperture approximately. The first approach is to densely pack sub-wavelength unit cells  to realize continuous or quasi-continuous apertures, which is referred to as reconfigurable holographic surface (RHS)\cite{HOL_Deng2021,HOL_Wan2021}. Utilizing tightly coupled arrays of broadband active antennas is a more promising approach\cite{HUAWEI}, which relies on a high-power uni-traveling-carrier (UTC) photodetector (PD)-coupled antenna array. The UTC-PD is bonded to a photodiode-coupled array antenna, which has the advantages of low costs and low power consumption\cite{HOL_Konkol2017,Zong2019_34}.

How to develop holographic radio communication theory, how to establish reliable channel models, and how to perform low latency and high reliability data processing on the massive data generated by holographic radio systems are still open~problems.

\subsection{More Capabilities}

\textcolor{black}{In this part, technologies that may strengthen 6G systems in all levels will be introduced. AI, ISAC, and blockchain can provide 6G systems with new capabilities of intelligence, sensing, and security, respectively. Semantic communication will expand communication systems capabilities greatly, making intelligent connection of everything a reality.}

\subsubsection{AI}
\

Over the last decade, AI has developed rapidly and shown its overwhelming advantages in a vast array of industries. \textcolor{black}{ML and DL are important subsets of AI, able to learn and develop over time. AI technologies have high robustness, adaptive learning ability, and strong understanding and reasoning ability, which equip them with great application potential in many aspects, especially for scenarios where significant amounts of data are available for training.} 
At the same time, the communication system is developing at a high speed, enabling larger throughput, lower latency, greater number of connections, and more intelligent services. The introduction of new demands and technologies has led to an increase in the data volume and complexity of communication networks, imposing serious limitations on  traditional communication algorithms.
AI is expected to be applied to all layers of 6G networks. It will simplify network management and optimization, making communication systems more efficient and intelligent \cite{AI_You2019,IMT-2030,IMT2030-Wireless_AI_WP,SEU}. \textcolor{black}{A similar outlook for AI was expressed in\cite{Shen2022_DT} and the authors proposed a four-level AI architecture for pervasive network intelligence.}

AI will bring a disruptive change to the traditional air interface design. \textcolor{black}{In the physical layer, AI is employed in wireless channel research to enable the modeling and prediction of complicated channels based on a large number of propagation environment parameters\cite{AI_OShea2017a, AI_Huang2020,AI_Yin2022}.} Due to the high computational complexity of conventional channel estimation techniques, there are many efforts attempting to perform channel estimation and signal detection utilizing AI\cite{AI_Ye2018,AI_Gao2022}. Besides, AI can also be applied to E2E transceiver design\cite{AI_Dorner2018,AI_Cammerer2020}, channel encoding and decoding techniques\cite{AI_Farsad2018,AI_Buchberger2021}, as well as modulation and waveform design\cite{AI_Zhao2021,AI_Aoudia2021}. AI has the ability to efficiently extract and express large dimensional feature space. Therefore, its application in the field of MIMO technology, where the antenna dimension has grown significantly, has also received a lot of attention\cite{AI_Gao2017,AI_Elbir2020}. In the medium access control (MAC) layer, AI is widely used for active user detection\cite{AI_ZhangZ2019,AI_Kim2020}, access control\cite{AI_Zhu2018}, and wireless link scheduling\cite{AI_Cui2019}. In addition, MAC protocols are expected to be designed automatically thanks to recent developments in deep multiagent reinforcement learning \cite{AI_Hoydis2021}. \textcolor{black}{A deep neural network-based transmission scheduling scheme was introduced in \cite{Sherman_new_4}.} \textcolor{black}{What's more, AI can play a part in achieving higher wireless positioning accuracy \cite{CJIT_Wang2021e}.}

In the upper layers of networks, AI is expected to be more widely used and bring disruptive changes to the network architecture, leading to pervasive network intelligence\cite{AI_Sun2019,Shen2022_DT, 6GANA}. First, AI can be used to design efficient resource allocation and interference management schemes. In \cite{AI_Sun2018} and \cite{AI_Liang2020}, deep neural network was utilized to solve the interference management problem, achieving better power control results while requiring less computational resources. In \cite{SPE_Hu2021}, a blockchain and AI-empowered dynamic resource sharing architecture was proposed to implement efficient pattern recognition and decision-making. In addition, as data traffic increases, network traffic control becomes an important issue. AI can be used to predict the network traffic, and then, the corresponding resource control algorithms are adopted to reduce congestion. In \cite{AI_Fu2018}, ML algorithms for traffic management of 5G networks were discussed. In \cite{AI_Zhang2019}, main AI algorithms applied to transport layer congestion control were summarized. Besides, AI can be used for network demand prediction and caching, helping to reduce latency and decrease operational costs\cite{AI_Jiang2019}. In the face of the explosive growth of application data, the AI-based distributed computing technology can further release the potential of computing and data resources of edge nodes\cite{AI_Qi2021}. In addition to applying AI to solve specific problems at each layer of the network, an intelligent-endogenous network architecture based on DL and knowledge graph was proposed\cite{AI_Zhou2022}. The network can automatically change in response to new service requirements, which is expected to further unlock the scalability, iterative enhancement, and model generalization application capabilities of AI. \textcolor{black}{Besides, a dynamic RAN slicing scheme based on a two-layer constrained reinforcement learning was introduced in \cite{Sherman_new_7}.}

A number of issues need to be addressed in the progress of applying AI to air interface design. AI cannot completely replace conventional methods and there is a need to identify the application area of learning methods \cite{AI_Zhang2020a}. Besides, future research needs to explore suitable AI learning techniques for specific problems \cite{AI_Jiang2017}. Challenges for network AI will be discussed later.

\subsubsection{ISAC}
\

\textcolor{black}{As one of the six visions of the 6G communication systems, ``Full Applications" puts forward the need for the integration of communication network and perceptive network.} In a broad sense, the perceptive network refers to a system that can perceive the attributes and states of all services, networks, users, terminals, and environmental objects \cite{IMT2030_ISAC}. ISAC is an important supporting technology for realizing 6G integrated network. Perception and communication systems are integrated to efficiently utilize congested wireless resources and/or hardware resources, and mutually assist each other to improve the efficiency of hardware, spectrum, time, and energy~\cite{ISAC_Cui2021}.

Over the past few decades, communication and perceptive technologies have been developed in parallel, and the two systems are relatively independent. However, with the development of communication technologies, the perceptive system and the communication system are more and more coupled. They both tend to utilize consistent high-frequency and large-aperture antennas, and are expected to use similar signals and data processing methods \cite{ISAC_Zhang2021}.

The emergence of the concept of ISAC can date back to the 1960s. In \cite{ISAC_Mealey1963}, R. M. Mealey \emph{et al.} used coded pulses to transmit information from ground-based radars to spacecraft. However, there were few further developments afterwards. Recently, with the development of related technologies, ISAC has received widespread attention. A large number of theoretical designs and system implementations of related technologies have been carried out by domestic and foreign scholars. C. Sturm \emph{et al.} discussed the waveform design and signal processing for ISAC and implemented the first ISAC system utilizing OFDM waveforms \cite{ISAC_Sturm2011}. The first information theoretical analysis of ISAC was conducted by A. R. Chiriyath \emph{et al.} \cite{ISAC_Chiriyath2016}. An ISAC implementation for vehicle communication scenarios was realized in \cite{ISAC_Kumari2018}. 
F. Liu \emph{et al.} provided an overview of the applications, research status, and future directions of ISAC design \cite{ISAC_Liu2020}. Focusing on signal processing, J. A. Zhang \emph{et al.} summarized the related technologies of ISAC \cite{ISAC_Zhang2021_signal}. Y. Cui \emph{et al.} provided a general survey of the progress of ISAC, listed the use cases of ISAC under the IoT architecture, and introduced several challenges and opportunities for future development \cite{ISAC_Cui2021}. \textcolor{black}{In \cite{ISAC_Ma2020}, joint radar-communication strategies for autonomous vehicle were surveyed. In \cite{SCIS_Xiao2022}, an overview research on integrated localization and communication was proposed.} What's more, IEEE 802.11bf was established in 2020, focusing on wireless local area network (WLAN) sensing. \textcolor{black}{Besides, 3GPP SA1 started a study item on ISAC in March 2022.}

\textcolor{black}{Research challenges include building high-precision ISAC measurement equipment, designing reasonable measurement scenarios, selecting efficient transmission frequency bands, evaluating the correlation between sensing channels and communication channels, and establishing accurate ISAC channel models. On this basis, it is necessary to consider how to integrate both communication and sensing requirements in terms of hardware architecture, system design, waveform design, and anti-jamming signal processing on the basis of avoiding interference and collisions\cite{IMT2030_ISAC}.}

\subsubsection{Blockchain}
\

Blockchain technique was first proposed for cryptocurrency in 2008 \cite{BC_Nakamoto2008}. Using distributed databases connected by hash pointers, blockchain has the characteristics of decentralization, transparency, anonymity, immutability, traceability, and resiliency \cite{BC_Wang2021}. Because of the flatter structure and more frequent data transformation in 6G network, the traditional centralized security authentication and access control mechanisms will no longer be fully applicable \cite{BC_Li2020}. Blockchain provides a promising solution to the trust-and security-related issues among distributed and heterogeneous network devices and infrastructures, and, is considered as one of the essential technologies for 6G communication systems \cite{IMT-2030,SEU,Chowdhury2020_6}. In 2018 Mobile World Congress Americas (MWCA), FCC outlined their vision for deploying blockchain in future 6G networks. In \cite{BC_Ling2019} a blockchain-RAN (B-RAN) architecture was proposed for decentralized secure radio access.

A series of studies on the application of blockchain in the field of communication are underway. Because of its characteristics of tamper resistance, decentralization, fine-grained auditability, and imbedded asymmetric encryption, blockchain is expected to provide both the data security and privacy protection. In \cite{BC_Aitzhan2018}, a blockchain-based distributed domain name system was proposed to defend against DDoS attacks. In \cite{BC_Wan2019}, blockchain was used to privacy protection of identity and confidential data of users in wireless networks. 
Besides, blockchain can promote more efficient resource sharing in a wide range of separated network entities. In \cite{BC_Weiss2019}, blockchain-based spectrum management was proposed to provide secure and highly efficient decentralized spectrum sharing. In \cite{BC_Backman2017}, blockchain was used in network slicing to promote slice leasing for the first time. In \cite{BC_Liu2021}, blockchain-empowered MEC was proposed to guarantee the security and traceability of computing and storage capacities sharing. \textcolor{black}{In \cite{BC_Shen2022}, blockchain solutions were investigated to balance transparency, efficiency, and privacy requirements in decentralized data management. }
Furthermore, blockchain plays an important role in massive IoT, ensuring the security and data privacy, traceability of massive data, and interoperability across devices. In \cite{BC_Ray2021}, blockchain and IoT were combined to store, access, and manage real-time sensory data from patients in a secure and efficient way. In \cite{BC_Kouicem2022}, a hierarchical and scalable blockchain-based trust management protocol was proposed for IoT systems, which was proved to be superior to existing solutions in terms of scalability, mobility support, communication, and computation costs. In \cite{BC_Mollah2021}, the adoption of blockchain for supporting the information exchange in internet of vehicles (IoV) was comprehensively surveyed. In \cite{BC_Hui2022}, a blockchain-based collaborative crowdsensing (BCC) scheme was proposed for IoV, promoting the security and efficiency of autonomous vehicles crowdsensing. 
In the field of smart grid, blockchain can be applied to promote privacy-preserving and efficient data aggregation \cite{BC_Zhuang2021}. 
One last application to note is the federated-style learning, which is inherently supported by blockchain. In \cite{BC_Lu2021}, a blockchain empowered federated learning framework was proposed to share intelligence in an open and compatible manner, thus improving the efficiency and security of digital twin wireless networks.

Although blockchain is expected to facilitate the development of 6G networks in various aspects, its decentralized characteristics will cause problems, such as relatively long latency, inefficient storage, and limited  throughput performance. Thus, it may only be useful for certain application scenarios. Besides, the current underlying technology platforms of blockchain show a fragmented state, and there is a need to establish unified standardization and regulation to promote the integration of individual systems.

\subsubsection{Semantic communication}
\

Semantic communication is a communication method in which semantic information is extracted from a source and encoded for transmission in a noisy channel. Rather than requiring error-free transmission at the bit level, semantic communication relies primarily on building a semantic knowledge base that is pervasive and comprehensible among a large number of human users and machines. It is expected to break through the bottleneck of transmission in classical communication systems and liberate communication networks from the traditional architecture based on data protocols and formats. Semantic communication is promising to further improve communication efficiency and reliability, enhance the quality of human-oriented services, and realize the true seamless intelligent connection of everything \cite{IMT2030_Network,SEM_Shi2021}.

\textcolor{black}{In 1948, classic information theory was proposed by Shannon \cite{SEM_Shannon1948}. Later, Weaver indicated that semantic problem concerns how precisely the transmitted symbols can convey the desired meaning, inspiring thinking and research of semantic information \cite{SEM_Shannon1949}.} The concept of semantic information theory was proposed in \cite{SEM_Bar-Hillel1953} and was further refined in\cite{SEM_Barwise1981,SEM_Floridi2004,SEM_Bao2011}. In \cite{SEM_Zhong2017_1}, the key point of semantic information theory was considered to be the understanding of content and the ability of logical deduction. In \cite{SEM_Zhong2017_2}, it was proved that the representation of semantic information is unique. In \cite{SEM_Shi2021}, the limitations of classic point-to-point semantic communication were analyzed, and a resource-efficient semantic-aware networking architecture based on federated edge intelligence was proposed to reduce resource consumption and improve communication efficiency. \textcolor{black}{In \cite{Xu2021_Security}, a new semantic representation framework was proposed to set up an intelligent and efficient semantic communication network architecture, which had a lower bandwidth requirement, less redundancy, and more accurate intent identification.}

Recently, semantic communications has been utilized in E2E communication systems to address the bottlenecks in traditional block communication systems. Several related studies have considered different types of sources, focusing on image and text transmission \cite{SEM_O'Shea2017,SEM_Guler2018,SEM_Xie2021_DL,SEM_Xie2021_IOT}. Semantic communication can also be applied in speech signal processing to convert speech signals into textual information in automatic speech recognition, where the characteristics of the speech signal, such as speech rate and intonation, are not concerned~\cite{SEM_Dahl2012}.

At present, semantic communication technology is still under rapid development, and many basic concepts need to be developed and improved\cite{IMT2030_Network,SEM_Shi2021}. First, the accuracy of semantic fundamental elements determines the reliability in practical applications and is the most fundamental issue in semantic communication. Second, how to design an effective semantic error-tolerant and error-correcting mechanism remains unknown. Moreover, a simple and versatile solution for fast semantic information detection and processing that can be implemented in resource-limited devices is still lacking. Semantic information models between different entities are difficult to share, exacerbating the challenge of adopting semantic communication in communication systems. Besides, semantic communication also puts forward completely new requirements for network security, which requires the establishment of a perfect censorship mechanism to prevent the semantic knowledge graph from being maliciously tampered and a safe and reliable storage and a recall mechanism to prevent the leakage of user's private information.

\subsubsection{Energy-neutral devices}
\

Passive (energy-neutral) devices rely on power harvested from the ambient environment to support their operation. The most important category is devices that harvest RF power, either from specific power transmitters or from ambient signals. Typically, these devices comprise a capacitor that is charged by the incoming RF field and communicate via backscattering. With backscattering, there is no RF chain, but the devices modulate the load impedance of their antenna to change the backscattered field; this change is then detected by the network node communicating with the device \cite{kimionis2014increased, stockman1948communication,kimionis2021printed, zawawi2018multiuser, duan2017achievable , fasarakis2015coherent, CJIT_Xu2021a}. From an electronics viewpoint, a load-modulated backscattering antenna is similar to an atom in a RIS (see Section~\ref{sec:RIS}). 

Energy-neutral devices will enable a host of new applications, from massive sensor telemetry to tracking of goods and people in factories, hospitals, and smart cities, for instance. Particularly, building and deploying massive amount of electronics without batteries is attractive, since no battery charging or replacement is required, and since toxic materials required for conventional batteries can be avoided. 

One of the main challenges with backscattering communication is the link budget. With $\beta$ being the path gain from the network node to the device, the total path gain from the network node to the device and back scales as $\beta^2$. To overcome the high path loss, either a directional antenna must be used (at the network node and/or the device), or one must rely on antenna arrays at the network side \cite{kashyap2016feasibility, boyer2013backscatter, yang2015multi , mishra2019optimal }. Massive MIMO, the core physical-layer technology in 5G and 6G, enables an array gain proportional to the number of antennas both in receive and transmit mode. When operating in reciprocity-based (TDD) mode, this scaling holds even with estimated channels \cite{marzetta2016fundamentals}. Exploiting this insight, massive MIMO technology is viewed as a fundamental enabler for communication with backscattering devices and technology is currently being developed, for example, in the European H2020-REINDEER project \cite{REINDEER}. Another challenge is the cost. More devices are deployed in large scale so that the overall cost will be magnified if the individual device takes on a high~cost.

\section{6G Testbeds}
\textcolor{black}{In Section IV--V, we have introduced a number of potential network architectures and technologies. Currently, standardization organizations have carried out a series of forward-looking works. We have investigated study items and work items in 3GPP (R18-R19), IEEE standards, technical groups in ETSI, as well as study groups in ITU-T and ITU-R\cite{3GPP_Study, 3GPP_Work, IEEE_Standard, ETSI_Groups, ITU_Homepage}. Details of the standardization efforts on the 6G network architecture and potential technologies are summarized in Table~\ref{table_stan_net} and Table~\ref{table_stan_tech}, respectively. The summary also provides readers with an overview of the standardization process. However, several technologies have not been investigated by standardization organizations and not been summarized in the tables. In order to accelerate the technical research as well as the subsequent standardization work, studying 6G testbeds is one of the indispensable tasks.}

\textcolor{black}{To investigate the developments of testbeds used to explore channel characteristics, verify the key technologies, and evaluate the whole communication system, different 6G testbeds have been given for different purposes in this section.} We will focus on three types of such testbeds, testbeds for 6G channels, testbeds for 6G key technologies, and testbeds for comprehensive 6G~system~verification.

\renewcommand{\arraystretch}{1.2}
\begin{table*}[!t]
	\centering
	\caption{\textcolor{black}{Standardization of 6G network architecture\cite{3GPP_Study, 3GPP_Work, IEEE_Standard, ETSI_Groups, ITU_Homepage}.}}
\begin{tabular}{|ll|ll|l|lll|ll|}
	\hline
	\multicolumn{2}{|c|}{\multirow{2}{*}{\textbf{Network architecture}}}                                                                                       & \multicolumn{2}{c|}{\textbf{3GPP (R18-R19)}}                                 & \multicolumn{1}{c|}{\textbf{IEEE}} & \multicolumn{3}{c|}{\textbf{ETSI}}                                                     & \multicolumn{2}{c|}{\textbf{ITU}}                         \\ \cline{3-10} 
	\multicolumn{2}{|c|}{}                                                                                                                            & \multicolumn{1}{c|}{Study Items} & \multicolumn{1}{c|}{Work Items} & \multicolumn{1}{c|}{Standard}      & \multicolumn{1}{c|}{OSG}     & \multicolumn{1}{c|}{TC}      & \multicolumn{1}{c|}{ISG} & \multicolumn{1}{c|}{ITU-T}   & \multicolumn{1}{c|}{ITU-R} \\ \hline
	\multicolumn{1}{|l|}{\multirow{4}{*}{5G Enhancement}}           & NFV                                                                                        & \multicolumn{1}{l|}{$\surd$}          & $\surd$                              & $\surd$                            & \multicolumn{1}{l|}{}        & \multicolumn{1}{l|}{}        & $\surd$                  & \multicolumn{1}{l|}{$\surd$} &                            \\ \cline{2-10} 
	\multicolumn{1}{|l|}{}                               & SDN                                                                                        & \multicolumn{1}{l|}{}                 &                                      & $\surd$                            & \multicolumn{1}{l|}{$\surd$} & \multicolumn{1}{l|}{}        &                          & \multicolumn{1}{l|}{$\surd$} &                            \\ \cline{2-10} 
	\multicolumn{1}{|l|}{}                               & SBA                                                                                        & \multicolumn{1}{l|}{$\surd$}          & $\surd$                              &                                    & \multicolumn{1}{l|}{}        & \multicolumn{1}{l|}{}        &                          & \multicolumn{1}{l|}{$\surd$} &                            \\ \cline{2-10} 
	\multicolumn{1}{|l|}{}                               & Network slicing                                                                            & \multicolumn{1}{l|}{$\surd$}          & $\surd$                              &                                    & \multicolumn{1}{l|}{}        & \multicolumn{1}{l|}{}        &                          & \multicolumn{1}{l|}{$\surd$} &                            \\ \hline
	\multicolumn{1}{|l|}{\multirow{4}{*}{\begin{tabular}[c]{@{}l@{}}3D multi-network\\ integration\end{tabular}}} & Satellite                                                                                  & \multicolumn{1}{l|}{$\surd$}          & $\surd$                              &                                    & \multicolumn{1}{l|}{}        & \multicolumn{1}{l|}{$\surd$} &                          & \multicolumn{1}{l|}{$\surd$} & $\surd$                    \\ \cline{2-10} 
	\multicolumn{1}{|l|}{}                               & Maritime                                                                                   & \multicolumn{1}{l|}{}                 &                                      &                                    & \multicolumn{1}{l|}{}        & \multicolumn{1}{l|}{}        &                          & \multicolumn{1}{l|}{$\surd$} &                            \\ \cline{2-10} 
	\multicolumn{1}{|l|}{}                               & UAV                                                                                        & \multicolumn{1}{l|}{$\surd$}          & $\surd$                              & $\surd$                            & \multicolumn{1}{l|}{}        & \multicolumn{1}{l|}{}        &                          & \multicolumn{1}{l|}{$\surd$} &                            \\ \cline{2-10} 
	\multicolumn{1}{|l|}{}                               & (IoT/NR) NTN                                                                               & \multicolumn{1}{l|}{$\surd$}          & $\surd$                              &                                    & \multicolumn{1}{l|}{}        & \multicolumn{1}{l|}{}        &                          & \multicolumn{1}{l|}{}        &                            \\ \hline
	\multicolumn{1}{|l|}{\multirow{3}{*}{Secure and trustworthy}}        & Blockchain                                                                                 & \multicolumn{1}{l|}{}                 &                                      & $\surd$                            & \multicolumn{1}{l|}{}        & \multicolumn{1}{l|}{}        &                          & \multicolumn{1}{l|}{$\surd$} &                            \\ \cline{2-10} 
	\multicolumn{1}{|l|}{}                               & \begin{tabular}[c]{@{}l@{}}Quantum communication/ \\Quantum key distribution\end{tabular} & \multicolumn{1}{l|}{}                 &                                      & $\surd$                            & \multicolumn{1}{l|}{}        & \multicolumn{1}{l|}{}        & $\surd$                  & \multicolumn{1}{l|}{$\surd$} &                            \\ \cline{2-10} 
	\multicolumn{1}{|l|}{}                               & Enhancement of network security                                                            & \multicolumn{1}{l|}{$\surd$}          & $\surd$                              & $\surd$                            & \multicolumn{1}{l|}{}        & \multicolumn{1}{l|}{$\surd$} &                          & \multicolumn{1}{l|}{$\surd$} &                            \\ \hline
	\multicolumn{1}{|l|}{\multirow{4}{*}{Integration of CSC}}          & \begin{tabular}[c]{@{}l@{}}Multi-layer ubiquitous\\ computing network\end{tabular}         & \multicolumn{1}{l|}{}                 &                                      &                                    & \multicolumn{1}{l|}{}        & \multicolumn{1}{l|}{}        &                          & \multicolumn{1}{l|}{$\surd$} &                            \\ \cline{2-10} 
	\multicolumn{1}{|l|}{}                               & Cloud computing                                                                            & \multicolumn{1}{l|}{}                 &                                      & $\surd$                            & \multicolumn{1}{l|}{}        & \multicolumn{1}{l|}{}        &                          & \multicolumn{1}{l|}{$\surd$} &                            \\ \cline{2-10} 
	\multicolumn{1}{|l|}{}                               & Edge computing                                                                         & \multicolumn{1}{l|}{$\surd$}          & $\surd$                              & $\surd$                            & \multicolumn{1}{l|}{}        & \multicolumn{1}{l|}{}        & $\surd$                  & \multicolumn{1}{l|}{$\surd$} &                            \\ \cline{2-10} 
	\multicolumn{1}{|l|}{}                               & Fog computing                                                                              & \multicolumn{1}{l|}{}                 &                                      & $\surd$                            & \multicolumn{1}{l|}{}        & \multicolumn{1}{l|}{}        &                          & \multicolumn{1}{l|}{}        &                            \\ \hline
	\multicolumn{1}{|l|}{Natively intelligent}                   & \begin{tabular}[c]{@{}l@{}}Enhancement of \\ network intelligence\end{tabular}             & \multicolumn{1}{l|}{$\surd$}          & $\surd$                              &                                    & \multicolumn{1}{l|}{}        & \multicolumn{1}{l|}{}        & $\surd$                  & \multicolumn{1}{l|}{$\surd$} &                            \\ \hline
	\multicolumn{10}{|l|}{\begin{tabular}[l]{@{}l@{}} *OSG: open source groups, TC: technical comittee, ISG: industry specification group, CSC: communications, sensing, and computing. \end{tabular}} \\ \hline
\end{tabular}
	\label{table_stan_net}
\end{table*}

\renewcommand{\arraystretch}{1.2}
\begin{table*}[!t]
	\centering
	\caption{\textcolor{black}{Standardization of 6G potential technologies\cite{3GPP_Study, 3GPP_Work, IEEE_Standard, ETSI_Groups, ITU_Homepage}.}}
\begin{tabular}{|l|ll|l|lll|ll|}
	\hline
	\multirow{2}{*}{\textbf{Technology}} & \multicolumn{2}{c|}{\textbf{3GPP (R18-R19)}}                                                                                                         & \multicolumn{1}{c|}{\textbf{IEEE}} & \multicolumn{3}{c|}{\textbf{ETSI}}                                            & \multicolumn{2}{c|}{\textbf{ITU}}                                                      \\ \cline{2-9} 
	& \multicolumn{1}{c|}{Study Items}                                                                                   & \multicolumn{1}{c|}{Work Items} & \multicolumn{1}{c|}{Standard}      & \multicolumn{1}{c|}{OSG} & \multicolumn{1}{c|}{TC} & \multicolumn{1}{c|}{ISG} & \multicolumn{1}{c|}{ITU-T}   & \multicolumn{1}{c|}{ITU-R}                              \\ \hline
	THz                                  & \multicolumn{1}{l|}{}                                                                                              &                                 & $\surd$                            & \multicolumn{1}{l|}{}    & \multicolumn{1}{l|}{}   & $\surd$                  & \multicolumn{1}{l|}{}        &                                                         \\ \hline
	OWCs                                 & \multicolumn{1}{l|}{}                                                                                              &                                 & $\surd$                            & \multicolumn{1}{l|}{}    & \multicolumn{1}{l|}{}   &                          & \multicolumn{1}{l|}{}        & \begin{tabular}[c]{@{}l@{}}$\surd$\\ (VLC)\end{tabular} \\ \hline
	Spectrum management                  & \multicolumn{1}{l|}{}                                                                                              &                                 & $\surd$                            & \multicolumn{1}{l|}{}    & \multicolumn{1}{l|}{}   &                          & \multicolumn{1}{l|}{}        &                                                         \\ \hline
	Ultra-massive MIMO                   & \multicolumn{1}{l|}{\begin{tabular}[c]{@{}l@{}}$\surd$\\ (NR MIMO evolution for downlink and uplink)\end{tabular}} & $\surd$                         &                                    & \multicolumn{1}{l|}{}    & \multicolumn{1}{l|}{}   &                          & \multicolumn{1}{l|}{}        &                                                         \\ \hline
	RIS                                  & \multicolumn{1}{l|}{}                                                                                              &                                 &                                    & \multicolumn{1}{l|}{}    & \multicolumn{1}{l|}{}   & $\surd$                  & \multicolumn{1}{l|}{}        &                                                         \\ \hline
	AI                                   & \multicolumn{1}{l|}{$\surd$}                                                                                       & $\surd$                         & $\surd$                            & \multicolumn{1}{l|}{}    & \multicolumn{1}{l|}{}   & $\surd$                  & \multicolumn{1}{l|}{$\surd$} &                                                         \\ \hline
	ISAC                                 & \multicolumn{1}{l|}{$\surd$}                                                                                       & $\surd$                         &                                    & \multicolumn{1}{l|}{}    & \multicolumn{1}{l|}{}   &                          & \multicolumn{1}{l|}{}        &                                                         \\ \hline
	Blockchain                           & \multicolumn{1}{l|}{}                                                                                              &                                 & $\surd$                            & \multicolumn{1}{l|}{}    & \multicolumn{1}{l|}{}   &                          & \multicolumn{1}{l|}{$\surd$} &                                                         \\ \hline
	Semantic communication               & \multicolumn{1}{l|}{}                                                                                              &                                 &                                    & \multicolumn{1}{l|}{}    & \multicolumn{1}{l|}{}   &                          & \multicolumn{1}{l|}{$\surd$} &                                                         \\ \hline
	Energy-neutral devices               & \multicolumn{1}{l|}{\begin{tabular}[c]{@{}l@{}}$\surd$\\ (Study on network energy savings for NR)\end{tabular}}    & $\surd$                         &                                    & \multicolumn{1}{l|}{}    & \multicolumn{1}{l|}{}   &                          & \multicolumn{1}{l|}{}        &                                                         \\ \hline
	\multicolumn{9}{|l|}{\begin{tabular}[l]{@{}l@{}} *OSG: open source groups, TC: technical comittee, ISG: industry specification group. \end{tabular}} \\ \hline
\end{tabular}
	\label{table_stan_tech}
\end{table*}

\subsection{Testbeds for 6G Channels}
Channel characterization, measurements, and modeling are the foundations of system design, theoretical analysis, performance evaluation, optimization, and deployment. Any new frequency bands, new scenarios, and new communication technologies will lead to new channel characteristics. Therefore, with the development and evolution of communication systems, channel research has played an indispensable role at all levels of the communication systems.
In general, the wireless signal includes large-scale fading and small-scale fading. Large-scale fading includes path loss and shadowing, while small-scale fading includes multipath fading. Large-scale fading plays an important role in wireless networking, including link budget calculation, network planning, optimization, and resource allocation, while small-scale fading is mainly utilized in wireless transmission, such as channel estimation, modulation, and coder design. Other applications, such as channel capacity analysis, localization, and positioning, consider both large-scale and small-scale fading.

With the emergence of new application scenarios and new technologies, 6G channels will have new characteristics. Therefore, it is essential to provide future 6G research with applicable channel testbeds. 
Existing testbeds for channels are divided into software channel simulators and hardware channel sounders. The latest developments will be introduced as follows.
\subsubsection{6G Pervasive Channel Simulator}
Channel simulators aim to simulate the propagation characteristics of real channel environments by using different channel models. The most critical issue for 6G channel modeling is to fully consider the characteristics of 6G all-spectra and all-scenario channels. A pervasive 6G wireless channel modeling theory was proposed by Wang \emph{et al.} in \cite{P6GCM}.
Using a unified channel modeling framework and method, the pervasive 6G wireless channel modeling theory adopts a unified channel impulse responses (CIR) expression and comprehensively considers integrating statistical properties of 6G channels for all spectra and all scenarios. Based on this, a 6G pervasive channel model applicable to 6G all-spectra and all-scenarios is constructed to accurately reflect the characteristics of different channels.

\begin{figure*}[tb]
	\centerline{\includegraphics[width=0.9\textwidth]{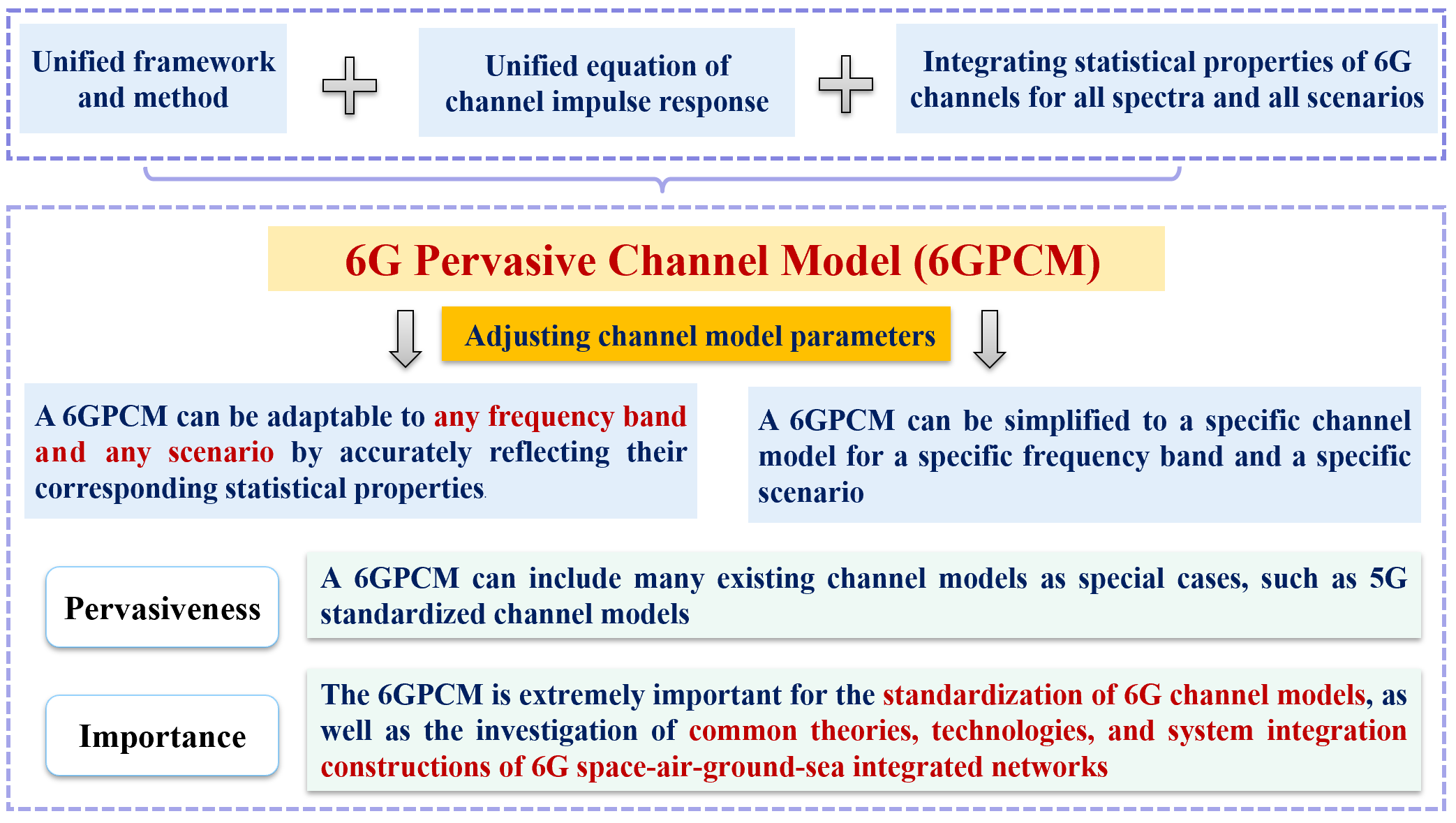}}
	\caption{A pervasive wireless channel modeling theory\cite{P6GCM}.}
	\label{fig_Channel_pervasive_method}
\end{figure*}

In Fig. \ref{fig_Channel_pervasive_method}, the 6G pervasive wireless channel modeling theory is illustrated,
which can be used to guide the construction of standardized 6G channel models. In terms of channel modeling framework and method, non-predictive models such as the geometry-based stochastic model (GBSM) and ray tracing (RT), as well as AI/ML-based predictive models can be adopted. 5G and previous standardized channel models such as 3GPP TR 38.901 and IMT-2020 mostly adopt the GBSM modeling method. It is expected that the standardized 6G channel models will mainly adopt the GBSM modeling method for the same reasons, while RT and AI/ML modeling methods can be used for individual frequency bands and scenarios as supplements.

Guided by the 6G pervasive channel modeling theory, a 6G pervasive channel model (6GPCM) based on a GBSM framework was first proposed in \cite{P6GCM}. It aims to construct a benchmark for 6G pervasive GBSM, which is expected to serve as a pioneer in the exploration of 6G standardized channel models. The proposed model is generally suitable for all spectra including sub-6 GHz, mmWave, THz, IR and VLC, global-coverage scenarios including low-earth-orbit (LEO) satellite, UAV, and maritime communication, and full-application scenarios such as ultra-massive MIMO, IIoT, and RIS channels in 6G communication systems. Parameter database for all spectra and scenarios can be constructed by using channel measurement fitting from existing standardized channel model documents. Considering the cost and limitations of channel measurements, RT and AI/ML-based methods can also be chosen to achieve channel parameter acquisition for different spectra and scenarios to complete the channel model parameter database. 
By adjusting the parameters of the channel model, the 6GPCM can be simplified to simulate channels for specific spectra and scenarios as special cases. By analyzing the 6GPCM, the complex mapping relationship among channel model parameters, channel characteristics and communication system performance can be investigated thoroughly. The 6GPCM is extremely important for the standardization of 6G channel models, as well as the investigation of theories, technologies, and system integration for the 6G space-air-ground-sea integrated network.

\subsubsection{Channel Sounders}
As a way to actively recognize channels, channel sounders are of great importance for establishing 6G all-spectra, global-coverage, and full-applications standardized channel models. A channel sounder is usually composed of a Tx, a Rx, and a data acquisition unit, which is a channel measurement platform to exploit the characteristics of unknown propagation environments \cite{Huang2019_mmWave}. In order to evaluate the performance of channel sounders, some key properties are defined, e.g., bandwidth, delay range, channel snapshot (CS) repetition rate, and dynamic range \cite{Nielsen2018}. First, the bandwidth and delay resolution are reciprocals of each other. The larger the bandwidth is, the higher the delay resolution will be, but usually the higher the cost or complexity of equipment as well. Second, the delay range refers to the maximum propagation distance detected by the channel sounder in the dynamic range. In general, the delay range required for indoor scenarios is relatively small, about a few hundred nanoseconds, while many microseconds for outdoor scenarios. Third, the CS repetition rate is an important indicator of channel measurements in mobile scenarios. Based on the Nyquist sampling theorem, the signal can be recovered perfectly when the sampling rate is higher than twice of the signal bandwidth. The bandwidth of a mobile channel equals to the maximum Doppler frequency, so the CS repetition rate needs to be greater than twice the maximum Doppler frequency. Last, the dynamic range is calculated by the difference between the strongest identifiable multipath component (MPC) power and the noise floor \cite{Karttunen2019}. The dynamic range can be improved by averaging over many CSs and increasing the length of probe waveform.

\begin{table*}[t]
	\centering
	\caption{Summary of 6G channel sounders for measurements.}
    \renewcommand\arraystretch{1.1}
    \scalebox{0.8}{
	\begin{tabular}{|m{1.5cm}<{\centering}|m{0.8cm}<{\centering}|m{5cm}<{\centering}|m{3cm}<{\centering}|m{4cm}<{\centering}|m{5cm}<{\centering}|}
		\hline
	\textbf{Category}	& \textbf{Ref.}  & \textbf{Organization} & \textbf{Frequency and (bandwidth)} & \textbf{Scenario/Application}  & \textbf{Antenna}  \\ \hline
		\multirow{10}{*}{ \rotatebox{270}{\textbf{All spectra}}}
        &\cite{sub-6_He2016}  &Beijing Jiaotong University & 3.5 GHz (30 MHz) &  Outdoor mobile scenarios & Tx\&Rx: single antenna  \\ \cline{2-6}
        &\cite{Sub-6_Chen2022}  &\textcolor{black}{SEU} & 0.7 GHz, 2.3 GHz, and 3.7 GHz (100 MHz) &  Urban Macro & Tx: single antenna;         Rx: 4×4 dual-polarization planar antenna array (0.7 GHz)        8×4 dual-polarization planar antenna array (2.3 GHz and 3.7 GHz)          \\ \cline{2-6}
        &\cite{sub-6_Zhang2022}  &\textcolor{black}{SEU} & 5.5 GHz (320 MHz) &  Indoor Office & SISO: single antennas; MIMO: Tx: 4$\times$4 dual-polarization planar antenna array,        Rx: 4$\times$8 dual-polarization cylindrical antenna array                 \\ \cline{2-6}
		&\cite{sub-6_Zhang2016}  &BUPT  & 3.5 GHz (100 MHz)  &Indoor (static, mobile)  & Tx: 32 antenna elements;  Rx: 56 antenna elements \\ \cline{2-6}
		&\cite{mmWave_Zhou2017}  &Tsinghua University & 28 GHz (1 GHz)   &Indoor office  &Virtual antenna array \\ \cline{2-6}
		&\cite{mmWave_Li2019}  & SEU & 40 GHz (500 MHz)  &Indoor and outdoor scenarios  &Virtual antenna array  \\ \cline{2-6}
		&\cite{THz_Abbasi2020}  &University of Southern California  & 140-220 GHz & Indoor office & Tx\&Rx: a horn antenna  \\ \cline{2-6}
		&\cite{THz_Tang2020}  & BUPT & 220-330 GHz &Indoor scenario  & Tx\&Rx: a horn antenna  \\ \cline{2-6}
		&\cite{CS_OWC_1}  &\"{O}zye\u{g}in University & 850nm (250 MHz) &Indoor (static, mobile)  & Tx: 4 IR LEDs; Rx: 5 PDs  \\ \cline{2-6}
		&\cite{CS_OWC_2}  &National Taiwan University  & -- & Outdoor V2V    & Tx: LED headlamp; Rx: a PD  \\ \hline
		\multirow{6}{*}{ \rotatebox{270}{\textbf{Global coverage}}} & \cite{maritime_Guerrero2014} &University of Cadiz & 5.8 GHz & Sea port & Tx\&Rx: an omnidirectional antenna  \\ \cline{2-6}
		&\cite{maritime_Yang2011}  & Super Radio AS &2.075 GHz (20 MHz) & Over sea & Tx\&Rx: a vertically antenna  \\ \cline{2-6}
		& \cite{UAV_Matolak2014} &University of South Carolina & C-band 5-5.1 GHz (50 MHz); L-band 960-977 MHz (5 MHz)  &A2G (820 m height) & Tx\&Rx: two monopole antennas  \\ \cline{2-6}
		&\cite{UAV_Khawaja2019}  &North Carolina State University & 3.1-4.8 GHz  & A2G, Open area (10-30 m) &Tx\&Rx: single dipole antenna  \\ \cline{2-6}
		& \cite{mine_Volos2007} &Mobile and Portable Radio Research Group & 3 GHz (900 MHz) &Room-and-pillar mine & Tx\&Rx: single antenna  \\ \cline{2-6}
		& \cite{mine_Ghaddar2016} &University of Quebec in Abitibi-Temiscamigue  & 60 GHz (7 GHz) &Tunnel mine & Tx\&Rx: multiple antennas  \\ \hline
		\multirow{6}{*}{ \rotatebox{270}{\textbf{Full applications}}} &\cite{Zheng2022_umMIMO}  & SEU & 5.3 GHz (160 MHz) &Urban scenarios/ultra-massive MIMO   &Tx: 128 elements ULA; Rx: 8 omnidirectional antennas  \\ \cline{2-6}
		&\cite{mmMIMO_Li2015}  &Beijing Jiaotong University & 1.47 GHz (91 MHz) &Outdoor stadium/massive MIMO   &Tx: 128 elements ULA/UCA; Rx: 2 antennas  \\ \cline{2-6}
		&\cite{RIS_Sun2022}  & SEU & 5.4 GHz (160 MHz) &Urban scenarios/RIS   &\begin{tabular}[c]{@{}c@{}} Tx: 32 elements UPA; \\ Rx: 64 elements cylindrical array\end{tabular} \\ \cline{2-6}
		&\cite{RIS_Alfred2021}  &France University of Grenoble-Alpes  & 28 GHz (4 GHz) &Indoor laboratory and office/RIS  &Tx: virtual antenna array; Rx: a monopole antenna \\ \cline{2-6}
		&\cite{IIoT_Zhang2022}  &\textcolor{black}{SEU} & 5.5 GHz (320 MHz) &Mechanical and automobile workshop  &Tx: a planar array with 32 uniformly spaced antenna elements;
		Rx: a cylindrical array with 64 antenna elements \\ \cline{2-6}
		&\cite{ISAC_Li2021}  &HUAWEI  & 140 GHz &Anechoic chamber/ISAC  &Tx: single horn antenna; Rx: vertical antenna array  \\ \cline{2-6}
		&\cite{ISAC_Pham2021}  &Barkhausen Institut & 26 GHz, 71-76 GHz   & Indoor/ISAC &Tx: a ULA with 16 antenna elements  \\ \hline
		\multicolumn{6}{|l|}{*BUPT: Beijing University of Posts and Telecommunications, IR: Infrared, LED: light-emitting diode, LD: laser diode, PD: photodiode, A2G: air-to-ground, V2V: vehicle-to-vehicle.} \\ \hline

	\end{tabular}}
\label{6G_channel_sounders}
\end{table*}

6G channel sounders for measurements can be used to investigate the properties of all-spectra, global-coverage, and full-applications channels. In Table \ref{6G_channel_sounders}, the specific configurations of these channel measurement campaigns are shown, and the detailed statements are as follows.

According to channel measurement frequency, channel sounders can be divided into sub-6 GHz, mmWave, THz, and OWC channel sounders. In \cite{sub-6_He2016}, a time domain channel sounder was used to conduct channel measurement at 3.5 GHz in outdoor mobile scenarios. \textcolor{black}{An omnidirectional antenna was used at both Tx and Rx sides. In \cite{Sub-6_Chen2022}, 0.7 GHz, 2.3 GHz, and 3.7 GHz cross-band channel measurements were conducted in Urban Macro scenarios by two sets of time-domain channel sounders. Two of these three frequency bands were measured simultaneously at each channel measurement. At the Rx side, 4 $\times$ 4 dual-polarization planar antenna array was used for 0.7 GHz channel measurements, and 8 $\times$ 4 dual-polarization planar antenna array was used for 2.3 GHz and 3.7 GHz channel measurements. At the Tx side, a single omnidirectional antenna was used for all frequency bands channel measurements. In \cite{sub-6_Zhang2022}, a time domain channel sounder was used to conduct SISO and MIMO channel measurements in a large indoor office environment. For SISO channel measurements, two omnidirectional antennas were used as the transmitting and receiving antennas. For MIMO channel measurements, the 4$\times$4 dual-polarization planar antenna array was used at the Tx side and the 4$\times$8 dual-polarization cylindrical antenna array was used at the Rx side.} In \cite{sub-6_Zhang2016}, a 32 $\times$ 56 MIMO channel measurement at 3.5 GHz was implemented via time-division multiplexing (TDM). The high-speed electrical switches were configured at both Tx and Rx sides. In order to investigate the path loss at mmWave, a 28 GHz channel sounder based on a vector network analyzer (VNA) was constructed in an outdoor scenario in \cite{mmWave_Zhou2017}. The rotating platforms were used at both the Tx and Rx sides to obtain spatial characteristics of channels. Moreover, a VNA based (i.e., frequency domain) channel sounder was presented in \cite{mmWave_Li2019}. Both the measurement frequency and bandwidth of this channel sounder are up to 40 GHz and 500 MHz, respectively. In \cite{THz_Abbasi2020}, a 140-220 GHz THz channel sounder based on a VNA and frequency extenders was constructed to conduct a LoS measurement campaign. By analyzing the power delay profile of measurements, multipaths were not neglected in the LoS channels. In \cite{THz_Tang2020}, a 220-330 GHz THz channel sounder with a bandwidth of 2 GHz was constructed to investigate the frequency dependence of path loss in indoor static scenarios. Both Tx and Rx of this sounder were equipped with high-gain directional horn antennas. In \cite{CS_OWC_1}, a frequency-domain channel sounder was constructed to conduct indoor channel measurements for LiFi communications using IR LEDs with centroid wavelength at 850 nm. Both transceivers have wide filed-of-view (FoV) and beamwidth, which can well support mobile scenarios. By using a time-domain channel sounder, an outdoor V2V channel measurement at the VL band was shown in \cite{CS_OWC_2}. A commercial car headlamp was used on the Tx side, and a PD installed on a car was used on the Rx side.

Also, based on the coverage scenarios of channel measurement, channel sounders can be mainly divided into UAV, maritime, and mine channel sounders. To study the wireless propagation channel in maritime communication scenarios, the corresponding channel sounders are needed for channel measurement. In \cite{maritime_Guerrero2014}, a time-domain channel sounder was used to measure the channel characteristics in a sea port scenario. The vertically polarized omnidirectional antennas were equipped at both Tx and Rx sides. A channel measurement campaign at an open sea environment was conducted by a time domain channel sounder \cite{maritime_Yang2011}. The channel sounder worked at 2.075 GHz with a bandwidth of 200 MHz. Two vertically polarized antennas were equipped at the Tx and Rx sides. UAV channel sounders are usually miniaturized and lightweight, as they need to be mounted on the UAV. In \cite{UAV_Matolak2014}, channel measurements at C-band and L-band in an air-to-ground scenario were conducted by a UAV channel sounder. The sounder measured the time-varying CIR via the collection of PDPs. In \cite{UAV_Khawaja2019}, a time domain UAV channel sounder was constructed to conduct channel measurement at 3.1-4.8 GHz in an air-to-ground scenario. Omnidirectional antennas were used at both the UAV and GS sides. In general, the structure of underground mines can be divided into two categories, i.e., tunnel mines and room-and-pillar mines. In \cite{mine_Volos2007}, a time domain channel sounder was used to conduct channel measurement at 3 GHz in a room-and-pillar mine. An omnidirectional biconical antenna was used at both Tx and Rx sides. In \cite{mine_Volos2007}, a frequency-domain channel sounder was used to conduct channel measurement at 60 GHz in a tunnel mine. Directional MIMO antennas were used at both Tx and Rx sides.

Finally, based on the application scenarios, current channel sounders can be divided into those for massive MIMO, RIS, \textcolor{black}{IIoT, and} ISAC. 
Massive MIMO channel sounders are usually equipped with large-scale antenna arrays. In \cite{Zheng2022_umMIMO} the channel measurements were conducted at 5.3 GHz by using the ultra-massive MIMO channel sounder. The Tx antenna array was equipped with 8 omnidirectional antennas. A uniform linear array (ULA) up to 4.3 m with antenna element spacing of 0.6 wavelengths was used at the Rx side. In \cite{mmMIMO_Li2015}, a massive MIMO time domain channel sounder was constructed to conduct channel measurements at 1.4725 GHz in an outdoor stadium. The Tx side was equipped with a virtual ULA with 128 antenna elements. The Rx side consists of 2-element antenna. In \cite{RIS_Sun2022}, RIS channel measurements are carried out at 5.4 GHz employing a RIS channel sounder. The whole measurement system consists of Tx, RIS hardware, and Rx. The Tx and Rx sides were equipped with a planar array with 32 uniformly spaced antenna elements and a cylindrical array with 64 antenna elements, respectively. RIS hardware consists of 9 same sub-arrays. In \cite{RIS_Alfred2021}, RIS-assisted mmWave channel measurements were conducted in the indoor laboratory and office environments by a RIS-enabled channel sounder. The channel sounder consists of a four-port VNA, the RIS, one monopole antenna, and an antenna positioner. The ISAC channel sounders consist of two parts: communication channel sounding and sensing channel sounding. \textcolor{black}{In \cite{IIoT_Zhang2022}, the authors conducted SISO and MIMO channel measurements in a mechanical and automobile workshop. The height of the Tx antenna is 4.5~m, which is approximately the height for the placement of wireless access points. The height of the Rx antenna is 0.3~m and 1~m to simulate the communication scenarios of the auto guided vehicle and wireless switches. The detailed description of the time-domain channel sounder can be referred to~\cite{sub-6_Zhang2022}.} In \cite{ISAC_Li2021}, a sensing channel measurement was conducted in an anechoic chamber by a THz ISAC channel sounder to verify the imaging accuracy of the sensing channel. The Tx antenna was a 140 GHz horn antenna, and Rx was a sampling surface. To validate the detect ability of the sensing targets, the channel measurement was conducted at 26 GHz by using an ISAC channel sounder \cite{ISAC_Pham2021}. A ULA equipped with 16 quasi-antenna elements was used on the Tx side. Then, the detection probability was calculated and analyzed.

\subsection{Testbeds for 6G Key Technologies}
In order to design and verify new technologies, several testbeds for 6G key technologies have been proposed by different organizations in recent years. Table \ref{Testbeds for 6G Key Technologies} gives an overview of representative testbeds. Details will be discussed as follows.
\renewcommand{\arraystretch}{1.3}
\begin{table*}[t]
	\centering
	\caption{Testbeds for 6G key technologies.}
	\begin{tabular}{|l|r|c|r|r|}
		\hline
		\textbf{Technology} & \textbf{Organization} &\textbf{Reference}  & \textbf{Time}  & \textbf{Testbed}  \\ \hline
		\multirow{3}{*}{mmWave}
		& University of Leuven & \cite{Anjos2022a} & 2022 & \begin{tabular}[r]{@{}r@{}} FORMAT (Flexible Organization and Reconfiguration \\ of Millimeter-wave Antenna Tiles) \end{tabular}
		  \\ \cline{2-5}
		& Lund University & \cite{Chung2021a} & 2021 & LuMaMi28 (Real-time millimeter-wave massive MIMO testbed) \\ \cline{2-5}
		& PML & \cite{Cai2022a} & 2022 & Real-time photonics-assisted mm-Wave communication system \\ \hline
		\multirow{2}{*}{THz}
		& Northeast University & \cite{Sen2020a} & 2020 & \begin{tabular}[r]{@{}r@{}} TeraNova (Integrated testbed for ultra-broadband \\ wireless communications at THz-bands) \end{tabular}
		  \\ \cline{2-5}
		& PML & \cite{Zhang2022b} & 2022 & Real-time transparent fiber-THz-fiber 2×2 MIMO transmission system \\ \hline
		\multirow{6}{*}{RIS}
		& HUST & \cite{pei2021ris} & 2021 & Prototype for RIS-aided wireless communication \\ \cline{2-5}
		& Princeton University & \cite{zhang2022intelligent}, \cite{zeng2022intelligent} & 2022 & Full-dimensional intelligent omni-surfaces \\ \cline{2-5}
		& Université Paris-Saclay & \cite{fara2022prototype} & 2022 & RIS prototype with continuous control of the reflection phase \\ \cline{2-5}
		& University of Surrey & \cite{testbed_Araghi2022} & 2021 & RIS prototype in the sub-6 GHz band  \\ \cline{2-5}
		& Sungkyunkwan University & \cite{testbed_Amri2021} & 2021 & 1-bit RIS testbed \\ \cline{2-5}
		& Tsinghua University & \cite{testbed_Dai2020} & 2020 & RIS-based wireless communication prototype \\  \hline
		\multirow{4}{*}{ISAC}
		& HUAWEI & \cite{Testbed20} & 2022 & \begin{tabular}[r]{@{}r@{}} 5G-advanced ISAC technology demonstration \\ of perceiving vehicles and people \end{tabular} \\ \cline{2-5}
		& HUAWEI & \cite{Testbed21} & 2022 & ISAC-OW prototype (ISAC with optical wireless) \\ \cline{2-5}
		& HUAWEI & \cite{Li2021h} & 2022 & THz-ISAC prototype (ISAC at THz band) \\ \cline{2-5}
		& University College London & \cite{xu2022experimental} & 2022 & OFDM-based MIMO SDR testbed \\ \hline
		\multirow{6}{*}{Cell-free}
		& Ericsson & \cite{Testbed23} & 2019 & Radio stripes \\ \cline{2-5}
		& Samsung & \cite{Yuan2022a} & 2022 & D-FD-MIMO (Distributed-Full Duplex-MIMO system) \\ \cline{2-5}
		& HUAWEI & \cite{Testbed25} & 2021 & User-centric 5G indoor distributed massive MIMO solution \\ \cline{2-5}
		& KU Leuven & \cite{callebaut2022techtile} & 2022 & Techtile (Open 6G modular testbed) \\ \cline{2-5}
		& SEU & \cite{ZC_Test_1} & 2020 & Cloud-based cell-free distributed massive MIMO system \\ \cline{2-5}
		& SEU & \cite{Testbed26} & 2022 & 6G-TK${\upmu}$ cell-free massive MIMO testbed \\ \hline
		\multirow{6}{*}{OWC}
		& University of Strathclyde & \cite{haas2020introduction} & 2020 & Real-world hybrid LiFi/Wi-Fi network deployment \\ \cline{2-5}
		& Kyocera Soraalaser & \cite{lee202126} & 2022 & 105 Gbps LiFi demonstration with WDM \\ \cline{2-5}
		& Mitsubishi Electric & \cite{matsuda2021field} & 2021 & Demonstration of real-time 14 Tb/s 220 m FSO transmission \\ \cline{2-5}
		& Graz University of Technology & \cite{ivanov2019evaluation} & 2019 & Testbed for deep space FSO \\ \cline{2-5}
		& The University of Edinburgh & \cite{haas2019solar} & 2019 & Testbed for solar cell receiver OWC technology \\ \cline{2-5}
		& YunTech & \cite{lain2019experimental} & 2019 & OCC testbed with DCO-OFDM \\ \hline
		\multicolumn{5}{|l|}{*HUST: Huazhong University of Science and Technology, YunTech: National Yunlin University of Science and Technology.} \\ \hline
	\end{tabular}
	\label{Testbeds for 6G Key Technologies}
\end{table*}

\subsubsection{Testbeds for mmWave}
It is an essential work to build the mmWave massive MIMO testbed for practical use. \textcolor{black}{During the 5G research era, various organizations including AT\&T, HUAWEI, NTT DoCoMo, New York University, and Intel/Fraunhofer have made extensive measurements of channel characteristics in the mmWave frequency band from 30 GHz to 100 GHz. According to the measurement and ray tracing results, \cite{docomo20165g} was developed to introduce corresponding channel characteristics and modeling.} Samsung \cite{Roh2014a} and Qualcomm \cite{Raghavan2018a} have presented 28 GHz mm-Wave MIMO Prototypes in 2014 and 2018, respectively, as well as corresponding measurement results. Recently, Anjos \emph{et al.} from the University of Leuven presented a reconfigurable millimeter-wave tile-based antenna array platform named Flexible Organization and Reconfiguration of Millimeter-wave Antenna Tiles (FORMAT) \cite{Anjos2022a}, aiming at offering an assembled hardware solution to demonstrate various antenna array concepts and thus providing valuable insights to beyond-5G. By employing FORMAT hardware at both base station and user terminal, a wireless link was set up to demonstrate a 4.8-Gbps downlink speed with 64 QAM modulation at 28.5 GHz. 

Another real-time millimeter-wave massive MIMO testbed named LuMaMi28 was presented in \cite{Chung2021a} by groups from Lund University. LuMaMi28 consists of a base station with 16 transceiver chains and multiple users equipped with a beam-switchable antenna array. Corresponding measurement results for mmWave massive MIMO performance with both static and mobile users in different actual scenarios were also provided in this paper.

In response to the limited coverage of mmWave due to its susceptibility to atmospheric attenuation, researchers demonstrated a promising real-time photonics-assisted mm-Wave communication technology to overcome the wall loss for mmWave and thus promote indoor coverage in \cite{Cai2022a}. This technology also incorporates Ka-band large-scale phased-array antenna and FPGA-based automatic beam tracking technique, which allows terminals to move freely within a max range of $\pm$50$^{\circ}$. It was also demonstrated that this photonics-assisted mm-Wave communication system can achieve 1.5 Gbps real-time bi-directional uncompressed high-definition video transmissions at 26.5--29.5 GHz.

\subsubsection{Testbeds for THz}
To tackle the emerging 6G testing challenges with up to 10 GHz of bandwidth, many instrument manufacturers have launched a variety of sub-THz testbed instruments to address a multitude of frequency bands, frequency bandwidths, and waveform types for experimental demonstration purpose, such as Keysight and National Instruments (NI). Researchers from Northeast University presented an integrated testbed for ultra-broadband wireless communications at true THz-band frequencies called TeraNova in \cite{Sen2020a}. The system consists of a transmitter and a receiver based on Schottky-diode frequency multiplying and mixing chains able to up- \& down-convert an intermediate frequency (IF) signal between 1 and 1.05 THz. Researchers then characterized the THz channel in the vicinity of the first absorption-defined window above 1 THz, as well as the thermal noise and the absorption noise in the TeraNova system. Through the analysis of the testing results, experiments using this platform also reveal several bottlenecks of THz researches at the physical layer to be~overcome.

Photonics-aided THz-wave can break the bottleneck of electronic devices and thus attracted dozens of attention. Zhang, \emph{et al.} from the Purple Mountain Laboratories \textcolor{black}{introduced their 352-Gbps THz wired transmission experiment at 325 GHz in \cite{SCIS_Zhu2022a}. Hollow-core fiber composed of a polycarbonate substrate tube and a silver film plated inner layer with 0.3$\upmu\text{m}$ thickness was exploited in this THz wired transmission, which realized the record-high 352 Gbps single line rate and 8.6 bps/Hz net spectrum efficiency by employing the 32 Gbaud PS-4096 QAM signals. Besides, Zhang, \emph{et al.} also showcased a real-time transparent fiber-THz-fiber 2$\times$2 MIMO transmission system based on photonic up-/down-conversion at 370 GHz THz band in \cite{Zhang2022b}}, which can offer a 100 GbE (103.125 Gbps) streaming service platform to play real-time movie and live surveillance video. Such Photonics-aided THz-wave architecture exhibits its superiority of high frequency, large bandwidth, and low transmission loss of optical devices, and can be seen as a promising solution for the seamless integration fiber-THz-fiber network in the future 6G.

\subsubsection{Testbeds for RIS}
In \cite{pei2021ris}, researchers conducted field trials to confirm that RIS is indeed a promising technology to improve communication performance. The developed RIS prototype is made of 1100 controllable elements working at 5.8 GHz band, and is configured by exploiting the geometrical array properties and a practical receiver-RIS feedback link. It can achieve 26dB power gain compared to the benchmark where the RIS is replaced by a copper plate for indoor scenarios. 

Normal RIS systems can only manipulate signals reflected on the same side, which to some extend restricts the service coverage. An intelligent omni-surfaces (IOS) was proposed in \cite{zhang2022intelligent}, which can support full-dimensional communications by employing its reflective and refractive properties. The physical structure of IOS and corresponding testbed were presented in \cite{zeng2022intelligent}. Another RIS prototype that can control the phase shifts of incident waves continuously was proposed in \cite{fara2022prototype}. Through experimental measurements, with the aid of full-wave simulations the properties of the proposed RIS prototype were also characterized. 

In \cite{testbed_Araghi2022}, groups from the University of Surrey presented a RIS testbed in the sub-6 GHz band, which was fabricated to operate at 3.5 GHz. The RIS testbed owns a surface composed of 2430 unit cells with conductive patches and a control unit that can control the response of the surface. It was demonstrated that, in the case of no LoS between the Tx and the Rx, the RIS can successfully configure itself to direct the reflected waves towards the target under different incident~angles. 

In \cite{testbed_Amri2021}, A 1-bit RIS testbed consisting of 16 $\times$ 16 unit cells was demonstrated. With a compressive sensing-based adaptive beamforming algorithm that can manipulate the beam towards the receiver, the RIS system can significantly improve the BER and SNR under different modulation schemes. Another RIS prototype considering mmWave frequency was presented in \cite{testbed_Dai2020}. Such prototype contains the hosts for parameter setting and data exchange, the universal software radio peripherals, and also the RIS having 256 2-bit elements. The measurement results have confirmed the effectiveness of this RIS prototype that it can achieve a 21.7 dBi antenna gain at 2.3 GHz and a 19.1 dBi antenna gain at 28.5 GHz.

\subsubsection{Testbeds for ISAC}
HUAWEI completed the world's first 5G-advanced ISAC technology verification at the Huairou Outfield in Beijing to validate the capability of ISAC of perceiving vehicles and people in various business scenarios, such as smart transportation and campus intrusion detection \cite{Testbed20}. The 3GPP 5G signal in the millimeter-wave band was employed in this test. Under the condition that the proportion of sensing resources not exceeding 15\%, the detection distance of the integrated 5G ISAC sensor is more than 500 meters, and the detection accuracy rate of vehicles and people reaches 100\%, which verifies the performance advantages of 5G-Advanced ISAC in KPIs such as detection distance and positioning accuracy over the mainstream traffic radar.

To meet the high-speed communication and high-precision sensing requirements of medical and industrial scenarios in the future, HUAWEI also proposed an ISAC-OW technology, and provided corresponding prototype verification \cite{Testbed21}. Such prototype can precisely sense and locate mobile robots through visible and infrared optical wireless links, while at the same time transmitting wireless real-time high-definition videos between mobile robots and the controller via an optical link. In addition, the ISAC-OW prototype can also monitor heart rate and breathing status in real-time without contact, with detection accuracy comparable to commercial smartwatches. Through the design of ISAC integrated waveform, hardware architecture and signal processing algorithm, the prototype achieves centimeter-level indoor positioning and high-speed wireless optical communication.

Due to the high accuracy of sensing and resolution, as well as the advantage in portability, the THz sensing has attracted massive research attention. HUAWEI has set up a THz-ISAC (Integrated Sensing and Communication at THz band) prototype suitable for the 100--300 GHz frequency band to explore and verify the technical feasibility of high-precision sensing and imaging on the terminal side and outdoor medium-distance ultra-high-speed transmission \cite{Li2021h}. Measurement results have demonstrated that with the assistance of virtual aperture, the prototype can realize millimeter-level resolution imaging of occluded objects and outdoor medium-distance and long-distance 240 Gbps high-speed line-of-sight over-the-air~transmission.

In order to verify that the dual-function radiation waveform can complete both radar sensing and communication functions simultaneously, researchers from University College London developed an OFDM-based MIMO software-defined radio (SDR) testbed in \cite{xu2022experimental}. By carrying out actual over-the-air experiments, they successfully demonstrated that the measured results of BER, which show the communication performance using dual-function waveform, can achieve comparable BER performance with that of the pure communication system, while keeping fine radar beampatterns at the same time. 

\subsubsection{Testbeds for Cell-free Systems}
As a strong candidate for 6G networking, cell-free systems need to be tested and verified to explore the limitations for  practical implementation. Ericsson developed a distributed MIMO (another name for cell-free MIMO systems) deployment named radio stripes in 2019 \cite{Testbed23} to serially process signals. Samsung presented a Distributed-Full Duplex-MIMO (D-FD-MIMO) system in \cite{Yuan2022a}, which employs a 2D planar antenna array at the base station to exploit the channel diversities in both elevation and horizontal domain. In 2021, HUAWEI deployed and tested a user-centric 5G indoor distributed massive MIMO solution \cite{Testbed25}, successfully increasing the user experience rate by 30\% while maintaining a stable Gbps rate experience during user movement. A smart connectivity platform project named REINDEER was jointly launched by KU Leuven, Linköpings universitet, Ericsson and other scientific research institutions in 2021 \cite{REINDEER}. The project is committed to designing cell-free protocols and real-time real-space interactive application processing algorithms with distributed intelligence, and to achieve energy-efficiency, scalability, and secure connectivity. Funded by this program, an open 6G modular testbed called Techtile for communication, sensing and federated learning was proposed in \cite{callebaut2022techtile} in 2022. Techtile owns 140 distributed computing resource units, as well as SDRs, sensors and LED resources. It can not only provide a platform to assess different networks and computing topologies (local-versus-central), but also support experimental research on hyper-connected interactive technologies. In order to address the scalability and synchronization issues of cell-free networks, all tiles of this testbed are connected, synchronized and powered based on Ethernet, and each tile is equipped with a SDR and a power supply. Such a flexible structure also allows the emulation of different distributed structures, such as mesh or tree.    

Groups from Southeast University built up a cloud-based cell-free distributed massive MIMO system in \cite{ZC_Test_1}, which supports demonstration in MIMO scenarios up to 128 $\times$ 128 antenna scale with 10.185 Gb/s throughput and more than 100 b/s/Hz spectrum utilization. Based on this testbed, a unified Bayes-network based baseband signal processor\cite{ZC_Test_2,ZC_Test_3} is designed, and the related application specific integrated circuit (ASIC) chip has been taped out and tested. Furthermore, they constructed a 6G-TK${\upmu}$ cell-free massive MIMO testbed \cite{Testbed26}, including a cell-free massive MIMO test system based on commercial 4.9 GHz remote radio units (RRU), and a scalable large-scale distributed phased array test system. By using the flexible frame structure, such testbed can achieve air interface calibration between the RRU and the terminal within 134 ${\upmu}$s. In addition, this testbed can also realize a parallel transmission of 16 data streams, with the spectral efficiency more than 200 bps/Hz.

\subsubsection{Testbeds for OWC}
OWC provides a direct path to leverage beyond RF spectrum for future 6G systems because of the availability of a wide range of optical transmitter and receiver devices. Testbeds in OWC can be classified in three categories: indoor mobile multiuser wireless networking referred to as LiFi, point-to-point FSO primarily for outdoor use cases, and OCC using embedded cameras as data detectors. 

Eindhoven University demonstrated 40 Gbps user data rate in a multiuser  scenario using a decentralised beam steering architecture \cite{zhang2020wide}. Fudan University has numerous VLC/LiFi testbeds and experimental systems. In a recent paper, this group have demonstrated a VLC MIMO link with Tomlinson-Harashima precoding \cite{niu2022phosphor}. pureLiFi Ltd in collaboration with the LiFi Research and Development Centre (LRDC) at the University of Strathclyde have developed a LiFi testbed for a classroom and demonstrated real-time handover, multiuser access and mobility support in a real-world environment \cite{haas2020introduction}. LiFi lends itself to simultaneous high speed data transmission and LiDAR. Kyocera Soraalaser (KSLD) has demonstrated this capability based on laserlight devices \cite{lee202126}. A collaboration between KSLD and the LRDC led to the development of a wavelength division multiplex (WDM) LiFi demonstrator capable of 105 Gbps using wavelengths in the visible and infrared spectrum. This demonstrator was showcased at CES 2022 \cite{M.Halper}. 

FSO can be used to build high speed terrestrial backhaul networks, and ensure that 6G is available everywhere. The German Aerospace Centre (DLR) in cooperation with ADVA Optical Networking showcased 13.16 Tb/s over a distance of 10.45 km using commercial coherent fiber optic transceivers \cite{dochhan201913}. Mitsubishi Electric in Japan demonstrated the transmission of 14 Tb/s over a 220 m \cite{matsuda2021field}. This demonstrator used a 2-dimensional optical transmitter array of size 3x3 array to guarantee that the system complied with class I eye safety regulations. A group at the Key Laboratory for Space Utilization in China demonstrated a real-time FSO link of 100 Gbps over a 2.1 km horizontal atmospheric link \cite{zhan2021demonstration}. In \cite{ivanov2019evaluation} an experimental platform for deep space FSO technology is shown. The testbed is based on a transmission link (operating on 1550 nm wavelength) realized with fibre optics technology. It contains a self-developed channel emulator, background noise module and a superconducting nanowire single-photon detector (SNSPD) as receiver. Recently, solar cells have been considered as high speed detectors for FSO systems to enable low cost systems because the massive MIMO solar panel constitutes a very large receiver aperture obviating the need for expensive beam acquisition and tracking units \cite{haas2019solar}.

OCC uses embedded cameras to establish a light-based wireless link, and can achieve data rates up to around 1 Mbps due to the nature of the optical sensor. The advantage of OCC is the availability of already integrated camera sensors in smartphones. OCC exploits the rolling shutter effect of a camera sensor to achieve orders of magnitude higher data rates than the typical framerates of a camera. Lain et al. have developed an OCC testbed using direct current optical orthogonal frequency division multiplexing (DCO-OFDM) \cite{lain2019experimental}. Signify has trialed OCC at Carrefour Lille in France and developed a precise indoor navigation system based on OCC.

\subsection{6G Comprehensive Verification Platforms}
In 6G, communication systems will not only undertake multi-band, multi-dimensional and high-performance communication needs, but also integrate new capabilities such as communication, sensing, computing and artificial intelligence \cite{PinTan2021}. To evaluate the key performances of 6G communication systems, such as throughout, spectral efficiency, and delay, the real data testbed for 6G is required. Hence, the Purple Mountain Laboratories is constructing a 6G comprehensive verification platform named as ``TK${\upmu}$" \cite{Testbed28}. This testbed mainly includes cell-free ultra-massive distributed MIMO communication, RIS-assisted 6G wireless communication, space-air-ground integrated communication, mmWave/THz communication, and grading intelligence part. It will be able to validate Tbps-level for peak rate, kbps/Hz-level for spectral efficiency, $ \upmu $s-level for delay, and endogenous security and intelligence. First, cell-free ultra-massive distributed MIMO adopts physical layer sinking, fronthaul networks, and decentralization merge to achieve cell-free networks, which is capable of breaking the boundary effect in traditional cellular architectures. Further, the cell-free architecture integrated with ultra-massive MIMO antennas is constructed to \textcolor{black}{guarantee the spectral efficiency of kbps/Hz-level}. Second, RIS-assisted 6G wireless communications are also incorporated into the ``TK${\upmu}$" 6G comprehensive verification platform to enhance the coverage of EM radiation and the rank of channel matrix, eliminate signal interference, focus energy, improve positioning accuracy, and improve information and power transmission, etc. Third, due to the limited coverage and network capacity for the ground network, it can no longer meet the explosive demand for high-speed and reliable network access anytime and anywhere on the earth. Hence, the space-air-ground integrated communication had been widely researched in the academia and industry \cite{Guo2022b}. The ``TK${\upmu}$" 6G comprehensive verification platform will combine space-air-ground-sea integrated \textcolor{black}{techniques} to achieve full coverage. Fourth, compared with sub-6 GHz, the mmWave and THz frequency can provide sufficient bandwidths to increase the peak rate for communication. Previous work shows that mmWave/THz bands are suitable for mobile, backhaul, and indoor wireless communications \cite{Huang2017,Rappaport2013}. Therefore, mmWave/THz \textcolor{black}{techniques} are further taken into account in the ``TK${\upmu}$" 6G comprehensive verification platform, which can guarantee \textcolor{black}{the peak rate of Tbps-level and delay of ${\upmu}$s-level}. Most importantly, the grading intelligence part is an integrated multi-layer and hierarchical intelligent network, which includes the data-driven edge intelligent center and the network intelligent cloud platform. By using the cloud, fog, and edge computing techniques, the data-driven edge intelligent center can directly interact with the communication networks and handle the generated data \textcolor{black}{to ensure endogenous security and intelligence}. In addition, the data-driven edge intelligent center will also share the communication data with the network intelligent cloud platform for further processing the data.

In addition, a joint research and design (R\&D) team formed by China Mobile Research Institute and Beijing University of Posts and Telecommunications designed and developed a prototype 6G universal prototype platform \cite{Testbed32}. The system adopts a universal baseband platform, supports multiple operating bands including visible light, and incorporates various capabilities such as communication and artificial intelligence, and supports flexible expansion and cloud-based technologies. Using this system, joint validation of various key 6G technologies can be performed, to help upgrade and iterate the technology and select technology solutions for 6G standardization. First, the system implements a fully open E2E link for each algorithm module, and improves the efficiency of the prototype system through algorithm structure optimization, AVX512 assembly instruction optimization, and multi-threading optimization. Second, the system achieves baseband heterogeneous acceleration and open module capability. Third, the system's visible light communication link has achieved multiple joint processing and aggregation capability. Next, the joint R\&D team will continue to upgrade the system and further explore the new software and hardware open architecture of the 6G common prototype verification system with multiple capability fusion. Moreover, it will also use the system as a public verification platform to cover the whole process of 6G key technology R\&D, standardization and industrialization to explore new paths for 6G development.

\textcolor{black}{HUAWEI 6G Research Team also posted their prototype for short-range communications at 70 GHz that could achieve ultra-low power consumption, ultra-high throughput and ultra-low latency \cite{HUAWEIimmersive}. Short-range communications exploits high frequency bands such as mmWave and THz to enable a truly immersive experience that allow free movements over the ``last meters" of the communication link, and thereby provide an extreme connection service for business scenarios such as immersive interactions based on XR and holographic communications. With several advanced technologies adopted, such as low-power polar encoding/decoding, low-power 1-bit ADC, and adaptive beam sweeping with a high-speed short-range phased array, HUAWEI 6G Research Team demonstrated a communication throughput over 10 Gbps with sub-millisecond latency, as well as 4K VR services in real time. The short-range transmission rate can achieve several times that of wired communication methods, while the overall power consumption of the prototype is less than 560mW.}

\section{Future Research Directions and Challenges for 6G}
\begin{figure*}[tb]
	\centerline{\includegraphics[width=0.99\textwidth]{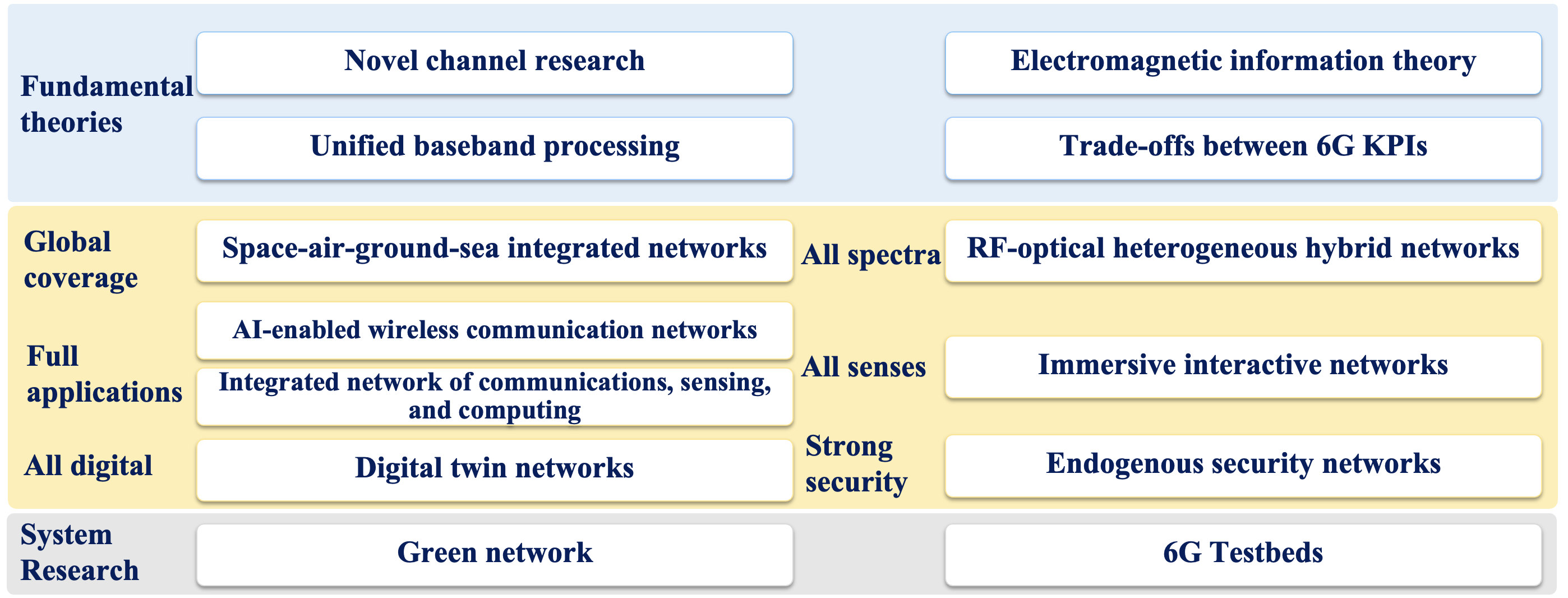}}
	\caption{\textcolor{black}{Future research directions and challenges for 6G.}}
	\label{fig_challenges}
\end{figure*}

In order to achieve the 6G vision of ``global coverage, full applications, strong security, all spectra, all senses, and all digital", a lot of issues and research directions need to be further explored. Moreover, 6G communication theories are in urgent need of breakthroughs. These research directions bring both challenges and new opportunities to 6G research, as summarized in Fig.\ref{fig_challenges}.
In this section, we will first discuss the challenges in fundamental theories, i.e., novel channel research, EM information theory, uniform baseband signal processing, and trade-off between 6G KPIs. 
Then, we will analyze and point out future research directions and key challenges to approach the 6G vision. 
\textcolor{black}{Finally, challenges for the overall system research will be introduced, including achieving green networks and establishing testbeds for the 6G developments.}

\subsection{\textcolor{black}{Fundamental Theories}}
\subsubsection{Novel Channel Research}
The traditional channel research often follows four steps: channel measurement, channel parameter estimation, channel characteristic analysis, and channel modeling. This passive way of recognizing channels has a number of limitations. The channel measurement is time-consuming, expensive, and labor-intensive. In addition, channel measurements in reality can never cover all frequency bands or scenarios. The large amount of data and high computation complexity also bring challenges to the channel parameter estimation. The channel characteristics can be analyzed only at known frequencies and in known scenarios, and it is unable to fully explore the complicated relationship between new characteristics and frequencies/scenarios. Finally, the traditional channel modeling is unable to predict channel characteristics of unknown frequency bands/scenarios in the future. For 6G, channel researches need to evolve from passively recognizing channels to actively recognizing and control channels, including 6G standardized pervasive channel modeling\cite{P6GCM}, AI-based predictive 6G channel modeling\cite{Yu2020_AI,Li2022_AI,Huang2022_AI_Survey1,Huang2022_AI_Survey2}, scenario adaptive channel modeling, and RIS channel modeling\cite{Sun2021_RIS,Huang2022_RIS}.

\subsubsection{EM Information Theory}
\begin{figure}[tb]
	\centerline{\includegraphics[width=0.49\textwidth]{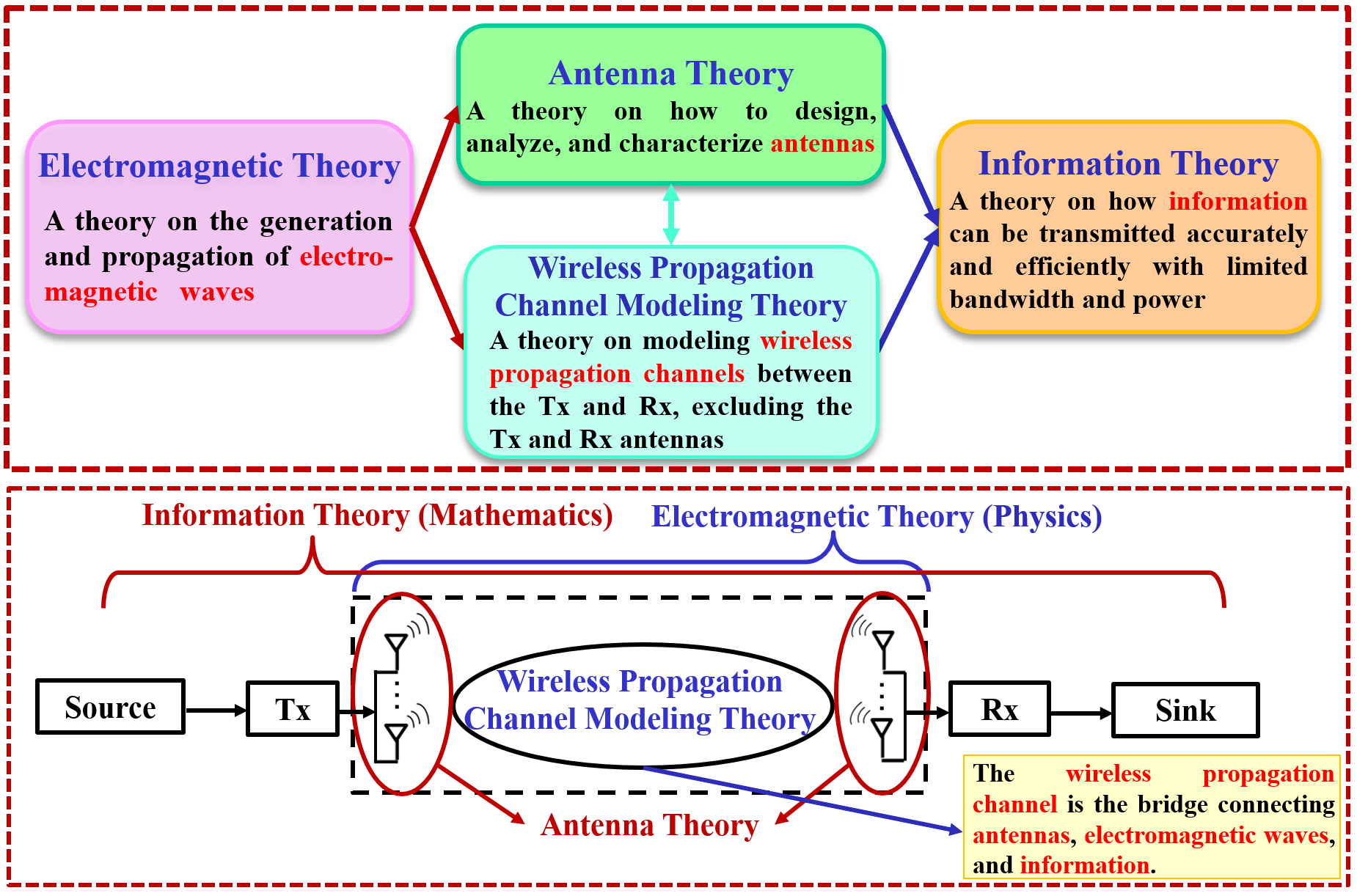}}
	\caption{Relationships between the EM theory, information theory, antenna theory, and wireless propagation channel modeling theory.}
	\label{fig_EIT}
\end{figure}
The EM theory and information theory are the two theoretical cornerstones of wireless communications. The EM theory is a theory on the generation and propagation of electromagnetic waves. The information theory is a theory on how information can be transmitted accurately and efficiently with limited bandwidth and power. The EM theory is based on continuous time and space, and can be used to obtain the continuous-space EM field distribution. However, using the EM theory alone cannot calculate the channel capacity. Information theory can be utilized to calculate the discrete-space channel capacity, but not the continuous-space channel capacity. The EM theory and information theory are connected by the antenna theory and wireless propagation channel modeling theory. In particular, the wireless propagation channel, originating from antennas and carries information via EM waves, serves as the bridge linking the EM theory and information theory. The relationships between these four classic theories are shown in Fig.~\ref{fig_EIT}. 

Note that with the development of 6G key technologies, some problems cannot be solved by the above individual theory, which brings both challenges and opportunities for the integration of the four theories. For example, as 6G wireless communication networks expand from local terrestrial coverage to global coverage of space-air-ground-sea integrated networks, the numbers of users, base stations, relays, and/or RISs continue to increase. The positions of base stations and users in moving networks such as LEO satellite, UAV, and vehicular networks tend to change continuously. Therefore, 6G wireless communication networks show a trend of evolution from discrete space to continuous space, requiring the acquisition of CSI and calculation of channel capacity at any position in the continuous space to facilitate system design. In this case, the integration of the above four theories is inevitable. Additionally, with the increase of antenna size and antenna elements in ultra-massive MIMO, Tx/Rx antennas become more closely related to the environment. In other words, Tx/Rx antennas and the underlying wireless propagation channel become inseparable and therefore, the integration of antenna theory and wireless propagation channel modeling theory is necessary. Furthermore, as the antenna element spacing decreases in holographic MIMO, the evolution of antenna arrays from discrete to continuous apertures brings new demands on channel characterization, antenna design, and continuous-space channel capacity calculation. Again, the integration of the above four theories is required. In summary, compared with 5G, 6G wireless communication networks put forward new applications and technical requirements, which go beyond the application scopes of every individual theory. The developments of classic theories have encountered bottlenecks. Thus, it is urgent to study the integration of EM theory and information theory, i.e., EM information theory \cite{Migliore2008,Migliore2019}, which will serve as a theoretical foundation for 6G wireless communication networks and help achieve new breakthroughs.

\subsubsection{Unified Baseband Processing}
With the error-correcting capability, channel coding has always played an indispensable role in communication systems. In order to meet flexible and various scenarios in 6G communication systems, enhanced coding schemes and simplified decoding algorithms, together with their unified implementations are essential. However, only focusing on optimizing a single channel coding module is sometimes not enough to meet the stringent requirements imposed by 6G, and a wider research view into the design and implementation for the whole baseband processing is much more important.

\paragraph*{Iterative and Joint Baseband Processing} Iterative receivers have long been regarded to enjoy higher system capacity and link reliability than existing separate baseband processing\cite{ZC_Chall_1}. They iteratively exchange soft information among two or more of the channel estimator, MIMO detector, NOMA detector, channel decoder, and source decoder\cite{ZC_Chall_2,ZC_Chall_3,ZC_Chall_4,ZC_Chall_5,ZC_Chall_6,ZC_Chall_7,ZC_Chall_8,ZC_Chall_9}. However, the increased usage of each module caused by iterations may burden the system with high latency and computational complexity. As most modules are dedicated to solving the maximum a posterior estimation problem, they can be processed in a unified Bayesian network by exploiting their similarities, therefore improving the EE of the iterative receiver. Since the signal flow is serial, another approach is to regard multiple modules as a whole with the final output as the estimation object for joint processing, such as joint detection and decoding \cite{ZC_Chall_10,ZC_Chall_11}. 

\paragraph*{Unified Architectures for Baseband Processing} Based on the iterative or joint baseband processing, efficient unified hardware architecture and corresponding implementations are highly essential to fulfill different practical applications. In pre-6G communication systems, the design and implementation for each module are isolated, and the lack of system-level design thinking would result in an additional cost of hardware resources. To improve the flexibility, compatibility, and hardware efficiency of the implementations for 6G baseband processing, very large scale integration (VLSI)-digital signal processing (DSP) methodology\cite{ZC_Chall_12}, involving retiming, folding, and unfolding, can be adopted to design a unified architecture with fixed processing elements, optimized network interconnection, and iteration timing schedules.

\paragraph*{Hardware Auto-generation for Baseband Processing} With much more diverse and complex applications in 6G communication systems, the customized electronic design automation (EDA) tools for baseband processing are essential, which can lower the entry barriers of circuit designs for various scenarios and realize the customer demands. A well-designed EDA tool can automatically generate the circuit design based on the performance required from the customer. Based on the unified Bayesian network, most of the processing units can be formalized to accelerate the auto-generation methodology, based on which a design space is created for configuring the best parameters in the hardware\cite{ZC_Chall_13,ZC_Chall_14}. With complicated requirements, AI can help the design space exploration. It can also help to determine the approximation-caused compensating parameters, quantization schemes, and iteration schemes in the auto-generation hardware design\cite{AI_You2019}.

\subsubsection{Trade-off between 6G KPIs}
Nowadays, the research on the trade-offs between KPIs has been widely carried out, but the existing research results mainly focus on SE, EE, capacity, delay and other metrics, while the research on the new KPIs proposed in 6G is limited. Therefore, it is a promising direction to investigate the intrinsic connection and trade-offs between all 6G metrics. In addition, 6G has numerous application scenarios, but the system resources are insufficient to let each application scenario perform well in all aspects. There have been several works that analyzed the technical requirements for KPIs in different 6G application scenarios, but there still exist research gaps in the trade-offs between KPIs for specific scenarios. In addition, as the 6G frequency band is getting higher and 6G technologies such as ultra-massive MIMO are maturing, the importance of Shannon information theory in system performance analysis is gradually coming to light. It is essential to analyze the impact of antennas, coding and other factors on the KPIs such as delay, reliability, and capacity from the information theory point of view to provide guidelines for the design and deployment of 6G communication systems.

\subsection{\textcolor{black}{Challenges to Approach the 6G Vision}}
\subsubsection{Space-air-ground-sea Integrated Networks}
The space-air-ground-sea integrated network is an inevitable trend and also a key technology to realize the vision of 6G global coverage. 
However, the space-air-ground-sea integrated network is still under development, and there are still many theoretical and engineering issues to be studied in network planning, construction, maintenance, and optimization. First of all, communication channels are the basis for subsequent research and development of communication systems and networks. Diverse frequency bands and scenarios bring great challenges to the channel measurement and modeling, and it is necessary to consider how to integrate unique channel characteristics in satellite\cite{Bai2019_satellite,Li2021_satellite}, UAV\cite{Chang2021_UAV,Bian2021_UAV}, ocean\cite{Liu2021_UAV_maritime,He2022_Maritime}, and ground scenarios in a pervasive channel model\cite{Wang2020_6GCMM-survey}. Furthermore, the architecture design, mobility management, network protocol, resource allocation, routing strategy, EE enhancement, and other issues in the space-air-ground-sea integrated network are also in urgent need of innovation. It is necessary to design a safe and efficient network architecture, so that diverse communication systems can be smoothly integrated to provide users with reliable and safe communication services\cite{Li2020_SAGS}. \textcolor{black}{Efficient, safe, and anonymous authentication protocols for the integrated networks are also of great importance\cite{Sherman_new_5}.} Considering different types of mobility in various scenarios, innovative mobility management solutions are required to achieve seamless mobility management between homogeneous and heterogeneous networks\cite{SEU}. In addition, the characteristics of high mobility and dynamic network topology in the space-air-ground-sea integrated network need to be considered in the design of network protocols, \textcolor{black}{resource allocation}, and routing strategies\cite{SEU,Liu2022_SAGS,Li2022_SAGS,Sherman_new_2}. In terms of network maintenance and optimization, it is necessary to consider the power supply of the communication platform and the load of UAVs and other equipment in the space-based and air-based networks, which has high requirements for the overall EE of the network. At the same time, it is necessary to consider how to make compromise between the network performance improvement and the cost\cite{Yan2018}. What's more, how to use AI, DL, and other intelligent technologies to optimize the network architecture and improve the overall network performance\cite{Kato2019,Liu2020_SAGS} is also one of the current~challenges.

\subsubsection{RF-optical Heterogeneous and Hybrid Networks}
To achieve the vision of 6G full spectrum, it is a promising development of 6G that systems of different frequency bands, including RF and the whole optical wireless bands, will be integrated to realize heterogeneous and hybrid systems and networks that support all frequency bands. 
The RF-optical heterogeneous hybrid networks cover a series of hybrid networks, and their applications cover various scenarios \textcolor{black}{e.g., indoor, outdoor V2V, free space, and underwater}\cite{Chowdhury2020}. There are mainly two categories of hybrid networks, RF/Optical hybrid systems, which include mmWave/VLC systems, indoor WiFi/LiFi, underwater acoustic/optical communication, and optical/optical hybrid systems, which include VLC/OCC and FSO/OCC. However, the realization of RF-optical heterogeneous systems and networks face a series of challenges. On one hand, like other heterogeneous networks (such as space-air-ground-sea integrated networks), RF-optical heterogeneous networks face common challenges of heterogeneous networks, including mobility management (network switching), transmission network protocols design, load balancing, heterogeneous network synchronization, resource allocation and EE improvement \cite{Kashef2016}, spectrum allocation \cite{Kim2022}, \textcolor{black}{as well as joint access points and power allocation\cite{Aboagye2021}}. On the other hand, due to the large frequency gap between the RF and optical bands, there are several special challenges that RF-optical heterogeneous networks need to solve. First of all, it is necessary to consider the transceivers with different characteristics in RF and optical communication networks. For instance, when users are moving, the heterogeneity of access points brings great challenges to frequent handovers\cite{Arshad2021}. The hybrid systems need to integrate the optoelectronic characteristics and meet the different bandwidth requirements of different transmission media, which brings great difficulty to the fusion of RF/optical hardware systems and channel models \cite{P6GCM}. Secondly, the network selection criteria of different optical wireless networks are usually different, which are also different from the existing RF communication networks. It is necessary to consider how to design the optimal network selection strategy \cite{Chowdhury2020}. In addition, due to the limited energy of terminal equipment, the large influence of mobility on uplink OWCs, and the interference of uplink OWCs on the lighting, there are still great limitations in the uplink OWCs\cite{Chowdhury2020}. Finally, the challenge of system security brought by the heterogeneous systems cannot be ignored\cite{Pan2017}.

\subsubsection{AI-enabled Wireless Communication Networks}
\textcolor{black}{In recent years, the rapid development of AI technologies has greatly pushed the development of B5G/6G wireless communications. In the 6G era, in order to realize AI-enabled intelligent wireless communication networks, the network AI is an important supporting technology, which provides a complete AI environment in the network, including AI infrastructure, AI workflow logic, data and model services, etc\cite{6GANA-NetworkAI}. }
\textcolor{black}{Note that although AI technologies are developing rapidly and some new techniques mentioned before may contribute to the network AI, it is still in the early stage of research, and there are still many problems to be explored and solved.} 
First of all, data is the foundation of wireless network AI research. Therefore, how to collect and use data in wireless networks and establish a shared data set for research is the primary problem to be solved\cite{IMT2030-Wireless_AI_WP}. When applying AI to the network, it is also necessary to pay attention to the performance of AI services, and better AI performance can be achieved through the selection of AI algorithms and network resource allocation. The key challenge is to model the relationship between the AI performance and the network configuration as well as an online network configuration scheme that adapts to network dynamics\cite{He2020_AI}. Besides, noting that more advanced and accurate AI models typically consume a lot of energy and incur significant environmental costs, there is an urgent need to study how to improve EE and reduce energy consumption and costs before large-scale deployment of AI services\cite{Thompson2020}. Finally, although AI technologies have many advantages, issues of privacy and data security also need to be seriously considered. In the future, the dynamic nature of data collection, data transmission, and data distribution in the network will lead to the risk of user information leakage, bringing great challenges to the privacy and security of the network\cite{Nguyen2021_COMST}.

\subsubsection{Integrated Communications, Sensing, and Computing Networks}
The 6G network will \textcolor{black}{realize the integration of} mobile communication, intelligent perception, and computing power services, with high-level integration and mutual enhancement\cite{INCSC_WP}. The integrated network of communication, sensing, and computing is a novel comprehensive network developed by integrating three technologies. The current research is mainly on the basic work of ISAC and computing network, such as waveform design and signal processing\cite{Sturm2011}, sensory mobile network\cite{Zhang2021_VTMag}, network level sensing\cite{Shi2022_JSAC} in ISAC, computing power offloading\cite{Du2018_Computing}, collaboration of multi-layer computing power resources\cite{Ghosh2020,Fantacci2020,Alnoman2021}, and cloud-edge-terminal resource allocation for the multi-layer ubiquitous computing network. For the integrated network of communication, sensing, and computing, there are still many challenges in the future, including development of ISAC technologies, networked sensing technologies and computing power network technologies enabled by \textcolor{black}{integration of communication, sensing, and computing}\cite{INCSC_WP}. In order to realize the networked sensing with communication, sensing, and extra computing, both the access network and the core network need to have  capabilities of communication, sensing, and computing. Therefore, requirements are put forward for the transceiver and processing modules, topology, MAC, routing algorithm, and resource allocation. In addition, multi-point networking collaborative sensing and customized sensing requirements are also issues that need to be considered. Finally, on the basis of the multi-layer ubiquitous computing network, it is also necessary to study how to provide rich prior information for the optimal scheduling decision of distributed computing power through the network sensing function. In turn, it is challenging to perform customized feature extraction and information fusion for the sensing data through distributed computing power shared in real time. Besides, how to improve the ubiquitous computing power through the enhanced communication performance is also an interesting issue that needs to be explored in the future.

\subsubsection{Endogenous Security Networks}
In the future 6G network, various security technologies, such as blockchain, physical layer security, mimic defense network, and secure multi-party computing, will be introduced to achieve the endogenous network security. These diverse technologies bring opportunities along with a number of challenges. 
\textcolor{black}{For example, there have been} a series of theoretical studies\cite{Zeng2018,Lv2018,Xu2019} on the physical layer security, this kind of technology is still far from practical applications on a large scale. The obstacles brought by the existing network framework, the scalability of the underlying air interface, the limitation of network resources, and diverse new scenarios have brought great challenges to the physical layer security technology\cite{Yan2020}. In recent years, mimic defense network has also attracted research interest, which uses dynamic heterogeneous redundancy architecture and negative feedback mechanism to improve the system's ability to deal with unknown threats\cite{Wu2018_chinese}. However, the wireless endogenous information security, wireless endogenous functional security, and other issues still need to be further explored\cite{Jin2021_Security}. In addition, the secure multi-party computation is a cutting-edge cryptographic technology. Its future research challenges include secure multi-party computation schemes suitable for different scenarios\cite{Zhang2016_chinese} and in malicious model environments\cite{Jiang2021_chinese}, as well as efficient and secure multi-party computation protocols, malicious security protocols, and special security protocols for specific applications\cite{Mimic_url_2019}.

\subsubsection{Immersive Interactive Network}
Providing applications using all-senses is an important vision of 6G network, and the immersive interactive network is an important technology to meet this vision. It will transmit multi-sensory information including the visual, auditory, tactile, taste, and smell information, thus providing users with a near-real virtual experience. The goal of immersive interactive network is to realize real-time control of novel applications such as immersive cloud XR, holographic communications, intelligent interaction, and sensory interconnection\cite{IMT-2030}. The current communication network mainly transmits vision and auditory information. The immersive interactive network in the 6G era needs to transmit the tactile, taste, and smell information, which has extremely high requirements on the transmission rate, reliability, and delay of the network. Currently, the research on immersive interactive networks is still in its infancy. Several existing studies include the network architecture, KPIs, and vision of the tactile Internet\cite{Scheuvens2019,Holland2019,Boabang2021}, enabling technologies of the human bond communication\cite{Dixit2019}, possible technologies and optimization schemes for immersive services\cite{Chakareski2020}, and the adaptation of high-speed wireless communication protocols for the haptic data transmission\cite{Oparin2017}. However, there are still a lot of blanks to be filled. In the future, the technical challenges of immersive interactive networks mainly include two aspects, i.e., the acquisition and transmission of multimodal information and the realization of high-performance intelligent network. On one hand, compared with vision and auditory information, it is still challenging to acquire, store, and transmit sensory information such as the touch, taste, and smell information. On the other hand, the immersive interactive network will transmit a large amount of multi-sensory data. Therefore, it is necessary to consider the coordinated control of the transmission of concurrent data streams according to specific scenarios and business model. Besides, issues such as ensuring network latency requirements, designing applicable network routing, and guaranteeing system security under large amounts of data are quite challenging. At present, it is also a hot issue to introduce edge computing\cite{Chakareski2020_ICME,Du2020}, ML\cite{Guo2020,Boabang2021}, and other technologies to effectively improve the intelligence level and performance of the network.

\subsubsection{Digital Twin Network}
Digital twin is a key technology to support the realization of 6G all-digital vision. On one hand, developing from 5G, the 6G network with greatly improved performance will provide a series of novel digital twin applications, such as the digital twin body area network\cite{SEU} and digital twin city \cite{Ivanov2020}. On the other hand, applying the digital twin technology to the communication network can accelerate the realization of a more secure, efficient, intelligent, and visualized 6G network through real-time mapping and interaction between the physical network and the twin network. The vision of network digital twin puts forward high requirements for the network, including holographic network virtual-real interaction mapping, full life cycle management, and real-time closed-loop control\cite{Sun2020_chinese}, which also brings a series of difficulties to the realization of network digital twin. Firstly, the 6G network will contain large-scale network elements connected with complex network topologies. Therefore, it is a huge challenge to model the real physical network in real time. In particular, the wireless channel digital twin is an indispensable part of the network digital twin. A real-time and accurate scenario-adaptive channel model needs to consider how to \textcolor{black}{characterize the propagation environment accurately in real time} and how to predict the possible future changes of wireless channels. In addition, considering the inconsistency in the technical implementation and supported functions of equipment from different manufacturers in the network, attention should be paid to the compatibility of network equipment in the process of data acquisition, processing, and modeling\cite{Sun2020_chinese}. Besides, the large-scale 6G network brings great challenges to the data collection, storage, management, and processing. It is also necessary to explore how to mathematically construct complex network topology relationships in large-scale networks\cite{Sun2020_chinese,Zhu2021_DTN}.

\subsection{\textcolor{black}{System Research}}
\subsubsection{Green Networks}
Since 5G, green communication and sustainable development have caught the attention of researchers on a global scale. With the further development of 6G technologies, applications, and social perceptions, the concept of green networks is gaining more attention. The development of energy-neutral devices is expected to make communication systems more energy efficient and enable new application scenarios where terminal power consumption is limited, such as IoT, satellite communication, and UAV communication. Besides, from the environmental and economic point of view, the development of green network is also extremely necessary. 
Green networks require the achievement of low EM fields, which aims to reduce the threat of EM emissions and avoid any health impact. Thus, effective evaluation, testing, and control of EMF security have become one of the pressing issues for 6G scale deployment.
What’s more, it is important to research and develop near zero power consumption technologies, such as RF energy harvesting, backscattering, and low-power computing, which help to break the battery capacity limitation and achieve environmentally friendly networks. 
\textcolor{black}{In addition, AI technologies use large amounts of data to train networks, consuming large amounts of computing resources and energy, against the need for sustainable development\cite{CJIT_Luan2022a}.} Therefore, green AI, which can obtain new results without increasing the computational cost, is getting more and more attention \cite{Green_Zhu2022}. It is essential to design energy efficient AI models for 6G green communication, considering computational complexity, hardware design and network deployment issues need to be considered \cite{Green_Mao2022}.

\subsubsection{6G Testbeds}
Some application scenarios of 6G, such as THz, optical communication, ultra-wideband, ultra-high speed transmission, and ultra-massive MIMO, bring challenges to the construction of channel sounders. Power amplifiers for high frequency bands, such as THz and optical frequency bands are very difficult to manufacture. In addition, the measurement bandwidth of channel sounders is limited by the sampling rate of the Analog to digital converter (ADC) and Digital to analog converter (DAC). At the same time, how to store the data generated by the high-speed ADC to the disk array in real time is another difficult problem. In ultra-high speed transmission scenarios, a high-speed CS repetition rate is required. In ultra-massive MIMO scenarios, channel sounders are required to support a large number of channels. A low-cost solution is channel expansion using high-speed solid-state switches. However, this method will prolong the measurement time of a single CS, and how to ensure the synchronization of switching needs to be considered. 

In addition, with the continuous development of 6G research, more and more simulation platforms have been set up to verify the underlying theory and practicability of various 6G key technologies, and to discover the defects and limitations of actual implementations. From the current testbeds of various 6G key technologies, it can be observed that fusion is an inevitable developing trend. ISAC combined with THz brings high communication data rate and high sensing accuracy, while the combination of mmWave and optical access network can overcome wall loss and thus brings a larger coverage area. Facing the high service quality requirements of 6G, different combinations of key technologies using different frequency bands and network architectures will become a promising, creative and challenging research direction. How to find the right combination and how to correctly build a testbed for these combined technologies will become a challenge for the future development.

Furthermore, 6G comprehensive testbed is a complex communication system, which is required to verify all 6G key technologies. Hence, the system needs to upgrade continuously according to the development the 6G key technologies. However, the research on key technologies applied to 6G, such as ISAC, RIS, ultra-massive MIMO antennas, OWC, AI, space-air-ground-sea integrated networks, and edge intelligent platform, is not yet clear. Therefore, the 6G comprehensive testbed needs to evolve with the evolution of key technologies. Moreover, a large amount of communication data puts forward higher requirements on the processing capability of the intelligent cloud platform. How to further explore and storage the new software and hardware open architecture of the 6G comprehensive testbed integrated with powerful data processing capabilities, and use the system as a public verification platform is an issue that needs to be considered in the future. 

\section{Lessons Learned and a Brief Summary}
\textcolor{black}{At the time of writing 6G research is still in its infancy. In this section, we will outline the lessons learned and conclude with a summary based on the critical appraisal of the literature. }

\subsection{Lessons Learned}
\begin{figure*}[tb]
	\centerline{\includegraphics[width=0.99\textwidth]{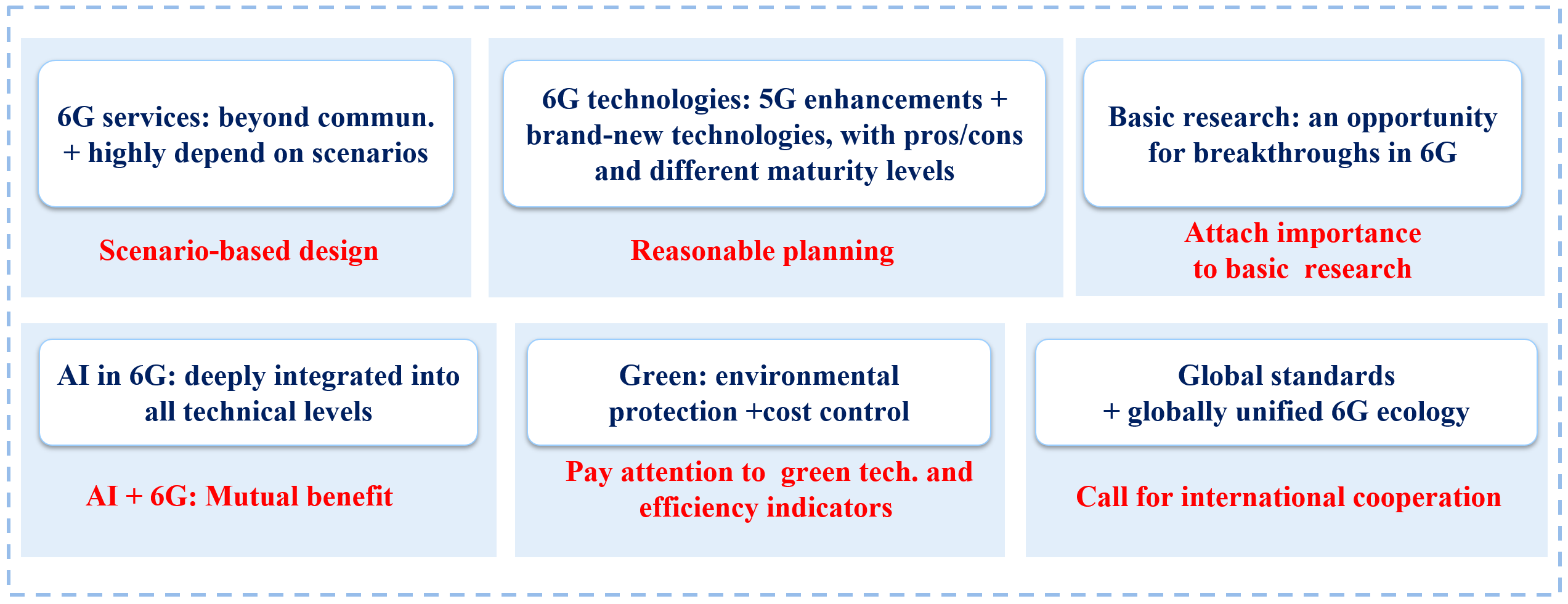}}
	\caption{Lessons Learned.}
	\label{fig_guidelines}
\end{figure*}
\textcolor{black}{Given the challenging vision of aiming for
  ``seamless global coverage, for the harmonization of heterogeneous
  frequency bands obeying widele different propagation properties,
  compelling telepresence-style meta-verse applications for the feast
  of all human senses, while maintaining both ultimate security ans
  well as anonymity of the communicating partie”, 6G networks will
  undoubtedly expand the conventional concept of communication
  services. As shown in Fig. \ref{fig_guidelines}, relying on the
  in-depth survey of 6G research and developments, we highlight some
  of the lessons learnt,
  as follows:}

\begin{enumerate}
\item {\emph{\textcolor{black}{Scenario-based design}}: 6G networks will provide substantial
    performance improvements in terms of the existing communication
    services. However, they will also enrich the services beyond the
    realms of communications and create new hitherto unexplored
    benefits in terms of sensing, localization and
    over-the-air-computing, just to name a few. These will pave the
    way for the seamless integration of the industrial `vertcal
    domains'. The 5G systems have already given cognizance to the
    widely heretogeneous requirements of sophisticated application
    scenarios and hence introduced the eMBB, mM2M and URLLC
    operational modes. In line with its extremely demanding
    specifications, the 6G system is expected to exhibit further
    proliferation of the operational modes in support of more specific
    scenarios and `vertical industrial' applications. This requires
    extremely refined multi-component optimizations tools, which are
    capable of finding all so-called Pareto-optimal operating
    points\cite{jingjing-wang+hanzo-ieee-cst-30-years-of-machine-learning}.
    More explicitly, instead of simply setting for example the maximum
    affordable transmit power and the maximum outage probability as a
    contraint and optimizing the sum-rate of a network as a
    single-component objective function, all three would have to be
    jointly optimized. This sophisticated stochastic optimization
    problem would then find all optimal operating points. As a result,
    none of the above-mentioned metrics can be improved without
    degrading at least one of the others. As a compelling benefit, the
    system controller would always activate the most appropriate
    operating mode for a specific application scenario.}
  
\item {\emph{\textcolor{black}{Reasonable planning}}: The 6G enabling technologies may be broadly
    divided into two categories, namely based on evolutionary
    enhancements of 5G and brand-new technologies. On one hand, each
    of these enabling technologies has its own strengths and
    weaknesses, which have to be critically appraised. On the other
    hand, a suite of novel 6G enabling technologies have been
    developed over the recent years, which have reached different stages. As it was discussed in
    Section~VI, technologies such as THz, RIS, and Cell-free solutions
    have already entered their technical verification stage, while
    technologies such as digital twins, semantic communication and
    meta-verse are still far away from reaching sufficient maturity
    for testbed-based verification, hence require radical further
    research. It is recommended to plan R\&D and deployment reasonably, which can reduce costs and maximize 6G network capabilities.}
  
\item {\emph{\textcolor{black}{Attach importance to basic research}}: Basic research 
    constitutes one of the key pillars of 6G research, paving the way
    for some breakthroughs in 6G. For example, wireless channel
    measurements and channel modeling for a wide range frequency bands
    and and scenarios will lay the foundations for the standardization
    of the 6G channel models, overall technical specifications and
    architectures.  Recent advances in EM information
    theory will also provide new foundations for the application of 6G
    technologies.}
  
\item {\emph{\textcolor{black}{Mutual benefit of AI+6G}}: AI technologies will be deeply integrated
    into all open system interconnection (OSI) layers of 6G networks and they will influence each
    other in a symbiotic manner. For example, on one hand, AI
    technologies can help improve the level of intelligence in 6G
    networks. On the other hand, the network will attain enhanced
    performance as a benefit of intelligent AI assistance, especially
    in the face of uncertainty, when its learning capability will come
    to rescue.  services. However, further research is required for
    their coordinated development.}
  
\item {\emph{\textcolor{black}{Pay attention to  green technologies and efficiency indicators}}: It is essential to aim for green and
    environmentally friendly designs all the way from research to
    pre-development and network roll-out. There is a general consensus
    that protecting the environment and cost control have never been
    more important, given the escalation of tele-traffic. On one hand,
    we should dedicate careful research attention to the conception of
    energy-neutral devices and green networks relying on sophisticated
    joint information and energy
    networking~\cite{hu-jie-kun-yang+hanzo-comst-paper}. On the other
    hand, efficiency indicators should be considered as an essential
    issue for the research and design of 6G~systems.}

\item {\emph{\textcolor{black}{Call for international cooperation}}: In the interest of seamless global roaming,
    we have to aim for global standards and a globally unified 6G
    ecology. This calls for international cooperation and development
    of 6G. }
\end{enumerate}

\subsection{A Brief Summary}
In summary, \textcolor{black}{6G will enrich the suite of global
  communication services by ushering in new application
  scenarios, bringing about fresh technological experiences and
  supporting economic growth.}  We have critically appraised the recent
solutions disseminated in a large body of the relevant literature, highlighting the associated developments and
challenges. We have discussed the associated vision for 6G, indicating
that 6G will be developed in six directions, aiming for global
coverage, relying on a wide range of spectral bands, attractive
applications, stimulating all human senses, while hinging on pervasive
digital intelligence, and strong security. Then, a discussion of the
6G KPIs and application scenarios offering exciting extensions of its
5G counterpart has been presented. The expected system performance
and the associated trade-offs between 6G KPIs have also been
discussed. Next, we have conducted an in-depth survey of the emerging
6G network architecture and technology developments. Following this,
recent efforts on 6G testbed development have been highlighted, with
special attention dedicated to the systems' critical elements.
%
\textcolor{black}{A host of open challenges facing 6G research and the
  corresponding research directions have also been analyzed from the
  perspective of fundamental research, \textcolor{black}{green networks} and the associated key technologies developed for
  supporting the proposed 6G vision, and 6G testbed
  developments. Finally, the associated lessons have been summarized. }
In conclusion, 6G research and standardization still face numerous open challenges. This work has revealed the envisioned appealing features of 6G and it is hoped that it has provided fresh motivation and inspiration for the community's 6G research.

\section*{APPENDIX}
\subsection*{\textcolor{black}{List of Abbreviations}}
\begin{tabbing}

	\hspace*{80bp}\=article\quad \=ÎÄÕÂÀà\kill

1G    \> the first generation  \\
2D	  \> two-dimensional   \\
2G    \> the second generation    \\
3D	  \> three-dimensional    \\
3G    \> the third generation   \\
3GPP  \>	3rd Generation Partnership Project  \\
4G	  \> the forth generation  \\
5G	  \> the fifth generation  \\
5GIA  \>	5G Infrastructure Association  \\
6G \>	the sixth generation \\
6GANA \>	6G Alliance of Network AI  \\
6G-IA \>	6G Smart Networks and Services \\ \> Industry Association  \\
6GPCM \>	6G pervasive channel model  \\
7G \>	the seventh generation  \\
ADC	 \> analog to digital converter   \\
AI	\> artificial intelligence   \\
AmBC \>	ambient backscatter communication    \\
AMF	\> access and mobility management function   \\
AR \>	augmented reality   \\
ASIC \>	application specific integrated circuit   \\
ATIS \>	alliance for telecommunications \\ \> industry solutions  \\
B5G	\> beyond 5G   \\
BCC	\> blockchain-based \\ \> collaborative crowdsensing  \\
BCJR \>	Bahl, Cocke, Jelinek, and Raviv  \\
BER	\> bit error rate   \\
BP	\> belief propagation  \\
B-RAN \> 	blockchain-RAN   \\
CCID \>	China Center for Information \\ \> Industry Development  \\
CIR \> 	channel impulse response  \\
CoMP \>	coordinated multiple points  \\
COVID-19 \> 	corona virus disease 2019   \\
CPUs \>	central processing units  \\
CR	\> cognitive radio  \\
CS \>	channel snapshot  \\
CSI	\> channel state information  \\
DAC \>	Digital to analog converter   \\
DDoS \>	distributed denial of service   \\
DDPG \>	deep deterministic policy gradient   \\
DEN2 \>	deep edge node and network   \\
DL \>	deep learning  \\
DRL	\> deep reinforcement learning \\
DSP	\> digital signal processing  \\
E2E	\> end-to-end  \\
ECCs \>	error-correcting codes  \\
EDA \>	electronic design automation   \\
EE \>	energy efficiency   \\
EM \>	electromagnetic    \\
eMBB \>	enhanced mobile broadband  \\
eni	\> experience network intelligence    \\
ETSI \>	European Telecommunications \\ \> Standards Institute   \\
EU \>	European Union   \\
euRLLC	\> enhanced-uRLLC   \\
FCC \>	Federal Communications Commission  \\
FDD	\> frequency-division duplex  \\
feMBB \>	further-eMBB   \\
FORMAT	\> Flexible Organization and Reconfiguration \\ \> of Millimeter-wave Antenna Tiles  \\
FSO	\> free space optical  \\
FTN	\> faster than nyquist   \\
GBSM \>	geometry-based stochastic model  \\
GDP	\> gross domestic product  \\
GRAND \>	guessing random additive noise decoding  \\
HD	\> high definition   \\
HST	\> high-speed train   \\
IBFD \>	in-band full-duplex    \\
ICT	\> information and \\ \> communications technology   \\
IF	\> intermediate frequency  \\
IIoT \>	industrial IoT  \\
IM	\> index modulation   \\
IMT-2020 \>	International Mobile \\ \> Telecommunications 2020  \\
IMT-2030 \>	International Mobile \\ \>  Telecommunications 2030   \\
IoE \>	internet of everything   \\
IoT	\> internet of things   \\
IoV	\> internet of vehicles    \\
IR	\> infrared   \\
ISAC \>	integrated sensing and communication   \\
ISAC-OW \>	ISAC with optical wireless   \\
ITU	\> International Telecommunications Union  \\
KPIs \>	key performance indicators   \\
KSLD \>	Kyocera Soraalaser   \\
LDPC \>	low density parity check code   \\
LEO	\> low-earth-orbit \\
LiDAR \>	light detection and ranging   \\
LiFi \> 	light fidelity  \\
LRDC \> 	the LiFi Research and \\ \> Development Centre   \\
LTE \>	long-term evolution  \\
MAC \>	medium access control   \\
MBRLLC	\> mobile broadband reliable and \\ \> low latency communication  \\
meMBB	\> massive eMBB   \\
MIMO	\> multiple-input multiple-output  \\
MISO	\> multiple-input single-output  \\
ML	\> machine learning     \\
mMTC	\> massive machine type communications    \\
mmWave	\> millimeter wave   \\
MPC \> multipath component \\
MR	\> mixed reality   \\
muRLLC	\> massive uRLLC    \\
MWCA	\> Mobile World Congress Americas  \\
NFV	\> network functions virtualization   \\
NFVI	\> NFV infrastructure   \\
NFV-MANO \>	NFV management and orchestration   \\
NGMN	\> Next Generation Mobile Networks   \\
NLOS	\> non-line-of-sight   \\
NOMA \>	non-orthogonal multiple access   \\
NR	\> new radio   \\
NTN	\> non-terrestrial networks   \\
NWDAF \>	network data analysis function   \\
OAM	\> orbital angular momentum  \\
OCC	\> optical camera communications   \\
OFDM	\> orthogonal frequency division \\ \>  multiplexing  \\
OFDMA	\> orthogonal frequency division \\ \> multiple access  \\
OLED	\> organic light emitting diode  \\
O-RAN \>	open-RAN  \\
OSD	\> statistics decoding   \\
OTFS \>	orthogonal time frequency space    \\
OVXDM \>	overlapped x domain multiplexing    \\
OWCs \>	optical wireless communications   \\
PAC	\> polarization-adjusted convolutional   \\
PAPR	\> peak to average power ratio   \\
PD	\> photodetector   \\
PDPs	\> power delay profiles    \\
PML	\> Purple Mountain Laboratory    \\
QAM	\> quadrature amplitude modulation    \\
QoS	\> quality of service     \\
R\&D \>	research and design    \\
RAN	\> radio access network    \\
RF	\> radio frequency    \\
RHS	\> reconfigurable holographic surface   \\
RIS \>	reconfigurable intelligent surfaces   \\
RT	\> ray tracing   \\
SBA	\> service-based architecture   \\
SC \>	successive cancellation   \\
SDN	\> software defined network   \\
SE	\> spectral efficiency   \\
SEFDM	\> spectrally efficient frequency \\ \> domain multiplexing   \\
SEU	\> Southeast University    \\
SIC	\> self-interference cancellation    \\
SMF	\> session management function      \\
SNR	\> signal to noise ratio        \\
SON	\> the self-organizing network     \\
SR	\> symbiotic radio      \\
SSN	\> self-sustaining network       \\
TCP	\> transmission control protocol    \\
TDD	\> time-division duplex               \\
TDMA  \> 	time division multiple access   \\
THz	\> terahertz    \\
UAV	\> unmanned aerial vehicle       \\
UDHN	\> ultra-dense heterogeneous network  \\
UDN	\> ultra-dense networking   \\
ULA \> 	uniform linear array   \\
umMTC \>	ultra-mMTC   \\
UPF	\> user plane function   \\
uRLLC	\> ultra-reliable and low \\ \> latency communications   \\
UTC	\> uni-traveling-carrier   \\
UV	\> ultraviolet        \\
V2V	\> vehicle-to-vehicle       \\
V2X	\> vehicle to everything        \\
VLC	\> visible light communications        \\
VLSI	\> very large scale integration      \\
VNA	\> vector network analyzer         \\
VNFs	\> virtual network functions       \\
VR	\> virtual reality                \\
WDM	\> wavelength division multiplex    \\
WiFi	\> wireless fidelity             \\
WLAN	\> wireless local area network     \\
XR	\> extended reality   \\

\end{tabbing}

\ifCLASSOPTIONcaptionsoff
\newpage
\fi
\iftrue

\bibliographystyle{IEEEtran}
\bibliography{6G_Ref.bib}

\begin{IEEEbiography}[{\includegraphics[width=1in,height=1.25in,clip,keepaspectratio]{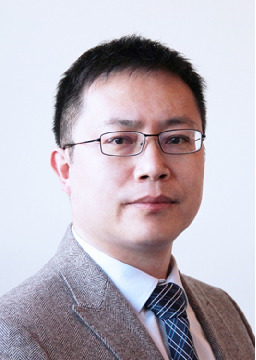}}]{Cheng-Xiang Wang}
	(Fellow, IEEE) received the B.Sc. and M.Eng. degrees in communication and information systems from Shandong University, China, in 1997 and 2000, respectively, and the Ph.D. degree in wireless communications from Aalborg University, Denmark, in 2004.
	
	He was a Research Assistant with the Hamburg University of Technology, Hamburg, Germany, from 2000 to 2001, a Visiting Researcher with Siemens AG Mobile Phones, Munich, Germany, in 2004, and a Research Fellow with the University of Agder, Grimstad, Norway, from 2001 to 2005. He has been with Heriot-Watt University, Edinburgh, U.K., since 2005, where he was promoted to a Professor in 2011. In 2018, he joined Southeast University, Nanjing, China, as a Professor. He is also a part-time Professor with Purple Mountain Laboratories, Nanjing. He has authored 4 books, 3 book chapters, and more than 480 papers in refereed journals and conference proceedings, including 26 highly cited papers. He has also delivered 24 invited keynote speeches/talks and 14 tutorials in international conferences. His current research interests include wireless channel measurements and modeling, 6G wireless communication networks, and electromagnetic information theory.
	
	Prof. Wang is a Member of the Academia Europaea (The Academy of Europe), a Member of the European Academy of Sciences and Arts (EASA), a Fellow of the Royal Society of Edinburgh (FRSE), IEEE, IET, and China Institute of Communications (CIC), an IEEE Communications Society Distinguished Lecturer in 2019 and 2020, a Highly-Cited Researcher recognized by Clarivate Analytics in 2017-2020, and one of the most cited Chinese Researchers recognized by Elsevier in 2021. He is currently an Executive Editorial Committee Member of the IEEE TRANSACTIONS ON WIRELESS COMMUNICATIONS. He has served as an Editor for over ten international journals, including the IEEE TRANSACTIONS ON WIRELESS COMMUNICATIONS, from 2007 to 2009, the IEEE TRANSACTIONS ON VEHICULAR TECHNOLOGY, from 2011 to 2017, and the IEEE TRANSACTIONS ON COMMUNICATIONS, from 2015 to 2017. He was a Guest Editor of the IEEE JOURNAL ON SELECTED AREAS IN COMMUNICATIONS, Special Issue on Vehicular Communications and Networks (Lead Guest Editor), Special Issue on Spectrum and Energy Efficient Design of Wireless Communication Networks, and Special Issue on Airborne Communication Networks. He was also a Guest Editor for the IEEE TRANSACTIONS ON BIG DATA, Special Issue on Wireless Big Data, and is a Guest Editor for the IEEE TRANSACTIONS ON COGNITIVE COMMUNICATIONS AND NETWORKING, Special Issue on Intelligent Resource Management for 5G and Beyond. He has served as a TPC Member, a TPC Chair, and a General Chair for more than 80 international conferences. He received 15 Best Paper Awards from IEEE GLOBECOM 2010, IEEE ICCT 2011, ITST 2012, IEEE VTC 2013Fall, IWCMC 2015, IWCMC 2016, IEEE/CIC ICCC 2016, WPMC 2016, WOCC 2019, IWCMC 2020, WCSP 2020, CSPS2021, WCSP 2021, and IEEE/CIC ICCC 2022. Also, he received the 2020-2022 “AI 2000 Most Influential Scholar Award Honorable Mention” in recognition of his outstanding and vibrant contributions in the field of Internet of Things.
	\end{IEEEbiography}
	
	\begin{IEEEbiography}[{\includegraphics[width=1in,height=1.25in,clip,keepaspectratio]{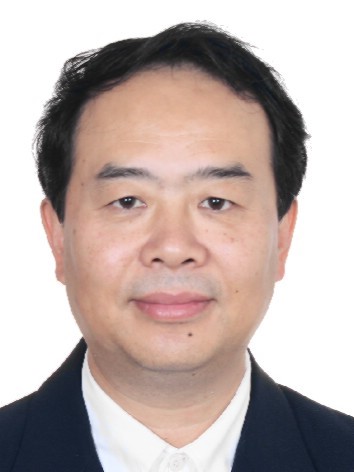}}]{Xiaohu You}
		(Fellow, IEEE) received his Master and Ph.D. Degrees from Southeast University, Nanjing, China, in Electrical Engineering in 1985 and 1988, respectively. Since 1990, he has been working with National Mobile Communications Research Laboratory at Southeast University, where he is currently professor and director of the Lab. He has contributed over 300 IEEE journal papers and 3 books in the areas of adaptive signal processing, neural network and wireless communications. From 1999 to 2002, he was the Principal Expert of China C3G Project. From 2001-2006, he was the Principal Expert of China National 863 Beyond 3G FuTURE Project. From 2013 to 2019, he was the Principal Investigator of China National 863 5G Project. His current research interests include wireless networks, advanced signal processing and its applications. 
		Dr. You was selected as IEEE Fellow in 2011. He was a recipient of China National First Class Invention Prize in 2011. He served as the General Chair for IEEE Wireless Communications and Networking Conference (WCNC) 2013, IEEE Vehicular Technology Conference (VTC) 2016 Spring, and IEEE International Conference on Communications (ICC) 2019. He is currently the Secretary General of the FuTURE Forum, and the Vice Chair of the China IMT-2020 (5G) Promotion Group. Dr. You won the IET Achievement Medal in 2021.
		
	\end{IEEEbiography}
	
	\begin{IEEEbiography}[{\includegraphics[width=1in,height=1.25in,clip,keepaspectratio]{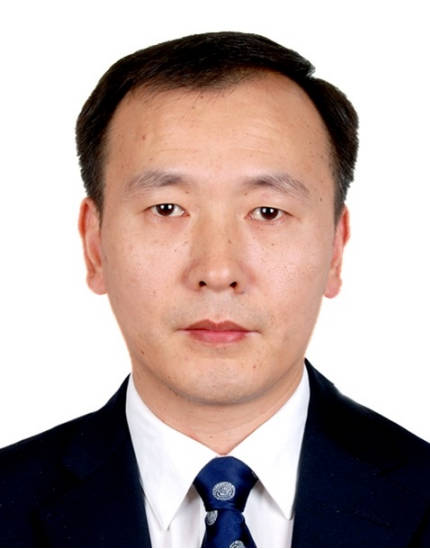}}]{Xiqi Gao}
		(Fellow, IEEE) received the Ph.D. degree in electrical engineering from Southeast University, Nanjing, China, in 1997. He joined the Department of Radio Engineering, Southeast University, in April 1992. Since May 2001, he has been a professor of information systems and communications. From September 1999 to August 2000, he was a visiting scholar at Massachusetts Institute of Technology, Cambridge, MA, USA, and Boston University, Boston, MA. From August 2007 to July 2008, he visited the Darmstadt University of Technology, Darmstadt, Germany, as a Humboldt scholar. His current research interests include broadband multicarrier communications, massive MIMO wireless communications, satellite communications, optical wireless communications, information theory and signal processing for wireless communications. From 2007 to 2012, he served as an Editor for the IEEE Transactions on Wireless Communications. From 2009 to 2013, he served as an Editor for the IEEE Transactions on Signal Processing. From 2015 to 2017, he served as an Editor for the IEEE Transactions on Communications. Dr. Gao was the recipient of the Science and Technology Awards of the State Education Ministry of China in 1998, 2006 and 2009, the National Technological Invention Award of China in 2011, the Science and Technology Award of Jiangsu Province of China in 2014, and the 2011 IEEE Communications Society Stephen O. Rice Prize Paper Award in the Field of Communications Theory.
	\end{IEEEbiography}
	
	\begin{IEEEbiography}[{\includegraphics[width=1in,height=1.25in,clip,keepaspectratio]{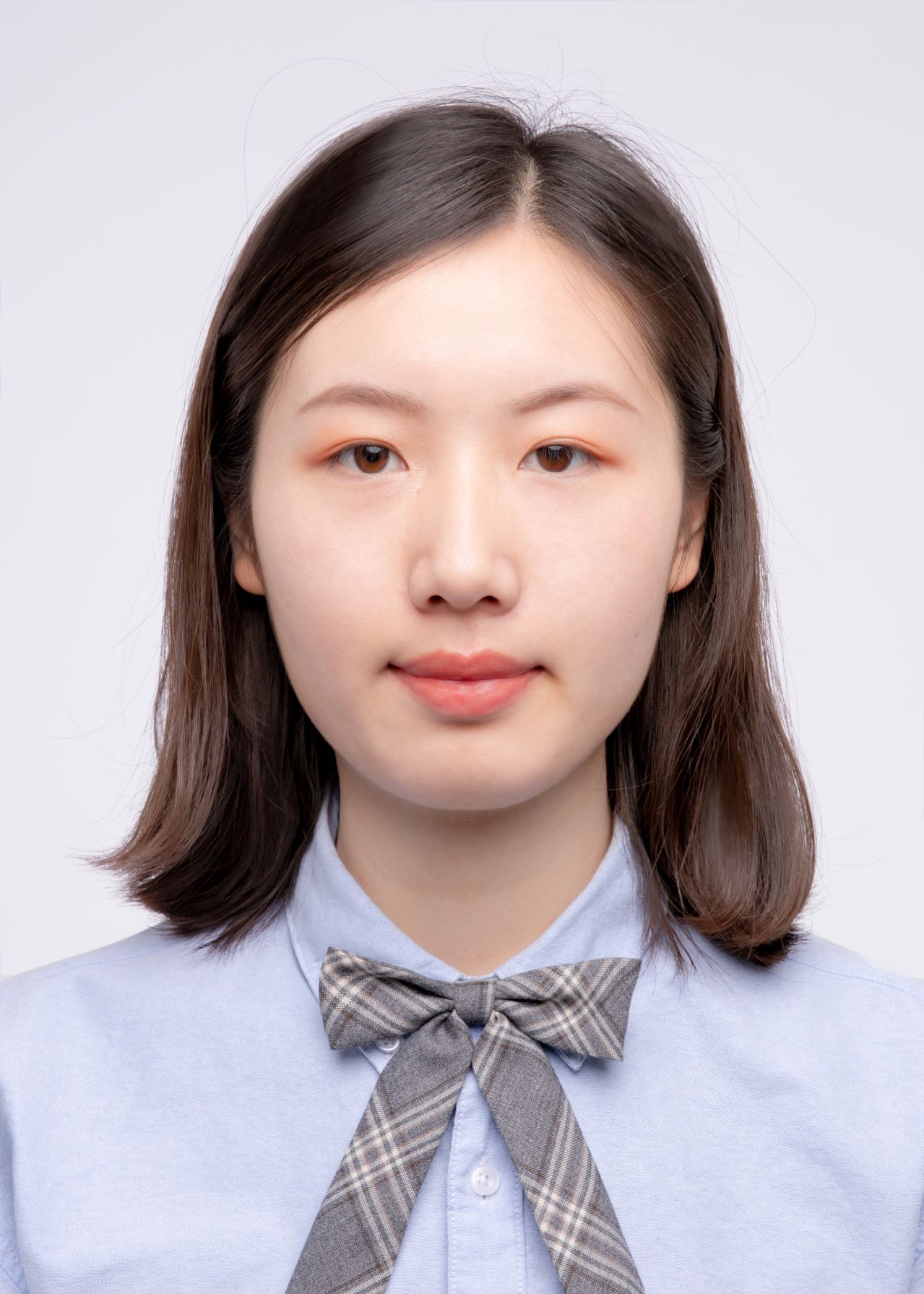}}]{Xiuming Zhu}
		received the B.E. degree in communication engineering from Harbin Institute of Technology, Weihai, China, in 2020. She is currently pursuing the M.Sc. degree in the National Mobile Communications Research Laboratory, Southeast University, China. Her research interests are optical wireless channel measurements and modeling.
	\end{IEEEbiography}
	
	\begin{IEEEbiography}[{\includegraphics[width=1in,height=1.25in,clip,keepaspectratio]{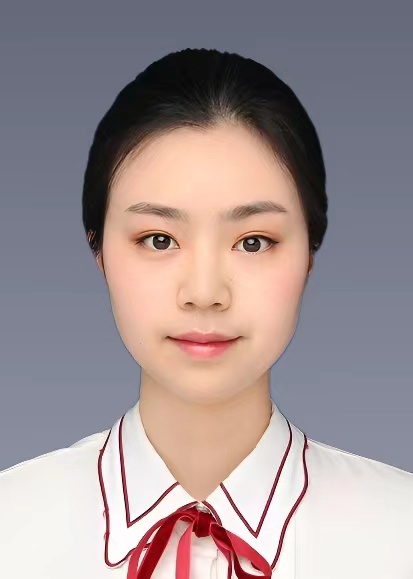}}]{Zixin Li}
		received the B.E. degree in information engineering from Southeast University, Nanjing, China, in 2020, where she is currently pursuing the M.Sc. degree with the National Mobile Communications Research Laboratory. Her research interests are satellite wireless channel measurements and modeling.
	\end{IEEEbiography}
	
	\begin{IEEEbiography}[{\includegraphics[width=1in,height=1.25in,clip,keepaspectratio]{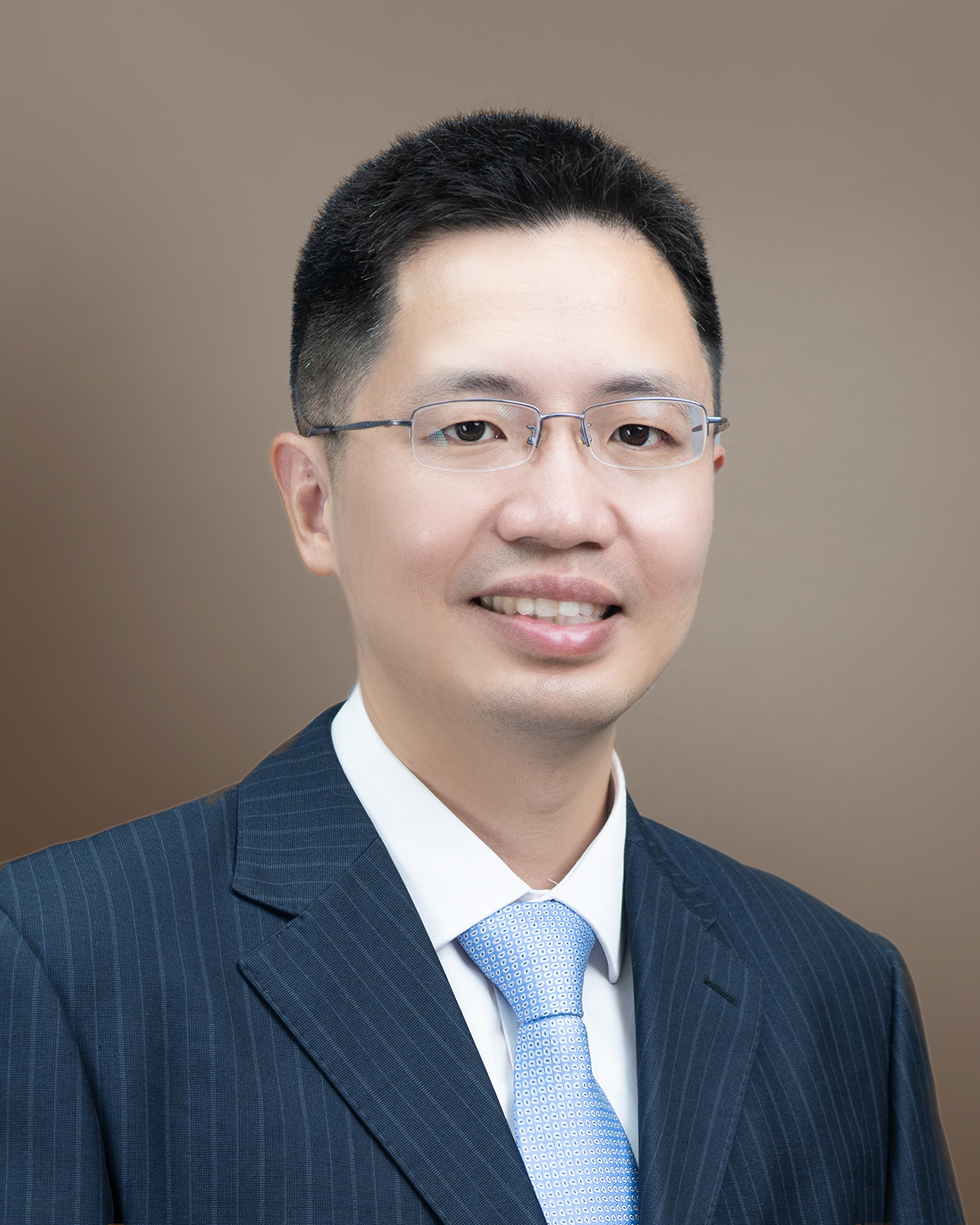}}]{Chuan Zhang}
		(Senior Member, IEEE) received the B.E. degree in microelectronics and the M.E. degree in very-large scale integration (VLSI) design from Nanjing University, Nanjing, China, in 2006 and 2009, respectively, and the Ph.D. degree from the Department of Electrical and Computer Engineering, University of Minnesota, Twin Cities (UMN), USA, in 2012.

		He is currently the Young Chair Professor of Southeast University. He is also with the LEADS, National Mobile Communications Research Laboratory, Frontiers Science Center for Mobile Information Communications and Security of MoE, Quantum Information Center of Southeast University, and the Purple Mountain Laboratories, Nanjing, China. His current research interests are algorithms and implementations for signal processing and communication systems.
	
		Dr. Zhang serves as an Associate Editor for the \textsc{IEEE Transactions on Circuits and Systems - II}. He served as an Associate Editor for the \textsc{IEEE Transactions on Signal Processing} and \textsc{IEEE Open Journal of Circuits and Systems}, and a Corresponding Guest Editor for the \textsc{IEEE Journal on Emerging and Selected Topics in Circuits and Systems} twice. He is a Distinguished Lecturer and the Vice Chair of the Circuits and Systems for Communications TC of the IEEE Circuits and Systems Society. He is also a member of the Applied Signal Processing Systems TC of the IEEE Signal Processing Society, and Circuits and Systems for Communications TC, VLSI Systems and Applications TC, and Digital Signal Processing TC of the IEEE Circuits and Systems Society. He received the Best Contribution Award of the IEEE Asia Pacific Conference on Circuits and Systems (APCCAS) in 2018, the Best Paper Award in 2016, the Best (Student) Paper Award of the IEEE International Conference on DSP in 2016, three Best (Student) Paper Awards of the IEEE International Conference on ASIC in 2015, 2017, and 2019, the Best Paper Award Nomination of the IEEE Workshop on Signal Processing Systems in 2015, three Excellent Paper Awards and two Excellent Poster Presentation Awards of the International Collaboration Symposium on Information Production and Systems from 2016 to 2018, the Outstanding Achievement Award of the Intel Collaborative Research Institute in 2018, and the Merit (Student) Paper Award of the IEEE APCCAS in 2008. He also received the Three-Year University-Wide Graduate School Fellowship of UMN and the Doctoral Dissertation Fellowship of UMN.
	\end{IEEEbiography}
	
	\begin{IEEEbiography}[{\includegraphics[width=1in,height=1.25in,clip,keepaspectratio]{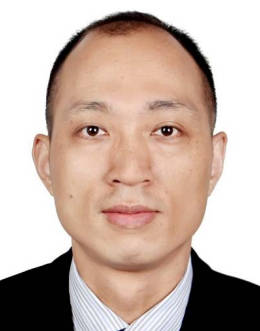}}]{Haiming Wang}
		(Member, IEEE) was born in 1975. He received the B.S., M.S., and Ph.D. degrees in electrical engineering from Southeast University, Nanjing, China, in 1999, 2002, and 2009, respectively. He joined the School of Information Science and Engineering and the State Key Laboratory of Millimeter Waves, Southeast University, in 2002, where he is currently a Distinguished Professor. He is also a Part-Time Professor with the Purple Mountain Laboratories, Nanjing, China. He has authored and coauthored over 50 technical publications in IEEE TRANSACTIONS ON ANTENNAS AND PROPAGATION and other peer-reviewed academic journals. He has authored and coauthored over more than 70 patents and 52 patents have been granted. His current research interests include AI-powered antenna and radio-frequency technologies (iART), AI-powered channel measurement and modeling technologies (iCHAMM), and AI-Powered Integrated Communications, Sensing, \& Positioning Technologies (iCSAP). 
		
		Dr. Wang was awarded twice for contributing to the development of IEEE 802.11aj by the IEEE Standards Association in 2018 and 2020.
	\end{IEEEbiography}
	
	\begin{IEEEbiography}[{\includegraphics[width=1in,height=1.25in,clip,keepaspectratio]{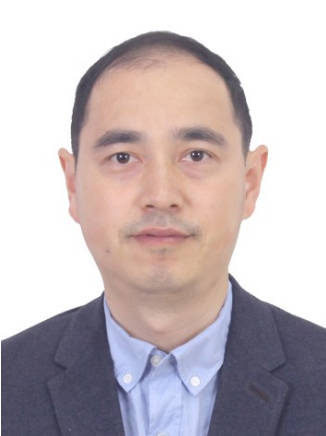}}]{Yongming Huang}
		(Senior Member, IEEE) received the B.S. and M.S. degrees from Nanjing University, Nanjing, China, in 2000 and 2003, respectively, and the Ph.D. degree in electrical engineering from Southeast University, Nanjing, in 2007. Since March 2007 he has been a faculty in the School of Information Science and Engineering, Southeast University, China, where he is currently a full professor. He has also been the Director of the Pervasive Communication Research Center, Purple Mountain Laboratories, since 2019. From 2008 to 2009, he was visiting the Signal Processing Lab, Royal Institute of Technology (KTH), Stockholm, Sweden. He has published over 200 peer-reviewed papers, hold over 80 invention patents. His current research interests include intelligent 5G/6G mobile communications and millimeter wave wireless communications. He submitted around 20 technical contributions to IEEE standards, and was awarded a certificate of appreciation for outstanding contribution to the development of IEEE standard 802.11aj. He served as an Associate Editor for the IEEE Transactions on Signal Processing and a Guest Editor for the IEEE Journal on Selected Areas in Communications. He is currently an Editor-at-Large for the IEEE Open Journal of the Communications Society.
	\end{IEEEbiography}
	
	\begin{IEEEbiography}[{\includegraphics[width=1in,height=1.25in,clip,keepaspectratio]{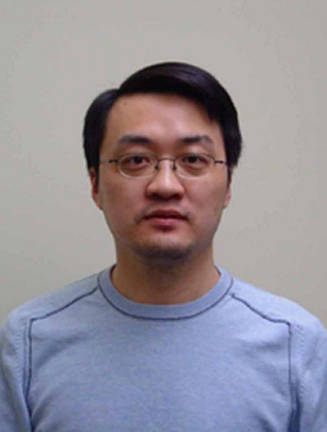}}]{Yunfei Chen}
		(Senior Member, IEEE) received his B.E. and M.E. degrees in electronics engineering from Shanghai Jiaotong University, Shanghai, P.R.China, in 1998 and 2001, respectively. He received his Ph.D. degree from the University of Alberta in 2006. He is currently working as a Professor in the Department of Engineering at the University of Durham, U.K. His research interests include wireless communications, performance analysis, joint radar communications designs, cognitive radios, wireless relaying and energy harvesting.
	\end{IEEEbiography}
	
	\begin{IEEEbiography}[{\includegraphics[width=1in,height=1.25in,clip,keepaspectratio]{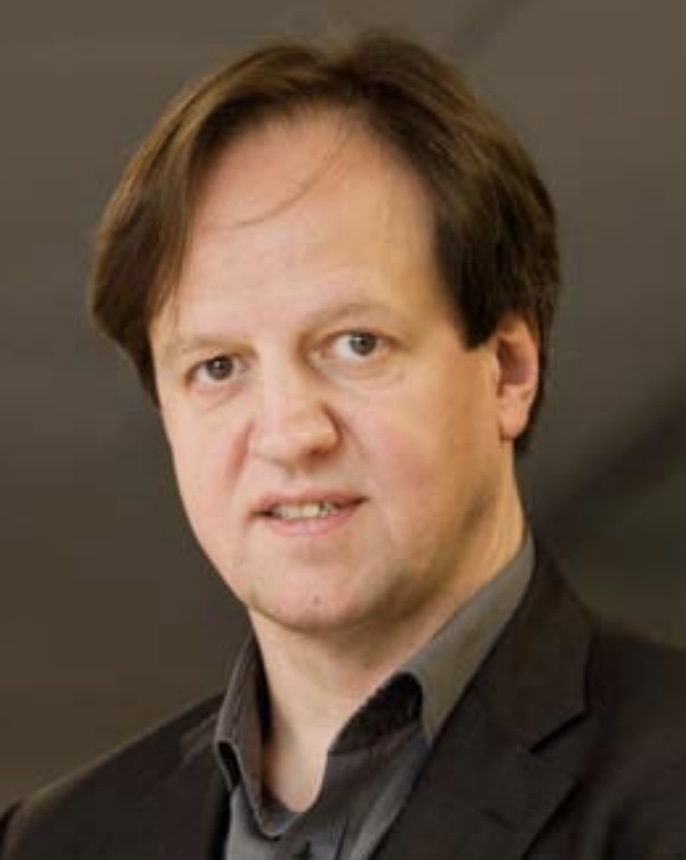}}]{Harald Haas}
		(Fellow, IEEE) received the Ph.D. degree from the University of Edinburgh, Edinburgh, U.K., in 2001.
		
		He is the Director of the LiFi Research and Development Center, University of Strathclyde, Glasgow, U.K. He is also the Initiator, the Co-Founder, and the Chief Scientific Officer of pureLiFi Ltd., Edinburgh. He has authored 550 conference and journal papers, including papers in \emph{Science and Nature Communications}. His main research interests are in optical wireless communications, hybrid optical wireless and RF communications, spatial modulation, and interference coordination in wireless networks.
		
		Dr. Haas received the Outstanding Achievement Award from the International Solid State Lighting Alliance in 2016 and the Royal Society Wolfson Research Merit Award. He was a recipient of the IEEE Vehicular Society James Evans Avant Garde Award in 2019. His team invented spatial modulation. He introduced LiFi to the public at an invited TED Global talk in 2011. This talk on Wireless Data from Every Light Bulb has been watched online over 2.72 million times. He was listed among the 50 best inventions in \emph{Time} in 2011. He gave a second TED Global lecture in 2015 on the use of solar cells as LiFi data detectors and energy harvesters. This has been viewed online over 2.75 million times. He was elected a Fellow of the Royal Society of Edinburgh in 2017. In 2018, he received a three-year EPSRC Established Career Fellowship extension and was elected Fellow of IET. He was elected a Fellow of the Royal Academy of Engineering in 2019.
	\end{IEEEbiography}
	
	\begin{IEEEbiography}[{\includegraphics[width=1in,height=1.25in,clip,keepaspectratio]{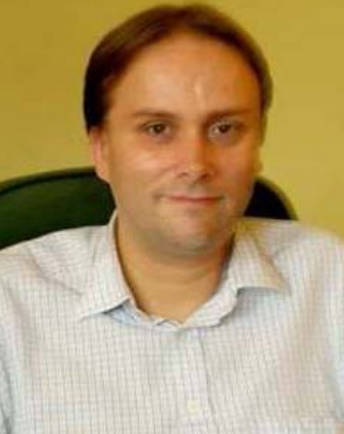}}]{John S. Thompson}
		(Fellow, IEEE) received the Ph.D. degree in electrical engineering from The University of Edinburgh, Edinburgh, U.K., in 1995.
		
		He is currently a Professor with the School of Engineering, the University of Edinburgh. His research interests include antenna array processing, cooperative communications systems, and energy-efficient wireless communications and their applications. He has published in excess of 350 papers on these topics. He currently participates in two U.K. research projects, which study new concepts for signal processing and for next-generation wireless communications. He was the Co-Chair of the IEEE SmartGridComm Conference held in Aalborg, Denmark, in 2018. In January 2016, he was elevated as a fellow of the IEEE for Contributions to Antenna Arrays and Multihop Communications. From 2015 to 2018, he was recognized by Thomson Reuters as a Highly Cited Researcher.
	\end{IEEEbiography}
	
	\begin{IEEEbiography}[{\includegraphics[width=1in,height=1.25in,clip,keepaspectratio]{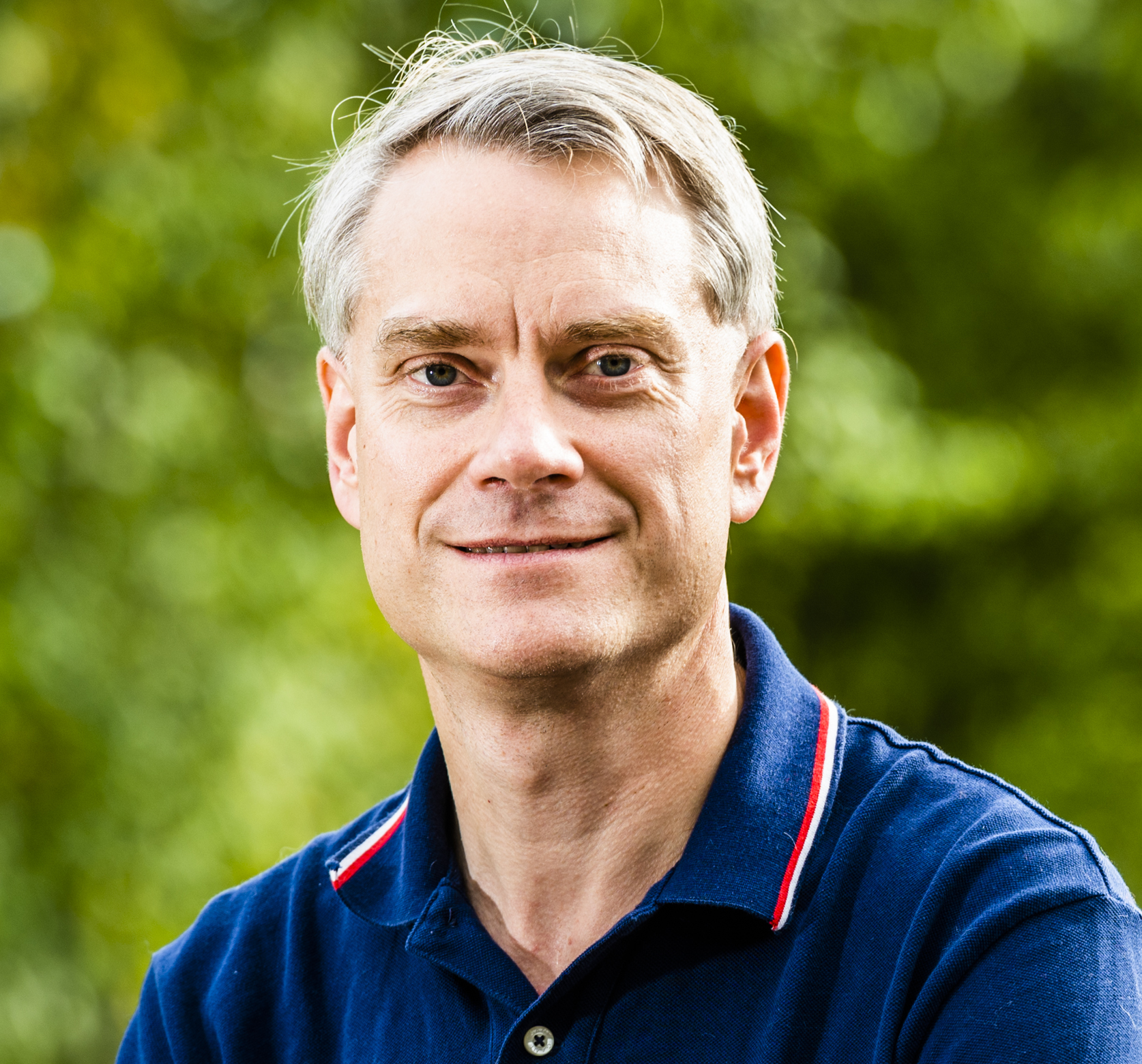}}]{Erik G. Larsson}
		(Fellow, IEEE) is Professor at Link\"oping University, Sweden, and Fellow of the IEEE. He co-authored the textbook \emph{Fundamentals of Massive MIMO} (Cambridge University Press, 2016). He received,	among others, the IEEE ComSoc Stephen O. Rice Prize in Communications Theory in 2015, the IEEE ComSoc Leonard G. Abraham Prize in 2017, the IEEE ComSoc Best Tutorial Paper Award in 2018, and the IEEE ComSoc Fred W. Ellersick Prize in 2019. His interests lie within wireless communications, statistical signal processing, and networks.
	\end{IEEEbiography}
	
	\begin{IEEEbiography}[{\includegraphics[width=1in,height=1.25in,clip,keepaspectratio]{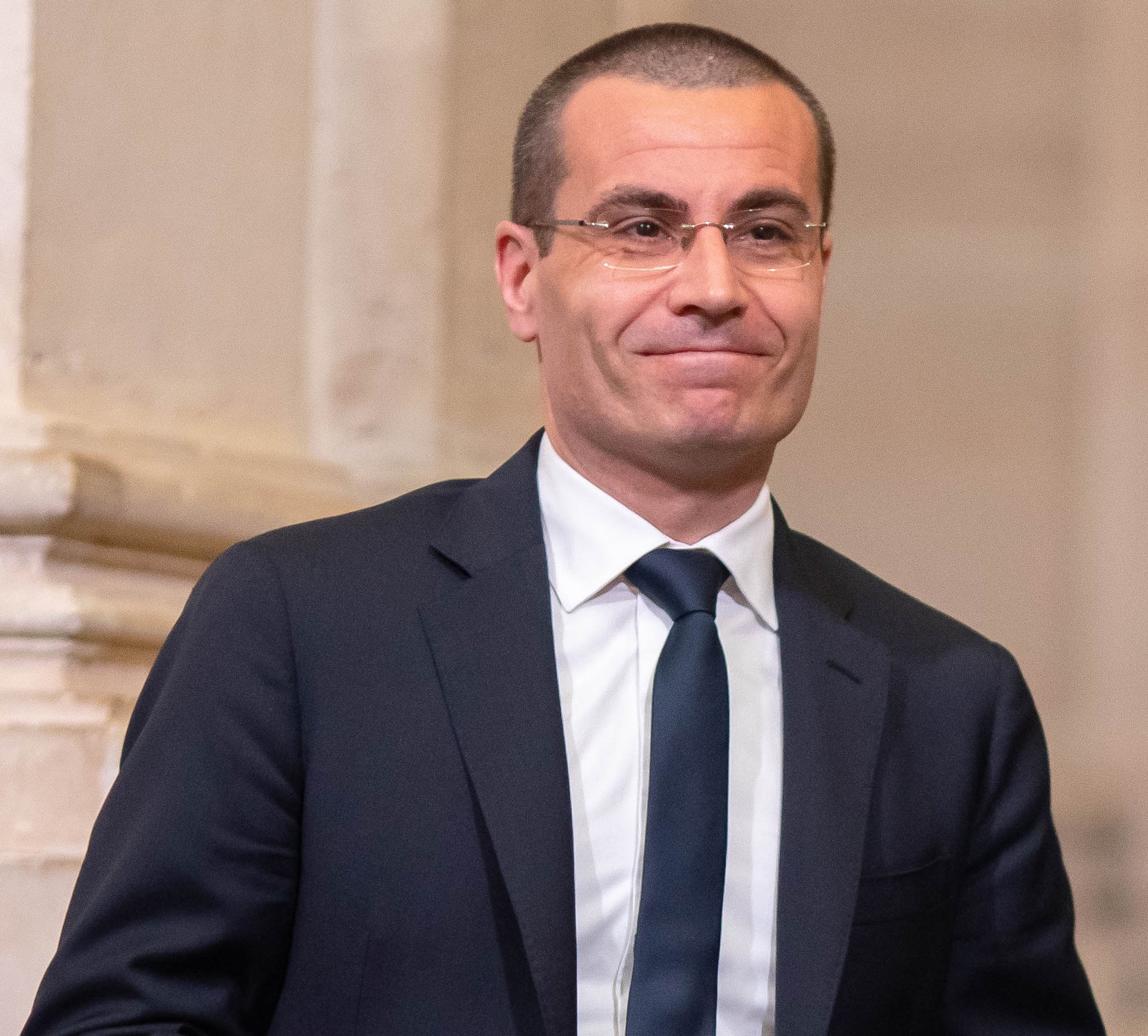}}]{Marco Di Renzo} 
		(Fellow, IEEE) received the Laurea (cum laude) and Ph.D. degrees in electrical engineering from the University of L'Aquila, Italy, in 2003 and 2007, respectively, and the Habilitation $\grave{a}$  Diriger des Recherches (Doctor of Science) degree from University Paris-Sud (now Paris-Saclay University), France, in 2013. Currently, he is a CNRS Research Director (Professor) and the Head of the Intelligent Physical Communications group in the Laboratory of Signals and Systems (L2S) at CentraleSupelec - Paris-Saclay University, Paris, France. In Paris-Saclay University, he serves as the Coordinator of the Communications and Networks Research Area of the Laboratory of Excellence DigiCosme, as a Member of the Admission and Evaluation Committee of the Ph.D. School on Information and Communication Technologies, and as a Member of the Evaluation Committee of the Graduate School in Computer Science. He is a Founding Member and the Academic Vice Chair of the Industry Specification Group (ISG) on Reconfigurable Intelligent Surfaces (RIS) within the European Telecommunications Standards Institute (ETSI), where he serves as the Rapporteur for the work item on communication models, channel models, and evaluation methodologies. He is a Fellow of the IEEE, IET, AAIA, and Vebleo; an Ordinary Member of the European Academy of Sciences and Arts, an Ordinary Member of the Academia Europaea; and a Highly Cited Researcher. Also, he is a Fulbright Fellow at City University of New York, USA, and was a Nokia Foundation Visiting Professor and a Royal Academy of Engineering Distinguished Visiting Fellow. His recent research awards include the 2021 EURASIP Best Paper Award, the 2022 IEEE COMSOC Outstanding Paper Award, and the 2022 Michel Monpetit Prize conferred by French Academy of Sciences. He serves as the Editor-in-Chief of IEEE Communications Letters.
	\end{IEEEbiography}
	
	\begin{IEEEbiography}[{\includegraphics[width=1in,height=1.25in,clip,keepaspectratio]{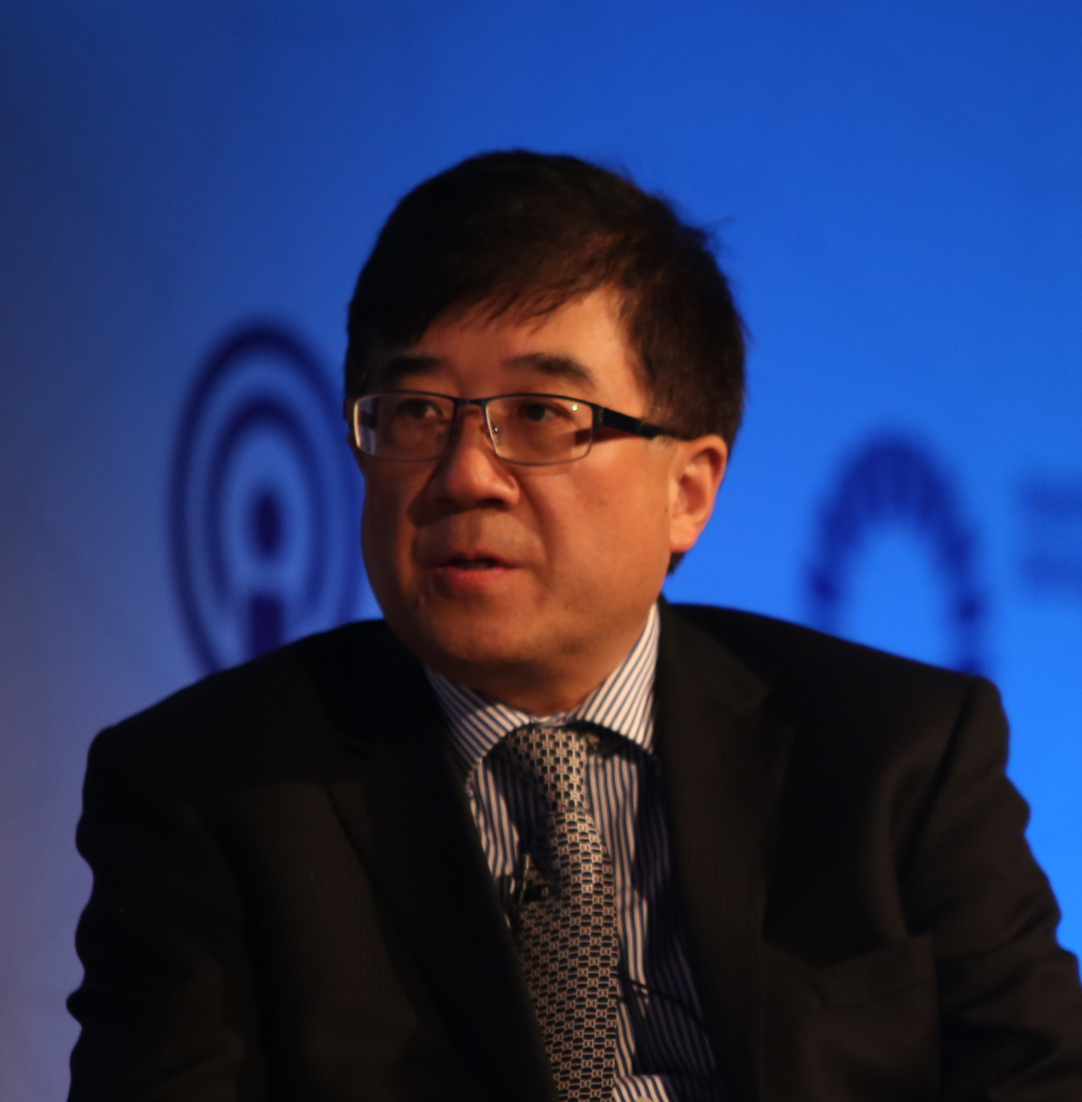}}]{Wen Tong}
		(Fellow, IEEE) received the B.S. degree from the Department of Radio Engineering, Nanjing Institute of Technology, Nanjing, China, in 1984, and the M.Sc. and Ph.D. degrees in electrical engineering from Concordia University, Montreal, QC, Canada, in 1986 and 1993, respectively. He joined the Wireless Technology Labs, Bell Northern Research, Canada, in 1995. In 2011, he was appointed as the Head of the Communications Technologies Labs, Huawei. He also spearheads and leads Huawei's 5G wireless technologies research and development. Prior to joining Huawei in 2009, he was the Nortel Fellow and the Head of the Network Technology Labs, Nortel. He is currently the Huawei Fellow and the CTO of Huawei Wireless. He is the Head of the Huawei Wireless Research. He pioneered fundamental technologies from 1G to 5G wireless with more than 500 awarded U.S. patents. He was elected as a Huawei Fellow. He was a recipient of the IEEE Communications Society Industry Innovation Award for ``the leadership and contributions in development of 3G and 4G wireless systems" in 2014, and the IEEE Communications Society Distinguished Industry Leader Award for ``pioneering technical contributions and leadership in the mobile communications industry and innovation in 5G mobile communications technology" in 2018. He is a fellow of the Canadian Academy of Engineering and a fellow of Royal Society of Canada. He also serves as the Board of Director for WiFi Alliance.
	\end{IEEEbiography}
	
	\begin{IEEEbiography}[{\includegraphics[width=1in,height=1.25in,clip,keepaspectratio]{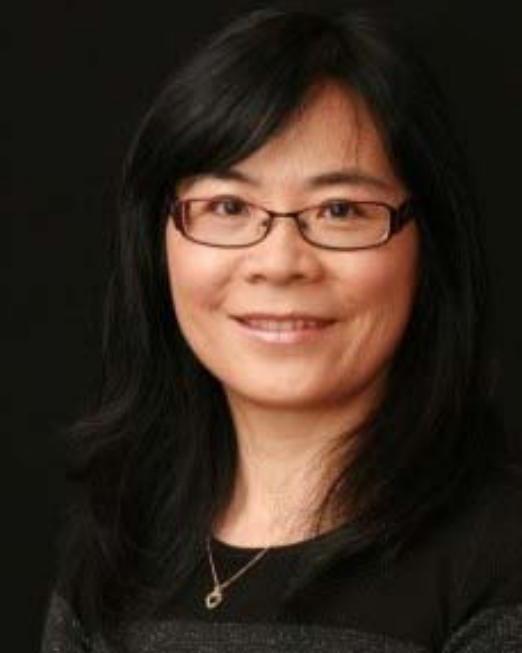}}]{Peiying Zhu}
		(Fellow, IEEE) received the M.Sc. degree from Southeast University in 1985 and the Ph.D. degree from Concordia University in 1993. Prior to joining Huawei in 2009, she was a Nortel Fellow and the Director of the Advanced Wireless Access Technology at the Nortel Wireless Technology Lab. She led the team and pioneered research and prototyping on MIMO-OFDM and Multi-hop relay. Many of these technologies developed by the team have been adopted into the LTE standards and 4G products. She is currently leading the 5G Wireless System Research in Huawei. The focus of her research is advanced wireless access technologies with over 200 granted patents. She has been regularly giving talks and panel discussions on 5G vision and enabling technologies. She is actively involved in 3GPP and IEEE 802 Standards Development. 
		
		Dr. Zhu is a Huawei Fellow. She is currently a WiFi Alliance Board Member. She has served as a Guest Editor for the IEEE Signal Processing Magazine Special Issue on the 5G revolution and the IEEE JSAC on Deployment Issues and Performance Challenges for 5G. She has co-chaired various 5G workshops in the IEEE GLOBECOM.
	\end{IEEEbiography}

	\begin{IEEEbiography}[{\includegraphics[width=1in,height=1.25in,clip,keepaspectratio]{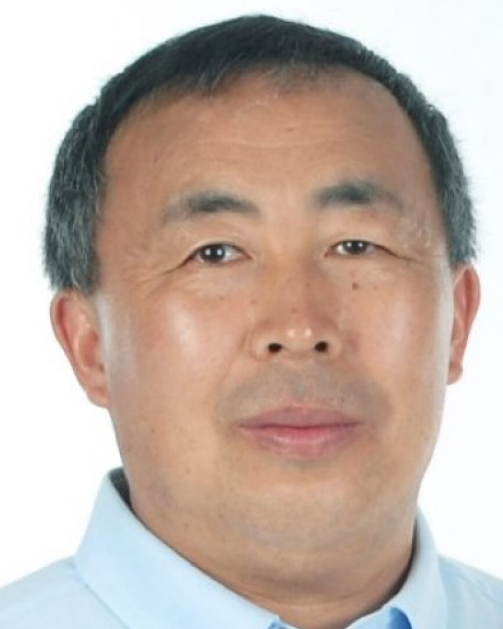}}]{Xuemin (Sherman) Shen}
		(Fellow, IEEE) received the Ph.D. degree in electrical engineering from Rutgers University, New Brunswick, NJ, USA, in 1990. He is a University Professor with the Department of Electrical and Computer Engineering, University of Waterloo, Canada. His research focuses on network resource management, wireless network security, Internet of Things, 5G and beyond, and vehicular ad hoc and sensor networks. Dr. Shen is a registered Professional Engineer of Ontario, Canada, an Engineering Institute of Canada Fellow, a Canadian Academy of Engineering Fellow, a Royal Society of Canada Fellow, a Chinese Academy of Engineering Foreign Member, and a Distinguished Lecturer of the IEEE Vehicular Technology Society and Communications Society. 
		
		Dr. Shen received the R.A. Fessenden Award in 2019 from IEEE, Canada, Award of Merit from the Federation of Chinese Canadian Professionals (Ontario) in 2019, James Evans Avant Garde Award in 2018 from the IEEE Vehicular Technology Society, Joseph LoCicero Award in 2015 and Education Award in 2017 from the IEEE Communications Society, and Technical Recognition Award from Wireless Communications Technical Committee (2019) and AHSN Technical Committee (2013). He has also received the Excellent Graduate Supervision Award in 2006 from the University of Waterloo and the Premier's Research Excellence Award (PREA) in 2003 from the Province of Ontario, Canada. He served as the Technical Program Committee Chair/CoChair for IEEE Globecom'16, IEEE Infocom'14, IEEE VTC'10 Fall, IEEE Globecom'07, and the Chair for the IEEE Communications Society Technical Committee on Wireless Communications. Dr. Shen is the President of the IEEE Communications Society. He was the Vice President for Technical \& Educational Activities, Vice President for Publications, Member-at-Large on the Board of Governors, Chair of the Distinguished Lecturer Selection Committee, Member of IEEE Fellow Selection Committee of the ComSoc. Dr. Shen served as the Editor-in-Chief of the IEEE IoT JOURNAL, IEEE Network, and IET Communications.
	\end{IEEEbiography}

	\begin{IEEEbiography}[{\includegraphics[width=1in,height=1.25in,clip,keepaspectratio]{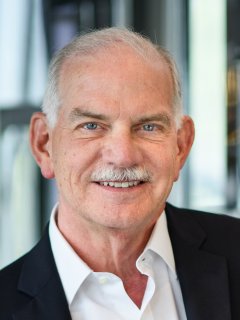}}]{H. Vincent Poor}
		(Life Fellow, IEEE) received the Ph.D. degree in EECS from Princeton University in 1977. From 1977 until 1990, he was on the faculty of the University of Illinois at Urbana-Champaign. Since 1990 he has been on the faculty at Princeton, where he is currently the Michael Henry Strater University Professor. During 2006 to 2016, he served as the dean of Princeton's School of Engineering and Applied Science. He has also held visiting appointments at several other universities, including most recently at Berkeley and Cambridge. His research interests are in the areas of information theory, machine learning and network science, and their applications in wireless networks, energy systems and related fields. Among his publications in these areas is the recent book \emph{Machine Learning and Wireless Communications}.  (Cambridge University Press, 2022). Dr. Poor is a member of the National Academy of Engineering and the National Academy of Sciences and is a foreign member of the Chinese Academy of Sciences, the Royal Society, and other national and international academies. He received the IEEE Alexander Graham Bell Medal in 2017.
	\end{IEEEbiography}
	
	\begin{IEEEbiography}[{\includegraphics[width=1in,height=1.25in,clip,keepaspectratio]{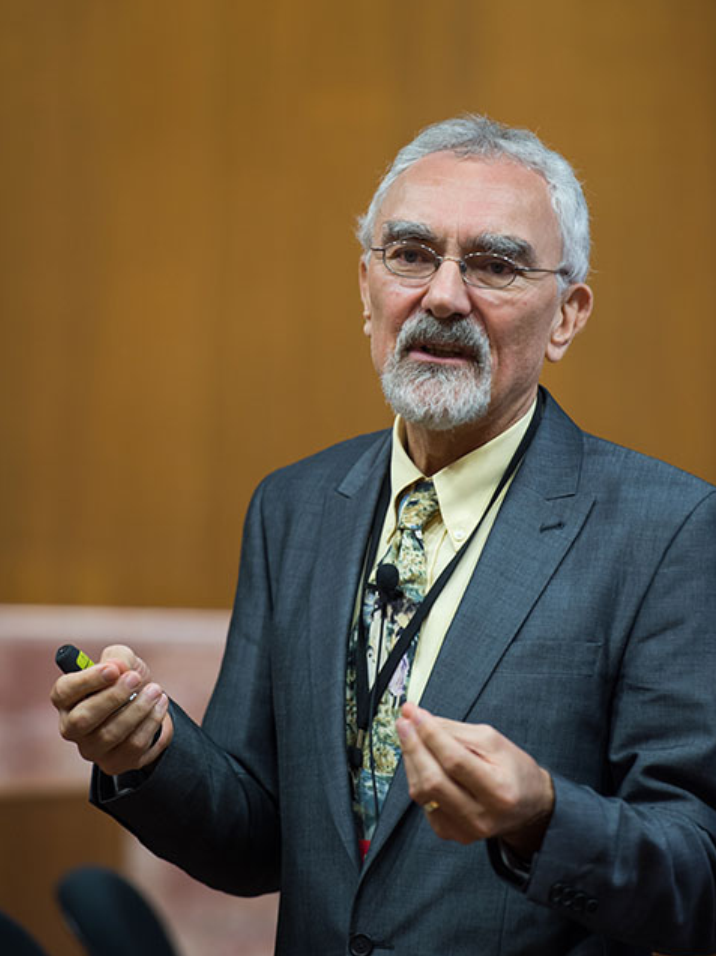}}]{Lajos Hanzo}
		(Life Fellow, IEEE) received his Master degree and Doctorate in 1976 and 1983, respectively from the Technical University (TU) of Budapest. He was also awarded the Doctor of Sciences (DSc) degree by the University of Southampton (2004) and Honorary Doctorates by the TU of Budapest (2009) and by the University of Edinburgh (2015). He is a Foreign Member of the Hungarian Academy of Sciences and a former Editor-in-Chief of the IEEE Press. He has served several terms as Governor of both IEEE ComSoc and of VTS. He has published widely at IEEE Xplore, coauthored 19 Wiley-IEEE Press books and has helped the fast-track career of 123 PhD students. He holds the Eric Sumner Technical Field Award. 
	\end{IEEEbiography}

	\end{document}